\documentclass[a4paper,11pt,english]{article}
\pdfoutput=1

\usepackage{jheppub}
\usepackage{amsmath,amssymb,amsfonts,amsthm}
\usepackage[framemethod=tikz]{mdframed}
\usepackage{placeins}
\usepackage{xcolor}
\usepackage{framed}
\definecolor{shadecolor}{gray}{0.95}
\usepackage{booktabs}
\usepackage{mathtools}
\definecolor{darkblue}{rgb}{0.1,0.1,.7}
\usepackage[]{amsmath}
\usepackage[]{graphicx}
\usepackage[]{latexsym}
\usepackage{amscd}
\usepackage[all,cmtip]{xy}
\usepackage{mathrsfs}
\usepackage{bbold}
\usepackage{ytableau}
\usepackage{youngtab}
\usepackage{simplewick}
\usepackage{changepage}
\usepackage[]{algorithm2e}
\usepackage{multirow}
\usepackage{slashed}
\usepackage[OT2,OT1]{fontenc}
\usepackage{braket}
\usepackage{tikz}

\theoremstyle{remark}

\makeatletter
\def\@fpheader{\ }
\makeatother

\title{QFT as a set of ODEs: higher dimensions}
\author{Fabiana De Cesare,}
\author{Manuel Loparco}

\affiliation{Istituto Nazionale di Fisica Nucleare, Sezione di Torino, and\\ 
Department of Physics, 
University of Turin,\\ 
	Via P. Giuria 1, 10125, Turin, Italy}

\emailAdd{fabiana.decesare@unito.it, manuel.loparco@gmail.com}
\abstract{Correlation functions of local operators in Quantum Field Theory (QFT) in Anti-de Sitter space (AdS) are completely fixed by the QFT data: the set of scaling dimensions $\Delta_i$ and OPE coefficients $C_{ijk}$ of the boundary operators, and the bulk-boundary (BOE) coefficients $b^{\hat\Phi}_i$ encoding how bulk fields decompose into boundary operators. In this work, we generalize the ordinary differential equations (ODEs) that govern the variation of the QFT data under a bulk relevant deformation, originally derived for AdS$_2$ \cite{Loparco:2026fki}, to the cases of AdS$_3$ and AdS$_4$. We demonstrate that these flow equations natively capture the mechanism of merger-annihilation when a boundary operator hits marginality, as well as level repulsion when different $\Delta_i$'s approach each other. Furthermore, we address the practical implementation of the framework: we propose substituting the ODE for the OPE coefficients with the crossing equation for greater efficiency, and we observe that Pad\'e approximants dramatically improve the convergence of the sums over boundary operators, at least in free theories. Altogether, these advances lay the groundwork for the future application of the flow equations to the study of strongly coupled QFTs in AdS and their flat space limits.}

\begin{document}
\maketitle

\newpage
\section{Introduction}
Many phenomena of Nature are believed to be described by strongly coupled quantum field theories (QFTs). Despite this, non-perturbative methods to compute observables in such theories remain scarce. The lattice approach \cite{Davoudi:2022Snowmass}, based on the discretization of spacetime, is the method that has yielded the most results, regarding for example the spectrum of many strongly coupled QFTs. However, it suffers from fundamental shortcomings, such as the sign problem and the inability to capture real-time dynamics. Hamiltonian truncation \cite{Fitzpatrick:2022dwq} and the numerical S-matrix bootstrap \cite{Kruczenski:2022lot} are two approaches that work directly in the continuum and can capture Lorentzian dynamics, but they come with their own limitations: Hamiltonian truncation often exhibits slow convergence with the UV cutoff, and the $2$-to-$2$ S-matrix bootstrap primarily yields bounds on Wilson coefficients rather than precise spectra\footnote{However, recently the S-matrix bootstrap has been succesfully utilized as a tool to compute observables in QCD \cite{He:2025fbk,He:2025gws,He:2023lyy,He:2024nwd,Guerrieri:2024jkn,Albert:2022oes,Albert:2023jtd,Albert:2026xyz}.}. In the absence of supersymmetry, conformal symmetry, integrability, or a strong-weak holographic duality, we are not left with many alternatives.

In \cite{Loparco:2026fki}, a program was initiated towards a new non-perturbative framework in the continuum. Let us describe the basic idea\footnote{The original idea presented in \cite{Loparco:2026fki} was itself inspired by the work on ODEs for the CFT data of 1D conformal manifolds by Behan \cite{Behan:2017mwi} and the study of the OPE in QFT by Hollands and Wald \cite{Hollands:2023txn}.}.

Consider placing a QFT on a $d+1$ dimensional rigid hyperboloid, which we will refer to as Anti-de Sitter (AdS) space. The isometries of AdS and its conformal boundary make it an ideal playground for studying strongly coupled QFTs \cite{Callan:1989em,Mazac:2018mdx,Hogervorst:2021spa,Antunes:2021abs,Antunes:2024hrt,Antunes:2025iaw,Bason:2025sxb,Carmi:2018qzm,Copetti:2023sya,Aharony:2012jf}. Crucially, these isometries imply the existence of a map between bulk QFT states and operators living on the boundary. These boundary operators are classified by their scaling dimensions $\Delta_i$ and their representations $\rho_i$ under $SO(d)$ (the group of rotations preserving the boundary)\footnote{Boundary operators may carry additional quantum numbers in the presence of global symmetries.}. Furthermore, the AdS isometries guarantee the existence of two types of convergent expansions: the operator product expansion (OPE) among boundary operators, and the boundary operator expansion (BOE), where a bulk operator is expanded in terms of boundary operators. Pictorially:
\begin{center}
  \includegraphics[scale=0.55]{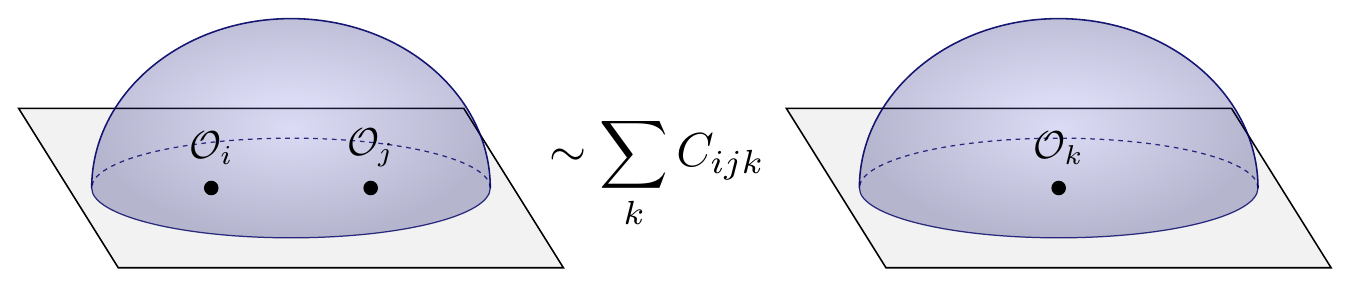}\\
  \includegraphics[scale=0.55]{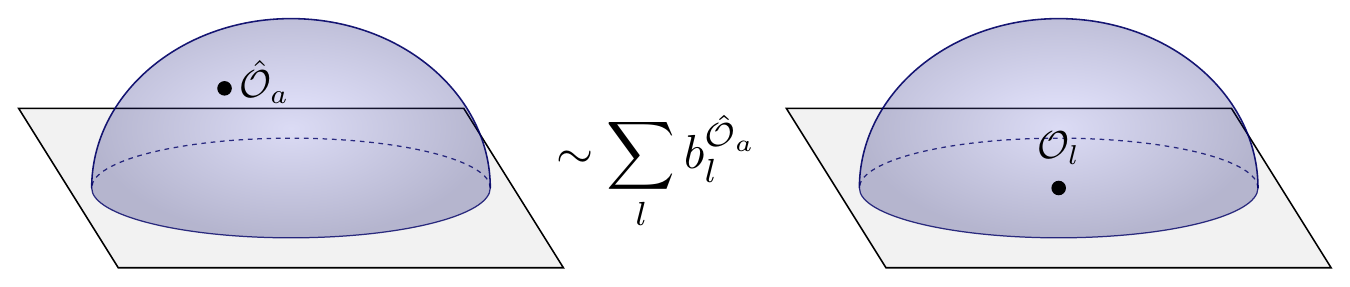}
\end{center}
where the light gray surface represents the boundary of AdS and the half-spheres are foliations of AdS in radial quantization (position dependence and the sum over descendants are suppressed). When operators carry non-trivial representations of $SO(d)$, there can be multiple OPE coefficients $C_{ijk}^{(n)}$ for a given triplet of boundary operators, and multiple BOE coefficients $b^{\hat{\mathcal{O}}_a(n)}_l$ for a given bulk-boundary pair. The OPE and BOE coefficients satisfy various non-trivial constraints, such as the crossing equations and the BOE locality sum rules \cite{Levine:2023ywq,Meineri:2023mps,Levine:2024wqn,Loparco:2025aag}.

Thanks to these convergent expansions, any correlation function of local operators in the bulk or on the boundary can be iteratively decomposed into lower-point correlators. Ultimately, all local observables are entirely determined by the set 
\begin{equation}
    \{\Delta_i,\rho_i,C_{ijk}^{(n)},b^{\hat{\mathcal{O}}_a(n)}_l\}
    \label{eq:QFTdata}
\end{equation}
which we refer to as the QFT data.

QFTs in AdS come in continuous families, parametrized by dimensionless variables $\lambda_i$ constructed from the dimensionful couplings $\bar\lambda_i$ of the relevant bulk operators $\hat\Phi_i$ in the action and the AdS radius $R$, such that $\lambda_i\equiv\bar\lambda_iR^{d+1-\Delta_{\hat\Phi_i}^{\text{UV}}}$, where $\Delta_{\hat\Phi_i}^{\text{UV}}$ is the dimension of $\hat\Phi_i$ in the UV CFT. For simplicity we will work with theories deformed by a single bulk relevant scalar operator $\hat\Phi$, and we will work in units where $R=1$. Under an infinitesimal variation of $\lambda$, the action changes as
\begin{equation}
    S(\lambda+\delta\lambda)=S(\lambda)+\delta\lambda\int_{\text{AdS}}dX\ \hat\Phi(X)\,,
    \label{eq:deformedaction}
\end{equation}
and consequently all QFT data vary continuously with $\lambda$. In \cite{Loparco:2026fki}, computing the universal effect of the deformation (\ref{eq:deformedaction}) on the QFT data in AdS$_2$ led to the derivation of an infinite set of coupled ODEs governing their evolution. 

In this paper, we generalize this framework to higher dimensions, motivated by the presence of many interesting strongly coupled QFTs in more than two dimensions. In $\text{AdS}_{d+1}$, our generalized ODEs take the form:
\begin{equation}
    \begin{aligned}
        \frac{d\Delta_i}{d\lambda}&=\sum_l b^{\hat\Phi}_l C_{lii}^{(0)}\mathcal{I}(\Delta_l)\,,\\
        \frac{db^{\hat\Phi}_i}{d\lambda}&=\sum_l\sum_j b^{\hat\Phi}_l b^{\hat\Phi}_j C_{lij}\mathcal{J}_{\Delta_i}(\Delta_l,\Delta_j)\,,
        \label{eq:flowequations}
    \end{aligned}
\end{equation}
where $\mathcal{I}$ and $\mathcal{J}$ are regulated integrals of conformal blocks, with explicit forms (\ref{eq:IDelta}) and (\ref{eq:JDelta}), $\Delta_i$ can be the scaling dimension of any boundary operator carrying a traceless symmetric representation of $SO(d)$, and $C_{lii}^{(0)}$ is the $n=0$ OPE coefficient in the basis defined in (\ref{eq:JJO3ptalt})\footnote{When $\mathcal{O}_i$ is a scalar, there is only one OPE coefficient and we omit the superscript.}. Crucially, the sums run over all scalar boundary operators excluding the identity.

We call these the \emph{flow equations}. Their practical applicability will rely heavily on the efficient numerical convergence of the sums in (\ref{eq:flowequations}). When utilizing standard conformal blocks, the sums do not generically converge. One known resolution is to instead use local blocks \cite{Levine:2023ywq,Meineri:2023mps,Levine:2024wqn}, a basis which possesses the physical analytic structure of a correlator block by block, altering the expressions of $\mathcal{I}$ and $\mathcal{J}$ to (\ref{eq:IDeltaalpha}) and (\ref{eq:Jalpha}) respectively. This resolves the issue for the first equation, while finding a basis of fully local blocks for the second equation remains an open question. In this work, we introduce an alternative method to render the sums of standard blocks convergent by employing Pad\'e approximants. We discuss this in section \ref{sec:pade}, where we show that the rate of convergence is drastically improved with respect to the local blocks method for both equations, at least in free theory.
\begin{figure}
    \centering
    \includegraphics[width=\linewidth]{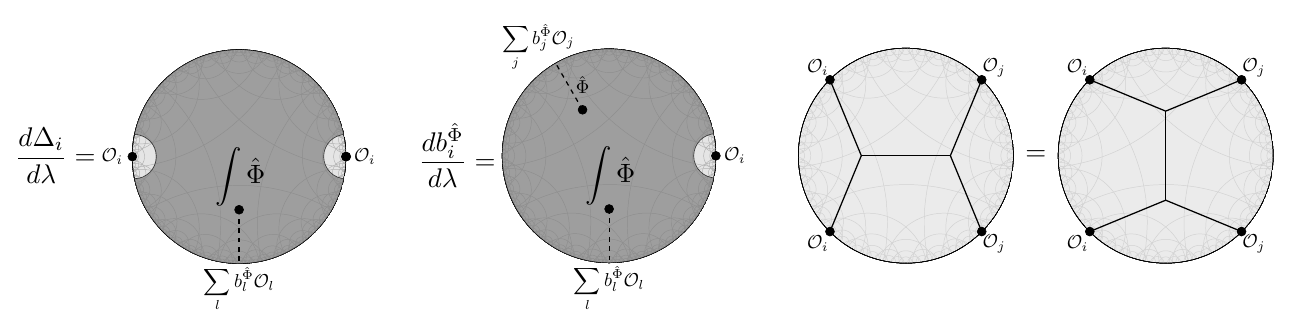}
    \caption{Pictorial representation of the flow equations. In this work, the crossing equations take the place of the ODE involving the derivative of the OPE coefficients.}
    \label{fig:enter-label}
\end{figure}

In AdS$_{d+1}$ with $d+1<5$, the flow equations close when supplemented with the crossing equations for a four-point function of the form $\langle\mathcal{O}_i\mathcal{O}_i\mathcal{O}_j\mathcal{O}_j\rangle$\footnote{In AdS$_{d+1}$ with $d\geq4$ mixed symmetry representations appear in the OPE expansion, hence one would have to derive the ODE for $\frac{d\Delta_i}{d\lambda}$ for operators carrying such representations in order to close the equations.}. In AdS$_3$, there is a single OPE coefficient for a triplet of boundary operators, so the crossing equations read (the notation is introduced in section \ref{subsec:crossing})
\begin{equation}
    \sum_m\Bigg(C_{mij}^2\mathcal{F}_{i,j,m}^{t}(\eta,\bar\eta)-C_{iim}C_{mjj}\mathcal{F}_{i,j,m}^{s}(\eta,\bar\eta)\Bigg)=\mathcal{F}_{i,j,\mathbb{1}}^{s}(\eta,\bar\eta)\,,
\end{equation}
where we took $i\ne j$.
In AdS$_4$, instead, considering a scalar $\mathcal{O}_i$ and a traceless symmetric tensor $\mathcal{O}_j$, crossing reads
\begin{equation}
    \sum_m\left(\sum_{a,b}C_{mij}^{(a)}C_{mij}^{(b)}\mathcal{F}_{i,j,m;I}^{t\,(a,b)}(v,u)-C_{mii}\sum_cC_{mjj}^{(c)}\mathcal{F}_{i,j,m;I}^{s\,(c)}(u,v)\right)=\mathcal{F}_{i,j,\mathbb{1};I}^{s\,(c)}(u,v)\,,
    \label{eq:crossing_intro}
\end{equation}
where again $i\ne j$.
We propose to use crossing rather than the generalization of the ODE involving $\frac{dC_{ijk}}{d\lambda}$ in \cite{Loparco:2026fki} because the integrated blocks $\mathcal{K}$ entering in that equation are only known through integral representations which are expensive to evaluate numerically, an issue that scales poorly in higher dimensions\footnote{We nevertheless discuss some properties of $\frac{dC_{ijk}}{d\lambda}$ in higher dimensions in appendix \ref{app:OPEflow}.}. Notice that solving crossing in this context is instead a relatively cheap numerical problem, in particular it is simpler than in the context of the conformal bootstrap: the spectrum, obtained by integrating (\ref{eq:flowequations}) over an infinitesimal $\delta\lambda$, is an input, and the OPE coefficients are the only unknowns.

Let us now explain in what sense equations (\ref{eq:flowequations}) and (\ref{eq:crossing_intro}) can provide a new non-perturbative method to extract observables of strongly coupled QFTs.  

\begin{figure}
    \centering
    \includegraphics[width=0.65\linewidth]{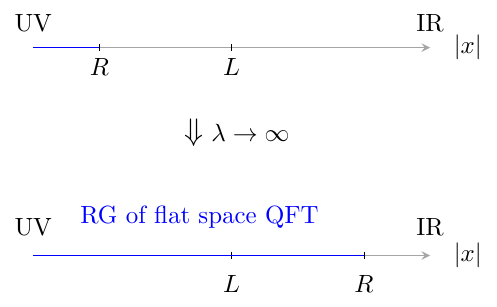}
    \caption{As the dimensionless coupling $\lambda\equiv \bar\lambda R^{d+1-\Delta_{\hat\Phi}^{\text{UV}}}$ grows, more of the flat space RG flow is revealed and imprinted on the QFT data. Here $L$ is a characteristic length scale of the QFT, $R$ is the AdS radius, and $|x|$ is the length scale at which we probe the theory. 
    }
    \label{fig:RGFlow}
\end{figure}

Consider a QFT in AdS defined as a relevant deformation of a conformal field theory (CFT) with conformal boundary conditions:
\begin{equation}
    S_{\text{QFT}}(\lambda)=S_{\text{CFT}}+\lambda\int_{\text{AdS}}\hat\Phi(X)\,.
\end{equation}
If the starting CFT is solvable (for example it is a free theory, or a 2D minimal model) its QFT data is known, or can be computed algorithmically. Then, this data furnishes the initial conditions of the ODEs. Because the coupling is measured in units of the AdS radius, increasing $\lambda$ by integrating the ODEs effectively evolves the data toward the flat space regime. When $\lambda\gg1$, the radius of AdS is much larger than the characteristic length scales of the theory, and the QFT data becomes sensitive to the flat space renormalization group (RG) flow (see figure \ref{fig:RGFlow} for a pictorial representation). 

In this regime, the ratios of the lightest scaling dimensions are expected to asymptote to the mass ratios of the stable particles in the flat space QFT\footnote{This identity applies to levels which are below the two-particle threshold of the lightest mass, e.g. when $\lim_{\lambda\to\infty}\frac{\Delta_i}{\Delta_1}<2$.}
\begin{equation}
    \lim_{\lambda\to\infty}\frac{\Delta_i(\lambda)}{\Delta_1(\lambda)}=\frac{m_i}{m_1}
\end{equation}
and specific combinations of OPE coefficients are expected to yield the flat space S-matrix \cite{Paulos:2016fap,Penedones:2010ue,Susskind:1998vk,Polchinski:1999ry,Hijano:2019qmi,Komatsu:2020sag,Li:2021snj,Cordova:2022pbl,vanRees:2022zmr,vanRees:2023fcf}.  Thus, finding the asymptotic solutions of the ODEs resolves the spectrum and dynamics of the flat space QFT.

We note that the concrete applicability of this non-perturbative method will rely on the development of efficient numerical algorithms that can implement the ODEs for interacting theories, while keeping the errors originating from the truncation of the sums in (\ref{eq:flowequations}) and (\ref{eq:crossing_intro}) under control. We envision that to be the main next step to be taken by this program.


\paragraph{Outline} We begin with section \ref{sec:preliminaries}, where we review the geometry of AdS and the classification of boundary operators in 3D and 4D. Then, we present the explicit form of standard conformal blocks and local blocks for bulk-boundary-boundary and bulk-bulk-boundary correlation functions, which we will need to derive the flow equations (\ref{eq:flowequations}). 

In section \ref{sec:derivationflowequations} we present the derivation of the flow equations (\ref{eq:flowequations}), which require the careful treating of divergences and the renormalization of boundary operators. 

In section \ref{sec:checks} we present checks of the flow equations in free theories, involving both scalar and spinning operators. We compare the performance of regular conformal blocks, local blocks and Pad\'e approximants of both and find that Pad\'e approximants of standard blocks converge fastest. We also present some rudimentary implementation of the crossing equations.

In section \ref{sec:pade} we discuss the details of the Pad\'e procedure, we explain why it works so well in free theory and compare analytic estimates of the error from theorems by Stahl with the errors observed in \ref{sec:checks} and find agreement. We then discuss possible generalizations to the interacting case.

In section \ref{sec:mergerrepulsion} we show that the mechanisms of level repulsion and merger-annihilation are implied by the flow equations, and derive some detailed features of these phenomena.

In section \ref{sec:discussion} we discuss how to move beyond some of our assumptions and outline some future directions and open questions.
\paragraph{Mathematica notebook} We attach to our ArXiv submission a Mathematica notebook where checks of the flow equations and the implementation of crossing from section \ref{sec:checks} can be reproduced.
\paragraph{Summary of assumptions}
Throughout most of this paper we assume we are working with a unitary and local QFT in AdS$_{d+1}$ with irrelevant boundary operators $\Delta_i>d$ and one relevant deformation in the bulk with UV scaling dimension $\Delta_{\hat\Phi}^{\text{UV}}<d+1$. We will assume the theory is parity invariant unless stated otherwise, and we will assume the convergence of the OPE and BOE and the absence of degeneracies. Finally, we assume that in the UV CFT the identity is the only scalar $\hat{\mathcal{O}}$ in the OPE of $\hat\Phi\times\hat\Phi$ with $\Delta_{\hat{\mathcal{O}}}^{\text{UV}}\leq2\Delta_{\hat\Phi}^{\text{UV}}-d-1$.
We will move past some of these assumptions in section \ref{sec:discussion}.
\section{Preliminaries}
\label{sec:preliminaries}

\subsection{Geometry}
\label{subsec:geometry}
\begin{figure}
    \centering
    \includegraphics[width=0.6\linewidth]{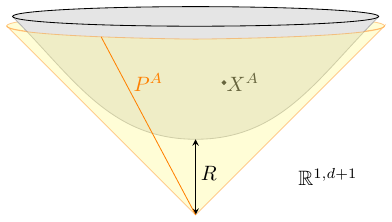}
    \caption{A representation of AdS (in gray) and the lightcone (in yellow) in embedding space $\mathbb{R}^{1,d+1}$. Points in the bulk of AdS are denoted by $X^A$. Lightrays, which are in one-to-one correspondence with points on the boundary of AdS, are indicated as $P^A$. $R$ is the radius of AdS. We slightly vertically displaced AdS from the lightcone for visualization purposes.}
    \label{fig:embeddingspace}
\end{figure}
We define $d+1$ dimensional Euclidean Anti-de Sitter space (for brevity AdS$_{d+1}$) through its embedding in $\mathbb{R}^{1,d+1}$. Points in embedding space belonging to AdS will be denoted as $X^A=(X^0,X^a,X^{d+1})\in\mathbb{R}^{1,d+1}$ where $a=1,\ldots d$, and they satisfy
\begin{equation}
    -(X^0)^2+\delta_{ab}X^aX^b+(X^{d+1})^2=-R^2\,,\qquad X^0>0\,,
    \label{eq:hyperboloid}
\end{equation}
where $X^0>0$ restricts us to one of the two branches of the hyperboloid. 

Points at the conformal boundary of AdS are embedding space lightrays, which we denote as $P^A\in\mathbb{R}^{1,d+1}$, satisfying
\begin{equation}
    -(P^0)^2+\delta_{ab}P^{a}P^b+(P^{d+1})^2=0\,,\quad P^0>0\,,
\end{equation}
with the identification $P^A\sim\lambda P^A$ with $\lambda>0$. The group of isometries preserving one branch of the hyperboloid is $SO^+(1,d+1)$, and it acts linearly on embedding space vectors. This implies in particular that all invariant cross ratios can be constructed from contractions of vectors $P^A$ and $X^A$ with the embedding metric $\eta_{AB}=\text{diag}(-1,1,\ldots,1)$ and with the Levi-Civita symbol $\epsilon_{A_1\ldots A_{d+2}}$.  We give a pictorial representation of the embedding space setup in figure \ref{fig:embeddingspace}.
\subsubsection{Poincar\'e half-plane coordinates}
We will frequently use Poincar\'e half-plane coordinates to chart AdS:
\begin{equation}
    \begin{aligned}
        &X^0=\frac{1+z^2+\mathbf{x}^2}{2z}\,,\qquad &&X^a=\frac{\mathbf{x}^a}{z}\,,\qquad &&&X^{d+1}=\frac{1-z^2-\mathbf{x}^2}{2z}\,,\\
        &P^0=\frac{1+\mathbf{x}^2}{2}\,,\qquad &&P^a=\mathbf{x}^a\,,\qquad &&& P^{d+1}=\frac{1-\mathbf{x}^2}{2}\,.
        \label{eq:poincare}
    \end{aligned}
\end{equation}
where we set $R=1$, as will be done for most of this paper. The metric then reads
\begin{equation}
    ds^2=\frac{\mathrm{d}z^2+\mathrm{d}\mathbf{x}^2}{z^2}\,,\qquad z>0\,,\qquad \mathbf{x}^i\in\mathbb{R}^d\,,
\end{equation}
where $z\to0$ is the limit to the conformal boundary. 
\subsubsection{Poincar\'e ball coordinates}
Alternatively, AdS can be represented as a ball. We can find the Poincar\'e ball coordinates by constructing the stereographic projection from $X^A=(-1,\mathbf{0},0)$
\begin{equation}
    \mathbf{y}^a\equiv \frac{X^a}{1+X^0}\,,\qquad u\equiv \frac{X^{d+1}}{1+X^0}\,.
\end{equation}
The hyperboloid constraint (\ref{eq:hyperboloid}) then forces $\mathbf{y}^2+u^2<1$, thus defining a $d+1$ dimensional ball $B^{d+1}$, with metric
\begin{equation}
    ds^2=\frac{4(\mathrm{d}u^2+\mathrm{d}\mathbf{y}^2)}{(1-u^2-\mathbf{y}^2)^2}\,,\qquad \mathbf{y}^2+u^2<1\,.
    \label{eq:ballcoordinates}
\end{equation}
\subsubsection{Boundary operators}
The geometry of AdS also dictates how we should organize states in the Hilbert space of our QFT, and in turn the way in which we should organize boundary operators. The two cases of interest to us, AdS$_3$ and AdS$_4$, benefit from a separate treatment.  The discussion for AdS$_4$ straightforwardly generalizes to traceless symmetric tensors in higher dimensions.
\paragraph{AdS$_3$} The group of isometries in the 3D case is $SO(3,1)$. Its associated Lie algebra, when complexified, decomposes as follows
\begin{equation}
    \mathfrak{so}(1,3)_{\mathbb{C}}\cong \mathfrak{sl}(2,\mathbb{C})\oplus\mathfrak{sl}(2,\mathbb{C})
    \label{eq:sl2Cisomorp}
\end{equation}
To make use of this fact, it is convenient to switch to complex coordinates
\begin{equation}
    \mathtt{z}=\mathbf{x}^1+i\mathbf{x}^2\,,\qquad \bar{\mathtt{z}}=\mathbf{x}^1-i\mathbf{x}^2\,,
    \label{eq:zzbarcoordinates}
\end{equation}
so that the associated metric becomes
\begin{equation}
    ds^2=\frac{\mathrm{d}z^2+\mathrm{d}\mathtt{z}\mathrm{d}\bar{\mathtt{z}}}{z^2}\,.
\end{equation}
Each boundary operator is then labeled by the eigenvalues $h$ and $\bar h$ of the two Cartans associated to the two copies of $\mathfrak{sl}(2,\mathbb{C})$, respectively $L_0$ and $\bar L_0$, realized as differential operators as
\begin{equation}
    L_0=-\mathtt{z}\partial_{\mathtt{z}}\,,\qquad \bar L_0=-\bar{\mathtt{z}}\partial_{\bar{\mathtt{z}}}\,.
\end{equation}
As is usually done, we will call $h$ and $\bar h$ the holomorphic and antiholomorphic weights. They are related to the scaling dimension $\Delta$ and the $SO(2)$ spin $J$ as
\begin{equation}
    h\equiv\frac{\Delta-J}{2}\,,\qquad \bar h\equiv\frac{\Delta+J}{2}\,.
\end{equation}
We will denote an operator with weights $(h_i,\bar h_i)$ as $\mathcal{O}_i(\mathtt{z},\bar{\mathtt{z}})$. It behaves under scalings and rotations as
\begin{equation}
\begin{aligned}
    \mathcal{O}_i(\lambda \mathtt{z},\lambda\bar{\mathtt{z}})&\to\lambda^{-h_i-\bar h_i}\mathcal{O}_i(\mathtt{z},\bar{\mathtt{z}})\,,\\
\mathcal{O}_i(e^{i\theta} \mathtt{z},e^{-i\theta} \bar{\mathtt{z}})&\to e^{-i\theta(h_i-\bar h_i)}\mathcal{O}_i(\mathtt{z},\bar{\mathtt{z}})\,.
\end{aligned}
\end{equation}
To relate these operators to their counterparts in cartesian coordinates, it suffices to know that 
\begin{equation}
    \mathcal{O}_i(\mathtt{z},\bar{\mathtt{z}})=\mathcal{O}^{\overbrace{\mathtt{z}\mathtt{z}\ldots\mathtt{z}}^{J_i}}_i(\mathtt{z},\bar{\mathtt{z}})\,.
\end{equation}
One can then change coordinates through (\ref{eq:zzbarcoordinates}). Notice that in complex coordinates it is manifest that $\mathcal{O}^{\mathtt{z}\mathtt{z}\ldots\mathtt{z}}_i$ and $\mathcal{O}^{\bar{\mathtt{z}}\bar{\mathtt{z}}\ldots\bar{\mathtt{z}}}_i$ are the only two independent components of an irreducible tensor: every mixed component would be proportional to a trace.

The action of parity on the coordinates is
\begin{equation}
    (z,\mathbf{x}^1,\mathbf{x}^2)\stackrel{\mathcal{P}}{\to}(z,\mathbf{x}^1,-\mathbf{x}^2)
    \label{eq:defparity}
\end{equation}
thus sending $\mathtt{z}\leftrightarrow\bar{\mathtt{z}}$. On operators, this has the effect of exchanging the holomorphic weights $( h_i, \bar h_i)\to(\bar h_i, h_i)$. We will denote the operator with weights $(\bar h_i, h_i)$ as $\mathcal{O}_{\bar i}(\mathtt{z},\bar{\mathtt{z}})$.

\paragraph{AdS$_4$} In four dimensions, boundary operators are organized into irreducible representations of $SO(4,1)$, which we classify by the scaling dimension $\Delta$ and the $SO(3)$ spin $J$. We will thus consider traceless symmetric tensors with $J$ indices, which we will represent in the usual index-free embedding formalism, by contracting the operator's indices with null vectors $Z^A\in\mathbb{C}^{5}$
\begin{equation}
    \mathcal{O}^{(J)}_i(P,Z)\equiv Z_{A_1}\cdots Z_{A_J}\mathcal{O}^{A_1\cdots A_J}_i(P)\,.
\end{equation}
Tangentiality and the nullness of $P$ imply the following properties for $Z^A$:
\begin{equation}
    P\cdot Z=0\,,\qquad Z\sim Z+\lambda P\,,\qquad \forall\,\lambda\,.
\end{equation}
In this form, scaling dimension and spin are encoded in the behavior of the operator under rescalings of $P^A$ and $Z^A$
\begin{equation}
    \begin{aligned}
        \mathcal{O}^{(J)}_i(\lambda P,Z)&\to\lambda^{-\Delta_i}\mathcal{O}^{(J)}_i(P,Z)\,,\\
        \mathcal{O}^{(J)}_i(P,\alpha Z)&\to\alpha^{J}\mathcal{O}^{(J)}_i(P,Z)\,.
    \end{aligned}
\end{equation}
The operator with indices can be retrieved through the action of a differential operator
\begin{equation}
    \mathcal{O}^{A_1\cdots A_J}_i(P)=\frac{1}{(\frac{1}{2})_JJ!}D_{Z_{A_1}}\cdots D_{Z_{A_J}}\mathcal{O}^{(J)}_i(P,Z)
\end{equation}
where $(a)_n\equiv\frac{\Gamma(a+n)}{\Gamma(a)}$ is the Pochhammer symbol and 
\begin{equation}
    D_{Z_A}\equiv \frac{1}{2}\partial_{Z_A}+(Z\cdot\partial_Z)\partial_{Z_A}-\frac{1}{2} Z_A(\partial_Z\cdot \partial_Z)\,.
\end{equation}
Sometimes it will be useful to consider the pullback to local coordinates
\begin{equation}
    \mathcal{O}_i^{a_1\cdots a_J}(\mathbf{x})=\frac{\partial P_{A_1}}{\partial\mathbf{x}_{a_1}}\cdots\frac{\partial P_{A_J}}{\partial\mathbf{x}_{a_J}}\mathcal{O}^{A_1\cdots A_J}(P)\,,
\end{equation}
and we will at times employ complex auxiliary null vectors $\mathbf{z}^a\in\mathbb{C}^3$ to contract indices in local coordinates
\begin{equation}
    \mathcal{O}_i^{(J)}(\mathbf{x},\mathbf{z})\equiv \mathbf{z}_{a_1}\cdots \mathbf{z}_{a_J}\mathcal{O}_i^{a_1\cdots a_J}(\mathbf{x})
\end{equation}
The explicit relation between $Z^A$ and $\mathbf{z}^a$ is
\begin{equation}
    Z^0=\mathbf{x}\cdot\mathbf{z}\,,\qquad Z^a=\mathbf{z}^a\,,\qquad Z^4=-\mathbf{x}\cdot\mathbf{z}\,.
\end{equation}
Sometimes we will need to strip off indices directly in local coordinates. That is done through the action of the differential operators
\begin{equation}
    D_{\mathbf{z}_a}\equiv \frac{1}{2}\partial_{\mathbf{z}_a}+(\mathbf{z}\cdot\partial_{\mathbf{z}})\partial_{\mathbf{z}_a}-\frac{1}{2} \mathbf{z}_a(\partial_{\mathbf{z}}\cdot \partial_{\mathbf{z}})\,.
    \label{eq:defza}
\end{equation}
\subsection{Operator expansions}
\begin{figure}
    \centering
    \includegraphics[width=\linewidth]{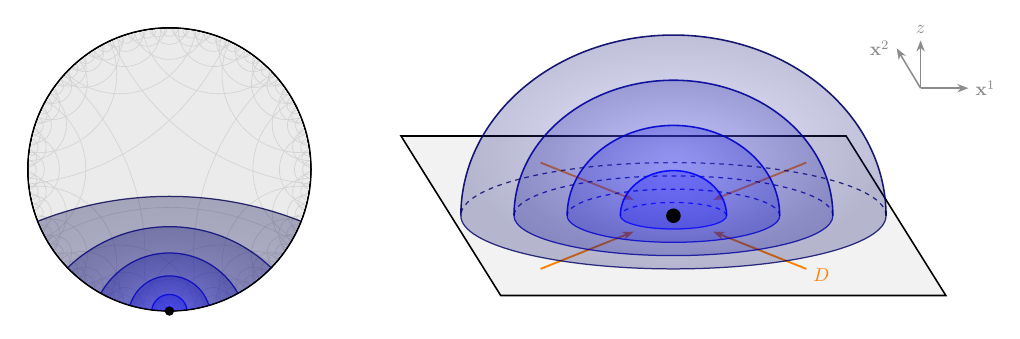}
    \caption{Radial quantization slices of AdS, represented on the left in the ball coordinates (\ref{eq:ballcoordinates}) and on the right in half-plane coordinates (\ref{eq:poincare}). The AdS isometries relate different slices to one another. In half-plane coordinates, it is clear that boundary dilatations relate to one another the spheres defined by the intersection between the slices and the boundary.}
    \label{fig:radialquantization}
\end{figure}
The main advantage of placing a QFT on AdS is the presence of a boundary on which the group of isometries acts like the conformal group. Moreover, in radial quantization, symmetries map the foliation surfaces into one another. On the conformal boundary, the edges of the foliation surfaces appear as spheres, which under dilatations can be shrunk down to points (see figure \ref{fig:radialquantization} for a pictorial representation). These facts ensure the existence of a map between states defined on the radial foliation surfaces and local operators on the boundary, which themselves can be classified in terms of Unitary Irreducible Representations (UIRs) of the conformal group. 

Here we will discuss two immediate consequences of the state-operator map for QFTs in AdS: the expansion of local bulk fields into a sum of local boundary operators -- the boundary operator expansion, or BOE -- and the expansion of a product of two boundary operators into a sum of boundary operators -- the operator product expansion, or OPE. First, let us discuss how we normalize boundary operators.
\subsubsection{Two-point functions and normalizations}
To begin, we need to define the normalization of our boundary operators. This is done by choosing a convention for the numerical factor appearing in the two-point function. 
\paragraph{AdS$_3$} We choose to normalize AdS$_3$ boundary operators as follows\footnote{Notice that in this normalization, the two-point function in cartesian coordinates reads
\begin{equation}
    \langle \mathcal{O}^{a_1 \dots a_J}_i(\mathbf{x}_1) \mathcal{O}_j^{b_1 \dots b_J}(\mathbf{x}_2) \rangle =\delta_{\Delta_i,\Delta_j} \frac{1}{(-2)^{J_i}}\frac{\mathcal{I}^{a_1 \dots a_J, b_1 \dots b_J}(\mathbf{x}_{12})}{(\mathbf{x}^2_{12})^{\Delta_i}}\,,
    \label{eq:2ptcartesian}
\end{equation}
where $\mathbf{x}^2\equiv\delta_{ab}\mathbf{x}^a\mathbf{x}^b$ and $\mathcal{I}^{a_1 \ldots a_J, b_1 \ldots b_J}(\mathbf{x})$ is the symmetric traceless tensor structure constructed from the product of $J$ fundamental inversion tensors $I^{ab}(\mathbf{x}) = \delta^{ab} - 2\frac{\mathbf{x}^a \mathbf{x}^b}{\mathbf{x}^2}$. To map this to complex coordinates, we use $I^{\mathtt{z}\mathtt{z}}(\mathtt{z},\bar{\mathtt{z}})  = -2{\mathtt{z}}/{\bar{\mathtt{z}}}$, $I^{\bar{\mathtt{z}}\bar{\mathtt{z}}}(\mathtt{z},\bar{\mathtt{z}})  = -2{{\bar{\mathtt{z}}/\mathtt{z}}}$, and that mixed components vanish.}
\begin{equation}
    \langle\mathcal{O}_i(\mathtt{z}_1,\bar{\mathtt{z}}_1)\mathcal{O}_j(\mathtt{z}_2,\bar{\mathtt{z}}_2)\rangle=\frac{\delta_{h_i,h_j}\delta_{\bar h_i,\bar h_j}}{\mathtt{z}_{12}^{2h}\bar{\mathtt{z}}_{12}^{2\bar h}}
    \label{eq:2ptcomplex}
\end{equation}
where, as in the rest of the paper, we use $\mathtt{z}_{ij}\equiv\mathtt{z}_i-\mathtt{z}_j$. Equivalently, we can write
\begin{equation}
\langle\mathcal{O}_i(\mathtt{z}_1,\bar{\mathtt{z}}_1)\mathcal{O}_j(\mathtt{z}_2,\bar{\mathtt{z}}_2)\rangle=\frac{\delta_{\Delta_i,\Delta_j}\delta_{J_i,J_j}}{|\mathtt{z}_{12}|^{2\Delta}}\left(\frac{\mathtt{z}_{12}}{\bar{\mathtt{z}}_{12}}\right)^{J_i}\,.
   \label{eq:2ptcomplexbis}
\end{equation}
From the two-point function it is immediate to see that complex conjugation relates an operator with holomorphic weights $(h,\bar h)$ to one with weights $(\bar h,h)$.

\paragraph{AdS$_4$} In four dimensions, we adopt the index-free embedding formalism and normalize boundary two-point functions as
\begin{equation}
    \langle\mathcal{O}_i^{(J_i)}(P_1,Z_1)\mathcal{O}_j^{(J_j)}(P_2,Z_2)\rangle=\delta_{\Delta_i,\Delta_j}\delta_{J_i,J_j}\frac{H_{1,2}^{J_i}}{P_{12}^{\Delta_i+J_i}}\,,\qquad P_{ij}\equiv-2P_i\cdot P_j
    \label{eq:twoPointJ}
\end{equation}
where $H_{1,2}$ is the invariant structure from \cite{Costa:2011mg}
\begin{equation}
    H_{i,j}\equiv(Z_i\cdot Z_j)(P_i\cdot P_j)-(Z_i\cdot P_j)(Z_j\cdot P_i)\,.
    \label{eq:Hijdef}
\end{equation}
Equation (\ref{eq:twoPointJ}) is also the form of two-point functions of symmetric traceless tensors in higher dimensions.
\subsubsection{Operator Product Expansion and three-point functions}
A powerful property of CFTs is the existence of a convergent and associative OPE. This is the fundamental fact that allows for the description of all local observables in terms of the CFT data, and it also holds for boundary operators in the context of QFT in AdS.

For two scalar operators, when the distance between their insertion points is smaller than their distance to any other operator insertion, the OPE takes the form
\begin{equation}
    \mathcal{O}_i(\mathbf{x}_1)\mathcal{O}_j(\mathbf{x}_2)=\sum_k \frac{C_{ijk}}{(\mathbf{x}^2_{12})^{\frac{\Delta_{ijk}}{2}}}\mathcal{C}_{a_1\ldots a_{J_k}}(\mathbf{x}_{12},\partial_2)\mathcal{O}^{a_1\ldots a_{J_k}}_k(\mathbf{x}_2)
    \label{eq:defOPE}
\end{equation}
where $C_{ijk}$ are called OPE coefficients, and the sum over descendants, encoded in the differential operator $\mathcal{C}$, is fully fixed by symmetry.

As a consequence of the OPE, three-point functions of primary scalar operators take the universal form
\begin{equation}
    \langle\mathcal{O}_i(\mathbf{x}_1)\mathcal{O}_j(\mathbf{x}_2)\mathcal{O}_k(\mathbf{x}_3)\rangle=\frac{C_{ijk}}{(\mathbf{x}_{12}^2)^\frac{\Delta_{ijk}}{2}(\mathbf{x}_{13}^2)^\frac{\Delta_{ikj}}{2}(\mathbf{x}_{23}^2)^\frac{\Delta_{jki}}{2}}\,,
    \label{eq:scalar3pt}
\end{equation}
where unitarity fixes $C_{ijk}\in\mathbb{R}$ and we are using the notation $\Delta_{ijk}\equiv\Delta_i+\Delta_j-\Delta_k$. 

For operators carrying spin, a separate treatment of AdS$_3$ and AdS$_4$ is in order.
\paragraph{AdS$_3$} 
If we classify boundary operators in AdS$_3$ by their holomorphic weights, each three-point function still depends on a single OPE coefficient \cite{Osborn:2012vt}
\begin{equation}
    \begin{aligned}
        \langle\mathcal{O}_i(\mathtt{z}_1,\bar{\mathtt{z}}_1)&\mathcal{O}_j(\mathtt{z}_2,\bar{\mathtt{z}}_2)\mathcal{O}_k(\mathtt{z}_3,\bar{\mathtt{z}}_3)\rangle=\frac{C_{ijk}}{\mathtt{z}_{12}^{h_{ijk}}\bar{\mathtt{z}}_{12}^{\bar h_{ijk}}\mathtt{z}_{13}^{h_{ikj}}\bar{\mathtt{z}}_{13}^{\bar h_{ikj}}\mathtt{z}_{23}^{h_{jki}}\bar{\mathtt{z}}_{23}^{\bar h_{jki}}}\\
        &=\frac{C_{ijk}}{|\mathtt{z}_{12}|^{\Delta_{ijk}}|\mathtt{z}_{13}|^{\Delta_{ikj}}|\mathtt{z}_{23}|^{\Delta_{jki}}}\left(\frac{\mathtt{z}_{12}}{\bar{\mathtt{z}}_{12}}\right)^{\frac{J_{ijk}}{2}}\left(\frac{\mathtt{z}_{13}}{\bar{\mathtt{z}}_{13}}\right)^{\frac{J_{ikj}}{2}}\left(\frac{\mathtt{z}_{23}}{\bar{\mathtt{z}}_{23}}\right)^{\frac{J_{jki}}{2}}
        \label{eq:ads33pt}
        \end{aligned}
\end{equation}
where $h_{ijk}\equiv h_i+h_j-h_k$ and $J_{ijk}\equiv J_i+J_j-J_k$. 

If we take a complex conjugate on each side of (\ref{eq:ads33pt}), we get
\begin{equation}
    C_{ijk}=\left(C_{\bar i \bar j \bar k}\right)^*\,.
\end{equation}
In a parity invariant theory, we can make further statements on these OPE coefficients. Consider the action of parity on an operator
\begin{equation}
    \mathcal{P}\mathcal{O}_i(\mathtt{z},\bar{\mathtt{z}})\mathcal{P}^{-1}=p_i\mathcal{O}_{\bar i}(\bar{\mathtt{z}},\mathtt{z})\,,\qquad p_i=\pm1\,,
\end{equation}
where $p_i=+$ for a parity even operator and $p_i=-1$ for a parity odd one. Importantly, notice that parity swaps both $\mathtt{z}\leftrightarrow\bar{\mathtt{z}}$ and $h_i\leftrightarrow\bar h_i$.

If we act with parity, defined as in (\ref{eq:defparity}), on both sides of (\ref{eq:ads33pt}), introducing the identity $\mathbb{1}=\mathcal{P}^{-1}\mathcal{P}$ in between operators, we get the equality
\begin{equation}
C_{ijk}=p_ip_jp_k C_{\bar i\bar j\bar k}\,.
\end{equation}
If one or all three operators are parity even, we have $C_{ijk}=C_{\bar i \bar j \bar k}$. If exactly two of them are parity even, $C_{ijk}=-C_{\bar i \bar j \bar k}$. 
\paragraph{AdS$_4$} Three-point functions of traceless symmetric boundary operators in AdS$_4$ depend on multiple independent OPE coefficients associated to different tensor structures. For a complete classification, see \cite{Costa:2011mg}. For the current purposes, we will need to study the case where we have two identical operators of spin $J$ and one parity even scalar operator and the case in which we have two parity even scalars and one spin $J$ operator. The complete decomposition for the first case includes $J+1$ independent OPE coefficients \cite{Costa:2011mg}
\begin{equation}
    \langle\mathcal{O}_i^{(J)}(P_1,Z_1)\mathcal{O}^{(J)}_i(P_2,Z_2)\mathcal{O}_l(P_3)\rangle=\sum_{n=0}^J\tilde C_{iil}^{(n)}\frac{V_{1,23}^nV_{2,31}^nH_{1,2}^{J-n}}{P_{12}^\frac{2\Delta_i-\Delta_l+2J}{2}P_{13}^\frac{\Delta_l}{2}P_{23}^\frac{\Delta_l}{2}}
    \label{eq:JJO3pt}
\end{equation}
where
\begin{equation}
\begin{aligned}
    V_{k,ij}&\equiv\frac{(Z_k\cdot P_i)(P_j\cdot P_k)-(Z_k\cdot P_j)(P_i\cdot P_k)}{P_i\cdot P_j}\,,
    \label{eq:defV}
\end{aligned}
\end{equation}
and $H_{i,j}$ was defined in (\ref{eq:Hijdef}).
We wrote tildes on the OPE coefficients because there is an alternative yet equivalent basis for the decomposition in (\ref{eq:JJO3pt}), which will be better suited to our purposes:
\begin{equation}
    \langle\mathcal{O}_i^{(J)}(P_1,Z_1)\mathcal{O}^{(J)}_i(P_2,Z_2)\mathcal{O}_l(P_3)\rangle=H_{1,2}^{J}\sum_{n=0}^JC_{iil}^{(n)}\frac{\mathcal{H}_n(\tau)}{P_{12}^\frac{2\Delta_i-\Delta_l+2J}{2}P_{13}^\frac{\Delta_l}{2}P_{23}^\frac{\Delta_l}{2}}
    \label{eq:JJO3ptalt}
\end{equation}
where $\mathcal{H}_n$ are a special kind of Jacobi polynomials
\begin{equation}
    \mathcal{H}_n(\tau)=\,_2F_1\left(\begin{matrix}-n, & n+\frac{1}{2}\\ & 1\end{matrix};2\tau\right)\,,\qquad \tau\equiv\frac{V_{1,23}V_{2,13}}{H_{1,2}}
    \label{eq:defHnv}
\end{equation}
The reasoning for using this basis becomes clear when deriving the spinning bulk-boundary-boundary conformal blocks, which we do in detail in appendix \ref{app:bulk-boundary-boundary}.

When instead we have only one operator with spin, there is only one OPE coefficient
\begin{equation}
    \langle\mathcal{O}_i^{(J)}(P_1,Z_1)\mathcal{O}_j(P_2)\mathcal{O}_l(P_3)\rangle=C_{ijl}\frac{V_{1,23}^J}{P_{12}^{\frac{\Delta_{ijl}+J}{2}}P_{23}^{\frac{\Delta_{jli}-J}{2}}P_{13}^{\frac{\Delta_{ilj}+J}{2}}}\,.
    \label{eq:JOOspinAdS4}
\end{equation}
Once again, the discussion in this paragraph generalizes straightforwardly to traceless symmetric operators in higher dimensions.
\subsubsection{Boundary Operator Expansion}
Radial quantization implies the existence of another kind of operator expansion, which involves bulk operators. In fact, any bulk local operator in AdS$_{d+1}$ can be expanded into a sum of boundary operators. For scalars, the explicit expansion reads \cite{Levine:2023ywq} 
\begin{equation}
    \hat{\Phi}(z,\mathbf{x})=\sum_l b^{\hat{\Phi}}_lz^{\Delta_l}\sum_{n=0}^\infty\frac{(-1)^n}{n!2^{2n}(\Delta_l-\frac{d-2}{2})_n}z^{2n}\Box^n\mathcal{O}_l(\mathbf{x})
    \label{eq:BOE}
\end{equation}
where $\Box\equiv\delta^{ab}\frac{\partial}{\partial\mathbf{x}^a}\frac{\partial}{\partial\mathbf{x}^b}$ and $(a)_n\equiv\frac{\Gamma(a+n)}{\Gamma(a)}$ are Pochhammer symbols. Since $\hat\Phi$ is a scalar, the index $l$ runs over scalar primary boundary operators, while the sum over $n$ is the sum over descendants, which is fixed by symmetry. We call $b^{\hat\Phi}_l$ the BOE coefficients. They are related to the form factors of the bulk field $\hat\Phi$ between the vacuum and the state created by $\mathcal{O}_l$ at the origin. Under this normalization of bulk operators, the bulk-boundary two-point function takes the form
\begin{equation}
    \langle\hat{\Phi}(z,\mathbf{x}_1)\mathcal{O}_l(\mathbf{x}_2)\rangle=b^{\hat\Phi}_l\left(\frac{z}{z^2+\mathbf{x}_{12}^2}\right)^{\Delta_l}\,.
\end{equation}
Unitarity fixes $b^{\hat\Phi}_l\in\mathbb{R}$.
\subsection{Conformal blocks}
Using the operator expansions we just discussed, we can derive compact representations for higher-point functions. In fact, iteratively applying the expansions of the previous section, one can express higher-point functions as sums of kinematic ``blocks", each of which is associated to specific representations of the AdS isometry group, and all dynamics are encoded in the QFT data. Let us discuss the various types of conformal blocks in order of complexity.
\subsubsection{Bulk-bulk}
\label{subsubsec:bulkbulk}
\begin{center}
    \includegraphics[]{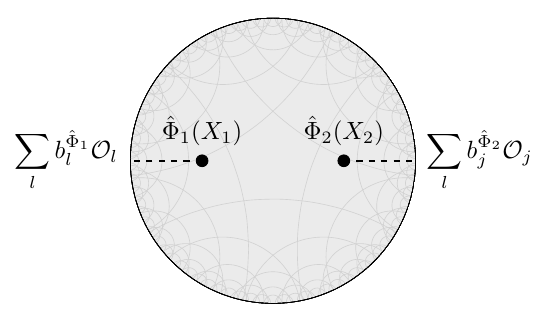}
\end{center}
Let us start from a bulk scalar two-point function. From two bulk points $X_1^A$ and $X_2^B$ one can construct the invariant 
\begin{equation}
    \sigma\equiv X_1\cdot X_2=-\frac{z_1^2+z_2^2+\mathbf{x}_{12}^2}{2z_1z_2}\,,
\end{equation}
which takes values $\sigma\leq-1$, with $\sigma=-1$ corresponding to coincident points configurations. The associated conformal blocks can be derived by applying the BOE (\ref{eq:BOE}) to both bulk fields and summing over the descendants. We do that explicitly in appendix \ref{app:bulk-bulk}. The result is 
\begin{equation}
    \langle\hat\Phi_1(X_1)\hat\Phi_2(X_2)\rangle=\sum_l b^{\hat\Phi_1}_lb^{\hat\Phi_2}_lG_{\text{BB}}^{\Delta_l}(\sigma)\,,
    \label{eq:BBdecompose}
\end{equation}
where
\begin{equation}
    G_{\text{BB}}^{\Delta}(\sigma)=\frac{1}{(-2-2\sigma)^{\Delta}}\,_2F_1\left(\begin{matrix}\Delta, & \Delta-\frac{d-1}{2},\\ & 2\Delta-d+1\end{matrix};\frac{2}{1+\sigma}\right)\,.
\end{equation}
Notice that this decomposition is valid for scalar bulk operators in any number of dimensions.
\subsubsection{Bulk-boundary-boundary}
\begin{center}
    \includegraphics[]{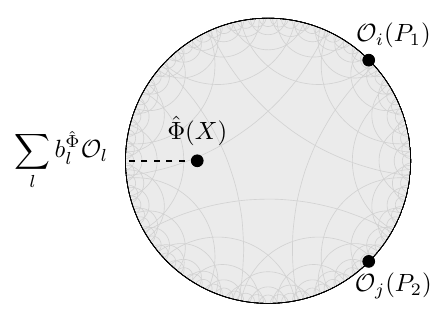}
\end{center}
From one bulk and two boundary points we can construct one invariant cross ratio 
\begin{equation}
    \chi\equiv-\frac{1}{2}\frac{P_1\cdot P_2}{(P_1\cdot X)(P_2\cdot X)}=\frac{z^2\mathbf{x}_{12}^2}{(z^2+(\mathbf{x}-\mathbf{x}_1)^2)(z^2+(\mathbf{x}-\mathbf{x}_2)^2)}\,.
    \label{eq:chimaindef}
\end{equation}
\paragraph{Scalars} Let us start by considering scalar operators only. Carrying out the BOE on the bulk field (see appendix \ref{app:bulk-boundary-boundary}), the conformal block decomposition takes the form 
\begin{equation}
    \langle\mathcal{O}_i(P_1)\mathcal{O}_j(P_2)\hat\Phi(X)\rangle=\frac{1}{P_{12}^\frac{{\Delta_i+\Delta_j}}{2}}\left(\frac{P_1\cdot X}{P_2\cdot X}\right)^{\frac{\Delta_j-\Delta_i}{2}}\sum_lb^{\hat\Phi}_lC_{ijl}G_{\text{Bbb}}^{\Delta_l,\Delta_j,\Delta_i}(\chi)
    \label{eq:Bbbdec}
\end{equation}
where 
\begin{equation}
   G_{\text{Bbb}}^{\Delta_l,\Delta_j,\Delta_i}(\chi)=\chi^\frac{\Delta_l}{2}\,_2F_1\left(\begin{matrix}
        \frac{\Delta_{lij}}{2}, & \frac{\Delta_{lji}}{2}\\
        &\Delta_l-\frac{d-2}{2}
    \end{matrix};\chi\right)\,,
    \label{eq:GBbb}
\end{equation}
where once again, this result is valid in any number of dimensions.

 We will also need the block decomposition for the case in which the two boundary operators are identical and carry a traceless symmetric spin $J$ representation of $SO(d)$. 
 
To derive it, we expand the bulk field to the boundary, reducing the correlator to a sum of CFT three-point functions. Here the dimension matters: for AdS$_3$, we must use (\ref{eq:ads33pt}), while for AdS$_4$ (\ref{eq:JJO3ptalt}). The result in AdS$_4$ easily generalizes to any number of dimensions. 

\paragraph{AdS$_3$}
In this case, the blocks are essentially the same as the scalar case with an overall tensor structure. When we have two identical spin $J_i$ boundary operators, the decomposition takes the form 
\begin{equation}
   \begin{aligned}\langle\mathcal{O}_i(\mathtt{z}_1,\bar{\mathtt{z}}_1)\mathcal{O}_i(\mathtt{z}_2,\bar{\mathtt{z}}_2)\hat\Phi(z,\mathtt{z}_3,\bar{\mathtt{z}}_3)\rangle=&\frac{1}{|\mathtt{z}_{12}|^{2\Delta_i}}\left(\frac{\mathtt{z}_{12}}{\bar{\mathtt{z}}_{12}}\right)^{J_i}\sum_lb^{\hat\Phi}_lC_{lii}G_{\text{Bbb}}^{\Delta_l,\Delta_i,\Delta_i}(\chi)
   \label{eq:ansatzblock2}
\end{aligned}
\end{equation}
where $G_{\text{Bbb}}^{\Delta_l,\Delta_i,\Delta_i}(\chi)$ is (\ref{eq:GBbb}) with $d=2$ and $\Delta_j\to\Delta_i$. See the detailed derivation in appendix \ref{app:bulk-boundary-boundary}.

\paragraph{AdS$_4$} In 4D (and higher) there are multiple OPE coefficients when two operators have spin. Explicitly, the decomposition reads
\begin{equation}
    \langle\mathcal{O}^{(J)}_i(P_1,Z_1)\mathcal{O}^{(J)}_i(P_2,Z_2)\hat\Phi(X)\rangle=\frac{(H_{1,2})^J}{P_{12}^{\Delta_i+J}}\sum_lb_l^{\hat\Phi}\sum_{n=0}^{J}C^{(n)}_{iil}\mathcal{H}_n(v)G^{\Delta_l,n} _{\text{Bbb}}(\chi)\,,
    \label{eq:BbbJ}
\end{equation}
where
\begin{equation}
    G^{\Delta_l,n} _{\text{Bbb}}(\chi)\equiv(1-\chi)^n\chi^{\frac{\Delta_l}{2}}\,_2F_1\left(\begin{matrix}
        \frac{\Delta_l}{2}+n, & \frac{\Delta_l}{2}+n\\
        & \Delta_l-\frac{d-2}{2}
    \end{matrix};\chi\right)\,
\end{equation}
and $\mathcal{H}_n$ was defined in (\ref{eq:defHnv}), but now the cross ratio $v$ is adapted to involve the bulk point $X$
\begin{equation}
    v\equiv\frac{1}{\chi-1}\frac{V_{1,23}V_{2,31}}{H_{1,2}}\,,\qquad P_3\to X
\end{equation}
The details of the derivation are in appendix \ref{app:bulk-boundary-boundary}.

Specifically for $\text{AdS}_4$, we also require the bulk-boundary-boundary block involving a spinning boundary operator, a boundary scalar, and a bulk scalar, see appendix \ref{app:OPEflow}. Using  (\ref{eq:JOOspinAdS4}), this decomposition gives
\begin{equation}
    \langle\mathcal{O}_i^{(J_i)}(P_1,Z_1)\mathcal{O}_j(P_2)\hat\Phi(X)\rangle=\frac{(V_{1,23})^{J_i}}{P_{12}^{\frac{\Delta_i+\Delta_j+J_i}{2}}}\left(\frac{P_1\cdot X}{P_2\cdot X}\right)^{\frac{\Delta_j-\Delta_i-J_i}{2}}\sum_lb^{\hat\Phi}_lC_{lij}G_{\text{Bbb},J_i}^{\Delta_l,\Delta_i,\Delta_j}(\chi)
    \label{eq:JOPhidec}
\end{equation}
where $V_{1,23}$ is as in (\ref{eq:defV}) but with $P_3\to X$, and 
\begin{equation}
    G_{\text{Bbb},J_i}^{\Delta_l,\Delta_i,\Delta_j}(\chi)=\left(-\frac{1}{2}\right)^{J_i}\chi^\frac{\Delta_l}{2}\ _2F_1\left(\begin{matrix}
        \frac{\Delta_{lij}}{2}+J_i, & \frac{\Delta_{lji}}{2}+J_i,\\
        &\Delta_l-\frac{d-2}{2}
    \end{matrix};\chi\right)\,.
    \label{eq:GBbbJOphi}
\end{equation}
\subsubsection{Bulk-bulk-boundary}
\begin{center}
    \includegraphics[]{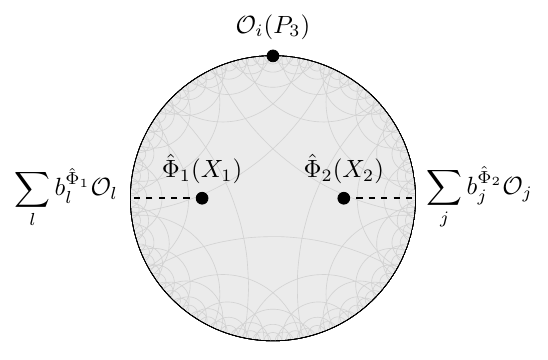}
\end{center}
We are interested in bulk-bulk-boundary three-point functions which involve only scalars.
When we have two points in the bulk and one on the boundary, there are two independent cross ratios, which we choose to be
\begin{equation}
    \xi=\left(\frac{P_3\cdot X_1}{P_3\cdot X_2}\right)^2\,,\qquad \rho=-1-\xi-2(X_1\cdot X_2)\sqrt{\xi}
    \label{eq:xirhodef}
\end{equation}
Notice that $\xi>0$ and $\rho\geq0$\footnote{That is because when $X_1\to X_2$ we have $\rho=-(1-\sqrt{\xi})^2$ but also $\xi=1$.}.
The conformal block expansion is obtained by expanding both bulk operators to the boundary. In appendix \ref{app:bulk-bulk-boundary} we get
\begin{equation}
    \langle\hat\Phi_1(X_1)\hat\Phi_2(X_2)\mathcal{O}_i(P_3)\rangle=\frac{1}{(-2P_3\cdot X_2)^{\Delta_i}}\sum_{j}\sum_lb^{\hat\Phi_1}_l b^{\hat\Phi_2}_jC_{lij}G_{\text{BBb}}^{\Delta_l,\Delta_j,\Delta_i}(\xi,\rho)\,.
    \label{eq:BBb3point}
\end{equation}
with 
\begin{equation}
    G_{\text{BBb}}^{\Delta_l,\Delta_j,\Delta_i}(\xi,\rho)=\frac{\xi^{\frac{\Delta_j-\Delta_i}{2}}}{\rho^{\frac{\Delta_{lji}}{2}}}F_4\left(\begin{matrix}\frac{\Delta_{lji}}{2} & \frac{\Delta_{lji}}{2}-\frac{d-2}{2}\\ \Delta_l-\frac{d-2}{2}, & \Delta_j-\frac{d-2}{2} \end{matrix}; -\frac{1}{\rho},-\frac{\xi}{\rho}\right)
    \label{eq:BBbdec}
\end{equation}
where $F_4$ is an Appell function, defined through the following series representation, convergent for $\sqrt{|x|}+\sqrt{|y|}<1$:
\begin{equation}
    F_4\left(\begin{matrix}a & b\\ c, & d \end{matrix};x,y\right)\equiv \sum_{m=0}^\infty\sum_{n=0}^\infty\frac{(a)_{m+n}(b)_{m+n}}{(c)_{m}(d)_nm!n!}x^my^n
    \label{eq:appellF4def}
\end{equation}
Notice that the domain of convergence does not include all possible values of the cross ratios. For an alternative (but more complicated) representation of this conformal block which converges for all AdS configurations, see equation (\ref{eq:convergentBBb}).
\subsubsection{Boundary 4-pt}
\label{subsubsec:bd4pt}
The last conformal blocks we need are those associated to four-point functions on the boundary. These are the standard CFT conformal blocks, and their explicit form depends on the number of dimensions. 

\paragraph{AdS$_2$} 
Four-point functions on the $1$D boundary can be decomposed in blocks as follows
\begin{equation}
    \langle\mathcal{O}_l(x_1)\mathcal{O}_i(x_2)\mathcal{O}_j(x_3)\mathcal{O}_k(x_4)\rangle=\left(\frac{x_{14}^2}{x_{13}^2}\right)^\frac{\Delta_{j}-\Delta_k}{2}\left(\frac{x_{24}^2}{x_{14}^2}\right)^\frac{\Delta_{l}-\Delta_i}{2}\frac{\mathcal{G}(\zeta)}{(x_{12}^2)^{\frac{\Delta_l+\Delta_i}{2}}(x_{34}^2)^{\frac{\Delta_j+\Delta_k}{2}}}
\end{equation}
where we used the fact that in 1D four-point functions can be fixed up to a function of a single cross ratio, which we take to be $\zeta\equiv\frac{x_{12}x_{34}}{x_{13}x_{24}}$. Moreover, operators are labeled only by their scaling dimension. 

The function $\mathcal{G}(\zeta)$ can be decomposed in various channels. The $s$-channel decomposition, defined by the pairings $(li)(jk)$, reads
\begin{equation}
    \mathcal{G}(\zeta)= \sum_{m}C_{lim}C_{mjk}G_{\Delta_m}^{lijk}(\zeta)
\end{equation}
where the explicit form of the block is
\begin{equation}
    G_{\Delta_m}^{lijk}(\zeta)=\zeta^{\Delta_{m}}\,_2F_1\left(\begin{matrix}
        \Delta_{mil}, & \Delta_{mjk}\\
        & 2\Delta_m
    \end{matrix};\zeta\right)
\end{equation}
where we use $lijk$ as a shorthand for $\Delta_l,\Delta_i,\Delta_j,\Delta_k$, but we emphasize the blocks only depend on the scaling dimensions of the operators.
\paragraph{AdS$_3$} 
In 2D CFT, the decomposition of a four-point function of operators with generic holomorphic weights reads \cite{Osborn:2012vt}
\begin{equation}
    \begin{aligned}
    \langle\mathcal{O}_l&(\mathtt{z}_1,\bar{\mathtt{z}}_1)\mathcal{O}_i(\mathtt{z}_2,\bar{\mathtt{z}}_2)\mathcal{O}_j(\mathtt{z}_3,\bar{\mathtt{z}}_3)\mathcal{O}_k(\mathtt{z}_4,\bar{\mathtt{z}}_4)\rangle\\
    &=\frac{1}{\mathtt{z}_{12}^{h_l+h_i}\mathtt{z}_{34}^{h_j+h_k}}\left(\frac{\mathtt{z}_{24}}{\mathtt{z}_{14}}\right)^{h_{li}}\left(\frac{\mathtt{z}_{14}}{\mathtt{z}_{13}}\right)^{h_{jk}}\frac{1}{\bar{\mathtt{z}}_{12}^{\bar h_l+\bar h_i}\bar{\mathtt{z}}_{34}^{\bar h_j+\bar h_k}}\left(\frac{\bar{\mathtt{z}}_{24}}{\bar{\mathtt{z}}_{14}}\right)^{\bar h_{li}}\left(\frac{\bar{\mathtt{z}}_{14}}{\bar{\mathtt{z}}_{13}}\right)^{\bar h_{jk}}\mathcal{G}(\eta,\bar{\eta})
    \end{aligned}
\end{equation}
where $h_{ij}\equiv h_i-h_j$ and we used the fact that in 2D we can fix a four-point function up to a function of two real variables, or one complex cross-ratio, which we take to be
\begin{equation}
    \eta\equiv \frac{\mathtt{z}_{12}\mathtt{z}_{34}}{\mathtt{z}_{13}\mathtt{z}_{24}}
\end{equation}
This function can then be decomposed into conformal blocks in the $s$-channel as follows
\begin{equation}
    \mathcal{G}(\eta,\bar{\eta})=\sum_m C_{lim}C_{mjk}G_{h_m,\bar h_m}^{lijk}(\eta,\bar{\eta})\,,
\end{equation}
Notice that the sum is taken to run over operators labeled by $(h_m,\bar h_m)$, hence the complex conjugate operators labeled by $(\bar h_m, h_m)$ must be included independently.

These blocks are known in closed form for every value of the external holomorphic weights \cite{Osborn:2012vt}
\begin{equation}
    G_{h_m,\bar h_m}^{lijk}(\eta,\bar{\eta})=\eta^{h_{m}}\bar{\eta}^{\bar h_m}\,_2F_1\left(\begin{matrix}
        h_{mil}, & h_{mjk}\\
        & 2h_m
    \end{matrix};\eta\right)\,_2F_1\left(\begin{matrix}
        \bar h_{mil}, & \bar h_{mjk}\\
        & 2\bar h_m
    \end{matrix};\bar{\eta}\right)
    \label{eq:3D4PT}
\end{equation}
where $h_{ijk}\equiv h_i+h_j-h_k$. With this normalization, the exchange of the identity gives $G^{iijj}_{0,0}=1$, as required by (\ref{eq:2ptcomplex}). When the operators $(h_m,\bar h_m)$ and $(\bar h_m,h_m)$ are collected into a single term, as we do in (\ref{eq:bbbb3D}), their blocks add up, and a factor $1/(1+\delta_{0,J_m})$ is needed to avoid double counting scalars.
\paragraph{AdS$_4$}
In 3D CFT, the situation is significantly more complicated. Let us focus on scalar operators. The four-point function can again be fixed up to a function of two cross-ratios
\begin{equation}
    \langle\mathcal{O}_l(\mathbf{x}_1)\mathcal{O}_i(\mathbf{x}_2)\mathcal{O}_j(\mathbf{x}_3)\mathcal{O}_k(\mathbf{x}_4)\rangle=\left(\frac{\mathbf{x}_{14}^2}{\mathbf{x}_{13}^2}\right)^\frac{\Delta_{j}-\Delta_k}{2}\left(\frac{\mathbf{x}_{24}^2}{\mathbf{x}_{14}^2}\right)^\frac{\Delta_{l}-\Delta_i}{2}\frac{\mathcal{G}(u,v)}{(\mathbf{x}_{12}^2)^{\frac{\Delta_l+\Delta_i}{2}}(\mathbf{x}_{34}^2)^{\frac{\Delta_j+\Delta_k}{2}}}
\end{equation}
which we take to be
\begin{equation}
    u=\frac{\mathbf{x}_{12}^2\mathbf{x}_{34}^2}{\mathbf{x}_{13}^2\mathbf{x}_{24}^2}\,,\qquad
v=\frac{\mathbf{x}_{14}^2\mathbf{x}_{23}^2}{\mathbf{x}_{13}^2\mathbf{x}_{24}^2}\,.
\end{equation}
The $s$-channel decomposition in this case reads
\begin{equation}
\begin{aligned}
   \mathcal{G}(u,v)&= \sum_{m}C_{lim}C_{mjk}G_{\Delta_m,J_m}^{lijk}(u,v)\,,
\end{aligned}
\end{equation}
where $\Delta_m$ and $J_m$ are the scaling dimension and spin of the exchanged operator $\mathcal{O}_m$, but the blocks are not known in closed form for generic $u$ and $v$. 

Complex cross ratios $(\eta,\bar\eta)$ can be used also in this case. They are related to $u$ and $v$ through
\begin{equation}
    u=\eta\bar\eta\,,\qquad v=(1-\eta)(1-\bar\eta)\,.
\label{eq:uvetabareta}
\end{equation}

The conformal block decomposition when external operators with spin are involved is significantly more complicated. We refer the reader to \cite{Kravchuk:2016qvl,Penedones:2015aga,Costa:2016xah,Erramilli:2019njx,Erramilli:2020rlr}.

\subsection{Local blocks}
\label{sec:local_blocks}
\begin{figure}
    \centering
    \includegraphics[width=0.4\linewidth]{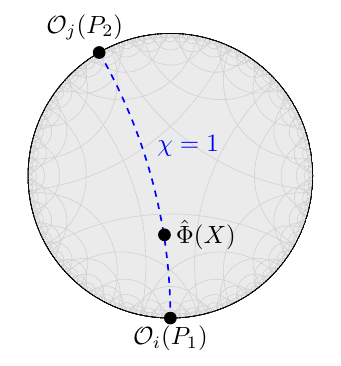}
    \includegraphics[width=0.4\linewidth]{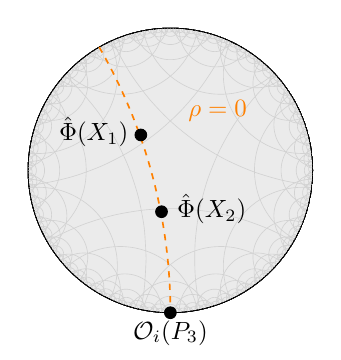}
    \caption{When a bulk operator is inserted on the geodesic connecting two other operators, the conformal block decompositions for bulk-boundary-boundary (\ref{eq:Bbbdec}) and bulk-bulk-boundary (\ref{eq:BBbdec}) correlation functions fail to converge. On the left, the geodesic in the bulk-boundary-boundary case, which corresponds to configurations where the cross ratio $\chi$ (\ref{eq:chimaindef}) is $1$. On the right, the geodesic in the bulk-bulk-boundary case, which corresponds to configurations where the cross-ratio $\rho$ (\ref{eq:xirhodef}) is equal to $0$ (and $\xi$ has any value). Local blocks solve the issue fully for the bulk-boundary-boundary case and only for $X_2$ on the portion of the geodesic between $X_1$ and $P_3$ in the bulk-bulk-boundary case.}
    \label{fig:geodesicfail}
\end{figure}
The conformal block decomposition of correlation functions involving bulk fields fails for certain configurations of points. Specifically, the bulk-boundary-boundary and bulk-bulk-boundary blocks present singularities whenever a bulk point lies on a geodesic which joins any two other insertion points, see figure \ref{fig:geodesicfail}. Take for example the decomposition of a bulk-boundary-boundary correlation function (\ref{eq:Bbbdec}). At the value of the cross ratio $\chi=1$, the blocks are singular:
\begin{equation}
    G_{\text{Bbb}}^{\Delta_l,\Delta_i,\Delta_j}(\chi)\stackrel{\chi\to1}{\sim}\begin{cases}\log(1-\chi)\,,\qquad d=2\\
    (1-\chi)^\frac{2-d}{2}\,,\qquad d>2
    \end{cases}
\end{equation}
At the same time, the analytic domain of a physical correlation function can be argued to be the cut complex plane $\chi\in\mathbb{C}\backslash(-\infty,0]$ \cite{Loparco:2025aag}, and so we do not expect a singularity at $\chi=1$.

This issue is fully resolved for the bulk-boundary-boundary case by introducing \textit{local blocks} \cite{Levine:2023ywq,Meineri:2023mps}, a basis which is composed of elements with the correct analytic structure term by term. We are going to limit ourselves here to report their form, and for in-depth discussions we direct the reader towards \cite{Levine:2023ywq,Meineri:2023mps}, and \cite{Loparco:2025aag} for the subtle case of AdS$_2$. For the bulk-bulk-boundary case, the currently available local blocks only cure the sum over one of the two bulk operators. Curing both, requires finding blocks that have the physical analytic structure in both cross ratios. This remains an open problem which we will not resolve here.
\subsubsection{Bulk-boundary-boundary}
The decomposition in terms of bulk-boundary-boundary local blocks mirrors the one in terms of regular conformal blocks (\ref{eq:Bbbdec})
\begin{equation}
    \langle\hat\Phi(X)\mathcal{O}_i(P_1)\mathcal{O}_j(P_2)\rangle=\frac{1}{P_{12}^\frac{{\Delta_i+\Delta_j}}{2}}\left(\frac{P_1\cdot X}{P_2\cdot X}\right)^{\frac{\Delta_j-\Delta_i}{2}}\sum_lb^{\hat\Phi}_lC_{ijl}G_{\text{Bbb}}^{\Delta_l,\Delta_i,\Delta_j,\alpha}(\chi)
    \label{eq:Bbbalphadec}
\end{equation}
where now \cite{Levine:2023ywq,Meineri:2023mps}
\begin{equation}
\begin{aligned}
    G_{\text{Bbb}}^{\Delta_l,\Delta_i,\Delta_j,\alpha}(\chi)&=G_{\text{Bbb}}^{\Delta_l,\Delta_i,\Delta_j}(\chi)-\chi^\alpha\frac{\Gamma(\alpha+\frac{\Delta_i-\Delta_j}{2})\Gamma(\alpha+\frac{\Delta_j-\Delta_i}{2})\Gamma(\Delta_l-\frac{d-2}{2})}{\Gamma(1+\alpha-\frac{\Delta_l}{2})\Gamma(\alpha+\frac{\Delta_l}{2}-\frac{d-2}{2})\Gamma(\frac{\Delta_{lij}}{2})\Gamma(\frac{\Delta_{lji}}{2})}\\
    &\quad\times\       
    _3F_2\left(\begin{matrix}
        1, & \alpha+\frac{\Delta_i-\Delta_j}{2}, & \alpha+\frac{\Delta_j-\Delta_i}{2}, \\
        & 1+\alpha-\frac{\Delta_l}{2}, & 1+\alpha-\frac{d-\Delta_l}{2}
    \end{matrix};\chi\right)
    \label{eq:Bbblocal}
\end{aligned}
\end{equation}
Notice the presence of an extra real parameter $\alpha$, which originates from the dispersion relation between local blocks and standard conformal blocks
\begin{equation}
    G_{\text{Bbb}}^{\Delta_l,\Delta_i,\Delta_j,\alpha}(\chi)=\chi^\alpha\int_{-\infty}^0\frac{d\chi'}{2\pi i}\frac{1}{\chi'-\chi}\text{Disc}_{\chi'\leq0}\left(\frac{G_{\text{Bbb}}^{\Delta_l,\Delta_i,\Delta_j}(\chi')}{\chi'^\alpha}\right)
    \label{eq:dispBbb}
\end{equation}
The validity of this dispersion relation and the convergence of the resulting block decomposition (\ref{eq:Bbbalphadec}) require a lower bound on $\alpha$ \cite{Meineri:2023mps,Loparco:2025aag}:
\begin{equation}
    \alpha>\frac{\Delta_i+\Delta_j+\Delta_{\hat\Phi}^{\text{UV}}}{2}
    \label{eq:alpha_lowerB}
\end{equation}
where $\Delta_{\hat\Phi}^{\text{UV}}$ is the scaling dimension of the bulk field $\hat\Phi$ at the UV fixed point of the bulk theory. Convergence of the integral at $\chi'=0$ also requires the upper bound $\alpha<1+\Delta_l/2$, but the result (\ref{eq:Bbblocal}) can be extended by analytic continuation to the whole interval (\ref{eq:alpha_lowerB}), as shown in \cite{Meineri:2023mps}.

It can be verified that now the blocks in (\ref{eq:Bbbalphadec}) are finite at $\chi=1$. Using this particular basis will be crucial for us to have convergent sums in the flow equations (\ref{eq:flowequations}).
\subsubsection{Bulk-bulk-boundary}
In the bulk-bulk-boundary case we expand one of the bulk operators to the boundary to then use the local block decomposition (\ref{eq:Bbbalphadec}) for the resulting correlation function. Schematically,
\begin{equation}
    \begin{aligned}\langle&\hat\Phi_1(z_1,\mathbf{x}_1)\hat\Phi_2(z_2,\mathbf{x}_2)\mathcal{O}_i(\mathbf{x}_3)\rangle=\sum_j b^{\hat\Phi_2}_jz^{\Delta_j}_2\langle\hat\Phi_1(z_1,\mathbf{x}_1)\mathcal{O}_j(\mathbf{x}_2)\mathcal{O}_i(\mathbf{x}_3)\rangle+\ldots\\
    &=\sum_{l,j} b^{\hat\Phi_1}_lb^{\hat\Phi_2}_jC_{lji}z^{\Delta_j}_2\frac{1}{(\mathbf{x}_{23}^2)^\frac{\Delta_i+\Delta_j}{2}}\left(\frac{z_1^2+\mathbf{x}_{12}^2}{z_1^2+\mathbf{x}_{13}^2}\right)^{\frac{\Delta_i-\Delta_j}{2}}G^{\Delta_l,\Delta_j,\Delta_i,\alpha} _{\text{Bbb}}(\chi_{(2,3)})+\ldots\\
    &\equiv\frac{1}{(-2P_3\cdot X_2)^{\Delta_i}}\sum_{l,j} b^{\hat\Phi_1}_lb^{\hat\Phi_2}_jC_{lji}G_{\text{BBb}}^{\Delta_l,\Delta_j,\Delta_i,\alpha}(\xi,\rho)
    \end{aligned}
    \label{eq:GBBbalpha1}
\end{equation}
where the dots stand for the sum over descendants of $\mathcal{O}_j$, which we omit to avoid clutter. Unlike the standard conformal block case, that sum cannot be carried out explicitly here. The result reads
\begin{align}
    &G_{\text{BBb}}^{\Delta_l,\Delta_j,\Delta_i,\alpha}(\xi,\rho)=G_{\text{BBb}}^{\Delta_l,\Delta_j,\Delta_i}(\xi,\rho)- \sum_{k,m,n}\frac{(-1)^{k+n}}{n!\ k!}\, \xi^{n-\frac{ \Delta_{ij}}{2}} \rho^{-k - m - n - \alpha +\frac{ \Delta_{ij}}{2}} \nonumber\\& \quad\times\frac{\Gamma\left(\alpha + \frac{\Delta_{ij}}{2}\right)\Gamma\left(k + \alpha - \frac{\Delta_{ij}}{2}\right)\Gamma\left(1 - \frac{d}{2} + \Delta_l\right)}{\Gamma\left(1 + \alpha - \frac{\Delta_l}{2}\right) \Gamma\left(1 - \frac{d}{2} + \alpha + \frac{\Delta_l}{2}\right) \Gamma\left(\frac{\Delta_{ij} + \Delta_l}{2}\right) \Gamma\left(\frac{-\Delta_{ij} + \Delta_l}{2}\right)} \nonumber\\& \quad\times\frac{\left(\alpha + \frac{\Delta_{ij}}{2}\right)_m \left(k + \alpha - \frac{\Delta_{ij}}{2}\right)_{m+n} \left(1 - \frac{d}{2} + k + \alpha - \frac{\Delta_{ij}}{2}\right)_{m+n}}{ \left(1 - \frac{d}{2} + k + \alpha - \frac{\Delta_{ij}}{2}\right)_m \left(1 - \frac{d}{2} + \Delta_j\right)_n \left(1 + \alpha - \frac{\Delta_l}{2}\right)_m \left(1 - \frac{d}{2} + \alpha + \frac{\Delta_l}{2}\right)_m}\,.
    \label{eq:GBBbalpha}
    \end{align}

\section{Derivation of the flow equations}
\label{sec:derivationflowequations}
Armed with the conformal block decompositions described in the previous section, we now present the derivation of our main results, which will be ODEs describing the variation of the QFT data in AdS$_3$ and AdS$_4$ under a bulk relevant deformation.

To derive the flow equations, we start from correlation functions at finite coupling $\lambda$, which are generically defined as
\begin{equation}
\langle\dots\rangle_{\lambda}=\frac{\int D{\phi} \ \dots\ e^{-S_0-\lambda\int_{AdS}\hat\Phi(z,\mathbf{x})}}{\int D{\phi}\ e^{-S_0-\lambda\int_{AdS}\hat\Phi(z,\mathbf{x})}}\,,
\end{equation}
where $S_0$ is the action of the undeformed bulk theory and $\phi$ schematically refers to all the fields involved in the theory. We consider correlation functions in which at least one operator is inserted at the boundary. 

Let us now introduce an infinitesimal shift in the bulk coupling, $\lambda \to \lambda + \delta\lambda$. As we will show shortly, the correction to the correlation functions in perturbation theory in $\delta\lambda$ generically has a divergence which arises from integrating over the bulk deformation close to the boundary, therefore corresponding to an IR divergence. To address this, we must renormalize the boundary theory by expressing the bare operators in terms of renormalized ones, as in\footnote{Bare primary operators can only mix with renormalized operators, either primaries or descendants, which are in the same $SO(d)$ representation. To obtain such operators we can either act with $\Box$ on operators in the same $SO(d)$ representation or with more generic combinations of derivatives on other kind of operators (e.g. a scalar operator can mix with the divergence of a vector). We here restrict to the first type of mixing because the second will not enter in the derivation of the flow equations.
 }
\begin{equation}
\mathcal{O}_i^{(\text{bare})}(\mathbf{x})=\sum_k\sum_{n=0}^\infty\mathcal{Z}_{ik;n}\Box^n\mathcal{O}^\mathrm{(ren)}_k(\mathbf{x})\,, \label{eq:barerenmatch}
\end{equation}
where the coefficients $\mathcal{Z}_{ik;n}$ are the field-strength renormalization constants, which can be written as
\begin{equation}
\mathcal{Z}_{ik;n} = \delta_{ik}\delta_{n,0} + \delta \mathcal{Z}_{ik;n}\,,
\end{equation}
where $\delta \mathcal{Z}_{ik;n}$ is of order $\delta\lambda$.

On the other hand, a shift in the coupling by $\delta\lambda$ corresponds to a shift in the bulk action of
\begin{equation}
\delta S_\mathrm{bulk}=\delta\lambda\int \frac{dz d^d\mathbf{x}}{z^{d+1}}\hat\Phi(z,\mathbf{x})\,.
\end{equation}
As a consequence, the change in correlation functions can be computed using bulk perturbation theory in the coupling $\delta\lambda$, which, at leading order, gives
\begin{equation}
\begin{aligned}
\langle\dots\rangle^{(\text{bare})}_{\lambda+\delta\lambda}= &\langle \dots\rangle_{\lambda} - \delta\lambda \int \frac{ dz d^d\mathbf{x}}{z^{d+1}}\langle \dots\hat\Phi(z,\mathbf{x})\rangle_\lambda     \\
&\ +\delta \lambda\langle \dots\rangle_{\lambda} \int \frac{ dz d^d\mathbf{x}}{z^{d+1}} \langle \hat\Phi(z,\mathbf{x})\rangle_\lambda +\mathcal{O}(\delta\lambda^2)\,,
\label{eq:corrPoint2}
\end{aligned}
\end{equation}
where we dropped the superscript on the right-hand side because these correlation functions are evaluated in the undeformed theory, where bare and renormalized operators coincide.

Let us first examine how this framework applies to the one-point function of a boundary operator. In the undeformed boundary theory, conformal invariance dictates that the one-point functions of all non-trivial primary operators vanish
\begin{equation}
\langle \mathcal{O}_i(\mathbf{x}_1)\rangle_{\lambda} =0\,. \label{eq:onePoint}
\end{equation}
When we shift the bulk coupling, the bare boundary operator must be renormalized according to \eqref{eq:barerenmatch}. Taking the expectation value of this expansion yields
\begin{equation}
\langle \mathcal{O}_i^{(\text{bare})}(\mathbf{x}_1)\rangle_{\lambda+\delta\lambda} =\sum_k\sum_{n=0}^\infty\mathcal{Z}_{ik;n}\Box^n\langle\mathcal{O}^\mathrm{(ren)}_k(\mathbf{x}_1)\rangle_{\lambda+\delta\lambda}\,.
\end{equation}
 Since the one-point functions vanish for all operators other than the identity, the only non-vanishing contribution in the sum comes from the identity operator. Using $\langle \mathbb{1} \rangle = 1$ and noting that derivatives of a constant vanish, the infinite sum collapses to a single term:
\begin{equation}
\langle \mathcal{O}_i^{(\text{bare})}(\mathbf{x}_1)\rangle_{\lambda+\delta\lambda} =\mathcal{Z}_{i\mathbb{1};0}=\delta\mathcal{Z}_{i\mathbb{1};0}\,.
\label{eq:ct_id1}
\end{equation}
Alternatively, we can compute this same shifted one-point function directly using bulk perturbation theory. By applying the perturbative expansion \eqref{eq:corrPoint2} to a single operator insertion and imposing the vanishing of the undeformed one-point function \eqref{eq:onePoint}, we are left with
\begin{equation}
\langle \mathcal{O}_i^{(\text{bare})}(\mathbf{x}_1)\rangle_{\lambda+\delta\lambda} =- \delta\lambda \int \frac{ dz d^d\mathbf{x}}{z^{d+1}} \langle \mathcal{O}_i(\mathbf{x}_1)\hat\Phi(z,\mathbf{x})\rangle_\lambda \,.
\label{eq:ct_id2}
\end{equation}
Comparing these two expressions allows us to identify the counterterm $\mathcal{Z}_{i\mathbb{1};0}$  with the integrated bulk-to-boundary correlation function. This result will be useful in the following.

In the remainder of this section, we apply this framework to compute the flow equations for the scaling dimensions and the BOE coefficients. We first focus on boundary two-point functions to extract the variation of the scaling dimensions $\Delta_i$. For scalar operators, we leave the spacetime dimension generic, while for spinning operators, we focus specifically on 3D and 4D, which require separate treatment. In both cases the result turns out to be closely related to the scalar case. Finally, we apply the framework to a bulk-to-boundary correlation function to derive the second ODE, which governs the flow of the BOE coefficients $b^{\hat{\Phi}}_i$.
\subsection{Flow of scaling dimensions}
\label{subsec:flowdeltas}
\paragraph{Scalars} We begin by examining the two-point function of scalar boundary primary operators at a finite coupling $\lambda$. Assuming the spectrum contains no degeneracies and choosing a diagonal basis for the operators, this reads 
\begin{equation}
    \langle \mathcal{O}_i(\mathbf{x}_1) \mathcal{O}_j(\mathbf{x}_2) \rangle_\lambda = \frac{\delta_{ij}}{(\mathbf{x}_{12}^2)^{\Delta_i(\lambda)}}\,.
\end{equation}
We now introduce an infinitesimal shift  $\delta\lambda$ to the bulk coupling. As a consequence, the operators renormalize following \eqref{eq:barerenmatch}.
Moreover, because we shifted the coupling in the bulk, the physical scaling dimensions of the operators generically shift by $\delta \Delta_i$. The two-point function of the renormalized operators then reads, imposing that they are canonically normalized and that they do not mix, 
\begin{equation}
    \langle \mathcal{O}^\mathrm{(ren)}_i(\mathbf{x}_1) \mathcal{O}^\mathrm{(ren)}_j(\mathbf{x}_2)\rangle_{\lambda+\delta\lambda} = \frac{\delta_{ij}}{(\mathbf{x}_{12}^2)^{\Delta_i(\lambda)+\delta\Delta_i}}\,.
\end{equation}
Combining this with the renormalization in \eqref{eq:barerenmatch}  and expanding for small deformations, we obtain
\begin{equation}
\begin{aligned}
    \langle\mathcal{O}^{(\text{bare})}_i&(\mathbf{x}_1)\mathcal{O}^{(\text{bare})}_j(\mathbf{x}_2)\rangle_{\lambda+\delta\lambda}=\delta_{ij}\left(\frac{1}{(\mathbf{x}_{12}^2)^{\Delta_i}}-\delta\Delta_i\frac{\log(\mathbf{x}_{12}^2)}{(\mathbf{x}_{12}^2)^{\Delta_i}}\right)\\
    &+\sum_{n=0}^\infty\frac{2^{2n}}{(\mathbf{x}_{12}^2)^n}\left(\frac{(\Delta_j)_n \left(\Delta_j - \frac{d-2}{2}\right)_n}{(\mathbf{x}_{12}^2)^{\Delta_j}}\delta\mathcal{Z}_{ij;n}+\frac{(\Delta_i)_n \left(\Delta_i - \frac{d-2}{2}\right)_n}{(\mathbf{x}_{12}^2)^{\Delta_i}}\delta\mathcal{Z}_{ji;n}\right)\\
    &+O(\delta\lambda^2)\,.
\end{aligned}
\label{eq:twoPoint1}
\end{equation}
On the other hand, applying \eqref{eq:corrPoint2} to this case gives
\begin{equation}
    \begin{aligned}
        \langle\mathcal{O}^{(\text{bare})}_i(\mathbf{x}_1)\mathcal{O}^{(\text{bare})}_j(\mathbf{x}_2)\rangle_{\lambda+\delta\lambda}= &\langle \mathcal{O}_i(\mathbf{x}_1) \mathcal{O}_j(\mathbf{x}_2)\rangle_{\lambda} - \delta\lambda \int \frac{ dz d^d\mathbf{x}}{z^{d+1}} \langle \mathcal{O}_i(\mathbf{x}_1) \mathcal{O}_j(\mathbf{x}_2) \hat\Phi(z,\mathbf{x})\rangle_\lambda \\&\ +\delta \lambda\langle \mathcal{O}_i(\mathbf{x}_1) \mathcal{O}_j(\mathbf{x}_2)\rangle_{\lambda} \int \frac{ dz d^d\mathbf{x}}{z^{d+1}} \langle \hat\Phi(z,\mathbf{x})\rangle_\lambda +\dots
    \end{aligned}
    \label{eq:twoPoint2}
\end{equation}
where we dropped the superscript on the right-hand side because these operators are evaluated in the undeformed theory, where bare and renormalized operators coincide.
Comparing this equation with \eqref{eq:twoPoint1}, we can extract a differential equation for the scaling dimensions $\Delta_i$, which will be the main result of this section. Before proceeding, we note that, as anticipated, the integral in \eqref{eq:twoPoint2} diverges as the bulk point approaches the boundary points $\mathbf{x}_1$ and $\mathbf{x}_2$. Crucially, two types of divergences arise, which can be regulated by introducing a cutoff $\epsilon$ near the boundary insertions. First, there are power-law divergences as $\epsilon\to0$, which are scheme-dependent and absorbable into the field-strength renormalization constants in \eqref{eq:twoPoint1}. 
Second, there are logarithmic divergences, which arise when $i=j$ and are instead universal. All these divergences must be perfectly canceled by the counterterms in \eqref{eq:twoPoint1}. Imposing this, we obtain
    \begin{equation}
        \frac{d\Delta_i}{d \lambda}=\frac{\left.\int \frac{ dz d^d\mathbf{x}}{z^{d+1}} \langle \mathcal{O}_i(\mathbf{x}_1) \mathcal{O}_i(\mathbf{x}_2) \hat\Phi(z,\mathbf{x})\rangle_\lambda\right|_\mathrm{log}}{\langle \mathcal{O}_i(\mathbf{x}_1) \mathcal{O}_i(\mathbf{x}_2) \rangle_\lambda }\,,
        \label{eq:firsteq1}
    \end{equation}
where we have denoted with the subscript $\log$ the terms proportional to $\log (\mathbf{x}_{12}^2)$.

\begin{figure}
\includegraphics[width=\linewidth]{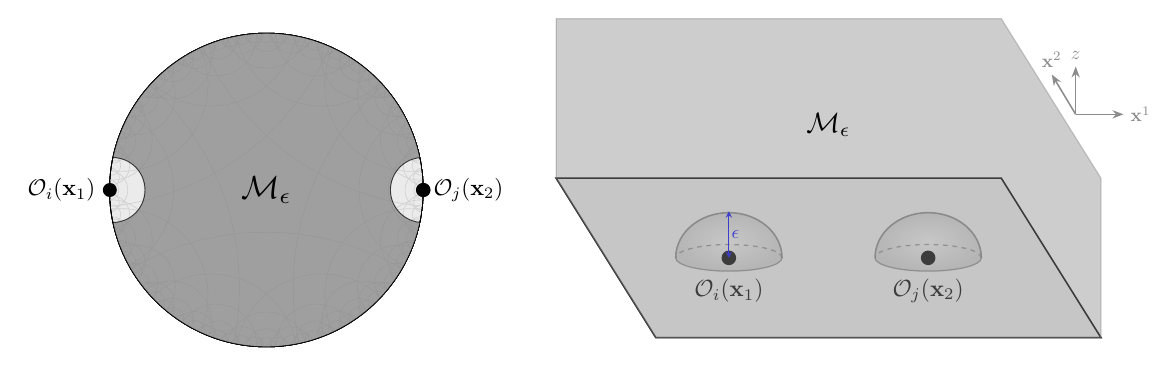}
    \caption{Domain of integration $\mathcal{M}_\epsilon$ in (\ref{eq:I_def}) in ball and half-plane coordinates. We remove the half-balls centered in the positions $\mathbf{x}_{1,2}$ of the boundary operators to regulate the IR divergences arising when the bulk point hits these points.}
    \label{fig:domain_first}
\end{figure}

We now select a regularization scheme and proceed with the computation of the first integral in  \eqref{eq:twoPoint2}. We choose to remove two half-balls centered in $\mathbf{x}_1$ and $\mathbf{x}_2$, as shown in figure~\ref{fig:domain_first}. The regularized integral reads
\begin{equation}
    I_{ij}^{\hat{\Phi}}(\mathbf{x}_1, \mathbf{x}_2, \epsilon) := \int_{\mathcal{M}_\epsilon} \frac{dz d^d\mathbf{x}}{z^{d+1}} \langle \hat{\Phi}(z, \mathbf{x}) \mathcal{O}_i(\mathbf{x}_1) \mathcal{O}_j(\mathbf{x}_2) \rangle_\lambda
    \label{eq:I_def}
\end{equation}
where $\mathcal{M}_\epsilon$ is defined by the restriction $z^2 + |\mathbf{x} - \mathbf{x}_k|^2 \ge \epsilon^2$ for $k=1,2$. The three point function inside the integral can be expressed in terms of the bulk-boundary-boundary block in equation \eqref{eq:Bbbdec}. The integral then reduces to 
\begin{equation}
    I_{ij}^{\hat{\Phi}}(\mathbf{x}_1, \mathbf{x}_2, \epsilon) = \sum_l b_l^{\hat{\Phi}}  C_{ijl} I^{\Delta_l,\Delta_i,\Delta_j}_{\text{Bbb}}(\mathbf{x}_1, \mathbf{x}_2, \epsilon)\,,
    \label{eq:I_decomp}
\end{equation}
where $I^{\Delta_l,\Delta_i,\Delta_j}_{\text{Bbb}}$ denotes the integrated conformal blocks over $\mathcal{M}_\epsilon$, which reads
\begin{equation}
    I^{\Delta_l,\Delta_i,\Delta_j}_{\text{Bbb}} := \int_{\mathcal{M}_\epsilon} \frac{dz d^d\mathbf{x}}{z^{d+1}} \frac{1}{|\mathbf{x}_{12}|^{\Delta_i+\Delta_j}} \left( \frac{|\mathbf{x}-\mathbf{x}_1|^2+z^2}{|\mathbf{x}-\mathbf{x}_2|^2+z^2} \right)^{\frac{\Delta_j-\Delta_i}{2}} G_{\text{Bbb}}^{\Delta_l,\Delta_i,\Delta_j}(\chi)\,.\label{eq:Y} 
\end{equation}
We note that in \eqref{eq:I_decomp} we exchanged the integral over the bulk point with the infinite sum of the block decomposition. Convergence of the block decomposition at every point of $\mathcal{M}_\epsilon$ is in general not sufficient for this: by the Fubini--Tonelli theorem, a sufficient condition is that the sum of the integrals of the absolute values of the summands is finite. For standard blocks, performing the integration term by term generally yields a divergent series. We will show at the end of this section that the decomposition in local blocks instead satisfies the Fubini-Tonelli condition. Moreover, as we will discuss in section \ref{sec:pade}, one can use Pad\'e approximants to resum the divergent series obtained by swapping sums and integrals of standard conformal blocks, at least in free theories. 

Let us now thus return to the computation of the integrated correlation function and postpone the issue of swapping integrals and sums over blocks to the end of this section. We first note that the sum in \eqref{eq:I_decomp} runs over all the operators in the BOE expansion of $\hat{\Phi}$, including  the identity operator ($l=\mathbb{1}$). We will now demonstrate that this specific contribution exactly cancels the disconnected piece in the second line of \eqref{eq:twoPoint2}. By isolating the identity exchange, the integrated block in \eqref{eq:I_decomp} becomes
\begin{equation}
    I_{ij}^{\hat{\Phi}}(\mathbf{x}_1, \mathbf{x}_2, \epsilon) = \langle \mathcal{O}_i(\mathbf{x}_1) \mathcal{O}_j(\mathbf{x}_2) \rangle_\lambda \int_{\mathcal{M}_\epsilon} \frac{dz d^d\mathbf{x}}{z^{d+1}} b_\mathbb{1}^{\hat{\Phi}}  + \sum_{l\ne \mathbb{1}} b_l^{\hat{\Phi}}  C_{ijl} I^{\Delta_l,\Delta_i,\Delta_j}_{\text{Bbb}}(\mathbf{x}_1, \mathbf{x}_2, \epsilon)\,,
\end{equation}
where we used the fact that the OPE coefficient is $C_{ij\mathbb{1}} =\delta_{ij}$ and the corresponding block is trivial, $G^{0,\Delta_i,\Delta_i}_{\text{Bbb}}(\chi)=1$. Recognizing that the bulk one-point function is precisely $\langle \hat{\Phi}(z, \mathbf{x}) \rangle = b_\mathbb{1}^{\hat{\Phi}}$ and removing the regulator, we find that  the first term on the right-hand side exactly matches and cancels the disconnected term in \eqref{eq:twoPoint2}. Consequently, the identity does not contribute to the shift in the correlation function\footnote{We thank Marco Meineri for pointing out to us the correct way to think of the exclusion of the identity in these sums.}. Thus from now on, for brevity, the sums over $l$ will implicitly exclude the identity. We now proceed with the computation of a generic term in the sum.
As explained in appendix \ref{app:first_eq_details}, equation \eqref{eq:Y} takes the form
\begin{equation}
   I^{\Delta_l,\Delta_i,\Delta_j}_{\text{Bbb}}(\mathbf{x}_1, \mathbf{x}_2, \epsilon) = \frac{1}{|\mathbf{x}_{12}|^{\Delta_i+\Delta_j}} \sum_{p=0}^\infty \frac{a_{2p}(\Delta_{ij}, \Delta_l)}{\Delta_{ij} - 2p} \left( \frac{\epsilon}{|\mathbf{x}_{12}|} \right)^{2p-\Delta_{ij}} + (\Delta_i \leftrightarrow \Delta_j)\,.
\end{equation}
The coefficients $a_{2p}(\Delta_{ij}, \Delta_l)$ are computed in closed form in appendix~\ref{app:first_eq_details}, see \eqref{eq:a2p}. For $p=0$, they read
\begin{equation}
    a_0(\Delta_{ij}, \Delta_l)=\frac{{\pi}^\frac{d}{2}\ \Gamma\left(\frac{\Delta_l-d}{2}\right)}{\Gamma\left(\frac{\Delta_l}{2}\right)} {}_3F_2\left( \frac{\Delta_{lij}}{2}, \frac{\Delta_{lji}}{2}, \frac{\Delta_l-d}{2} ; \Delta_l +1- \frac{d}{2}, \frac{\Delta_l}{2} ; 1 \right)\,.
\end{equation}
Let us now replace the result for $I^{\Delta_l,\Delta_i,\Delta_j}_{\text{Bbb}}$ in \eqref{eq:I_decomp} and then in \eqref{eq:twoPoint2}, and compare it with  \eqref{eq:twoPoint1}. We begin by considering the case $\Delta_i\ne\Delta_j$. We see that matching term by term in $\mathbf{x}_{12}$ we find the explicit expression for the counterterms in this regularization scheme, which reads
\begin{equation}
    \delta \mathcal{Z}_{ij;p} = - \frac{\delta\lambda\, \epsilon^{2p-\Delta_{ij}}}{2^{2p}(\Delta_j)_{p}(\Delta_j-\frac{d-2}{2})_{p}} \sum_l b_l^{\hat{\Phi}} \frac{C_{jil} a_{2p}(\Delta_{ij}, \Delta_l) }{\Delta_{ij} - 2p} \quad \text{for} \quad i \neq j
    \label{eq:counterterm}
\end{equation}
Eq.~\eqref{eq:counterterm} appears singular when $\Delta_{ij}=2p$ with $p\geq1$, i.e.\ when $\mathcal{O}_i$ has the same scaling dimension as the descendant $\Box^p\mathcal{O}_j$; such resonances are not excluded by the assumption of a non-degenerate spectrum. A pole at $\Delta_{ij}=2p$ would be the analytic continuation of a term proportional to $\log(\epsilon/|\mathbf{x}_{12}|)/(\mathbf{x}_{12}^2)^{\Delta_i}$ in the bare two-point function. Such a term cannot be matched: the renormalized side \eqref{eq:twoPoint1} contains no $\log|\mathbf{x}_{12}|$ for $i\neq j$, and the counterterms, which multiply pure powers of $|\mathbf{x}_{12}|$, cannot generate one. Indeed, the pole is absent term by term: $a_{2p}(2p,\Delta_l)=0$ for every $\Delta_l$, see appendix \ref{app:first_eq_details}. Each term in \eqref{eq:counterterm} thus has a finite limit at the resonance, which is scheme dependent and drops out of the flow equations, as all counterterms do.

Now consider the case $\Delta_i=\Delta_j$, which instead contains the universal logarithmic divergence. Isolating the $p=0$ term, as higher-order terms vanish in the limit $\epsilon \to 0$, and taking the limit $\Delta_i\rightarrow\Delta_j$ the integrated block becomes
\begin{equation}
    I^{\Delta_l,\Delta_i,\Delta_i}_{\text{Bbb}}(\mathbf{x}_1, \mathbf{x}_2, \epsilon)= \frac{ a_0(0, \Delta_l) }{|\mathbf{x}_{12}|^{2\Delta_i}} \log\left(\frac{\mathbf{x}_{12}^2}{\epsilon^2} \right)\,,
\end{equation}
where we used that 
\begin{equation}
    \left. \frac{d a_0(\Delta_{ij}, \Delta_l)}{d\Delta_{ij}} \right|_{\Delta_{ij}=0} = 0\,.
\end{equation}
Inserting this into the bare two-point function in \eqref{eq:twoPoint2} and comparing again with \eqref{eq:twoPoint1}, we obtain the following matching condition
\begin{equation}
    - \delta\lambda \sum_l b_l^{\hat{\Phi}}  C_{iil}I^{\Delta_l,\Delta_i,\Delta_i}_{\text{Bbb}}(\mathbf{x}_1, \mathbf{x}_2, \epsilon) = \frac{1}{(\mathbf{x}_{12}^2)^{\Delta_i}}\left(2\delta\mathcal{Z}_{ii;0} - \delta \Delta_i \log(\mathbf{x}_{12}^2)\right)
\end{equation}
This equation is satisfied by imposing 
\begin{align}
    &\delta\mathcal{Z}_{ii;0} =  \delta\lambda \sum_l b_l^{\hat{\Phi}} C_{iil} a_0(0, \Delta_l) \log \epsilon\,, \\
    &\delta \Delta_i = \delta\lambda \sum_l b_l^{\hat{\Phi}} C_{iil} a_0(0, \Delta_l)\,.
\end{align}
The second relation gives the main result of this section, which reads
\begin{mdframed}[backgroundcolor=shadecolor,linewidth=0pt]
\begin{equation}
    \frac{d\Delta_i}{d\lambda} =  \sum_l b_l^{\hat{\Phi}} C_{iil} \mathcal{I}(\Delta_l)\,,
    \label{eq:first_eq_scalar}
\end{equation}
\end{mdframed}
where
\begin{equation}
    \mathcal{I}(\Delta_l)\equiv a_0(0, \Delta_l) = \frac{2 \pi ^{d/2} \Gamma (\Delta_l-\frac{d-2}{2})}{\Gamma (\frac{\Delta_l}{2})\Gamma (\frac{\Delta_l+2}{2}) (\Delta_l-d)}\,.
    \label{eq:IDelta}
\end{equation}

\paragraph{AdS$_3$} In AdS$_3$, the analysis generalizes easily to the case of spinning operators. Following the same logic as the scalar case, we can compute the flow equation for the scaling dimension of operators with generic spin, which reads 
\begin{equation}
        \frac{d\Delta_i}{d \lambda}=\frac{\left.\int \frac{{dz d\mathtt{z} d\mathtt{\bar{z}}}}{2z^3}\langle \mathcal{O}_i(\mathtt{z}_1,\mathtt{\bar{z}}_1) \mathcal{O}_i(\mathtt{z}_2,\mathtt{\bar{z}}_2) \hat\Phi(z, \mathtt{z},\mathtt{\bar{z}})\rangle_\lambda\right|_\mathrm{log}}{\langle \mathcal{O}_i(\mathtt{z}_1,\mathtt{\bar{z}}_1) \mathcal{O}_i(\mathtt{z}_2,\mathtt{\bar{z}}_2)\rangle_\lambda }\,,
        \label{eq:firsteq12d}
    \end{equation}
    where now the subscript log refers to terms proportional to $\log|\mathtt{z}_{12}|^2$. Looking at the two point function in \eqref{eq:2ptcomplexbis} and at the bulk-boundary-boundary block in \eqref{eq:ansatzblock2}, we see that the overall spin structure simplifies and we get 
    \begin{equation}
        \frac{d\Delta_i}{d \lambda}=\sum_lb^{\hat\Phi}_lC_{iil} \ \left.\int_{\mathcal{M}_\epsilon} d X\ G_{\text{Bbb}}^{\Delta_l,\Delta_i,\Delta_i}(\chi)\right|_{\mathrm{log}}\,,
        \label{eq:first_eq_3}
    \end{equation}
    where the identity is again excluded from the sum. The integral over AdS is again generically divergent and needs to be regulated with some cutoff $\epsilon$.  Adopting the same regularization scheme that we used for the scalar case, we identify the integral in \eqref{eq:first_eq_3} with  $I^{\Delta_l,\Delta_i,\Delta_i}_{\text{Bbb}}$ in \eqref{eq:Y}, when the limit $\Delta_i\rightarrow\Delta_j$ is taken. Using the results of the previous section, the flow equation for the scaling dimension then reads
     \begin{equation}
        \frac{d\Delta_i}{d \lambda}=\sum_lb^{\hat\Phi}_lC_{iil}\mathcal{I}(\Delta_l)
    \end{equation}
with $\mathcal{I}(\Delta_l)$ the same as in (\ref{eq:IDelta}) with $d=2$.
\paragraph{AdS$_4$}  In AdS$_4$, the case of spinning operators is treated by generalizing equation \eqref{eq:firsteq1} into
\begin{equation}
    \frac{d\Delta_i}{d \lambda}=\frac{\left.\int_X\ \langle \mathcal{O}^{(J)}_i(P_1,Z_1) \mathcal{O}^{(J)}_i(P_2,Z_2)\Phi(X)\rangle\right|_\mathrm{log}}{\langle \mathcal{O}^{(J)}_i(P_1,Z_1) \mathcal{O}^{(J)}_i(P_2,Z_2)  \rangle_\lambda }\,,
\end{equation}
where for convenience we used embedding coordinates. The denominator is given by the two-point function in equation \eqref{eq:twoPointJ}, while the numerator can be expressed in terms of the bulk-boundary-boundary block in equation \eqref{eq:BbbJ}. The latter then becomes
\begin{equation}
    \int_X\frac{(H_{1,2})^J}{P_{12}^{\Delta_i+J}}\ \left.\sum_lb_l^{\hat\Phi}\sum_{n=0}^{J}C^{(n)}_{iil}\mathcal{H}_n(v)G^{\Delta_l,n} _{\text{Bbb}}(\chi)\right|_\mathrm{log}\,,
    \label{eq:firsteqdenJ}
\end{equation}
where the sum over $l$ again excludes the identity operator.
It seems that, in principle, we get contributions from each tensor structure $\mathcal{H}_n(v)$. However, crucially, the only relevant contribution comes from the $n=0$ term, as all other structures vanish when the integral over $X$ is computed. A simple way to see this is by going to the special configuration in which we place one boundary operator at the origin in Poincar\'e coordinates and the other at infinity, while the bulk point remains generic. This corresponds to 
\begin{equation}
\begin{aligned}
    P_1&=\left(\frac{1}{2},\mathbf{0},\frac{1}{2}\right),\quad P_2=\left(\frac{1}{2},\mathbf{0},-\frac{1}{2}\right),\quad Z_1=(0,\mathbf{z}_1,0),\quad Z_2=(0,\mathbf{z}_2,0)\\
    X&=\left(\frac{1+z^2+\mathbf{x}^2}{2z},\frac{\mathbf{x}}{z},\frac{1-z^2-\mathbf{x}^2}{2z}\right)
\end{aligned}
\end{equation}
With this choice of external points, equation \eqref{eq:firsteqdenJ} becomes
\begin{equation}
    \int\frac{  dz d^3\mathbf{x}}{z^4}(\mathbf{z}_1\cdot\mathbf{z}_2)^J\sum_lb_l^{\hat\Phi}\sum_{n=0}^{J}C^{(n)}_{iil}\mathcal{H}_n\left(\frac{(\mathbf{z}_1\cdot\mathbf{x})(\mathbf{z}_2\cdot\mathbf{x})}{\mathbf{x}^2(\mathbf{z}_1\cdot\mathbf{z}_2)}\right)G^{\Delta_l,n} _{\text{Bbb}}\left(\frac{z^2}{\mathbf{x}^2+z^2}\right)\Bigg|_{\text{log}}\,.
\end{equation}
In this configuration, all the terms in the expansion have the property of being traceless with respect to the full cross-trace, except for the $n=0$ term:
\begin{equation}
    (D_{\mathbf{z}_1}\cdot D_{\mathbf{z}_2})^J\left((\mathbf{z}_1\cdot\mathbf{z}_2)^J\mathcal{H}_n\left(\frac{(\mathbf{z}_1\cdot\mathbf{x})(\mathbf{z}_2\cdot\mathbf{x})}{\mathbf{x}^2(\mathbf{z}_1\cdot\mathbf{z}_2)}\right)\right)=0\,,\quad n\ne 0\,.
\end{equation}
This implies that, once indices are opened, each of these terms is proportional to one or more terms in the form
\begin{equation}
    \left( \delta^{a_1 a_2}-d \frac{\mathbf{x}^{a_1}\mathbf{x}^{a_2}}{\mathbf{x}^2}\right)\,,
\end{equation}
which give vanishing contribution when integrated over $\mathbf{x}$ by $SO(d)$ invariance\footnote{The regulator we chose, see Figure \ref{fig:domain_first}, preserves $SO(d)$, hence the above statement remains true in the regulated domain.}. Then, we deduce that the only non-vanishing contribution after integration comes from the $n=0$ term. This property clearly needs to be true in any configuration.\footnote{We explicitly checked that this is true in a generic frame by performing the integral in embedding coordinates in the cases of $J=1$, $J=2$ and $J=3$.} equation \eqref{eq:firsteqdenJ} then reduces to
\begin{equation}
    \frac{(H_{1,2})^J}{P_{12}^{\Delta_i+J}}\left.\sum_lb_l^{\hat\Phi}C^{(0)}_{iil}\ \int_XG^{\Delta_l,\Delta_i,\Delta_i} _{\text{Bbb}}(\chi)\right|_\mathrm{log}\,,
\end{equation}
where we used
\begin{equation}
    \mathcal{H}_0(v)=1\,,\quad  G^{\Delta_l,0} _{\text{Bbb}}(\chi)=G^{\Delta_l,\Delta_i,\Delta_i} _{\text{Bbb}}(\chi)\,,
\end{equation}
where $G^{\Delta_l,\Delta_i,\Delta_i} _{\text{Bbb}}(\chi)$ is the scalar bulk-boundary-boundary block. The tensor structure then simplifies with that of the two-point function in the denominator, and we recover again the integral in $I^{\Delta_l,\Delta_i,\Delta_i}_{\text{Bbb}}$. This implies 
\begin{equation}
    \frac{d\Delta_i}{d \lambda}=\sum_lb^{\hat\Phi}_lC^{(0)}_{iil} \ \mathcal{I}(\Delta_l)\,,
    \label{eq:firstflowspinj}
\end{equation}
with $\mathcal{I}(\Delta)$ given in equation \eqref{eq:IDelta}.

\paragraph{The Fubini-Tonelli theorem and local blocks} As anticipated, the sum in the flow equation with integrated standard conformal blocks generically diverges due to the illegitimate swapping of the sum over blocks and the integral over AdS. Here we argue that, as reported in \cite{Loparco:2026fki}, local blocks solve this issue.  We focus on the case $i=j$, which computes the variation of the dimension, and assume the $i\neq j$ case, which computes counterterms, works by similar arguments. We start from (\ref{eq:firsteq1}) and define
\begin{equation}
    I\equiv\int_{\mathcal{M}_\epsilon}\frac{dzd^d\mathbf{x}}{z^{d+1}}\frac{\langle\mathcal{O}_i(\mathbf{x}_1)\mathcal{O}_i(\mathbf{x}_2)\hat\Phi(z,\mathbf{x})\rangle}{\langle\mathcal{O}_i(\mathbf{x}_1)\mathcal{O}_i(\mathbf{x}_2)\rangle}= \int_{\mathcal{M}_\epsilon}\frac{dzd^d\mathbf{x}}{z^{d+1}}\mathcal{G}_{\hat\Phi ii}(\chi)\,,
\end{equation}
where, in the Poincar\'e frame $\mathbf{x}_1=\mathbf{0}$ and $|\mathbf{x}_2|=\infty$, $\chi=\frac{z^2}{z^2+\mathbf{x}^2}$ and $\mathcal{G}_{\hat\Phi ii}$ is defined in (\ref{eq:genframeBbb}). If we go to polar coordinates centered around $\mathbf{x}_1$
\begin{equation}
    z=r\cos\theta\,,\qquad |\mathbf{x}|=r\sin\theta\,,\qquad \hat{\mathbf{x}}\in S^{d-1}\,,\qquad\theta\in[0,\frac{\pi}{2})\,,
\end{equation}
we have that $\chi=\cos^2\theta$, and we can carry out the rest of the integrals, which only involve the measure. Based on the discussion in the previous parts of this section, we can retrieve the flow equation for the scaling dimension by extracting the coefficient of $-\log\epsilon^2$. We get
\begin{equation}
    \frac{d\Delta_i}{d\lambda}=\frac{\pi^\frac{d}{2}}{\Gamma(\frac{d}{2})}\int_0^1d\chi\ \frac{(1-\chi)^\frac{d-2}{2}}{\chi^\frac{d+2}{2}}\mathcal{G}_{\hat\Phi ii}(\chi)\,.
\end{equation}
Notice that now the right hand side is a finite integral since the reduced correlator behaves as $\mathcal{G}\stackrel{\chi\to0}{\sim}\chi^\frac{\Delta_1}{2}$ with $\Delta_1>d$ being the scaling dimension of the lightest operator in the BOE of $\hat\Phi$, and it is regular at $\chi=1$ as discussed in section \ref{sec:local_blocks}.

Now we want to decompose the reduced correlator in blocks with (\ref{eq:Bbbalphadec}) and swap the sum and integral. The Fubini-Tonelli theorem ensures that a sufficient condition for this move to be allowed is 
\begin{equation}
    \sum_l|b^{\hat\Phi}_lC_{lii}|\frac{\pi^\frac{d}{2}}{\Gamma(\frac{d}{2})}\int_0^1d\chi\ \frac{(1-\chi)^\frac{d-2}{2}}{\chi^\frac{d+2}{2}}|G_{\text{Bbb}}^{\Delta_l,\Delta_i,\Delta_i,\alpha}(\chi)|<\infty
\end{equation}
The rest of this section is devoted to proving that this condition is verified for a certain regime of $\alpha$.

First of all, we observe that local blocks have a constant sign in $\chi\in(0,1)\,.$ We prove this fact in Appendix \ref{subsec:signlocal}.  We can thus bring the absolute value out of the integral and find the equivalent condition
\begin{equation}
    \sum_l|b^{\hat\Phi}_lC_{lii}|\frac{\pi^\frac{d}{2}}{\Gamma(\frac{d}{2})}\left|\int_0^1d\chi\ \frac{(1-\chi)^\frac{d-2}{2}}{\chi^\frac{d+2}{2}}G_{\text{Bbb}}^{\Delta_l,\Delta_i,\Delta_i,\alpha}(\chi)\right|<\infty\,.
\end{equation}
The integral can now be carried out in total analogy to what is done in Appendix \ref{app:first_eq_details} by expanding the hypergeometric function, integrating term by term, and resumming. The result is 
\begin{align}
    \frac{\pi^\frac{d}{2}}{\Gamma(\frac{d}{2})}\int_0^1d\chi\ \frac{(1-\chi)^\frac{d-2}{2}}{\chi^\frac{d+2}{2}}G_{\text{Bbb}}^{\Delta_l,\Delta_i,\Delta_i,\alpha}(\chi)&=\frac{\pi^\frac{d}{2}\Delta_l\Gamma(\alpha)\Gamma(\alpha-\frac{d}{2})\Gamma(\Delta_l-\frac{d-2}{2})}{(\Delta_l-d)\Gamma(\alpha-\frac{\Delta_l}{2})\Gamma(\frac{\Delta_l}{2}+1)^2\Gamma(\alpha-\frac{d-\Delta_l}{2})}\nonumber\\
    &\equiv\mathcal{I}^\alpha(\Delta_l)\,.
    \label{eq:IDeltaalpha}
\end{align}
The sufficient condition for the Fubini-Tonelli theorem thus becomes
\begin{equation}
    \sum_l|b^{\hat\Phi}_lC_{lii}\mathcal{I}^\alpha(\Delta_l)|<\infty\,.
\end{equation}
The large $\Delta_l$ regime of the integrated local block can be studied explicitly, and is
\begin{equation}
    \mathcal{I}^\alpha(\Delta)=c_{\alpha,d}\sin(\pi(\alpha-\Delta/2))2^{\Delta}\Delta^{\frac{1}{2}-2\alpha}\left(1+O(\Delta^{-1})\right)
\end{equation}
hence a sufficient condition is 
\begin{equation}
    \sum_l|b^{\hat\Phi}_lC_{lii}|2^{\Delta_l}\Delta_l^{\frac{1}{2}-2\alpha}<\infty\,.
\end{equation}
To make progress, we study the tail of the sum above a truncation $\Delta_{l_{\text{max}}}$ and use the Cauchy-Schwarz inequality to put a bound on it in terms of sums of BOE and OPE coefficients separately
\begin{equation}
    \sum_{l>l_{\text{max}}}|b^{\hat\Phi}_lC_{lii}|2^{\Delta_l}\Delta_l^{\frac{1}{2}-2\alpha}\leq\sqrt{\left(\sum_{l>l_{\text{max}}}\left(b^{\hat\Phi}_l\right)^2\Delta_l^{-\alpha_1}\right)\left(\sum_{l>l_\text{max}}C_{lii}^24^{\Delta_l}\Delta_l^{-\alpha_2}\right)}\,,
\end{equation}
where $\alpha_1$ and $\alpha_2$ are constants obeying $\alpha_1+\alpha_2=4\alpha-1$. We are almost done since we can use the bounds on summed OPE and BOE coefficients discussed in Appendix \ref{subsec:weightedconvergence}, which state that there exist constants $K_b$, $K_C$ and $\Delta_0$ for which, for every $\Delta_{l_{\text{max}}}>\Delta_0$,
\begin{equation}
    \sum_{l>l_{\text{max}}}\left(b^{\hat\Phi}_l\right)^2\Delta_l^{-\alpha_1}\leq K_b\Delta_{l_{\text{max}}}^{2\Delta^{\text{UV}}_{\hat\Phi}-\alpha_1}\,,\qquad \sum_{l>l_{\text{max}}}C_{lii}^24^{\Delta_l}\Delta_l^{-\alpha_2}\leq K_C\Delta_{l_{\text{max}}}^{4\Delta_i-\alpha_2}\,,
\end{equation}
provided that $\alpha_1>2\Delta^{\text{UV}}_{\hat\Phi}$ and $\alpha_2>4\Delta_i$. Putting all the pieces together, the Fubini-Tonelli condition is satisfied if
\begin{equation}
    \alpha>\Delta_i+\frac{\Delta_{\hat\Phi}^{\text{UV}}}{2}+\frac{1}{4}\,,
\end{equation}
and for such a range of $\alpha$, the summation and integration of the local blocks can be swapped in the derivation of the first flow equation, leading to an absolutely convergent sum:
\begin{equation}
    \frac{d\Delta_i}{d\lambda}=\sum_lb^{\hat\Phi}_lC_{lii}\mathcal{I}^\alpha(\Delta_l)
\end{equation}
with $\mathcal{I}^\alpha(\Delta_l)$ given in (\ref{eq:IDeltaalpha}). We expect the same to be true for the spinning case (\ref{eq:firstflowspinj}) by substituting $\mathcal{I}(\Delta_l)\to\mathcal{I}^\alpha(\Delta_l)$. We verify numerically that the sums are convergent in free theories with spinning boundary operators in section \ref{sec:checks}.

\subsection{Flow of BOE coefficients}
\begin{figure}
    \centering
    \includegraphics[width=\linewidth]{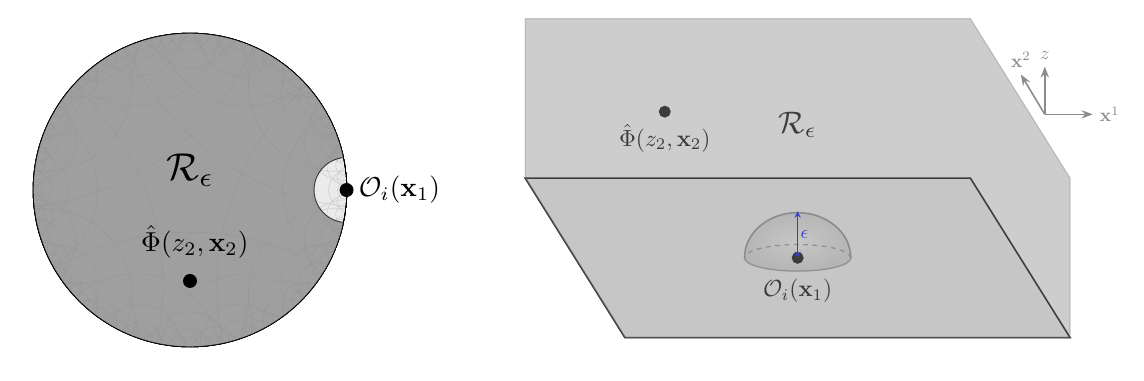}
    \caption{The domain of integration $\mathcal{R}_\epsilon$ in (\ref{eq:IBBb_def}) in ball and half-plane coordinates. We remove the half-ball centered at the position $\mathbf{x}_1$ of the boundary operator to regulate the IR divergence arising when the integrated bulk field hits this point. There is no UV divergence when the integration point hits the other bulk insertion when assumption (\ref{eq:nobulkuv}) is satisfied. We discuss how to move past this assumption in section \ref{sec:discussion}}
    \label{fig:Repsilondef}
\end{figure}

Let us now examine the bulk-boundary two-point function 
\begin{equation}
    \langle \mathcal{O}_i(\mathbf{x}_1) \hat{\Phi}(\mathbf{x}_2, z_2) \rangle_\lambda = b_i^{\hat{\Phi}}(\lambda) \left( \frac{z_2}{\mathbf{x}_{12}^2  + z_2^2} \right)^{\Delta_i(\lambda)}\,,
\end{equation}
to extract the flow equation for the BOE coefficients $b_i^{\hat{\Phi}}$. 
We only restrict to the case in which the bulk operator is the deforming operator $\hat{\Phi}$ because this is sufficient to obtain a closed system of differential equations describing the RG flow. Now, following the logic of the previous section, we introduce an infinitesimal shift $\delta\lambda$ to the bulk coupling. Under this shift, both the BOE coefficient and the dimension of the boundary operator get shifted. Moreover, while both bulk and boundary operators generically renormalize, we focus here on the scenario where only the boundary operators do so. Below, we outline the conditions required for this realization, postponing the analysis of the general case to section \ref{subsubsec:bulkrelevants}. The shift then gives
\begin{equation}
\begin{aligned}
    \langle \mathcal{O}^\mathrm{(bare)}_i(\mathbf{x}_1)\hat{\Phi}( z_2,\mathbf{x}_2)  \rangle_{\lambda+\delta\lambda}  &= \sum_{j} \sum_{n=0}^\infty \mathcal{Z}_{ij;n}\Box_{\mathbf{x}_1}^n \langle \mathcal{O}^\mathrm{(ren)}_j(\mathbf{x}_1)\hat{\Phi}( z_2,\mathbf{x}_2) \rangle_{\lambda+\delta\lambda} \\
    &= \left(\frac{z_2}{\mathbf{x}_{12}^2+z_2^2}\right)^{\Delta_i(\lambda)} \left[b_i^{\hat{\Phi}} + b_i^{\hat{\Phi}}\delta\Delta_i\log\left(\frac{z_2}{\mathbf{x}_{12}^2+z_2^2}\right) + \delta b_i^{\hat{\Phi}}\right] \\
    &\quad + \sum_{j} \sum_{n=0}^\infty b_j^{\hat{\Phi}}\delta\mathcal{Z}_{ij;n}\Box_{\mathbf{x}_1}^n \left(\frac{z_2}{\mathbf{x}_{12}^2+z_2^2}\right)^{\Delta_j(\lambda)} \,.
    \label{eq:Bbprop1}
\end{aligned}
\end{equation}
On the other side, we can express the change in the bulk-boundary two-point function in perturbation theory as
\begin{equation}
    \begin{aligned}
    \langle\mathcal{O}^\mathrm{(bare)}_i(\mathbf{x}_1)  \hat{\Phi}( z_2,\mathbf{x}_2) \rangle_{\lambda+\delta\lambda}  &=\langle   \mathcal{O}_i(\mathbf{x}_1)\hat{\Phi}(z_2,\mathbf{x}_2)\rangle_{\lambda} - \delta\lambda \int \frac{ dz d^d\mathbf{x}}{z^{d+1}} \langle\mathcal{O}_i(\mathbf{x}_1) \hat{\Phi}(z_2,\mathbf{x}_2)  \hat\Phi(z,\mathbf{x})\rangle_\lambda\\&\ +\delta \lambda\langle \mathcal{O}_i(\mathbf{x}_1)  \hat{\Phi}( z_2,\mathbf{x}_2) \rangle_{\lambda} \int \frac{ dz d^d\mathbf{x}}{z^{d+1}} \langle \hat\Phi(z,\mathbf{x})\rangle_\lambda +\dots\,.
    \label{eq:bulkbulkint1}
    \end{aligned}
\end{equation}
The integral in the first line generically diverges as the integrated bulk point approaches the boundary point $\mathbf{x}_1$. Similarly to what we did in the previous section, we then regularize the integral by removing the half-ball around $\mathbf{x}_1$, which gives
\begin{equation}
    I_{i}^{\hat{\Phi}\hat{\Phi}}(\mathbf{x}_1,\mathbf{x}_2, z_2, \epsilon) := \int_{\mathcal{R}_\epsilon} \frac{dz d^d\mathbf{x}}{z^{d+1}} \langle\mathcal{O}_i(\mathbf{x}_1)\hat{\Phi}(z_2, \mathbf{x}_2) \hat{\Phi}(z, \mathbf{x})  \rangle_{\lambda}\,.
    \label{eq:IBBb_def}
\end{equation}
where $\mathcal{R}_\epsilon$ is the region defined by $(\mathbf{x}-\mathbf{x}_1)^2+z^2>\epsilon^2$, represented in figure \ref{fig:Repsilondef}.

As anticipated, this is not the only divergence of this integral. Indeed, the integral might also diverge when the integrated point approaches the other bulk point $(z_2,\mathbf{x}_2)$. This limit is controlled by the UV bulk OPE:
\begin{equation}
    \hat\Phi(z,\mathbf{x})\hat\Phi(z_2,\mathbf{x}_2)\approx\sum_{\hat{\mathcal{O}}}C^{\text{UV}}_{\hat{\mathcal{O}}\hat\Phi\hat\Phi}\sum_{n=0}^\infty c_n\left(\frac{zz_2}{(z-z_2)^2+(\mathbf{x}-\mathbf{x}_2)^2}\right)^{\Delta_{\hat\Phi}^{\text{UV}}-\frac{\Delta^{\text{UV}}_{\hat{\mathcal{O}}}}{2}-n}\hat{\mathcal{O}}(z_2,\mathbf{x}_2)\,,
\end{equation}
where $\Delta_{\hat{\mathcal{O}}}^{\text{UV}}$ and $C^{\text{UV}}_{\hat{\mathcal{O}}\hat\Phi\hat\Phi}$ are the CFT data of the bulk UV theory, and $n$ labels curvature corrections to the flat space OPE. The sum over $\hat{\mathcal{O}}$ is taken to run over both primaries and descendants, but we focus on scalars for simplicity. The contribution of spinning operators in the bulk OPE would be killed by choosing a rotationally invariant regulator around $(z_2,\mathbf{x}_2)$. Notice that the contribution of the identity is zero because $\langle\mathcal{O}_i\rangle=0$.

Now, using this OPE in (\ref{eq:bulkbulkint1}), let us examine the region of integration near $(z_2,\mathbf{x}_2)$. For that purpose, we change integration variables to $r=\sqrt{(z-z_2)^2+(\mathbf{x}-\mathbf{x}_2)^2}$ and angular variables which will not be important. The integral in the region close to $(z_2,\mathbf{x}_2)$ is dominated by the contribution of the bulk operator with the minimal scaling dimension $\Delta_{\hat{\mathcal{O}}}^{\text{UV}}$ and in particular by the $n=0$ term
\begin{equation}
    \sim\int_0 r^{d-2\Delta_{\hat\Phi}^{\text{UV}}+\text{min}(\Delta_{\hat{\mathcal{O}}}^{\text{UV}})}\,.
\end{equation}
We can thus state that this integral does not suffer from bulk UV divergences if
\begin{equation}
    \Delta_{\hat\Phi}^{\text{UV}}<\frac{1}{2}\left(d+1+\min_{\hat{\mathcal{O}}\in\hat\Phi\times\hat\Phi}\Delta^{\text{UV}}_{\hat{\mathcal{O}}}\right)\,.
    \label{eq:nobulkuv}
\end{equation}
\begin{figure}
    \centering
    \includegraphics[width=0.6\linewidth]{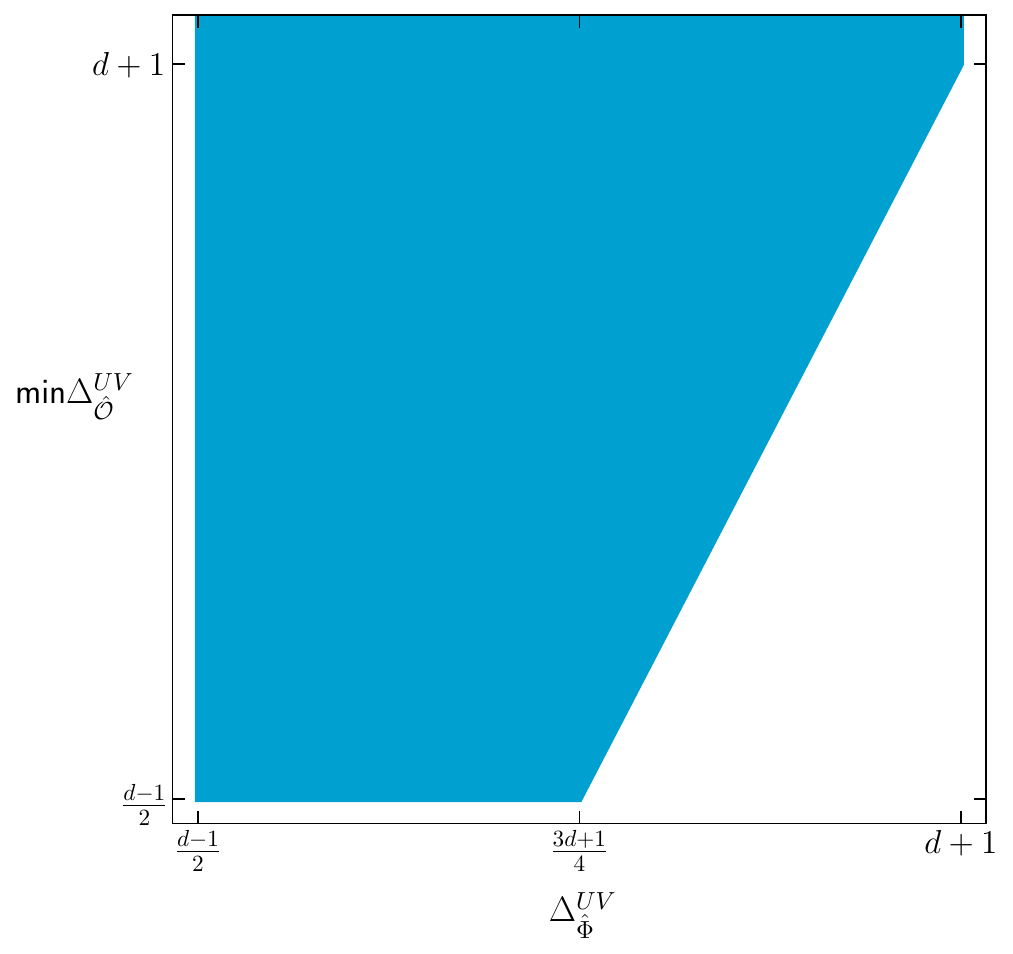}
    \caption{When the lightest bulk relevant operator of the theory appearing in the bulk OPE of the deforming operator $\hat\Phi\times\hat\Phi$ has UV scaling dimension $\Delta_{\hat{\mathcal{O}}}^{\text{UV}}$ that lies in the blue shape, there are no UV divergences in the derivation of the BOE flow equation. The lower bound $\frac{d-1}{2}$ is the UV unitarity bound. In this section, we assume our theory lies in the blue shape. We discuss how to drop this assumption in section \ref{subsubsec:bulkrelevants}.}
    \label{fig:NoBulkUV}
\end{figure}
We note that if all bulk operators in the UV other than $\hat\Phi$ are irrelevant, this inequality is satisfied. If there are some relevant operators other than $\hat\Phi$, this is a nontrivial constraint. In figure \ref{fig:NoBulkUV} we plot the values of $\Delta_{\hat\Phi}^{\text{UV}}$ and $\min(\Delta^{\text{UV}}_{\hat{\mathcal{O}}})$ for which this integral is UV finite. In the rest of this section, we will assume our QFT lives in that region. We will discuss extensions of this assumption in section \ref{subsubsec:bulkrelevants}.

Let us now go back to the integral in \eqref{eq:IBBb_def}. We begin by re-expressing the integrand in terms of the bulk-bulk-boundary block introduced in \eqref{eq:BBb3point}. We then obtain
\begin{equation}
 I_{i}^{\hat{\Phi}\hat{\Phi}}(\mathbf{x}_1,\mathbf{x}_2, z_2, \epsilon)=\left(\frac{z_2}{\mathbf{x}_{12}^2+z_2^2}\right)^{\Delta_i} \int_{\mathcal{R}_\epsilon} \frac{dz d^d\mathbf{x}}{z^{d+1}}\sum_{j,l}b^{\hat\Phi}_l b^{\hat\Phi}_jC_{lij}G_{\text{BBb}}^{\Delta_l,\Delta_j,\Delta_i}(\xi,\rho)\,.
\end{equation}
We will now exchange summation and integration, as done for the scaling dimension flow equation. This is not justified when using standard conformal blocks, for the reasons established in section \ref{sec:local_blocks}. It also is not justified using the currently available local blocks in the bulk-bulk-boundary case, as they render convergent only one of the two BOE sums. We nevertheless shall perform the exchange and we will show afterwards that, at least in free theory, the sequence of Pad\'e approximants of the partial sums is convergent. Proceeding, we thus get
\begin{equation}
    I_{i}^{\hat{\Phi}\hat{\Phi}}(\mathbf{x}_1,\mathbf{x}_2, z_2, \epsilon)=\left(\frac{z_2}{\mathbf{x}_{12}^2+z_2^2}\right)^{\Delta_i} \sum_{j,l}b^{\hat\Phi}_l b^{\hat\Phi}_jC_{lij}I_\mathrm{BBb}^{\Delta_l,\Delta_j,\Delta_i}(\mathbf{x}_1,\mathbf{x}_2, z_2, \epsilon)\,,
    \label{eq:I_second_eq}
\end{equation}
where we have introduced the bulk-bulk-boundary integrated block
\begin{equation}
    I_\mathrm{BBb}^{\Delta_l,\Delta_j,\Delta_i}(\mathbf{x}_1,\mathbf{x}_2, z_2, \epsilon)=\int_{\mathcal{R}_\epsilon} \frac{dz d^d\mathbf{x}}{z^{d+1}}G_{\text{BBb}}^{\Delta_l,\Delta_j,\Delta_i}(\xi,\rho)\,.
    \label{eq:IBBbnew}
\end{equation}
As before, the sums in \eqref{eq:I_second_eq} include contributions from the identity operator. However, because two bulk operators are now expanded to the boundary, there are two distinct identity contributions to consider. The contribution arising from the BOE of the integrated operator $\hat{\Phi}(\mathbf{x},z)$ cancels exactly against the disconnected piece in \eqref{eq:bulkbulkint1}, following the same mechanism described in the previous section. Conversely, the identity contribution arising from the expansion of the unintegrated operator $\hat{\Phi}(\mathbf{x}_2,z_2)$ evaluates to
\begin{equation}
    I_{i}^{\hat{\Phi}\hat{\Phi}}(\mathbf{x}_1,\mathbf{x}_2, z_2, \epsilon)\Big|_\text{id.} = \int_{\mathcal{R}_\epsilon} \frac{dz d^d\mathbf{x}}{z^{d+1}} b^{\hat{\Phi}}_\mathbb{1} \langle \mathcal{O}_i(\mathbf{x}_1)\hat{\Phi}(\mathbf{x},z)\rangle_\lambda\,.
\end{equation}
Using \eqref{eq:ct_id1} and \eqref{eq:ct_id2} and taking $\epsilon$ to zero, we can directly identify this integral with the identity counterterm:
\begin{equation}
    I_{i}^{\hat{\Phi}\hat{\Phi}}(\mathbf{x}_1,\mathbf{x}_2, z_2, \epsilon)\Big|_\text{id.} = -  b^{\hat{\Phi}}_\mathbb{1}\frac{\delta\mathcal{Z}_{i\mathbb{1};0}}{\delta\lambda}\,.
\end{equation}
This term exactly cancels the identity contribution present in the last line of \eqref{eq:Bbprop1}. We thus conclude that all identity contributions systematically cancel out of the variation. Consequently, we can safely exclude the identity operator from all subsequent sums.

Let us now go back to the integral in \eqref{eq:IBBbnew}. This is finite for $\Delta_j>\Delta_i$, so we can compute it explicitly in this regime and then analytically continue the result to other values of $\Delta_i,\Delta_j$. 
We refer to appendix~\ref{sec:sec_eq} for the details,  and report here the final result, which reads
\begin{equation}
\begin{aligned}
    I_{i}^{\hat{\Phi}\hat{\Phi}}(\mathbf{x}_1,\mathbf{x}_2, z_2, \epsilon)= &-\left(\frac{z_2}{\mathbf{x}_{12}^2+z_2^2}\right)^{\Delta_i} \sum_{j,l}b^{\hat\Phi}_l b^{\hat\Phi}_jC_{lij}[\mathcal{J}_{\Delta_i}(\Delta_l,\Delta_j)]_\mathrm{reg}\, \\&- b^{\hat{\Phi}}_i\left(\frac{z_2}{\mathbf{x}_{12}^2+z_2^2}\right)^{\Delta_i} \sum_l b^{\hat{\Phi}}_l C_{iil} 
\mathcal{I}( \Delta_l)  \log\left(\frac{z_2 \epsilon}{\mathbf{x}_{12}^2+z_2^2}\right)  \\&-\sum_{j\ne i} \sum_{n=0}^\infty b_j^{\hat{\Phi}}\frac{\delta\mathcal{Z}_{ij;n}}{\delta\lambda}\Box_{\mathbf{x}_1}^n \left(\frac{z_2}{\mathbf{x}_{12}^2+z_2^2}\right)^{\Delta_j(\lambda)} \,.
\label{eq:IBBb _result}
    \end{aligned}
\end{equation}
where $[\mathcal{J}_{\Delta_i}(\Delta_l,\Delta_j)]_\mathrm{reg}$ denotes the quantity
\begin{equation}
   \mathcal{J}_{\Delta_i}(\Delta_l,\Delta_j)=-\frac{\pi ^{d/2} \Gamma \left(\frac{\Delta_{ji}}{2}\right) \Gamma \left(\Delta_l-\frac{d-2}{2}\right) \Gamma \left(\Delta_j-\frac{d-2}{2}\right) \Gamma \left(\frac{\Delta_l-d}{2}\right)}{\Delta_l \Gamma \left(\frac{\Delta_l}{2}\right) \Gamma \left(\frac{\Delta_{lji}}{2}\right) \Gamma \left(\frac{\Delta_i+\Delta_j-d+2}{2} \right) \Gamma \left(\frac{\Delta_{lji}-d+2}{2} \right)}\,,
   \label{eq:JDelta}
\end{equation}
with the pole at $\Delta_j=\Delta_i$ subtracted:
\begin{equation}
    [\mathcal{J}_{\Delta_i}(\Delta_l, \Delta_j)]_{\mathrm{reg}} := \oint_{\Delta_j} \frac{d\Delta}{2\pi i} \frac{\mathcal{J}_{\Delta_i}(\Delta_l, \Delta)}{\Delta - \Delta_j} \,.
    \label{eq:regJ}
\end{equation}
From here on, we drop the subscript ``reg"  and always understand the quantity $\mathcal{J}_{\Delta_i}(\Delta_l, \Delta_j)$ to be regularized in this manner.
Replacing everything in \eqref{eq:bulkbulkint1} and comparing with \eqref{eq:Bbprop1}, we find that the counterterm contributions cancel each other out, while the variation of the scaling dimension is canceled by the sum over $l$ in the second row of \eqref{eq:IBBb _result}, by means of the first flow equation in \eqref{eq:first_eq_scalar}. The final result is a differential equation for the BOE coefficients, which reads

\begin{mdframed}[backgroundcolor=shadecolor,linewidth=0pt]
\begin{equation}
    \frac{db_i^{\hat{\Phi}}}{d\lambda} = \sum_{l,j} b_l^{\hat{\Phi}} b_j^{\hat{\Phi}} C_{ijl} \mathcal{J}_{\Delta_i}(\Delta_l, \Delta_j) \,.
    \label{eq:second_eq_final}
\end{equation}
\end{mdframed}
The equation involves only scalars, because $\hat\Phi$ is a bulk scalar.

Note that the above equation holds for all boundary operators, even for the identity $\mathcal{O}_i=\mathbb{1}$, for which the equation provides the evolution of the vacuum expectation value (vev) of $\hat\Phi$. In that case, however, the condition for no bulk divergences (\ref{eq:nobulkuv}) must be adapted. The evolution of the vev of $\hat\Phi$ in fact requires the computation of $\int_X\langle\hat\Phi(X)\hat\Phi(X_1)\rangle$. The bulk OPE of the two deforming operators contains the identity, which, in contrast to the generic case, leads to a non-vanishing contribution that goes like $\sim\int_0 dr\ r^{d-2\Delta_{\hat\Phi}^{\text{UV}}}$. The correct condition for no UV divergences in the BOE flow of $b^{\hat\Phi}_{\mathbb{1}}$ thus becomes
\begin{equation}
    2\Delta_{\hat\Phi}^{\text{UV}}<d+1\,.
\end{equation}


\paragraph{Local blocks} As for the first flow equation, we can replace standard blocks with local blocks to obtain a generalized version of the equation in \eqref{eq:second_eq_final} with better convergence properties. The result is the same, with $\mathcal{J}_{\Delta_i}(\Delta_l,\Delta_j) $ replaced with
\begin{equation}
\begin{aligned}
\mathcal{J}^{\alpha}_{\Delta_i}(\Delta_l,\Delta_j) = -&\frac{
    2 \pi^{d/2} 
    \Gamma\left(\alpha - \frac{d}{2}\right) 
    \Gamma\left(\alpha + \frac{\Delta_{ij}}{2}\right) 
    \Gamma\left(1 - \frac{d}{2} + \Delta_j\right)
}{
    \Delta_{ji}
    (\Delta_l - d) 
    \Gamma\left(\frac{2 - d + \Delta_{j} + \Delta_i}{2}\right) 
    \Gamma\left(\alpha - \frac{\Delta_l}{2}\right) 
    \Gamma\left(\frac{\Delta_{lji} }{2}\right) 
} \\
&\quad \times \frac{\Gamma\left(1 - \frac{d}{2} + \frac{\Delta_l}{2}\right) }{  \Gamma\left(\alpha -\frac{d}{2}+ \frac{\Delta_{ilj}}{2}\right)} \, {}_3F_2\left( 
    \begin{matrix} 
        \frac{\Delta_l - d}{2}, \, 
        \alpha + \frac{\Delta_l - d}{2}, \, 
        \frac{\Delta_{ilj}}{2} 
        \\ 
        \alpha + \frac{\Delta_{ilj}-d}{2}, \, 
        1 - \frac{d}{2} + \Delta_l 
    \end{matrix} 
    ;\ 1 \right) \,,
    \label{eq:Jalpha}
\end{aligned}
\end{equation}
which is meant to be regularized as explained in \eqref{eq:regJ}. These local blocks render the sum over $l$ convergent for $\alpha$ large enough, while the sum over $j$ is generically divergent. We will show in section \ref{sec:pade} that the sequence of Pad\'e approximants of the partial sums is convergent, at least in free theories. The issue of finding a method which is guaranteed to provide convergent sums in interacting theories for this flow equation remains an open problem.

\paragraph{Singularities at $\Delta_{ij}=2p$} The singularities at $\Delta_{ij}=2p$ with $p\geq1$, discussed below \eqref{eq:counterterm}, also affect the flow of BOE coefficients, where they produce poles through the factor $\Gamma(\Delta_{ji}/2)$ in \eqref{eq:JDelta}. Unlike in \eqref{eq:counterterm}, these poles are present in the individual terms, and cancel only after summing over $l$, see appendix~\ref{app:resonances}. The finite remainder, however, cannot be obtained by discarding the poles term by term, i.e.\ by applying \eqref{eq:regJ} around $\Delta_j=\Delta_i-2p$: this gives a well-defined but incorrect result. The correct prescription is to apply the regularization \eqref{eq:regJ} to a rescaled function,
\begin{equation}
    \sum_l b_l^{\hat{\Phi}}C_{ijl}\,\mathcal{J}_{\Delta_i}(\Delta_l,\Delta_j)\Big|_{\Delta_{ij}=2p}=\sum_l b_l^{\hat{\Phi}}C_{ijl}\left[\frac{\Gamma\left(\frac{\Delta_l+\Delta_{ij}}{2}\right)\Gamma\left(\frac{\Delta_l-\Delta_{ij}}{2}\right)}{\Gamma\left(\frac{\Delta_l}{2}+p\right)\Gamma\left(\frac{\Delta_l}{2}-p\right)}\,\mathcal{J}_{\Delta_i}(\Delta_l,\Delta_j)\right]_{\mathrm{reg}}
    \label{eq:Jresonant}
\end{equation}
with the contour in \eqref{eq:regJ} now encircling $\Delta_j=\Delta_i-2p$. The rescaling factor equals one at $\Delta_{ij}=2p$, but its derivative there multiplies the residue of each term, and thus shifts its finite part. Since this derivative depends on $\Delta_l$, these shifts do not cancel in the sum over $l$, unlike the residues themselves.  Their role is to compensate for the violation of the locality sum rules that the naive continuation introduces away from $\Delta_{ij}=2p$. We derive \eqref{eq:Jresonant} in appendix~\ref{app:resonances}.

This procedure can be avoided altogether by working with local blocks: since $\mathcal{J}^{\alpha}_{\Delta_i}$ in \eqref{eq:Jalpha} has no pole at $\Delta_j=\Delta_i-2p$, it can be evaluated there directly.

Finally, we note that such singularities are ubiquitous in free theories, where for instance $\Delta_{[\phi^2]_n}-\Delta_{[\phi^2]_m}=2(n-m)$ for all $\lambda$, whereas in interacting theories  they generically occur only at isolated values of $\lambda$.
\subsection{Closing the system with crossing}
\label{subsec:crossing}
In order to have a closed set of equations, \cite{Loparco:2026fki} also derived a flow equation for $\frac{dC_{ijk}}{d\lambda}$ in AdS$_2$ by studying how the three-point function of boundary operators varies under a change in coupling of the bulk relevant deformation. The result depended on a kinematic function $\mathcal{K}$ which is the regulated integral of a bulk-boundary-boundary-boundary conformal block, and which is only known through a complicated integral representation.  Here, we do not follow that route, due to the increased difficulty in higher dimensions and the inefficacy of the numerical evaluation of the final result, even in 2D. Nevertheless, we study some properties of that equation in higher dimensions in appendix \ref{app:OPEflow}. 

Instead, we propose using crossing. The crossing equation, at the basis of the modern conformal bootstrap \cite{Poland:2018epd}, is obtained by decomposing four-point functions of conformal primaries by taking the OPE with two different pairings and equating the result.  Let us discuss the AdS$_3$ and AdS$_4$ cases separately
\paragraph{AdS$_3$} In AdS$_3$, as discussed in the preliminaries, there is only one OPE coefficient for a triplet of boundary operators. In the flow equation for the scaling dimensions, the OPE coefficients $C_{lii}$ appear, with $\mathcal{O}_l$ being a scalar operator and $\mathcal{O}_i$ can be a scalar or a spinning operator. The flow equation for the BOE coefficients (\ref{eq:second_eq_final}) instead features the OPE coefficients $C_{lij}$, where all three operators are scalars. Since the OPE of two scalars contains spinning operators, the equations do not close if we only consider crossing with external scalars. In fact, we need to know the scaling dimension of the exchanged spinning operators, which would be evolved by the first flow equation (\ref{eq:firsteq1}). This then requires knowing the OPE coefficients with two spinning operators and a scalar. Hence, the minimal way to close the system of equations is to consider crossing for the four-point function $\langle\mathcal{O}_i(0,0)\mathcal{O}_i(\eta,\bar\eta)\mathcal{O}_j(1,1)\mathcal{O}_j(\infty,\infty)\rangle$ in the case where $\mathcal{O}_i$ and $\mathcal{O}_j$ are both scalars, and the case where one of them has spin. Following the discussion in section \ref{subsubsec:bd4pt}, taking into account all prefactors, the associated crossing equation becomes, for $i\neq j$,
\begin{equation}
    \sum_l\Bigg(C_{lij}^2\mathcal{F}_{i,j,l}^{t}(\eta,\bar\eta)-C_{iil}C_{ljj}\mathcal{F}_{i,j,l}^{s}(\eta,\bar\eta)\Bigg)=\mathcal{F}_{i,j,\mathbb{1}}^{s}(\eta,\bar\eta)
\end{equation}
where
\begin{equation}
\begin{aligned}
    \mathcal{F}_{i,j,l}^{s}(\eta,\bar\eta)&\equiv G^{iijj}_{h_l,\bar h_l}(\eta,\bar\eta)\\
    \mathcal{F}_{i,j,l}^{t}(\eta,\bar\eta)&\equiv \frac{\eta^{2h_i}\bar\eta^{2\bar h_i}}{(1-\eta)^{h_i+h_j}(1-\bar\eta)^{\bar h_i+\bar h_j}}G^{jiij}_{h_l,\bar h_l}(1-\eta,1-\bar\eta)
\end{aligned}
\end{equation}
and the conformal blocks $G^{lijk}_{h_m,\bar h_m}$ are reported explicitly in (\ref{eq:3D4PT}).

\paragraph{AdS$_4$} In AdS$_4$, we have more complications. Triplets of boundary operators have many independent OPE coefficients. The first equation involves $C_{lii}^{(0)}$, where $\mathcal{O}_i$ can have spin and $\mathcal{O}_l$ is a scalar. The BOE flow equation involves only scalars. Analogously with what we discussed for AdS$_3$, the minimal way to close the flow equations is to consider crossing for four-point functions of the kind $\langle\mathcal{O}_i\mathcal{O}_i\mathcal{O}_j\mathcal{O}_j\rangle$ in cases where $\mathcal{O}_i$ and $\mathcal{O}_j$ are both scalars, and cases where one of them has spin. Here we write the equations for the case where $\mathcal{O}_j$ has spin. Crossing then reads, for $i\ne j$,
\begin{equation}
    \sum_l\left(\sum_{a,b}C_{lij}^{(a)}C_{lij}^{(b)}\mathcal{F}_{i,j,l;I}^{t\,(a,b)}(v,u)-C_{lii}\sum_cC_{ljj}^{(c)}\mathcal{F}_{i,j,l;I}^{s\,(c)}(u,v)\right)=\mathcal{F}_{i,j,\mathbb{1};I}^{s\,}(u,v)\,,
    \label{eq:crossing}
\end{equation}
where the index $I$ labels 4-point tensor structures, while the latin indices $a,b,c$ label 3-point tensor structures. The blocks $\mathcal{F}$ are not known in closed form, but they can be computed through recursion relations \cite{Penedones:2015aga,Costa:2016xah,Erramilli:2019njx,Erramilli:2020rlr}.

We stress that crossing closes the system only formally. Since the spectrum is known, crossing is linear in products of OPE coefficients, but linear independence of the blocks is not sufficient to determine them uniquely and stably. Crossing involves infinitely many blocks, and in practice it can only be imposed up to a small residual, due to truncation, finite precision, and errors in the spectrum. Combinations of blocks that nearly cancel, i.e.\ whose sum is smaller than this residual, then leave the corresponding combinations of OPE coefficients undetermined. As a result, the truncated system is exponentially ill-conditioned, and only the lightest OPE coefficients are reliably fixed, see figure~\ref{fig:opeads2}. Another problem arises for degenerate operators, whose blocks coincide, so that crossing fixes their OPE coefficients only up to a rotation in the degenerate subspace, which must instead be fixed by degenerate perturbation theory. Degeneracies are common in free theories and generically lifted by the interaction, and level repulsion generically prevents them from reappearing, see section~\ref{subsec:levelrepulsion}. Near an avoided crossing,
\begin{equation}
    C_{1ij}^2G_{\Delta_1}+C_{2ij}^2G_{\Delta_2}=\left(C_{1ij}^2+C_{2ij}^2\right)G_{\bar\Delta}+\frac{\Delta_{21}}{2}\left(C_{2ij}^2-C_{1ij}^2\right)\partial_{\Delta}G_{\bar\Delta}+O(\Delta_{21}^2)\,,\qquad\bar\Delta=\frac{\Delta_1+\Delta_2}{2}\,,
\end{equation}
so crossing fixes $C_{1ij}^2+C_{2ij}^2$ and $\Delta_{21}(C_{2ij}^2-C_{1ij}^2)$, and separating the two amplifies errors by $1/\Delta_{21}$, which is finite by level repulsion but large close to the free theory and at narrow avoided crossings. Finally, crossing does not fix the sign of $C_{lij}$, which must be determined by smoothness along the flow or by the locality sum rules \eqref{eq:linearsumrules}.

\subsection{Convergence rate of sums in the flow equations}
\label{subsec:convergence}
Any practical application of the flow equations (\ref{eq:flowequations}) will have to deal with the issue that the sums over boundary operators must be truncated. The convergence of these sums is thus an important aspect that needs to be studied. As discussed before, the scaling dimension flow equation (\ref{eq:firsteq1}) involves sums which are proven to be absolutely convergent when $\mathcal{I}$ is an integrated local block. For the BOE flow equation (\ref{eq:second_eq_final}) we will show here that the sum over $\Delta_l$ for fixed $\Delta_j$ is absolutely convergent for some choice of $\alpha$ when $\mathcal{J}$ are integrated local blocks. In the free theory checks described in section \ref{sec:checks}, we will show that suitably defined sequences of Pad\'e approximants of the partial sums over standard integrated blocks converge fastest for both flow equations. We will explain this fact in section \ref{sec:pade}, where we also  discuss possible generalizations of this approach to interacting theories.

For our discussion, it will be useful to recall the universal bounds on sums of BOE and OPE coefficients which we derive in appendix \ref{app:convergence}. The final result of that appendix is that there exist constants $K_b$, $K_C$ and $\Delta_0$ for which, for all $\Delta_{l_{\text{max}}}\geq\Delta_0$,
\begin{equation}
    \sum_{l>l_{\text{max}}}\left(b^{\hat\Phi}_l\right)^2\Delta_l^{-\alpha_1}\leq K_b\ \Delta_{l_{\text{max}}}^{2\Delta^{\text{UV}}_{\hat\Phi}-\alpha_1}\,,\qquad \sum_{l>l_{\text{max}}}C_{lij}^2\,4^{\Delta_l}\Delta_l^{-\alpha_2}\leq K_C\ \Delta_{l_{\text{max}}}^{2(\Delta_i+\Delta_j)-\alpha_2}\,,
    \label{eq:universalbC}
\end{equation}
with $\alpha_1>2\Delta^{\text{UV}}_{\hat\Phi}$ and $\alpha_2>2(\Delta_i+\Delta_j)$.
\subsubsection{Flow of scaling dimensions}
The flow equation describing the evolution of the scaling dimensions reads
\begin{equation}
    \frac{d\Delta_i}{d\lambda}=\sum_lb^{\hat\Phi}_lC_{lii}\mathcal{I}(\Delta_l)\,.
\end{equation}
Studying the explicit expression of $\mathcal{I}$ (\ref{eq:IDelta}), we find that for large $\Delta_l$ it behaves as
\begin{equation}
    \mathcal{I}(\Delta)=c_d\ 2^\Delta\Delta^{-\frac{d+1}{2}}\left(1+O(\Delta^{-1})\right)
\end{equation}
for some unimportant constant $c_d$.
We can thus bound the tail of the sum of absolute values of the summands above a certain truncation
using the Cauchy-Schwarz inequality
\begin{equation}
  \sum_{l>l_{\text{max}}}\left|b_l^{\hat\Phi}C_{lii}\mathcal{I}(\Delta_l)\right|\leq  c_d\sqrt{\left(\sum_{l>l_{\text{max}}}\left(b^{\hat\Phi}_l\right)^2\Delta_l^{-\alpha_1}\right)\left(\sum_{l>l_\text{max}}C_{lii}^24^{\Delta_l}\Delta_l^{-\alpha_2}\right)}\,,
  \label{eq:asymptotic1stflow}
\end{equation}
with $\alpha_1+\alpha_2=d+1$. If $\alpha_1>2\Delta^{\text{UV}}_{\hat\Phi}$ and $\alpha_2>4\Delta_i$, we can use the bounds (\ref{eq:universalbC}) and obtain
\begin{equation}
    \sum_{l>l_{\text{max}}}\left|b_l^{\hat\Phi}C_{lii}\mathcal{I}(\Delta_l)\right|\leq c_d\sqrt{K_bK_C}\Delta_{l_{\text{max}}}^{\Delta^{\text{UV}}_{\hat\Phi}+2\Delta_i-\frac{d+1}{2}}
\end{equation}
which implies that using standard blocks ensures a convergent sum if
\begin{equation}
    2\Delta_i+\Delta_{\hat\Phi}^{\text{UV}}<\frac{d+1}{2}\,,
\end{equation}
a condition which is violated by most boundary operators $\mathcal{O}_i$.

One way to render these sums absolutely convergent is to use local blocks, leading to a modification of $\mathcal{I}$ into expression (\ref{eq:IDeltaalpha}). The tail can then be bounded as
\begin{equation}
    \sum_{l>l_{\text{max}}}\left|b_l^{\hat\Phi}C_{lii}\mathcal{I}^\alpha(\Delta_l)\right|\leq c_{d,\alpha}\sqrt{K_bK_C}\Delta_{l_{\text{max}}}^{\Delta^{\text{UV}}_{\hat\Phi}+2\Delta_i-2\alpha+\frac{1}{2}}
\end{equation}
the sum is thus ensured to be convergent if we choose
\begin{equation}
\alpha>\Delta_i+\frac{\Delta_{\hat\Phi}^{\text{UV}}}{2}+\frac{1}{4}\,.
\end{equation}
We can choose the coefficient $\alpha$ to depend on the truncation $\Delta_{l_{\text{max}}}$. Following appendix F.3.2 of \cite{Loparco:2026fki}, we find the same optimal relation found there
\begin{equation}
    \alpha_{\text{optimal}}=\frac{\Delta_{l_{\text{max}}}}{2}\,.
\end{equation}
\subsubsection{Flow of BOE coefficients}
The flow equation describing the evolution of the BOE coefficients reads
\begin{equation}
    \frac{db^{\hat\Phi}_i}{d\lambda}=\sum_l\sum_j b^{\hat\Phi}_l b^{\hat\Phi}_j C_{lij}\mathcal{J}_{\Delta_i}(\Delta_l,\Delta_j)\,.
\end{equation}
We start by studying the sums over $l$ for fixed $j$ and sum over $j$ for fixed $l$ using standard integrated blocks. We use the asymptotics of the explicit expression of $\mathcal{J}$ as well as the bounds (\ref{eq:universalbC}) as we did for the scaling dimension flow, to put a bound on the tails:
\begin{equation}
  \sum_{l>l_{\text{max}}}\left| b^{\hat\Phi}_lC_{lij}\mathcal{J}_{\Delta_i}(\Delta_l,\Delta_j)\right|\leq c_{\Delta_i,\Delta_j,d}\Delta_{l_{\text{max}}}^{\Delta_{\hat\Phi}^{\text{UV}}+2\Delta_i-\frac{d+1}{2}}\,,
   \label{eq:largelBBb}
\end{equation}
provided that $2\Delta_i+\Delta_{\hat\Phi}^{\text{UV}}<\frac{d+1}{2}$, and
\begin{equation}
    \sum_{j>j_{\text{max}}}\left| b^{\hat\Phi}_jC_{lij}\mathcal{J}_{\Delta_i}(\Delta_l,\Delta_j)\right|\leq c_{\Delta_i,\Delta_l,d}\Delta_{j_{\text{max}}}^{\Delta_{\hat\Phi}^{\text{UV}}+\Delta_i+\frac{d-1}{2}}\,,
    \label{eq:largejBBb}
\end{equation}
provided that $\Delta_i+\Delta^{\text{UV}}_{\hat\Phi}<\frac{1-d}{2}$.
Notice that both conditions are generically violated. To improve the situation, we use the local blocks $\mathcal{J}^\alpha$ (\ref{eq:Jalpha}). We obtain the bounds
\begin{equation}
  \sum_{l>l_{\text{max}}}\left| b^{\hat\Phi}_lC_{lij}\mathcal{J}^\alpha_{\Delta_i}(\Delta_l,\Delta_j)\right|\leq c_{\Delta_i,\Delta_j,d,\alpha}\Delta_{l_{\text{max}}}^{\Delta_{\hat\Phi}^{\text{UV}}+\Delta_i+\Delta_j+\frac{1}{2}-2\alpha}\,,
   \label{eq:largelBBbalpha}
\end{equation}
provided $\alpha>\frac{\Delta_{\hat\Phi}^{\text{UV}}+\Delta_i+\Delta_j}{2}+\frac{1}{4}$, and 
\begin{equation}
  \sum_{j>j_{\text{max}}}\left| b^{\hat\Phi}_jC_{lij}\mathcal{J}^\alpha_{\Delta_i}(\Delta_l,\Delta_j)\right|\leq c_{\Delta_i,\Delta_l,d,\alpha}\Delta_{j_{\text{max}}}^{\Delta_{\hat\Phi}^{\text{UV}}+\Delta_i+\alpha-\frac{3}{2}}\,,
   \label{eq:largejBBbalpha}
\end{equation}
provided $\Delta_{\hat\Phi}^{\text{UV}}+\Delta_i+\alpha<\frac{3}{2}$.

Notice that $\alpha$ can be chosen to increase with $\Delta_j$, leading to a convergent sum over $\Delta_l$, while the sum over $\Delta_j$ cannot be proven to converge. In practice, we observe that choosing a truncation such that $l_{\text{max}}\gg j_{\text{max}}$ leads to convergent sums in our checks in section \ref{sec:checks}, while the opposite ordering of truncations does not. As we will discuss in section \ref{sec:pade}, using Pad\'e approximants  gives convergent sums even with symmetric truncations in free theories.
\section{Checks in free theories}
\label{sec:checks}
The flow equations (\ref{eq:flowequations}) apply to any QFT in AdS realized as a relevant deformation of a UV CFT. In free theories, we can check their validity explicitly at any arbitrary value of the coupling, which will be taken to be the mass of the free field in units of the AdS radius. Here, we quantify the relative errors obtained by comparing the truncated flow equations to the expected analytic results for the derivatives of the CFT data for some value of the mass. We will compare the usage of standard conformal blocks, local blocks, and Pad\'e approximants of both. Surprisingly we find that, despite the fact that the sequence of partial sums of integrated standard blocks does not converge, the sequences of their Pad\'e approximants are the ones that converge fastest. We explain this fact in section \ref{sec:pade}.

In this section, for any partial sum $S_a$ defined by truncating the flow equations (\ref{eq:flowequations}), we quantify the relative error from the analytic expression of the derivative of a piece of QFT data $a$ as
\begin{equation}
\text{err}_a(l_{\text{max}})\equiv\left|\frac{\frac{da}{d\lambda}-S_a(l_{\text{max}})}{\frac{da}{d\lambda}}\right|\,,\qquad a=\Delta_i\ \text{or}\ b^{\hat\Phi}_i\,.
\label{eq:deferror}
\end{equation}
Throughout this section, we will use $\mathcal{I}$ and $\mathcal{J}$ to indicate integrated blocks, whether standard (\ref{eq:IDelta}), (\ref{eq:JDelta}) or local (\ref{eq:IDeltaalpha}), (\ref{eq:Jalpha}). The plot legends will state whether a certain curve was obtained with standard or local blocks. All plots can be reproduced with the attached Mathematica notebook.
\subsection{Flow of scaling dimensions}
Here we present checks of the first flow equation; for conciseness we focus on AdS$_4$
\begin{equation}
    \frac{d\Delta_i}{d\lambda}=\sum_lb^{\hat\Phi}_l C_{lii}\mathcal{I}(\Delta_l)\,.
\end{equation}
We specifically compare the convergence of the sum over boundary operators when using integrated standard conformal blocks and local blocks. We will indicate the partial sums as
\begin{equation}
    S_i(l_{\max})\equiv\sum_l^{l_{\max}}b^{\hat\Phi}_l C_{lii}\mathcal{I}(\Delta_l)\,.
\end{equation}
We also compare these to Pad\'e approximants, which we discuss in detail in section \ref{sec:pade}. Considering that in free theories the spectrum of scalars created by the bulk deformation is integer spaced $\Delta_{l}=2\Delta_1+2l$, we define the function
\begin{equation}
    f_{i,l_{\text{max}}}(x)\equiv\sum_{l=0}^{l_{\max}}b^{\hat\Phi}_l C_{lii}\mathcal{I}(\Delta_l)x^l
\end{equation}
and consider the sequence of its diagonal Pad\'e approximants (for a definition, see (\ref{eq:defdiagpade})) at $x=1$
\begin{equation}
    S_i^{\text{Pad\'e}}(l_{\text{max}})\equiv\left[\left\lfloor \frac{l_{\text{max}}}{2}\right\rfloor\right]_{f_{i,l_{\text{max}}}}(1)
\end{equation}

\subsubsection{Free scalar}
Consider a free massive scalar in AdS$_{d+1}$
\begin{equation}
    S=\frac{1}{2}\int\frac{dzd^d\mathbf{x}}{z^{d+1}}\left[\partial_\mu\hat\phi\partial^\mu\hat\phi+m^2\hat\phi^2\right]\,,
    \label{eq:scalarAction}
\end{equation}
where we absorbed the coupling between the scalar and the AdS curvature in the definition of $m^2$. We will consider how the QFT data of this theory varies under an infinitesimal change of the dimensionless coupling 
\begin{equation}
\lambda\equiv\frac{1}{2}m^2R^2
\end{equation}
around an arbitrary value. This theory in particular includes the boundary primary operator $\phi$, appearing in the BOE of the bulk operator $\hat\phi$, with scaling dimension
\begin{equation}
    \Delta_\phi(\lambda)=\frac{d}{2}\pm\sqrt{\frac{d^2}{4}+2\lambda}\,.
\end{equation}
The choice of sign corresponds to the choice of boundary condition for the scalar (Dirichlet: $+$, Neumann: $-$), where Neumann exists and is unitary only for $-\frac{d^2}{8}<\lambda<\frac{4-d^2}{8}$\footnote{This upper bound assumes $d\geq2$. For AdS$_2$, Neumann exists and is unitary for $-\frac{1}{8}<\lambda<0$}, while Dirichlet exists for all $\lambda>-\frac{d^2}{8}$. We will also consider the boundary spin $J$ primaries $[\phi^2]_{n,J}$ appearing for example in the OPE between two $\phi$ operators. Their scaling dimensions are 
\begin{equation}
    \Delta_{[\phi^2]_{n,J}}(\lambda)=2\Delta_\phi(\lambda)+2n+J
\end{equation}
The flow equation for the scaling dimensions of $\phi$ reads
\begin{figure}
    \centering
    \includegraphics[width=0.75\linewidth]{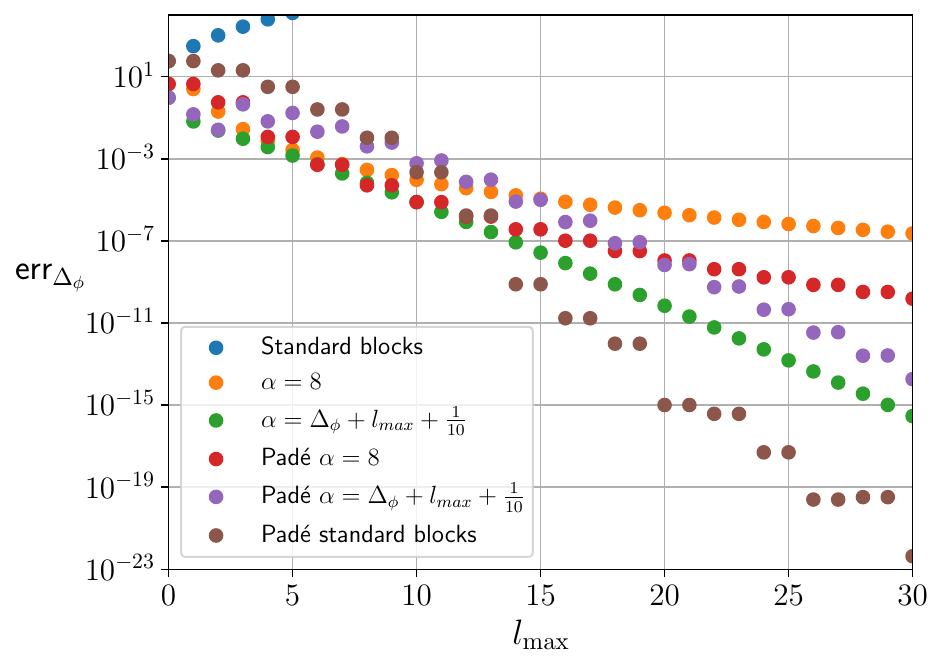}
    \caption{The relative error (\ref{eq:deferror}) of the partial sums in the scaling dimension flow equation reproducing the derivative of $\Delta_\phi$ in free scalar theory in AdS$_4$ at the point $\lambda=\frac{1}{2}m^2R^2=4$. We compare the partial sum over integrated standard blocks, local blocks with fixed $\alpha$, local blocks with $\alpha\propto l_{\max}$, and Pad\'e approximants of all previous cases. The Pad\'e approximants of the integrated standard blocks partial sums perform best.}
    \label{fig:DeltaError}
\end{figure}
\begin{equation}
    \frac{d\Delta_\phi}{d\lambda}=\frac{2}{2\Delta_\phi-d}=\sum_{n=0}^\infty b^{\hat\phi^2}_{[\phi^2]_{n,0}}C_{\phi\phi[\phi^2]_{n,0}}\mathcal{I}(\Delta_{[\phi^2]_{n,0}})\,,
    \label{eq:flowdDeltaphi}
\end{equation}
where the coefficients $b^{\hat\phi^2}_{[\phi^2]_{n,0}}$ and $C_{\phi\phi[\phi^2]_{n,0}}$ are given explicitly in (\ref{eq:bphi2phi2}) and (\ref{eq:Cphiphiphi2}).

Notice that only the scalar double trace operators appear, due to the fact that the equation originates from the BOE of a scalar, $\hat\phi^2$.

In figure \ref{fig:DeltaError} we plot the relative errors of the partial sums for this flow equation in the case of AdS$_4$. We make the following observations:
\begin{itemize}
    \item The sequence of partial sums over standard integrated blocks diverges, as expected.
    \item Pad\'e approximants improve the behavior of the partial sums over local integrated blocks when $\alpha$ is fixed, but worsen it when $\alpha$ scales with $l_{\max}$.
    \item The sequence of partial sums which converges fastest is the one of Pad\'e approximants of standard integrated blocks. Even though this is counterintuitive, we explain this fact in section \ref{sec:pade}.
\end{itemize}
In the free scalar theory we checked also $\frac{d\Delta_{[\phi^2]_{n,0}}}{d\lambda}$ for some values of $n$ and $\frac{d\Delta_{\phi^3}}{d\lambda}$, and obtained the same qualitative behavior with comparable precision. The OPE coefficients relevant for these cases are reported in appendix \ref{app:QFTdata}.
\subsubsection{Free traceless symmetric tensor}
Let us consider the flow of the scaling dimension of a boundary operator with spin. To exemplify this, we study the theory of a free traceless symmetric tensor in AdS$_{4}$
\begin{figure}
    \centering
    \includegraphics[width=0.75\linewidth]{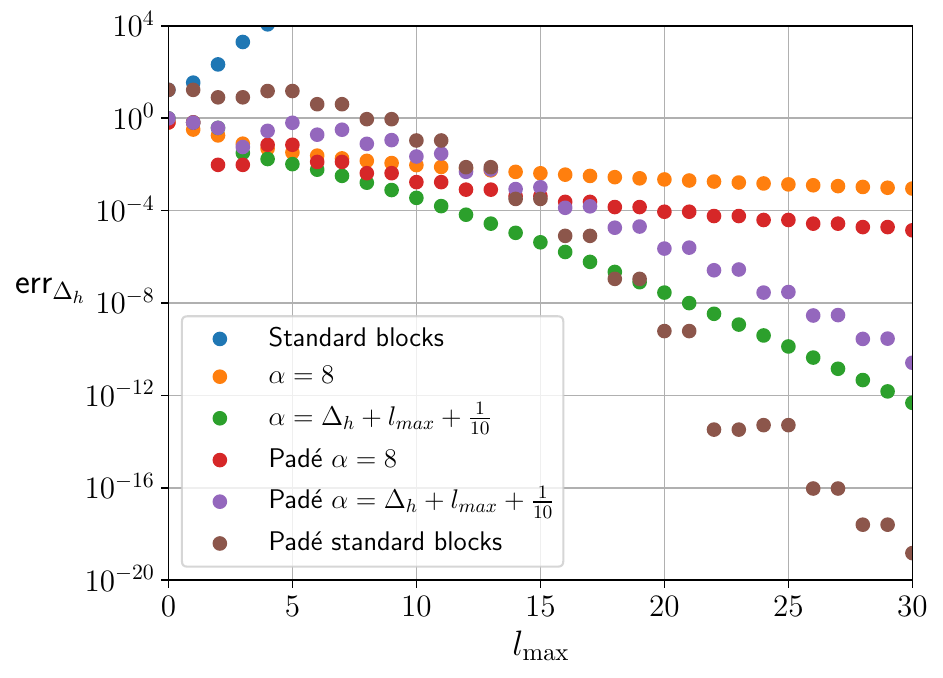}
    \caption{The relative error (\ref{eq:deferror}) of the partial sums in the scaling dimension flow equation reproducing the derivative of $\Delta_h$ in the theory of a free massive traceless symmetric tensor in AdS$_4$ at the point $\lambda=\frac{1}{2}m^2R^2=4$. We compare the partial sum over integrated standard blocks, local blocks with fixed $\alpha$, local blocks with $\alpha\propto l_{\max}$, and Pad\'e approximants of all previous cases. The Pad\'e approximants of the integrated standard blocks partial sums perform best.}
    \label{fig:DeltaTError}
\end{figure}
\begin{equation}
    S=\frac{1}{2}\int\frac{dzd^3\mathbf{x}}{z^{4}}\left[\nabla_\mu\hat h^{\nu\alpha}\nabla^\mu\hat h_{\nu\alpha}-2\nabla^\mu\hat h^{\nu\alpha}\nabla_\alpha\hat h_{\mu\nu}+(m^2+6)\hat h_{\mu\nu}\hat h^{\mu\nu}\right]\,,
    \label{eq:actiontensor}
\end{equation}
where we chose the definition of $m^2$ such that in the limit $m^2\to0$ we get the action of the AdS linearized graviton (although one should restore terms proportional to $\hat h_{\mu}^{\ \mu}$ for gauge invariance).

Once again we will consider the derivatives of the QFT data as we vary the dimensionless mass $\lambda\equiv\frac{1}{2}m^2R^2$. In particular, we will focus on the spin 2 boundary operator $h_{ab}$ appearing in the BOE of $\hat h_{\mu\nu}$, with scaling dimension
\begin{equation}
    \Delta_h(\lambda)=\frac{3}{2}\pm\sqrt{\frac{9}{4}+2\lambda}
\end{equation}
where once again the sign corresponds to the choice of Dirichlet vs Neumann boundary conditions. We will focus on Dirichlet. We will need to consider also the boundary scalar primaries $[h^2]_{n,0}$ associated to the BOE of $\hat h_{\mu\nu}\hat h^{\mu\nu}$ with scaling dimensions
\begin{equation}
    \Delta_{[h^2]_{n,0}}=2\Delta_h(\lambda)+2n\,.
\end{equation}
The flow equation for the scaling dimensions of $h_{ab}$ reads
\begin{equation}
    \frac{d\Delta_h}{d\lambda}=\frac{2}{2\Delta_h-3}=\sum_{n=0}^\infty b^{\hat h^2}_{[h^2]_{n,0}}C_{hh[h^2]_{n,0}}^{(0)}\mathcal{I}(\Delta_{[h^2]_{n,0}})\,,
\end{equation}
where the product of the coefficients $b^{\hat h^2}_{[h^2]_{n,0}}C_{hh[h^2]_{n,0}}^{(0)}$ is given explicitly in (\ref{eq:BCtensor}).

Analyzing the sequence of partial sums, we find analogous behaviors as in the scalar case. We plot them in figure \ref{fig:DeltaTError}.

\subsection{Flow of BOE coefficients}
Now let us consider the second flow equation
\begin{equation}
    \frac{d b^{\hat\Phi}_i}{d\lambda}=\sum_{l}\sum_j b^{\hat\Phi}_l b^{\hat\Phi}_jC_{ilj}\mathcal{J}_{\Delta_i}(\Delta_l,\Delta_j)
\end{equation}
Since the deformation is taken to be a bulk scalar, only boundary scalar operators are involved in this equation. When only one of the sums is involved, like in the case where we evolve for example $b^{\hat\phi}_\phi$, the partial sums and their Pad\'e approximants are defined as in the previous section. When two sums are involved, we empirically observe that having a higher cutoff for the sum over $l$ as we increase $j$ leads to convergence, at least in the checks we ran. For example, we take
\begin{equation}
    S_i(j_{\text{max}})\equiv\sum_j^{j_{\text{max}}}\sum_l^{j+j_{\text{max}}}b^{\hat\Phi}_l b^{\hat\Phi}_jC_{ilj}\mathcal{J}_{\Delta_i}(\Delta_l,\Delta_j)\,.
    \label{eq:truncatelocal}
\end{equation}
For local blocks, this choice is compatible with (\ref{eq:largelBBbalpha}), where we compensate the weaker bound on the tail of the sum over $l$ as $\Delta_j$ increases by simultaneously increasing $\alpha$. At the same time, we have no convergent bound on the tail of the sum over $j$. Nevertheless, the sum converges in the example of the free scalar.

The sequence of Pad\'e approximants of the partial sums with standard blocks, instead, converges even when we choose the same truncation for $\Delta_l$ and $\Delta_j$. Since we will be considering free theories we have $\Delta_l=2\Delta_1+2l$ and similarly for $\Delta_j$. We will proceed as follows: we define the functions
\begin{equation}
    f_{i,j,l_{\text{max}}}(x)\equiv\sum_{l=0}^{l_{\text{max}}}b^{\hat\Phi}_l b^{\hat\Phi}_jC_{ilj}\mathcal{J}_{\Delta_i}(\Delta_l,\Delta_j)x^l\,,
\end{equation}
and take their diagonal Pad\'e approximants at $x=1$, defined in (\ref{eq:defdiagpade})
\begin{equation}
    S_{i,j}^{\text{Pad\'e}}(l_{\text{max}})=\left[\left\lfloor\frac{l_{\text{max}}}{2}\right\rfloor\right]_{f_{i,j,l_{\text{max}}}}(1)\,.
\end{equation}
Then, we do the same again for the sum over $j$. We construct a function $g_{i,l_{\text{max}}}(y)$ and construct its Pad\'e approximants around $y=0$, then evaluated at $y=1$
\begin{equation}
    g_{i,l_{\text{max}}}(y)\equiv\sum_j^{l_{\text{max}}}S_{i,j}^{\text{Pad\'e}}(l_{\text{max}})y^j\,, \qquad S_i^{\text{Pad\'e}}(l_{\text{max}})=\left[\left\lfloor\frac{l_{\text{max}}}{2}\right\rfloor\right]_{g_{i,l_{\text{max}}}}(1)\,.
    \label{eq:padeBOE}
\end{equation}
These are the partial sums of which we plot the error for example in figure \ref{fig:bphi2Error}.
\subsubsection{Free scalar}
\begin{figure}
    \centering
    \includegraphics[width=0.75\linewidth]{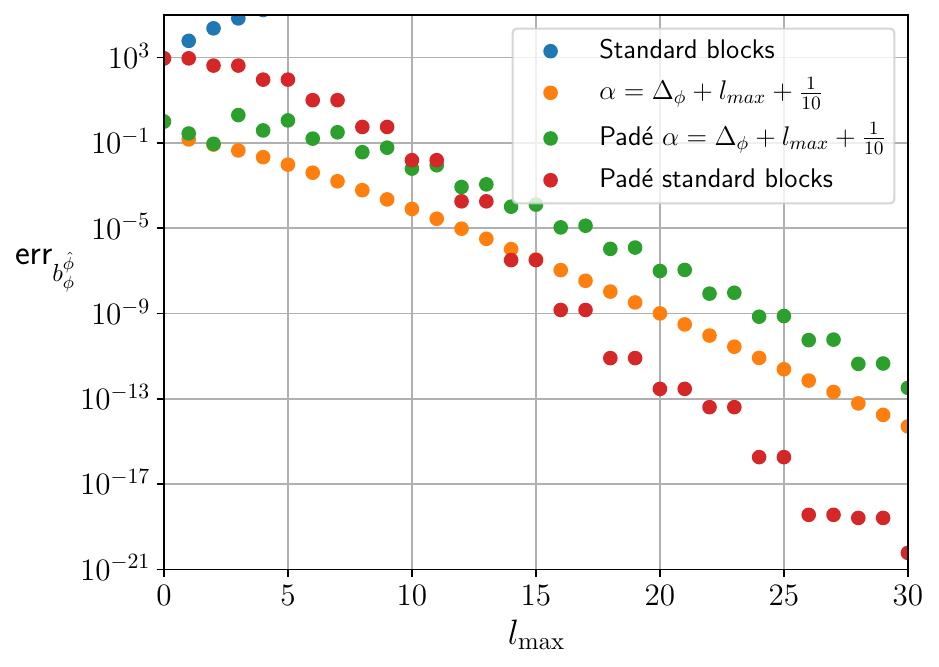}
    \caption{The relative error (\ref{eq:deferror}) of the partial sums in the BOE coefficient flow equation reproducing the derivative of $b^{\hat\phi}_\phi$ in the theory of a free massive scalar in AdS$_4$ at the point $\lambda=\frac{1}{2}m^2R^2=4$. We compare the partial sum over integrated standard blocks, local blocks with $\alpha\propto l_{\max}$, and Pad\'e approximants of all previous cases. The Pad\'e approximants of the integrated standard blocks partial sums perform best.}
    \label{fig:bphiError}
\end{figure}
In the theory of a free scalar in AdS$_{d+1}$, with action (\ref{eq:scalarAction}), we can consider the evolution of the BOE coefficients of the bulk primaries $\hat\phi$ and $\hat\phi^2$. Since the BOE of $\hat\phi$ involves a single boundary primary operator, the associated flow equation only involves one sum
\begin{equation}
    \frac{d b^{\hat\phi}_\phi}{d\lambda}=b^{\hat\phi}_\phi\sum_{l=0}^\infty b^{\hat\phi^2}_{[\phi^2]_{l,0}}C_{\phi\phi[\phi^2]_{l,0}}\mathcal{J}_{\Delta_\phi}(\Delta_{[\phi^2]_{l,0}},\Delta_\phi)
\end{equation}
We plot the relative errors (defined in (\ref{eq:deferror})) of the partial sums
in figure \ref{fig:bphiError}. 
In this case standard blocks, local blocks and Pad\'e approximants all behave similarly to the flow equation of the scaling dimensions.

\begin{figure}
    \centering
    \includegraphics[width=0.75\linewidth]{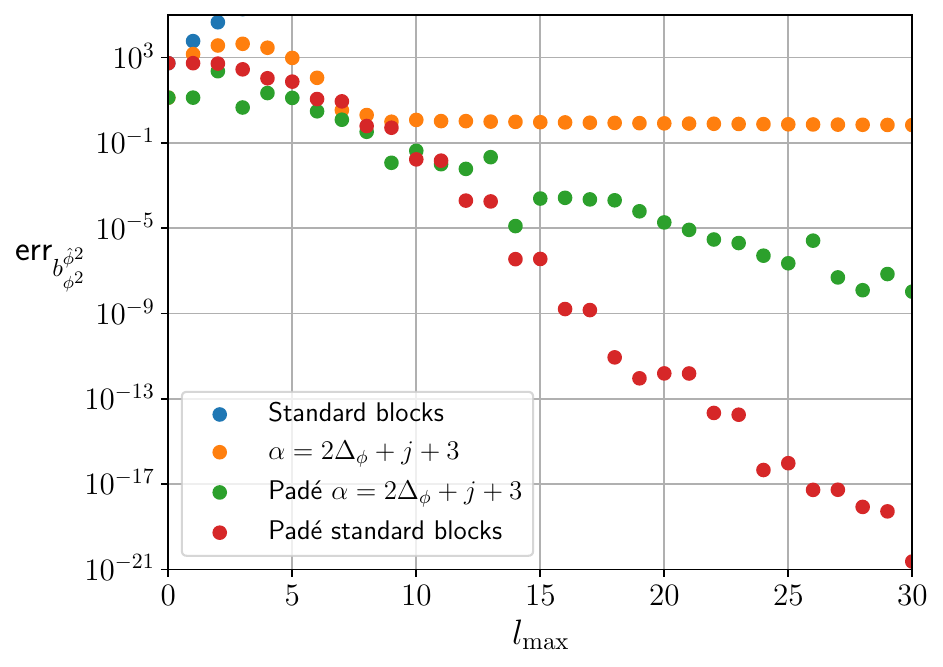}
    \caption{The relative error (\ref{eq:deferror}) of the partial sums in the BOE coefficient flow equation reproducing the derivative of $b^{\hat\phi^2}_{\phi^2}$ in the theory of a free massive scalar in AdS$_4$ at the point $\lambda=\frac{1}{2}m^2R^2=4$. We compare the partial sum over integrated standard blocks, local blocks with $\alpha\propto j$, and their Pad\'e approximants. Despite appearances, the orange dots are slowly decaying. It is evident that Pad\'e approximants introduce a significant improvement over local blocks. Pad\'e approximants of standard blocks perform best here too.}
    \label{fig:bphi2Error}
\end{figure}

A more emblematic example is that of the flow of $b^{\hat\phi^2}_{\phi^2}$. In this case, two sums are involved
\begin{equation}
    \frac{db^{\hat\phi^2}_{\phi^2}}{d\lambda}=\sum_{l=0}^\infty\sum_{j=0}^\infty b^{\hat\phi^2}_{[\phi^2]_{l,0}}b^{\hat\phi^2}_{[\phi^2]_{j,0}}C_{\phi^2[\phi^2]_{l,0}[\phi^2]_{j,0}}\mathcal{J}_{\Delta_{\phi^2}}(\Delta_{[\phi^2]_{l,0}},\Delta_{[\phi^2]_{j,0}})\,.
    \label{eq:flowbphi2phi2}
\end{equation}
 with the BOE and OPE coefficients given in (\ref{eq:bphi2phi2}) and (\ref{eq:cphi2phi2phi2}) respectively. This case is where the Pad\'e approximants bring the biggest improvement over local blocks.  In figure \ref{fig:bphi2Error} we show the error, defined in (\ref{eq:deferror}), of the partial sums truncated as described in (\ref{eq:truncatelocal}) and of the Pad\'e approximants (\ref{eq:padeBOE}). We explain the reason why Pad\'e approximants work this well in these examples in section \ref{sec:pade}.

\subsection{OPE coefficients from crossing}
To fix the OPE coefficients at a given step in a numerical application of the flow equations, we advocate for the use of the crossing equation. Concretely, one can devise a numerical algorithm where the scaling dimensions of the boundary operators are evolved in an infinitesimal step from the first flow equation, and then crossing is solved with the updated spectrum. Here we exemplify that, when the spectrum is known, crossing can be reliably and cheaply solved with various numerical methods in this setting with good precision. Of course, signs cannot be determined this way. One can argue that the signs are fixed by the previous step in the iterative solution of the ODEs, or that they can be checked, for example, using the linear sum rules (\ref{eq:linearsumrules}). For now, we focus on determining the absolute values of the coefficients, and study cases with two identical operators. We postpone the development of more sophisticated numerical methods for non-identical operators and operators with spin to future works.

\paragraph{AdS$_2$}
In the 2D case, using the expression of the blocks in \ref{subsubsec:bd4pt},  defining for convenience 
\begin{equation}
    \hat G^{\Delta_i}_{\Delta_m}(\zeta)\equiv\zeta^{-2\Delta_i}G^{iiii}_{\Delta_m}(\zeta)\,,
\end{equation}
focusing on the case $\mathcal{O}_i=\mathcal{O}_j$, the crossing equation takes the form
\begin{equation}
    \sum_m C_{iim}^2\left(\hat G_{\Delta_m}^{\Delta_i}(\zeta)-\hat G_{\Delta_m}^{\Delta_i}(1-\zeta)\right)=\hat G_{0}^{\Delta_i}(1-\zeta)-\hat G_{0}^{\Delta_i}(\zeta)
    \label{eq:crossing53}
\end{equation}
where we've separated the contribution of the identity by hand to highlight that we fixed $C_{ii\mathbb{1}}=1$ in line with our normalization for boundary two-point functions.
\begin{figure}
    \centering
    \includegraphics[width=0.47\linewidth]{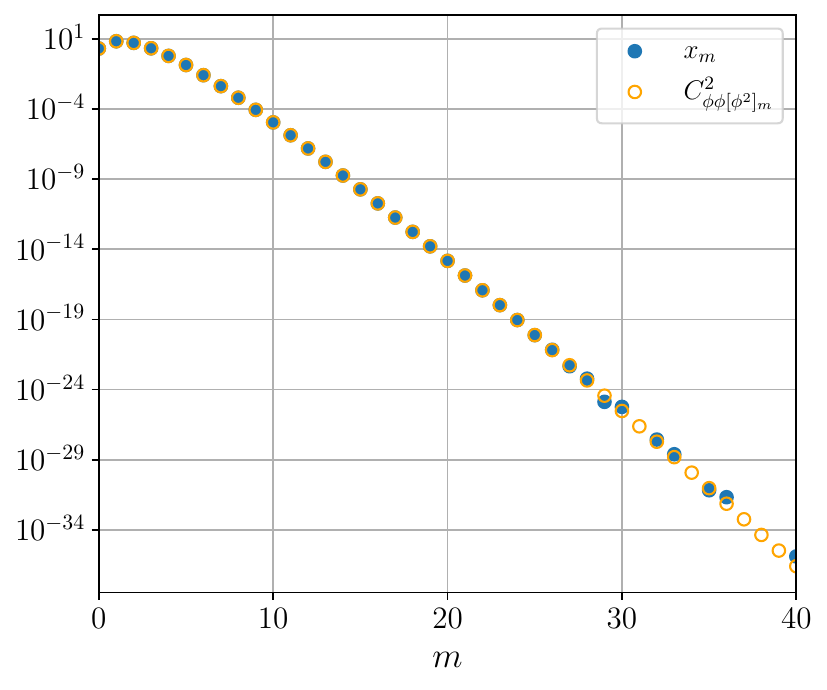}
    \includegraphics[width=0.47\linewidth]{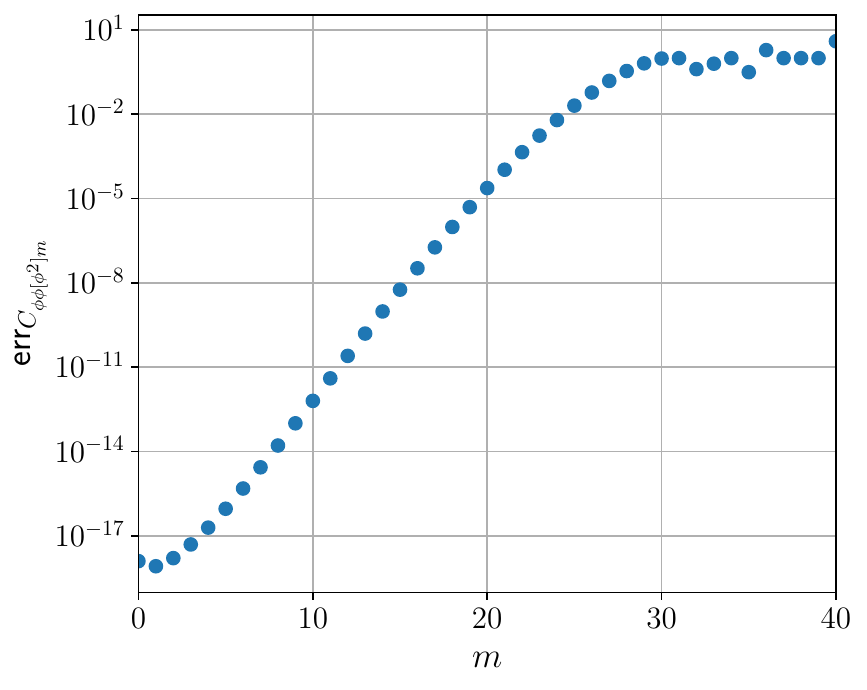}
    \caption{We show the numerical estimate $x_m$ of the OPE coefficients $C_{\phi\phi[\phi^2]_m}$ in free scalar theory in AdS$_2$ obtained by solving the linear optimization problem (\ref{eq:linearoptimize}) with $m_{\text{max}}=40$ states and $\Delta_\phi=\frac{27}{11}$ compared to the analytic value. On the left we compare the absolute values, on the right we show the relative error (\ref{eq:errorOPE}). We used Mathematica's \texttt{LinearOptimization} function. The coefficients determined with most accuracy are those of the lighest operators. The missing blue dots on the left are $x_m=0$ within the numerical accuracy we used.}
    \label{fig:opeads2}
\end{figure}

To solve for the OPE coefficients, we set up the following optimization problem. We choose 41 equally spaced points $\zeta_n$, $\zeta_n\in(\frac{1}{4},\frac{1}{2})$. Then, we build a matrix $M$ and a vector $b$ by keeping 41 double trace operators $\mathcal{O}_m$ to sum over with $\Delta_m=2\Delta_i+2m$
\begin{equation}
    M_{mn}\equiv \hat G_{\Delta_m}^{\Delta_i}(\zeta_n)-\hat G_{\Delta_m}^{\Delta_i}(1-\zeta_n)\,,\qquad b_n\equiv \hat G_{0}^{\Delta_i}(1-\zeta_n)-\hat G_{0}^{\Delta_i}(\zeta_n)\,.
\end{equation}
Finally, we ask a linear optimizer (\texttt{LinearOptimization} on Mathematica) to find the vector $x_m\geq0$ which minimizes the residuals, parametrized by a positive number $\delta$
\begin{equation}
    \left|(M\cdot x)_n-b_n\right|\leq\delta\,,\qquad \forall n\,.
    \label{eq:linearoptimize}
\end{equation}
We then identify the solution $x_m$ with the vector of approximate squared OPE coefficients $C_{iim}^2$.

In figure \ref{fig:opeads2} we show the error with which we determine $x_m$ in the free scalar theory, defined as
\begin{equation}
    \text{err}_{C_{\phi\phi[\phi^2]_m}}=\left|\frac{x_m-C_{\phi\phi[\phi^2]_m}^2}{C_{\phi\phi[\phi^2]_m}^2}\right|
    \label{eq:errorOPE}
\end{equation}
where $x_m$ is the numerical vector determined by the optimizer. 
We chose $\Delta_\phi=\frac{27}{11}$. The lightest OPE coefficients are determined with the greatest relative accuracy.

\paragraph{AdS$_3$} In the 3D case, following the discussion in section \ref{subsubsec:bd4pt}, the crossing equation for identical external operators reads
\begin{equation}
    \sum_m C_{iim}^2\left(\tilde G^{\Delta_i}_{\Delta_m,J_m}(\eta,\bar\eta)-\tilde G^{\Delta_i}_{\Delta_m,J_m}(1-\eta,1-\bar\eta)\right)=\tilde G^{\Delta_i}_{0,0}(1-\eta,1-\bar\eta)-\tilde G^{\Delta_i}_{0,0}(\eta,\bar\eta)
\end{equation}
where we are using that, when external operators are scalars, the OPE coefficients of operators related by parity can be collected, since $C_{ii\bar m}=C_{iim}$ in that case, and the blocks simplify as follows\footnote{The Kronecker delta $\delta_{0,J_m}$ is there to avoid double counting of scalar operators which have $h_m=\bar h_m$.}
\begin{equation}
\begin{aligned}
    &\tilde G^{\Delta_i}_{\Delta_m,J_m}(\eta,\bar\eta)=\frac{1}{1+\delta_{0,J_m}}\frac{1}{(\eta\bar\eta)^{\frac{2\Delta_i-\Delta_m+J_m}{2}}}\\
    &\times\left[\eta^{J_m}\,_2F_1\left(\begin{matrix}\frac{\Delta_m-J_m}{2}, & \frac{\Delta_m-J_m}{2}\\ &\Delta_m-J_m\end{matrix};\bar\eta\right)\,_2F_1\left(\begin{matrix}\frac{\Delta_m+J_m}{2}, & \frac{\Delta_m+J_m}{2}\\ &\Delta_m+J_m\end{matrix};\eta\right)+(\eta\leftrightarrow\bar\eta)\right]\,.
    \label{eq:bbbb3D}
\end{aligned}
\end{equation}
Now we need to choose a numerical method to solve for the OPE coefficients. It turns out that the linear optimizer we used in AdS$_2$ is much slower here due to spin causing a proliferation of states when choosing a certain cutoff in $\Delta_m$. A faster method is to construct an auxiliary quadratic function which we ask Mathematica's \texttt{FindMinimum} to minimize. 

The steps are the following. We choose a finite set of pairs of points $\{(\eta,\bar\eta)_n\}$ over which we scan and a truncation $\Delta_{m_{\text{max}}}$. Then, we build the matrix and vector
\begin{equation}
\begin{aligned}
    M_{mn}&\equiv \tilde G^{\Delta_i}_{\Delta_m,J_m}(\eta_n,\bar\eta_n)-\tilde G^{\Delta_i}_{\Delta_m,J_m}(1-\eta_n,1-\bar\eta_n)\,,\\ b_{n}&\equiv\tilde G^{\Delta_i}_{0,0}(1-\eta_n,1-\bar\eta_n)-\tilde G^{\Delta_i}_{0,0}(\eta_n,\bar\eta_n)
\end{aligned}
\end{equation}
and construct the function
\begin{figure} 
    \centering
    \includegraphics[width=0.48\linewidth]{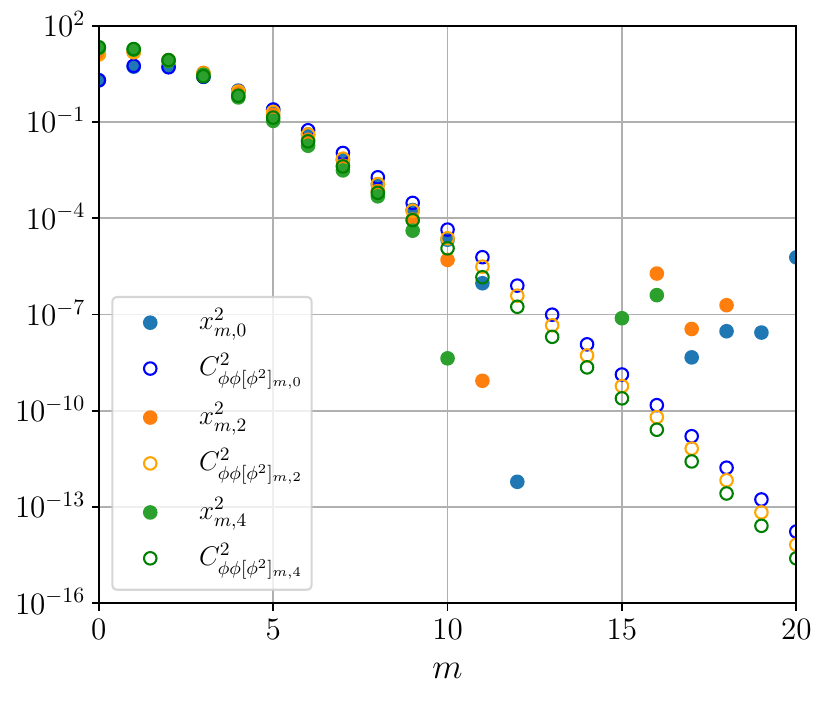}
    \includegraphics[width=0.48\linewidth]{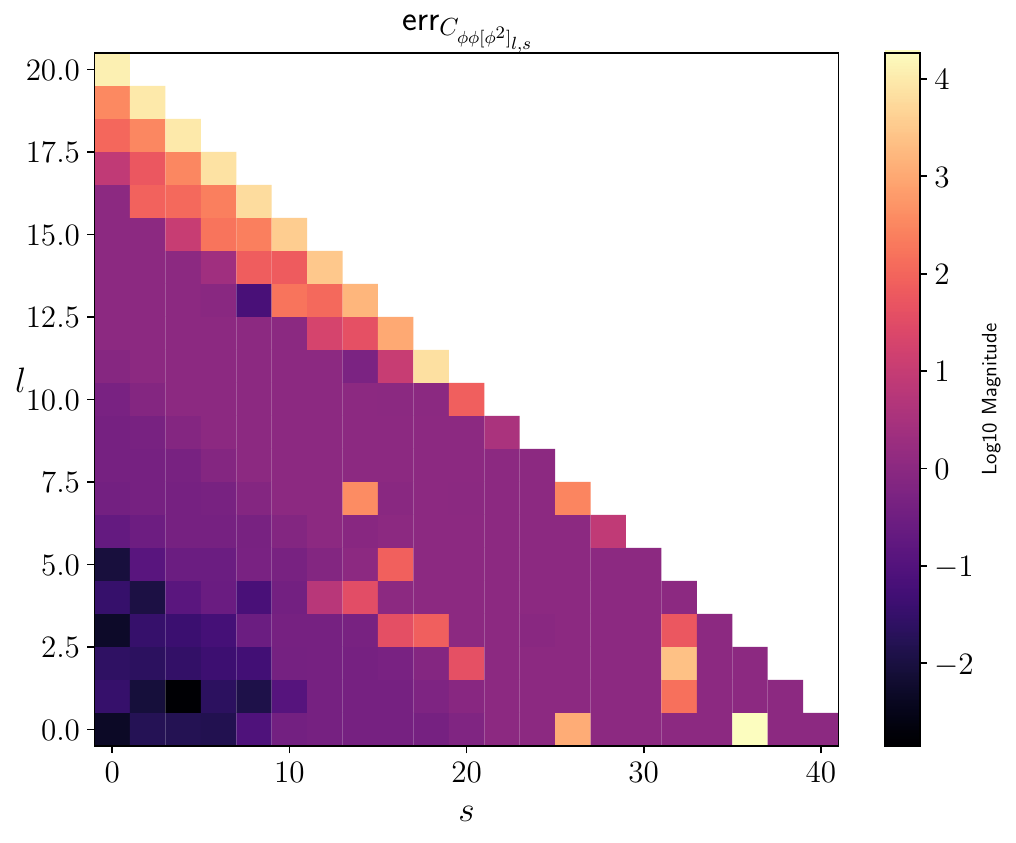}
        \caption{Numerical estimate of the OPE coefficients $C_{\phi\phi[\phi^2]_{m,J}}$ using Mathematica's \texttt{FindMinimum} on the function (\ref{eq:f(x)OPE}) in free scalar theory in AdS$_3$ with $\Delta_\phi=\frac{37}{11}$. On the left, we compare the numerical values with the analytic ones in the cases $J=0,2,4$. On the right, the relative errors (\ref{eq:errorOPE}). The lightest states are determined with the highest relative precision. The missing blue dots on the left are zero up to numerical precision.}
    \label{fig:OPEError3D}
\end{figure}
\begin{equation}
    f(x)=\sum_{n}\left((M\cdot x^2)_{n}-b_{n}\right)^2
    \label{eq:f(x)OPE}
\end{equation}
where here $x^2\equiv(x_0^2,x_1^2,\ldots x_{m_{\text{max}}}^2)$. Finally, we feed the function to Mathematica's \texttt{FindMinimum} and ask it to find the $x^2$ which minimizes (\ref{eq:f(x)OPE}) through a Levenberg-Marquardt algorithm. The resulting values will be identified with approximate OPE coefficients squared $x^2_m\equiv C_{iim}^2$.

In practice, we check this method in free scalar theory. We choose $\Delta_\phi=\frac{37}{11}$, and truncate the double trace operators at $\Delta_{\text{max}}=2\Delta_\phi+40$. Since the dimensions of the double trace operators are $\Delta_{[\phi^2]_{m,J}}=2\Delta_\phi+2m+J$, we have to implement the truncation by summing over $m$ up to 20 and over even $J$ up to $40-2m$, so that we have 231 unknowns. We have to choose at least 231 pairs of points $\{(\eta,\bar\eta)_n\}$, we pick 256. Finally, this sort of algorithm requires an initial guess of $x$ from which the minimization procedure starts. To simulate the ignorance we would have of the OPE coefficients from a previous step in an iterative solution of the flow equations, we generate the initial condition by using the actual analytic expression of $C_{\phi\phi[\phi^2]_{m,J}}$ and multiplying each element by a $10\%$ random noise. We repeat the numerical reconstruction five times with independent random initializations and average the results, to avoid selecting a particularly favorable realization. We show the errors, defined as in (\ref{eq:errorOPE}), of an application of this algorithm in figure \ref{fig:OPEError3D}. More details and the whole code to reproduce this plot are in the ancillary Mathematica notebook. We note that increasing the spectral cutoff improves the numerical accuracy of the reconstructed OPE coefficients. Moreover, the choice of sampling points is relevant because the conformal block expansion converges more rapidly in some regions than in others, affecting the truncation error at a fixed spectral cutoff. These tests use exact spectra: stability against spectral errors accumulated along the flow remains to be investigated. 

Let us make a final comment. Of course, these are rudimentary methods to determine OPE coefficients from crossing. More sophisticated methods achieving higher precision can be used when one will attempt a numerical application of the flow equations to an interacting theory. Our point is simply to demonstrate the kind of numerical accuracy one can achieve with fast evaluations on a standard laptop.

\section{The (un)reasonable effectiveness of Padé}
\label{sec:pade}
As we showed in section \ref{subsec:convergence}, the convergence of the sums in the flow equations (\ref{eq:flowequations}) cannot be established with standard blocks (and they diverge in our free-theory examples), while local blocks give absolutely convergent sums with power-law tails, except for the sum over j in the BOE flow. Here we explain why our alternative approach, based on Pad\'e approximants, leads to a drastic improvement in the convergence of the sums in the free theory checks in section \ref{sec:checks}. We then discuss a proposal on how to extend the applicability of these improvements to interacting theories.

Let us start with a lightning introduction to Pad\'e approximants. See \cite{Baker_Graves-Morris_1996} for a textbook on the subject.

\subsection{Quick intro to Pad\'e approximants}
Given a function of a single real variable $f(x)$, and two integers $m\geq0 $ and $n\geq1$, the $\left[\frac{m}{n}\right]$ Pad\'e approximant of $f(x)$ is defined as
\begin{equation}
    \left[\frac{m}{n}\right]_f(x)\equiv\frac{\sum_{i=0}^ma_ix^i}{1+\sum_{j=1}^nb_jx^j}\,,
    \label{eq:padedef}
\end{equation}
where the coefficients $a_i$ and $b_j$ are uniquely fixed by requiring that all derivatives at a certain point (here taken to be $x=0$) match
\begin{equation}
\frac{d^k}{dx^k}\left[\frac{m}{n}\right]_f(x)\Bigg|_{x=0}=\frac{d^k}{dx^k}f(x)\Bigg|_{x=0}\,,\quad k=0,\ldots,m+n
\end{equation}
This ensures that the Taylor series of the function and its Padé approximant agree up to order $m+n$.
We will in particular use diagonal Pad\'e approximants, where $m=n$, and we will indicate them as
\begin{equation}
    [m]_f(x)\equiv\left[\frac{m}{m}\right]_f(x)\,.
    \label{eq:defdiagpade}
\end{equation}
The power of Pad\'e approximants lies in the fact that, by allowing a more general ansatz than just simple powers, they can extend the domain of convergence of a series. Consider the series representation of the logarithm
\begin{equation}
    f(x)\equiv\log(1+x)=\sum_{n=1}^\infty\frac{(-1)^{n-1}}{n}x^n\,.
    \label{eq:log(1+x)}
\end{equation}
This series has radius of convergence $|x|<1$, bounded by the singularity at $x=-1$. Attempting to use the truncated Taylor series to reproduce $\log(2)$ is highly inefficient\footnote{The Taylor series still converges conditionally at $x=1$, but slowly (the error goes to zero as $n_{\text{max}}^{-1}$).}, for example
\begin{equation}
   \left|\sum_{n=1}^8\frac{(-1)^{n-1}}{n}-\log(2)\right|\sim 6\times 10^{-2}\,.
\end{equation}
On the other hand, the sequence of Pad\'e approximants converges much faster. The error obtained by using the $4$-th order Pad\'e approximant is
\begin{equation}
    \left|[4]_{f}(1)-\log(2)\right|\sim 7\times 10^{-7}
    \label{eq:logerr}
\end{equation}
 To understand why this happens, it is instructive to look at the analytic structure of Padé approximants in the complex plane. Since the discussion below is somewhat technical, readers primarily interested in the final results may skip it and proceed directly to the error estimate in \eqref{eq:errPade}.

Being rational functions, Pad\'e approximants only possess poles and cannot contain genuine branch cuts. However, when approximating a function with a branch cut, such as $\log(1+x)$ for $x \in (-\infty,-1]$, the poles and zeros of the diagonal approximants $[m]_f(x)$ accumulate densely along the cut. The Pad\'e approximant effectively mimics the discontinuity of the branch cut by condensing its poles along that line.

An approximant of order $m$ may also feature a few \emph{spurious} poles inside the domain of analyticity of the full function, each accompanied by a nearby zero. The rigorously relevant notion is therefore \emph{convergence in capacity}: a sequence $f_m$ converges in capacity to $f$ in a domain $D$ if, for every compact set $V\subset D$ and every $\varepsilon>0$, the logarithmic capacity (defined below) of the set $\{x\in V:\ |f(x)-f_m(x)|>\varepsilon\}$ tends to zero as $m\to\infty$. The set where the approximation fails thus shrinks, but it is allowed to move with $m$. In particular, convergence in capacity does not guarantee that the full sequence converges at a given point, since spurious poles may approach that point for infinitely many $m$. The branch cut itself is not part of this exceptional set: it has non-zero capacity, and it is excluded from the domain $D$ from the outset.

The branch cuts of a multi-valued function can be placed arbitrarily. It is thus natural to ask which specific choice of branch cut the Pad\'e approximants simulate. 

The answer is given by Stahl's theorems \cite{Stahl1997TheCO,Stahl1986OrthogonalPW}. Suppose that $f$ is analytic at the expansion point and can be continued analytically along every path that avoids a compact set $M$ of zero capacity, for instance a finite number of branch points and poles. The continuation may be multivalued: paths are allowed to cross branch cuts, which are not singularities of $f$, so only their endpoints belong to $M$. This assumption excludes natural boundaries. Making $f$ single-valued requires removing a cut that joins its branch points, and such a cut generally has non-zero capacity, unlike $M$. Stahl's theorem states that there is a unique compact set $B$, the cut of \emph{minimal} logarithmic capacity, such that $f$ is single-valued in $\Omega=\mathbb{C}\backslash B$, and the diagonal Pad\'e approximants converge to $f$ in capacity in $\Omega$. Moreover, all but $o(m)$ of the poles of $[m]_f$ accumulate on $B$. For instance, for $\log(1+x)$ expanded around $x=0$, $M=\{-1,\infty\}$ and $B=(-\infty,-1]$.

Let us expand on the concept of minimal capacity, as it will also be related to the rate of convergence of the Pad\'e approximants.

\paragraph{Logarithmic capacity and minimal cuts} Logarithmic capacity is a concept that originates in 2D electrostatics, where the term ``logarithmic" is used because the electric potential between two point charges in 2D scales like $-\log r$. 

Given a set of points and extended objects, treat them as conductors and imagine populating them with charges with total charge 1. Then, the logarithmic capacity of the set is $e^{-E}$ where $E$ is the electrostatic energy of the configuration in which all charges have settled at equilibrium. In the context of the Pad\'e approximants, we are interested in the logarithmic capacity of the branch cuts in the frame where the expansion point is at infinity. Given two branch points, the statement is that \emph{the poles of the Pad\'e approximants will accumulate on the cut which joins the two branch points in a way which minimizes the logarithmic capacity in this frame. }

Stahl also determined the rate of this convergence:
\begin{equation}
    \lim_{m\to\infty}|f(x)-[m]_f(x)|^\frac{1}{2m}=e^{-g_{\Omega}(x,x_0)}\quad\text{in capacity in }\Omega\,,
\end{equation}
where $g_\Omega(x,x_0)$ is the Green's function of the 2D Laplacian on the domain $\Omega=\mathbb{C}\backslash B$, where $B$ is the set of minimal capacity branch cuts of $f$ and $x_0$ is the expansion point (until now we had set $x_0=0$). The Green's function is defined to have the following boundary conditions:
\begin{equation}
    \begin{aligned}
        g_{\Omega}(x,x_0)&\stackrel{x\to x_0}{\approx}-\log|x-x_0|+O(1)\,, \\
        g_{\Omega}(x,x_0)&=0\quad \text{for}\ x\in B\,.
        \label{eq:greensbc}
    \end{aligned}
\end{equation}
For concreteness, let us see how this works in the example of the logarithm (\ref{eq:log(1+x)}).

\begin{figure}
    \centering
    \includegraphics[width=0.95\linewidth]{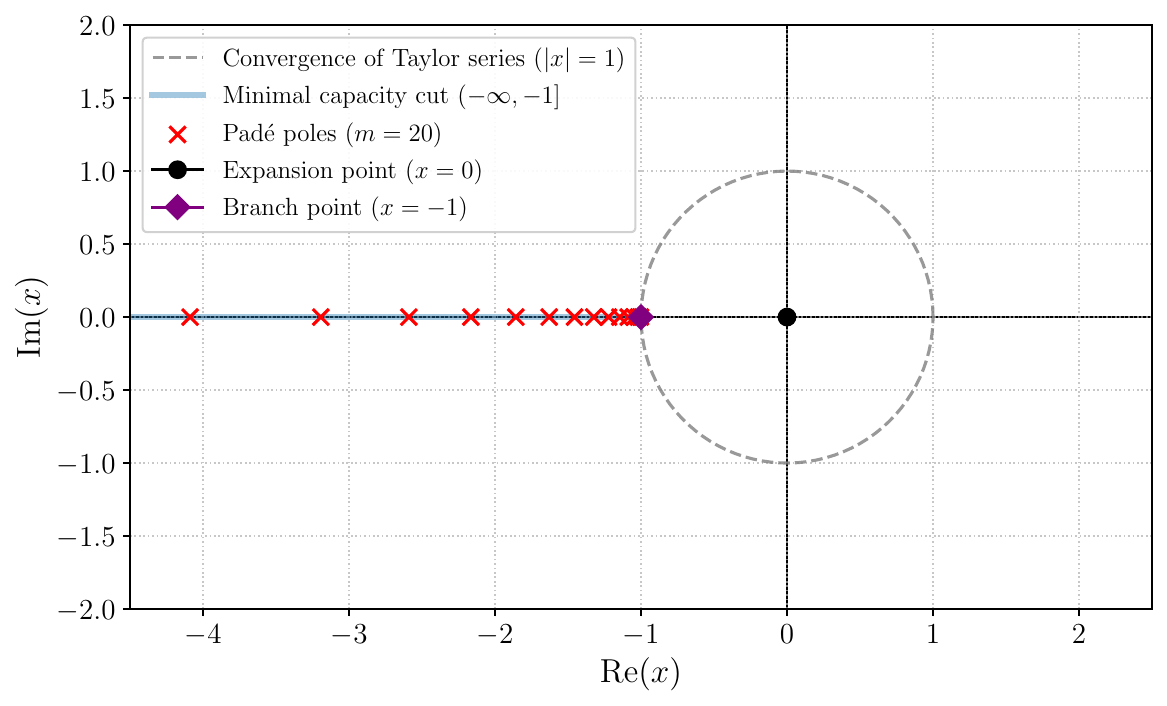}
    \caption{The poles of the Pad\'e approximant of order $m=20$ of $\log(x+1)$ around $x=0$ accumulate around the minimal capacity cut, $(-\infty,-1]$.}
    \label{fig:padepoles}
\end{figure}

First, we determine the minimal capacity cut. We thus map the expansion point to infinity by taking $y=\frac{1}{x}$, keeping one of the  branch points at $y=-1$ and placing the other at the origin $y=0$. Geometrically, the configuration that minimizes capacity between two points is simply the straight line segment connecting them; any curved path would increase the spatial extent, consequently lowering the energy and increasing the capacity. Thus, the minimal cut in the $y$-plane is the straight segment $[-1, 0]$. Mapping it back to the original coordinate $x$, this corresponds to the straight line $(-\infty, -1]$ on the real axis, hence this is where we expect the poles to accumulate. We confirm this in Figure \ref{fig:padepoles}.

Then, to study the error we look for the Green's function. For simplicity, we map the cut plane to the unit disk; the Green's function will transform like a scalar under this conformal mapping\footnote{A way to see this is to consider the fact that this is the Green's function of a free massless scalar in 2D, which has null scaling dimension.}. 
\begin{equation}
    w=\frac{\sqrt{1+x}-1}{\sqrt{1+x}+1}\,.
\end{equation}
Under this mapping, the expansion point remains the origin $w=0$, the branch cut is the circumference of unit radius $|w|=1$, and the whole domain $\Omega=\mathbb{C}\backslash(-\infty,-1]$ is mapped to the unit disk $\tilde\Omega=\{|w|<1\}$. In the $w$-plane, the Green's function with boundary conditions (\ref{eq:greensbc}) is then trivially
\begin{equation}
    \tilde g_\Omega(w,0)=-\log|w|\,.
\end{equation}
It is harmonic everywhere in $\Omega$, it has the right singularity and it vanishes on the cut. By conformal invariance, the Green's function in the $x$-plane is thus 
\begin{equation}
    g_\Omega(x,0)=-\log|w(x)|=-\log\left|\frac{\sqrt{1+x}-1}{\sqrt{1+x}+1}\right|
\end{equation}
Substituting this back into Stahl's formula for the error, we get
\begin{equation}
    \lim_{m\to\infty}|f(x)-[m]_f(x)|^\frac{1}{2m} = \left|\frac{\sqrt{1+x}-1}{\sqrt{1+x}+1}\right|\,, \qquad \text{in capacity in $\mathbb{C}\backslash(-\infty,-1]$}\,.
    \label{eq:errPade}
\end{equation}
This implies that the error of the diagonal Pad\'e approximant decays asymptotically as $|w(x)|^{2m}$.  Plugging in $x=1$ and $m=4$, we match the error we found in (\ref{eq:logerr}):
\begin{equation}
    \left|\frac{\sqrt{2}-1}{\sqrt{2}+1}\right|^{8}\sim 7\times 10^{-7}\,.
\end{equation}
We discussed some generalities and then studied the example of the logarithm. In the following we will show that the sums in the flow equations, at least in free theory, behave quite similarly to the example of the logarithm. 

\subsection{Pad\'e for exponential convergence in free theory}
\subsubsection{Flow of scaling dimensions}
In free theories in AdS, the scaling dimensions of scalar boundary operators take the form
\begin{equation}
    \Delta_{q,n}=q\Delta_1+2n\,,\qquad q,n\in\mathbb{N}
    \label{eq:qnspectrum}
\end{equation}
where $\Delta_1$ is the lightest operator above the identity and generically there are multiple degenerate operators with the same $\Delta_{q,n}$. Since the values of $q$ appearing in the flow equations depend on which bulk deformation we are considering, in free theories we only have to deal with $q=2$. We are thus left with sums over $n$, which as we discussed are divergent when using standard blocks. We thus introduce the following function $f(x)$ constructed from the summands in the scaling dimension flow equation (\ref{eq:first_eq_scalar}) with standard integrated blocks
\begin{equation}
    f_i(x)\equiv \sum_{n=0}^\infty x^n b_{[\phi^2]_{n,0}}^{\hat\phi^2}C_{[\phi^2]_{n,0}ii}\mathcal{I}(2\Delta_\phi+2n)\,.
    \label{eq:qndefinition}
\end{equation}
We are interested in the analytic structure of $f_i(x)$. In section \ref{sec:checks} we studied the specific cases of the flow equations evolving $\Delta_\phi$, $\Delta_{\phi^2}$ and $\Delta_{\phi^3}$ in free scalar theory. In all of them, the OPE coefficients $C_{ii[\phi^2]_{n}}$ exhibit alternating signs. In appendix \ref{app:QFTdata}, we found that $C_{\phi^m\phi^m[\phi^2]_n}=mC_{\phi\phi[\phi^2]_n}$, hence at least for every operator that is a pure monomial of the boundary $\phi$, this property holds. Remarkably, for this class of operators, $f_i(x)$ can be resummed exactly:
\begin{equation}
\begin{aligned}
    f_{\phi^m}(x)=\frac{m\Gamma(2\Delta_\phi-\frac{d}{2})}{\Gamma(\Delta_\phi+1)\Gamma(\Delta_\phi-\frac{d-2}{2})}\Bigg(&\ _2F_1\left(\begin{matrix}
        \Delta_\phi, & 2\Delta_\phi-\frac{d}{2}\\
        & \Delta_\phi+1
    \end{matrix};-x\right)\\
    &+\frac{2\Delta_\phi}{2\Delta_\phi-d}\ _2F_1\left(\begin{matrix}
        \Delta_\phi-\frac{d}{2}, & 2\Delta_\phi-\frac{d}{2} \\ & \Delta_\phi-\frac{d-2}{2}
    \end{matrix};-x\right)\Bigg)
\end{aligned}
\end{equation}
showcasing a branch cut along the negative real axis. $f_i(x)$ is thus analytic at the physical evaluation point $x=1$, where it gives 
\begin{equation}
    f_{\phi^m}(1)=\frac{2m}{2\Delta_\phi-d}=\frac{d\Delta_{\phi^m}}{d\lambda}\,.
\end{equation}
\begin{figure}
    \centering
    \includegraphics[width=0.75\linewidth]{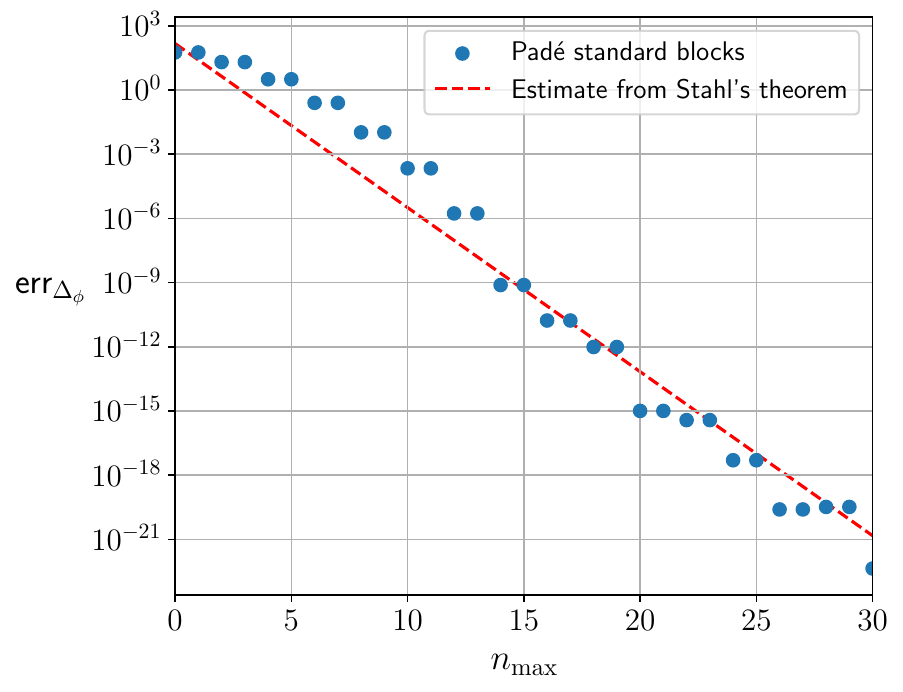}
    \caption{Comparison of the error (defined in (\ref{eq:deferror})) of the Pad\'e approximants of the partial sums in the flow equation (\ref{eq:flowdDeltaphi}) and the estimate from Stahl's theorem (\ref{eq:estimateerror}).}
    \label{fig:padefit}
\end{figure}
While the series expansion around $x=0$ diverges at $x=1$, due to the cut at $x=-1$, the underlying function is finite there. We are exactly in a situation analogous to the toy example (\ref{eq:log(1+x)}), where an originally divergent sum is analytically continued by Pad\'e approximants.  The error of the Pad\'e approximants should scale the same way
\begin{equation}
    \text{err}_{\Delta_\phi}\sim |w(1)|^{l_{\text{max}}}=\left|\frac{\sqrt{2}-1}{\sqrt{2}+1}\right|^{l_{\text{max}}}\sim e^{-1.765 l_{\text{max}}}\,.
    \label{eq:estimateerror}
\end{equation}
We compare this expectation with numerical results in figure \ref{fig:padefit}, and find that they approximately saturate this estimate.

We can ask ourselves whether Pad\'e approximants also improve the convergence of this sum when using local blocks. In that case, the explicit expression of $\mathcal{I}^{\alpha}(\Delta_{[\phi^2]_n})$ happens to compensate the alternation of signs in $C_{ii[\phi^2]_n}$, leading to a linear combination of hypergeometric functions with argument $+x$, with an explicit form that is unimportant and cumbersome. The function thus has a branch point at $x=1$, but it is also finite there. Stahl's theorem thus cannot guarantee exponential convergence at $x=1$, explaining why we observed no improvement when using Pad\'e approximants on partial sums of integrated local blocks, for example in figure \ref{fig:DeltaError}.
\subsubsection{Flow of BOE coefficients}
Now consider the sums involved in the flow equation of the BOE coefficients (\ref{eq:flowequations}).
The formula involves a double series, and the theory of Pad\'e approximants of two variables is not fully developed. Instead, we will take the approach of using Pad\'e sequentially. We will introduce two auxiliary variables $x$ and $y$
\begin{equation}
\begin{aligned}
    f_i(x,y)&\equiv\sum_{m=0}^\infty b^{\hat\phi^2}_{[\phi^2]_{m,0}}y^{m}f_{i,m}(x)\,,\\
    f_{i,m}(x)&\equiv\sum_{n=0}^\infty b^{\hat\phi^2}_{[\phi^2]_{n,0}}C_{i,[\phi^2]_n[\phi^2]_m}\mathcal{J}_{\Delta_i}(2\Delta_\phi+2n,2\Delta_\phi+2m)x^{n}
\end{aligned}
\end{equation}
We will first compute the Pad\'e approximant of $f_{i,m}(x)$ around $x=0$. Then, after taking $x\to1$, we do the Pad\'e approximant of $f_i(1,y)$ around $y=0$ and finally take $y\to1$.

Let us focus on the case $\mathcal{O}_i=\phi^2$, which we studied in section \ref{sec:checks}. In this case, the elements with $m=0$ have to be treated carefully through the prescription outlined in (\ref{eq:regJ}). Let us first write in closed form the rest of the sum: 
\begin{align}
    &\frac{f_{\phi^2}(x,y)-b^{\hat\phi^2}_{\phi^2}f_{\phi^2,0}(x)}{\pi^{-\frac{d}{2}}\Gamma(2\Delta_\phi-\frac{d}{2})y}\sqrt{2}\Gamma\left(\Delta_\phi-\frac{d-2}{2}\right)\Gamma\left(\Delta_\phi-\frac{d-4}{2}\right)\nonumber\\
    &=\Bigg(\ _2F_1\left(\begin{matrix}
        \Delta_\phi, & 2\Delta_\phi-\frac{d}{2}\\
        & \Delta_\phi+1
    \end{matrix};-x\right)
    +\frac{2\Delta_\phi}{2\Delta_\phi-d}\ _2F_1\left(\begin{matrix}
        \Delta_\phi-\frac{d}{2}, & 2\Delta_\phi-\frac{d}{2} \\ & \Delta_\phi-\frac{d-2}{2}
    \end{matrix};-x\right)\Bigg)\\
    &\ \times\Bigg(\ _3F_2\left(\begin{matrix}
        1, & 1, & \Delta_\phi+1\\
        & 2, & \Delta_\phi-\frac{d-4}{2}
    \end{matrix};-y\right)
    +\frac{1}{2\Delta_\phi-\frac{d-2}{2}}\ _3F_2\left(\begin{matrix}
        1, &\Delta_\phi+1, & 2\Delta_\phi-\frac{d-2}{2} \\ & \Delta_\phi-\frac{d-4}{2}, & 2\Delta_\phi-\frac{d-4}{2}
    \end{matrix};-y\right)\Bigg)\nonumber
\end{align}
\begin{figure}
    \centering
    \includegraphics[width=0.75\linewidth]{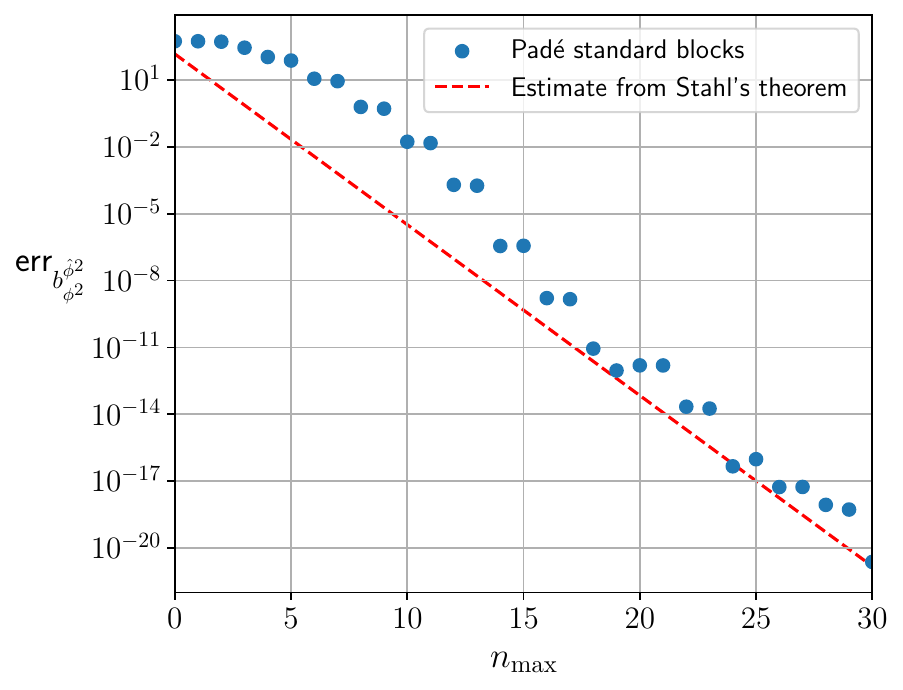}
    \caption{Comparison of the error (defined in (\ref{eq:deferror})) of the Pad\'e approximants of the partial sums in the flow equation (\ref{eq:flowbphi2phi2}) and the estimate from Stahl's theorem (\ref{eq:estimateerror}).}
    \label{fig:padefitb}
\end{figure}
We are not able to find a closed form for $f_{\phi^2,0}(x)$. Assuming it does not spoil the analytic structure of $f_{\phi^2}(x,y)$, we observe that, the full function seems to have separate cuts at $x\in(-\infty,-1]$ and $y\in(-\infty,-1]$, and that the two series in $m$ and $n$ factorize (except for $m=0$ terms). This suggests that the Pad\'e procedure can factorize on the two sums. 
 Due to the analytic structure in $x$ and $y$ being completely analogous to the scaling dimension case, we expect similar convergence with the truncation for both Pad\'e sums. We choose equal truncations on $n$ and $m$ and compare the error of the double truncated Pad\'e sum with the estimate from Stahl's theorem in figure \ref{fig:padefitb}, verifying these claims.

\subsection{Proposal for Pad\'e in interacting theories}
\label{subsec:criteria}
In this section we will propose a Pad\'e procedure for interacting theories which are deformations of free theories
\begin{equation}
    S(\lambda)=S_{\text{free}}+\lambda\int_{\text{AdS}}\hat\Phi\,.
    \label{eq:freedeform}
\end{equation}
We have no proof that the proposed procedure converges in interacting theories, hence the contents of this section should be taken as a conjecture to be tested.

In theories of the type (\ref{eq:freedeform}), the spectrum at $\lambda=0$ is organized as in (\ref{eq:qnspectrum}). For finite $\lambda$, degeneracies are lifted and dimensions are not integer spaced, but each state can be continuously traced back as originating from a certain $(q,n)$ family at $\lambda=0$. The generalization of (\ref{eq:qndefinition}) we propose is thus to group operators by their original free theory family and introduce two auxiliary variables
\begin{equation}
    f_i(x,y)\equiv \sum_{n=0}^\infty x^n\sum_{q=1}^\infty y^q\sum_{l\in(q,n)\ \text{family}}b_{l}^{\hat\Phi}C_{lii}\mathcal{I}(\Delta_{l})\,,
    \label{eq:interactqndef}
\end{equation}
where ``$l\in(q,n)$ family" means $\Delta_l(\lambda\to0)=q\Delta_1+2n$.  Notice that for finite $\lambda$ the sum over $q$ does not truncate. For small $\lambda$, BOE coefficients that were zero in free theory will be highly suppressed, possibly favoring the convergence of the sum over $q$, and signs will keep alternating in $n$ by continuity. But at large values of $\lambda$, nothing guarantees these facts. If the signs are distributed like independent random signs, theorems by Paley and Zygmund \cite{Paley_Zygmund_1930,Paley_Zygmund_1932} state that the function would almost surely\footnote{Here ``almost surely" should be intended in the sense of probability theory.} have natural boundaries at $|x|=1$ and $|y|=1$. In that case it would be impossible for Pad\'e to analytically continue the function to $x=1,y=1$\footnote{These theorems technically apply to functions of a single variable. We expect the situation can only be worse with two variables.}.

In a generic CFT, the eigenstate thermalization hypothesis (ETH) \cite{Deutsch:2018ulr} would lead us to expect this pseudo-random behavior for the signs of the OPE coefficients. However, we are studying in particular CFTs that live on the boundary of AdS. Bulk locality has been shown to imply nontrivial constraints on CFT data \cite{Maldacena:2015iua,Li:2017lmh,Levine:2023ywq}. So we do not exclude the possibility that the summands in (\ref{eq:interactqndef}) will present a structured sign modulation for high values of $q$ and $n$. If, in addition, $f_i$ could be analytically continued to $x=y=1$ with no singularity there, Stahl's theorem would apply to each step of the sequential procedure. Whether this happens in interacting theories, is an open question.

\section{Merger-annihilation and level repulsion}
\label{sec:mergerrepulsion}
The flow equations (\ref{eq:flowequations}) exhibit many interesting properties. In 2D \cite{Loparco:2026fki} it was shown that they generically imply level repulsion, the phenomenon by which different curves of the QFT data, when approaching each other, effectively repel; and merger-annihilation, where one boundary operator hitting marginality $(\Delta_i=d)$ generically implies all the QFT data exhibits a square root-like behavior and  becomes complex after a critical value of $\lambda$. Here we will generalize these facts to higher dimensions, closely following the arguments in \cite{Loparco:2026fki}.

The equations in this section are valid when the sums converge. That means they are valid when using local blocks $\mathcal{I}^{\alpha}, \mathcal{J}^{\alpha}$ or the Pad\'e procedure. We simply indicate $\mathcal{I}$ and $\mathcal{J}$, omitting the superscripts. We also focus specifically on scalar states, but we expect the generalization to spinning states to be straightforward.

\subsection{Merger-annihilation}
Throughout the paper, we have assumed all boundary operators are irrelevant, $\Delta_i>d$. Now let us imagine that the lightest operator above the identity, with dimension $\Delta_1(\lambda)$, approaches $d$ from above at some critical $\lambda_c$. Here we assume this happens when we approach $\lambda_c$ from below, but identical formulae with adjusted signs can be derived in the scenario where marginality is reached as $\lambda\to\lambda_c$ from above.

\subsubsection{Scaling dimensions near merger-annihilation}
We start by observing that the explicit forms of $\mathcal{I}(\Delta)$ (\ref{eq:IDelta}) and $\mathcal{I}^\alpha(\Delta)$ (\ref{eq:IDeltaalpha})  have a pole at $\Delta=d$ with residue Vol$(S^{d-1})$. When $\Delta_1\to d$, the whole sum over boundary operators is thus dominated by the lightest term. If we study the derivative of $\Delta_1$ we find (assuming $b^\Phi_1\neq0$ and $C_{111}\neq0$)
\begin{equation}
    \frac{d\Delta_1}{d\lambda}\approx\frac{\text{Vol}(S^{d-1})b^{\hat\Phi}_1C_{111}}{\Delta_1-d}\quad \implies\quad \Delta_1(\lambda)\approx d\pm\sqrt{2(\lambda-\lambda_c)\text{Vol}(S^{d-1})b^{\hat\Phi}_1C_{111}}
\end{equation}
where the QFT data on the right-hand side is evaluated at $\lambda=\lambda_c$. Notice that $b^\Phi_1C_{111}$ is negative by construction because $\Delta_1$ decreases as $\lambda$ increases in this setup.  

The square root behavior is typical of fixed-point annihilation \cite{Kaplan:2009kr, Gorbenko:2018ncu,Gorbenko:2018dtm}. We see that for $\lambda<\lambda_c$ there are two sets of consistent QFT data, merging at $\lambda=\lambda_c$. One of the two branches involves an operator that is relevant, going beyond our initial assumptions. We will discuss theories with boundary relevant operators in section \ref{subsubsec:boundaryrelevants}. For $\lambda>\lambda_c$, the data becomes complex, so there is no unitary QFT that is described by the analytic continuation of the QFT data curves to this regime of the coupling.

The rest of the scaling dimensions generically inherits the same square root behavior, even though they're far from marginality. That is because $\mathcal{O}_1$ appears in the sums in the other flow equations as well, leading to
\begin{equation}
    \frac{d\Delta_i}{d\lambda}\approx\frac{\text{Vol}(S^{d-1})b^{\hat\Phi}_1C_{1ii}}{\Delta_1-d}\quad \implies\quad \Delta_i(\lambda)\approx \Delta_i\pm C_{1ii}\sqrt{2(\lambda-\lambda_c)\frac{\text{Vol}(S^{d-1})b^{\hat\Phi}_1}{C_{111}}}\,.
\end{equation}
In a theory with a global symmetry we expect merger-annihilation only when a singlet hits marginality.
This is consistent with our results, since $b^{\hat\Phi}_1=0$ if $\mathcal{O}_1$ is not a singlet  because $\hat\Phi$ only creates singlets. 
\subsubsection{BOE coefficients near merger-annihilation}
The same phenomenon takes place for the BOE coefficients. The coefficients $\mathcal{J}_{\Delta_i}(\Delta_l,\Delta_j)$ (\ref{eq:JDelta}) also have a pole at $\Delta_l=d$. This time, the residue nontrivially depends on $\Delta_i$ and $\Delta_j$
\begin{equation}
\begin{aligned}
    F_{\Delta_i}(\Delta_j)&\equiv\underset{\Delta_l=d}{\text{Res}}\mathcal{J}_{\Delta_i}(\Delta_l,\Delta_j)\\
    &=-\oint_{\Delta_j}\frac{d\Delta}{2\pi i}\frac{1}{\Delta-\Delta_j}\frac{2 \pi ^{d/2} \Gamma \left(\Delta-\frac{d-2}{2}\right)}{(\Delta-\Delta_i) \Gamma \left(\frac{d-\Delta_i+\Delta}{2}\right) \Gamma \left(\frac{\Delta_i+\Delta+2-d}{2}\right)}
\end{aligned}
\end{equation}
where the contour integral here has nothing to do with the residue at $\Delta_l=d$, but rather the regulated form of $\mathcal{J}$ (\ref{eq:regJ}).

We thus get, in the limit where $\Delta_1\to d$,
\begin{equation}
    \frac{db^{\hat\Phi}_i}{d\lambda}\approx\frac{b^{\hat\Phi}_1}{\Delta_1-d}\sum_j b^{\hat\Phi}_jC_{ij1}F_{\Delta_i}(\Delta_j)
\end{equation}
implying
\begin{equation}
    b^{\hat\Phi}_i(\lambda)\approx b^{\hat\Phi}_i(\lambda_c)\pm \sqrt{\frac{2(\lambda-\lambda_c)b^{\hat\Phi}_1}{\text{Vol}(S^{d-1})C_{111}}}\sum_j b^{\hat\Phi}_jC_{ij1}F_{\Delta_i}(\Delta_j)\,,
\end{equation}
where again all the QFT data on the right hand side is evaluated at $\lambda=\lambda_c$.
\subsubsection{OPE coefficients near merger-annihilation}
We expect the same phenomenon to take place for OPE coefficients. In AdS$_2$, this was shown explicitly in \cite{Loparco:2026fki}. We did not derive an ODE that couples the derivative of the OPE coefficients to the rest of the QFT data in higher dimensions, instead we use the crossing equation to fix OPE coefficients independently. 

In appendix \ref{app:OPEflow}, we nevertheless study certain properties of such an ODE. Concretely, we find that for triplets $ijk$ of scalar operators,
\begin{equation}
    \frac{dC_{ijk}}{d\lambda}=\sum_lb^{\hat\Phi}_l\sum_mC_{lim}C_{mjk}\mathcal{K}_{\Delta_i\Delta_j\Delta_k}(\Delta_l,\Delta_m)+\text{perms.}+\text{spin}
    \label{eq:CijkODE}
\end{equation}
where ``perms." are the cyclic permutations of the $i,j,k$ indices, and ``spin" indicates contributions from exchanged spinning operators, which we will not need to consider here, and $\mathcal{K}$ is an integrated conformal block which we do not compute explicitly.

The main property we need here is that, as we argued in appendix \ref{app:OPEflow}, the kinematic coefficients $\mathcal{K}$ must have a pole at $\Delta_l=d$. Following the discussion for scaling dimensions and BOE coefficients, this immediately implies
\begin{equation}
    C_{ijk}(\lambda)\approx C_{ijk}(\lambda_c)\pm\sqrt{\frac{2(\lambda-\lambda_c)b^{\hat\Phi}_1}{\text{Vol}(S^{d-1})C_{111}}}\sum_mC_{lim}C_{mjk}H_{\Delta_i\Delta_j\Delta_k}(\Delta_m)+\text{perms.}+\text{spin}
\end{equation}
where we do not know the explicit expression of $H\equiv\underset{\Delta_l=d}{\text{Res}}\mathcal{K}$ in $d+1>2$.

\begin{figure}
    \centering
    \includegraphics[width=0.75\linewidth]{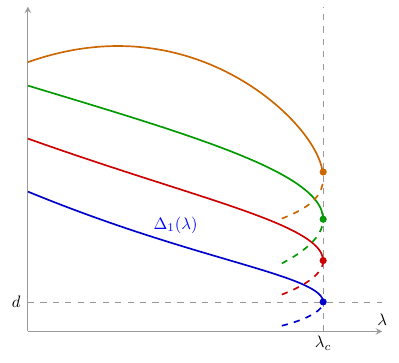}
    \caption{When the scaling dimension $\Delta_1$ of the lightest boundary operator above the identity hits marginality, generically all the QFT data has a square root behavior. The curves can in principle be prolonged along the other branch of the square root, and a new consistent set of QFT data is found. They can also be prolonged for $\lambda>\lambda_c$ leading to complex QFT data, we do not show that in this figure. The curves above $\Delta_1(\lambda)$ can represent any other piece of QFT data: $\Delta_i(\lambda)$, $b^{\hat\phi}_i(\lambda)$, $C_{ijk}(\lambda)$.}
    \label{fig:mergerannihilate}
\end{figure}
\subsection{Level repulsion}
\label{subsec:levelrepulsion}
When a Hamiltonian depends on a continuous parameter $\lambda$, its eigenvalues are known to generically experience level repulsion: when two eigenvalues get close they repel each other, and never cross\footnote{They can cross if the two energy eigenstates belong to different charge sectors of the theory.} \cite{1929PhyZ...30..467V}. The scaling dimensions of boundary operators in QFT in AdS are the eigenvalues of the Hamiltonian in cylinder coordinates, and as $\lambda$ varies, they experience level repulsion too. It requires a few steps to prove this. We follow the arguments in \cite{Loparco:2026fki}, themselves inspired by \cite{Behan:2017mwi}.

Consider two boundary operators $\mathcal{O}_1$ and $\mathcal{O}_2$, which have scaling dimensions $\Delta_1$ and $\Delta_2$ such that $0<\Delta_{21}\ll|\Delta_{ij}|$ for all other pairs $ij$ (here we use the notation $\Delta_{ij}\equiv\Delta_i-\Delta_j$). We want to show that 
\begin{equation}
    \frac{d^2\Delta_{21}}{d\lambda^2}=\frac{4c^2}{\Delta_{21}(\lambda_0)^3}+O(\Delta_{21}^{-2})\,, \qquad c\in\mathbb{R}\,.
\end{equation}
This is enough to argue for level repulsion: as $\Delta_{21}\to0$ from above, the singularity on the right hand side of this equation acts as a potential barrier which prevents the difference of scaling dimensions from ever becoming $0$ and flipping sign.

We start by taking a derivative of the scaling dimension flow equation for $\Delta_2$
\begin{equation}
    \frac{d^2\Delta_{2}}{d\lambda^2}=\sum_l\left(b_l^{\hat\Phi}\frac{dC_{22l}}{d\lambda}\mathcal{I}(\Delta_l)+\frac{db^{\hat\Phi}_l}{d\lambda}C_{22l}\mathcal{I}(\Delta_l)+b^{\hat\Phi}_lC_{22l}\frac{d\mathcal{I}(\Delta_l)}{d\lambda}\right)\,.
    \label{eq:scaling2ndder}
\end{equation}
Let us study each term in the limit $\Delta_1\to\Delta_2$. 

For the first term, we once again appeal to the fact that an ODE for $\frac{dC_{ijk}}{d\lambda}$ would have the form (\ref{eq:CijkODE}). As stated before, we do not have the explicit expression of $\mathcal{K}_{\Delta_i\Delta_j\Delta_k}(\Delta_l,\Delta_m)$ in $d+1>2$, but we do not need it for our arguments. The only fact we need is that $\mathcal{K}$ must possess a pole at $\Delta_m=\Delta_i$, with residue proportional to $\mathcal{I}$. We prove this in appendix \ref{app:OPEflow}. Near this pole, it must behave as
\begin{equation}
    \mathcal{K}_{\Delta_i\Delta_j\Delta_k}(\Delta_l,\Delta_m)\sim-\frac{\mathcal{I}(\Delta_l)}{\Delta_m-\Delta_i}+O(1)\sim\mathcal{K}_{\Delta_i\Delta_k\Delta_j}(\Delta_l,\Delta_m)\,.
\end{equation}
Now, the sum over $\Delta_m$ contains both $\Delta_1$ and $\Delta_2$. The exact pole at $\Delta_m=\Delta_2$ would be removed by renormalization, just like it happens for $\mathcal{J}$ (explained in appendix \ref{sec:sec_eq}). The approximate singularity as $\Delta_1\to\Delta_2$, though, is physical and dominates $\frac{dC_{22l}}{d\lambda}$. Taking this into account, we get
\begin{equation}
\begin{aligned}
    \sum_lb^{\hat\Phi}_l\frac{dC_{22l}}{d\lambda}\mathcal{I}(\Delta_l)&=  \sum_lb^{\hat\Phi}_l\mathcal{I}(\Delta_l)\sum_p\sum_m b^{\hat\Phi}_p\Bigg[2C_{p2m}C_{m2l}\mathcal{K}_{\Delta_2\Delta_2\Delta_l}(\Delta_p,\Delta_m)\\
    &\qquad\qquad\qquad\qquad\qquad\qquad+C_{plm}C_{m22}\mathcal{K}_{\Delta_l\Delta_2\Delta_2}(\Delta_p,\Delta_m)\Bigg]\\
    &\approx \frac{1}{\Delta_{21}}\Omega_{12}\left[2\Omega_{12}+b^{\hat\Phi}_2C_{122}\mathcal{I}(\Delta_2)-b^{\hat\Phi}_1C_{222}\mathcal{I}(\Delta_1)\right]
    \label{eq:1sttermomega}
\end{aligned}
\end{equation}
where we defined the shorthand
\begin{equation}
    \Omega_{12}\equiv\sum_l b^{\hat\Phi}_lC_{12l}\mathcal{I}(\Delta_l)\,.
\end{equation}
In the following we will assume that $\Omega_{12}$ is nonzero.

Now let us focus on the second term.
\begin{equation}
    \sum_l\frac{db^{\hat\Phi}_l}{d\lambda}C_{22l}\mathcal{I}(\Delta_l)=\sum_lC_{22l}\mathcal{I}(\Delta_l)\sum_{j}\sum_k b^{\hat\Phi}_jb^{\hat\Phi}_kC_{ljk}\mathcal{J}_{\Delta_l}(\Delta_j,\Delta_k)
\end{equation}
The coefficient $\mathcal{J}_{\Delta_l}(\Delta_j,\Delta_k)$ (\ref{eq:JDelta}) has the following analytic structure near $\Delta_k=\Delta_l$:
\begin{equation}
    \mathcal{J}_{\Delta_l}(\Delta_j,\Delta_k)=-\frac{\mathcal{I}(\Delta_j)}{\Delta_k-\Delta_l}+O(1)
\end{equation}
so that in the limit $\Delta_1\to\Delta_2$, the sums are dominated by $l=1,k=2$ and $l=2,k=1$, giving 
\begin{equation}
    \sum_l\frac{db^{\hat\Phi}_l}{d\lambda}C_{22l}\mathcal{I}(\Delta_l)\approx-\frac{\Omega_{12}}{\Delta_{21}}\left(b^{\hat\Phi}_2C_{221}\mathcal{I}(\Delta_1)-b^{\hat\Phi}_1C_{222}\mathcal{I}(\Delta_2)\right)
    \label{eq:2ndtermomega}
\end{equation}
Finally, the third term in (\ref{eq:scaling2ndder}) is not divergent as $\Delta_1\to\Delta_2$, hence it is subleading to the first two. 

Now, plugging (\ref{eq:2ndtermomega}) and (\ref{eq:1sttermomega}) into (\ref{eq:scaling2ndder}), and slightly rearranging terms, we get
\begin{equation}
    \frac{d^2\Delta_2}{d\lambda^2}\approx\frac{1}{\Delta_{21}}\left[2\Omega_{12}^2+\Omega_{12}\left(b^{\hat\Phi}_2C_{221}+b^{\hat\Phi}_1C_{222}\right)\left(\mathcal{I}(\Delta_2)-\mathcal{I}(\Delta_1)\right)\right]\approx\frac{2\Omega_{12}^2}{\Delta_{21}}\,.
\end{equation}
Repeating again for $\frac{d^2\Delta_1}{d\lambda^2}$ and taking the difference, we get
\begin{equation}
    \frac{d^2\Delta_{21}}{d\lambda^2}\approx\frac{4\Omega_{12}^2}{\Delta_{21}}\,,
    \label{eq:d2delta21}
\end{equation}
\begin{figure}
    \centering
    \includegraphics[width=0.6\linewidth]{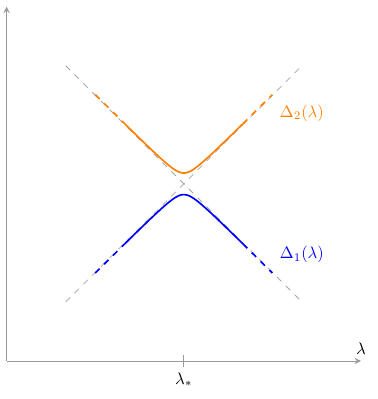}
    \caption{When two scaling dimension get close to each other, they repel. This is the phenomenon of level repulsion, predicted by the flow equations, and for $\lambda\sim \lambda_*$ it is captured by equation (\ref{eq:levelrepulsion}). After the repulsion event, the two curves exchange slopes.}
    \label{fig:levelrepulsion}
\end{figure}
and what remains to be done is to determine the scaling of $\Omega_{12}^2$ as $\Delta_{21}\to0$. To do that, consider that
\begin{equation}
\begin{aligned}
    \frac{d\Omega_{12}}{d\lambda}&=\sum_l\left[\frac{db^{\hat\Phi}_l}{d\lambda}C_{12l}\mathcal{I}(\Delta_l)+b^{\hat\Phi}_l\frac{dC_{12l}}{d\lambda}\mathcal{I}(\Delta_l)+b^{\hat\Phi}_lC_{12l}\frac{d\mathcal{I}(\Delta_l)}{d\lambda}\right]\\
    &\approx  \frac{\Omega_{12}}{\Delta_{21}}\left[(b^{\hat\Phi}_1C_{122}+b^{\hat\Phi}_2C_{112})(\mathcal{I}(\Delta_2)-\mathcal{I}(\Delta_1))-\frac{d\Delta_{21}}{d\lambda}\right]\\
    &\approx-\frac{\Omega_{12}}{\Delta_{21}}\frac{d\Delta_{21}}{d\lambda}\,,
    \label{eq:dOmega12}
\end{aligned}
\end{equation}
where to go from the first to the second line we repeated the steps in (\ref{eq:1sttermomega}) but now for $C_{12l}$ instead of $C_{22l}$, and we used the scaling dimension flow equation to repackage some of the terms as $\frac{d\Delta_{21}}{d\lambda}$.

Formula (\ref{eq:dOmega12}) implies that $\Omega_{12}\sim c/\Delta_{21}$. Plugging it into (\ref{eq:d2delta21}) we get the differential equation we wanted to prove
\begin{equation}
    \frac{d^2\Delta_{21}}{d\lambda^2}\approx\frac{4c^2}{\Delta_{21}^3}
\end{equation}
This differential equation can be solved, leading to an approximate solution valid when $0<\Delta_{21}\ll|\Delta_{ij}|$:
\begin{equation}
    \Delta_{21}(\lambda)\approx\sqrt{c_1(\lambda-\lambda_*)^2+\frac{4c^2}{c_1}}
    \label{eq:levelrepulsion}
\end{equation}
where $c_1$ and $\lambda_*$ are integration constants. This implies that, in the vicinity of $\lambda\approx \lambda_*$, the two curves exchange slopes after experiencing level repulsion.
\section{Discussion}
\label{sec:discussion}
Let us discuss some further properties of the flow equations and some open research directions which shall be explored in future work.
\subsection{Beyond our assumptions}
Let us start by discussing extensions of our framework beyond the original assumptions. In particular, we will discuss the effect of having relevant boundary operators and bulk operators which violate assumption (\ref{eq:nobulkuv}).

As in the previous section, we use $\mathcal{I}$ and $\mathcal{J}$ to indicate the integrated blocks. In practice one should use their local versions $\mathcal{I}^\alpha$ and $\mathcal{J}^\alpha$ or Pad\'e approximants to make the sums convergent, and any equation we write is intended implicitly with one of the two methods applied.
\subsubsection{Relevant boundary operators}
\label{subsubsec:boundaryrelevants}
Throughout this work we have assumed that all boundary operators have irrelevant scaling dimensions $\Delta_i>d$. In section \ref{sec:mergerrepulsion}, we studied what happens when the lightest operator hits marginality, $\Delta_i\to d$. More generally, one can ask whether the flow equations are valid when there are multiple relevant operators. The discussion is totally analogous with the 2D case \cite{Loparco:2026fki}:

Consider the integrals of correlation functions which we computed in section \ref{sec:derivationflowequations} to derive the flow equations. The presence of relevant boundary operators would introduce power-law IR divergences. Introducing a cutoff $\delta$ near the boundary of AdS in Poincar\'e coordinates, we would have
\begin{equation}
    \int_{z>\delta}\frac{dzd^d\mathbf{x}}{z^{d+1}}\langle\hat\Phi(z,\mathbf{x})\cdots\rangle=\sum_{\Delta_l<d}\frac{\delta^{\Delta_l-d}}{d-\Delta_l}b^{\hat\Phi}_l\int d^d\mathbf{x}\langle\mathcal{O}_l(\mathbf{x})\cdots\rangle+\text{finite as }\delta\to0\,.
    \label{eq:IRdeltadivergences}
\end{equation}
These divergences can be removed by adding a boundary action with appropriate counterterms involving these relevant operators. Physically, this corresponds to considering a fine tuned RG flow in which conformal symmetry on the boundary is preserved. 

    Analytically continuing the scaling dimensions in our flow equations precisely corresponds to throwing away the divergences in (\ref{eq:IRdeltadivergences}) when deriving the integrals of blocks $\mathcal{I}$ and $\mathcal{J}$ appearing in (\ref{eq:flowequations}). Since, as was shown in \cite{Lauria:2023uca}, the AdS isometries prevent us from turning on relevant operators on the boundary independently of the bulk deformation, we are led to state that our flow equations also describe these types of fine-tuned flows in AdS\footnote{Examples of fine-tuned flows with relevant boundary operators are free theories with Neumann boundary conditions and more in general the flows defined by the prolongation of QFT data curves beyond merger-annihilation (like the dashed lines in figure \ref{fig:mergerannihilate}). A physically interesting example would be the so-called ``Dirichlet$^*$" boundary theory which merges and annihilates with the Dirichlet boundary conditions of Yang-Mills in four dimensions. \cite{Ciccone:2024guw,Ciccone:2025dqx}}.
\subsubsection{Bulk relevant operators beyond assumption (\ref{eq:nobulkuv})}
\label{subsubsec:bulkrelevants}
When deriving the BOE flow equation, in order to avoid extra UV divergences we made the assumption
\begin{equation}
    \Delta_{\hat\Phi}^{\text{UV}}<\frac{1}{2}\left(d+1+\min_{\hat{\mathcal{O}}\in\hat\Phi\times\hat\Phi}\Delta^{\text{UV}}_{\hat{\mathcal{O}}}\right)\,.
    \label{eq:UVassumption}
\end{equation}
We can give up this assumption by introducing a UV cutoff $\delta$ in the integral in (\ref{eq:bulkbulkint1}) and carefully renormalizing the bulk operators.  This would lead to flow equations involving the UV cutoff. 
\begin{equation}
    \frac{db^{\hat\Phi}_i}{d\lambda}=\lim_{\delta\to0}\left[\sum_{l,j}b^{\hat\Phi}_lb^{\hat\Phi}_jC_{lji}\mathcal{J}_{\Delta_i}^{(\delta)}(\Delta_l,\Delta_j)-\sum_{\hat{\mathcal{O}}}b^{\hat{\mathcal{O}}}_iC_{\hat\Phi\hat\Phi\hat{\mathcal{O}}}^{\text{UV}}\sum_{n=0}^\infty\frac{a_n}{\delta^{2\Delta_{\hat\Phi}^\text{UV}-\Delta_{\hat{\mathcal{O}}}^\text{UV}-d-1-2n}}\right]
    \label{eq:UVfloweq}
\end{equation}
where the sum over $\hat{\mathcal{O}}$ runs over bulk operators violating (\ref{eq:UVassumption}), $a_n$ are some kinematic coefficients and $\mathcal{J}^{(\delta)}$ is a regulated version of $\mathcal{J}$ with a UV cutoff $(z-z_2)^2+(\mathbf{x}-\mathbf{x}_2)^2>\delta^2$. In the edge case $\Delta_{\hat\Phi}^{\text{UV}}=\frac{1}{2}\left(d+1+\min_{\hat{\mathcal{O}}\in\hat\Phi\times\hat\Phi}\Delta^{\text{UV}}_{\hat{\mathcal{O}}}\right)$, the equation (\ref{eq:UVfloweq}) will also involve logarithms of the UV cutoff $\delta$.

Notice that now the flow equations do not close anymore. One would in fact need to flow the BOE coefficients of the bulk operators $\hat{\mathcal{O}}$ appearing in the sum in (\ref{eq:UVfloweq}). This can be remedied quickly, by running again our arguments in section \ref{sec:derivationflowequations}, to obtain
\begin{equation}
    \frac{db^{\hat{\mathcal{O}}}_i}{d\lambda}=\sum_l\sum_jb^{\hat\Phi}_lb^{\hat{\mathcal{O}}}_jC_{lji}\mathcal{J}_{\Delta_i}(\Delta_l,\Delta_j)
\end{equation}
where this time UV convergence is ensured if the lightest operator in the UV OPE of $\hat\Phi\times\hat{\mathcal{O}}$ is not lighter than $\hat{\mathcal{O}}$. 

In 2D \cite{Loparco:2026fki}, another possibility was presented, which would avoid the appearance of the UV cutoff in the flow equations.  The prescription is explicitly presented in section 5.3 there, we summarize it here and refer the reader to that section for a detailed justification.

The reasoning starts by observing that only a finite set of bulk operators $\hat{\mathcal{O}}$ will violate (\ref{eq:UVassumption}), and will thus appear in the sum in (\ref{eq:UVfloweq}). We can choose a basis of bulk operators $\hat{\mathcal{O}}_a$ that are ordered by their UV scaling dimensions $\Delta_a^{\text{UV}}\leq\Delta_{a+1}^{\text{UV}}$ and one for boundary operators such that they are ordered by boundary scaling dimensions $\Delta_i\leq\Delta_{i+1}$. Moreover, we are free to fix $b^{\hat{\mathcal{O}}_a}_i=0$ if $i<a$ and $b^{\hat{\mathcal{O}}_i}_i=1$. Now, the prescription is to shift the variations of the BOE coefficients iteratively as follows
\begin{equation}
    \delta b^{\hat{\mathcal{O}}_a}_i\to\delta b^{\hat{\mathcal{O}}_a}_i-\delta b_j^{\hat{\mathcal{O}}_a}b_i^{\hat{\mathcal{O}}_j}\,,\qquad \forall a\geq j\,,\quad \forall j\,.
    \label{eq:iterativestep}
\end{equation}
To give a concrete example, consider the case $a=1$, the lightest operator:
\begin{equation}
    \frac{db^{\hat{\mathcal{O}}_1}_i}{d\lambda}\to\frac{db^{\hat{\mathcal{O}}_1}_i}{d\lambda}-\frac{db^{\hat{\mathcal{O}}_1}_1}{d\lambda}b_i^{\hat{\mathcal{O}}_1}
\end{equation}
Using the flow equation (\ref{eq:UVfloweq}) for $\frac{db^{\hat{\mathcal{O}}_1}_1}{d\lambda}$ and the renormalization condition $b^{\hat{\mathcal{O}}_1}_1=1$ we obtain
\begin{equation}
    \frac{db^{\hat{\mathcal{O}}_1}_i}{d\lambda}=\sum_l\sum_j b^{\hat\Phi}_lb^{\hat{\mathcal{O}}_1}_j\left[C_{lij}\mathcal{J}_{\Delta_i}(\Delta_l,\Delta_j)-b_i^{\hat{\mathcal{O}}_1}C_{lj1}\mathcal{J}_{\Delta_1}(\Delta_l,\Delta_j)\right]
\end{equation}
which is now finite and does not involve the UV cutoff.

For the second lightest bulk operator, we must do
\begin{equation}
    \frac{db^{\hat{\mathcal{O}}_2}_i}{d\lambda}\to\frac{db^{\hat{\mathcal{O}}_2}_i}{d\lambda}-\frac{db^{\hat{\mathcal{O}}_2}_2}{d\lambda}b_i^{\hat{\mathcal{O}}_2}\to\frac{db^{\hat{\mathcal{O}}_2}_i}{d\lambda}-\frac{db^{\hat{\mathcal{O}}_2}_2}{d\lambda}b_i^{\hat{\mathcal{O}}_2}-\frac{db^{\hat{\mathcal{O}}_2}_1}{d\lambda}b_i^{\hat{\mathcal{O}}_1}+\frac{db^{\hat{\mathcal{O}}_2}_1}{d\lambda}b_2^{\hat{\mathcal{O}}_1}b_i^{\hat{\mathcal{O}}_2}
\end{equation}
obtaining again a finite flow equation
\begin{equation}
\begin{aligned}
    \frac{db^{\hat{\mathcal{O}}_2}_i}{d\lambda}=\sum_l\sum_j b^{\hat\Phi}_lb^{\hat{\mathcal{O}}_2}_j\Big[&C_{lij}\mathcal{J}_{\Delta_i}(\Delta_l,\Delta_j)-b_i^{\hat{\mathcal{O}}_2}C_{lj2}\mathcal{J}_{\Delta_2}(\Delta_l,\Delta_j)\\
    &\quad\quad-(b_i^{\hat{\mathcal{O}}_1}-b^{\hat{\mathcal{O}}_1}_2b^{\hat{\mathcal{O}}_2}_i)C_{lj1}\mathcal{J}_{\Delta_1}(\Delta_l,\Delta_j)\Big]
\end{aligned}
\end{equation}
This procedure can be applied to any bulk operator which violates assumption (\ref{eq:UVassumption}), and leads to finite flow equations free of the UV cutoff.
\subsubsection{Marginally relevant bulk deformation}
\label{subsubsec:marginallyrelevant}
Many interesting QFTs, such as the $O(N)$ model in 2D or Yang-Mills and QCD in 4D, have a deforming operator which is classically marginal $\Delta_{\hat\Phi}^{\text{UV}}=d+1$. In particular, in these examples the deforming operator is the lightest relevant scalar in the UV theory\footnote{For QCD, this is true for the specific case of exactly massless quarks.}. Such theories violate assumption (\ref{eq:nobulkuv}), leading to logarithmic UV divergences in the BOE flow equations. These can be cured with the same methods discussed in section \ref{subsubsec:bulkrelevants}, but this time we need to apply the iterative procedure (\ref{eq:iterativestep}) only once. We normalize $\hat\Phi$ by the condition $b^{\hat\Phi}_1=1$ where $\mathcal{O}_1$ is the lightest boundary operator in the BOE of $\hat\Phi$ above the identity. Then, the flow equations for the BOE coefficients of $\hat\Phi$ can be written as
\begin{equation}
    \frac{db^{\hat\Phi}_i}{d\lambda}=\sum_l\sum_j b^{\hat\Phi}_lb^{\hat{\Phi}}_j\Big[C_{lij}\mathcal{J}_{\Delta_i}(\Delta_l,\Delta_j)-b_i^{\hat{\Phi}}C_{lj1}\mathcal{J}_{\Delta_1}(\Delta_l,\Delta_j)\Big]\,,\qquad i\geq2\,.
\end{equation}
once again giving finite flow equations with no need to introduce a UV cutoff.
\subsection{Flow equations with no coupling}
The coupling $\lambda$ appearing in our flow equations is not a physical observable. To phrase our framework in a more scheme-independent way, we could express all the QFT data as curves parametrized by the lightest scaling dimension above the identity
    \begin{equation}
        \{
\Delta_i(\Delta_1),\rho_i, C_{ijk}^{(n)}(\Delta_1),b^{\hat\Phi(n)}_i(\Delta_1)\}
    \end{equation}
    If $\Delta_1$ evolves monotonically, the flow equations can be rephrased by a simple use of the chain rule 
\begin{equation}
\begin{aligned}
    \frac{d\Delta_i}{d\Delta_1}&=\frac{\sum_lb^{\hat\Phi}_lC_{lii}\mathcal{I}(\Delta_l)}{\sum_lb^{\hat\Phi}_lC_{l11}\mathcal{I}(\Delta_l)}\\
    \frac{db^{\hat\Phi}_i}{d\Delta_1}&=\frac{\sum_l\sum_jb^{\hat\Phi}_lb^{\hat\Phi}_jC_{lij}\mathcal{J}_{\Delta_i}(\Delta_l,\Delta_j)}{\sum_lb^{\hat\Phi}_lC_{l11}\mathcal{I}(\Delta_l)}
\end{aligned}
\end{equation}
while the crossing equations are already independent of $\lambda$ and implicitly dependent on $\Delta_1$.
\subsection{Open questions}
\begin{itemize}
    \item What is the minimal set of QFT data? The locality sum rules of \cite{Levine:2023ywq} imply nontrivial constraints among the BOE and OPE coefficients
        \begin{equation}
       \sum_l b^{\hat\Phi}_lC_{lij}\theta^{\alpha,n}_{\Delta_l\Delta_i\Delta_j}=0\,,\qquad \forall\mathcal{O}_i,\mathcal{O}_j\,,\alpha>\frac{\Delta_i+\Delta_j+\Delta_{\hat\Phi}^{\text{UV}}}{2}\,, n\in\mathbb{N}_0
        \label{eq:linearsumrules}
        \end{equation}
        where $\theta^{\alpha,n}_{\Delta_l\Delta_i\Delta_j}$ is a kinematic coefficient which may be found explicitly in \cite{Levine:2023ywq,Loparco:2025aag}. Can these sum rules be used to, for example, express all BOE coefficients in terms of OPE coefficients and scaling dimensions?
    \item Can we prove that a generic interacting QFT in AdS has QFT data with an asymptotically ordered sign modulation for large scaling dimensions, as discussed in section \ref{subsec:criteria}? That would ensure that Pad\'e approximants are a useful tool to provide good convergence of the sums in the flow equations also for interacting examples.
    \item To get a closed set of flow equations and crossing equations for AdS$_{d+1}$ with $d+1\geq5$, one has to derive the ODE involving $\frac{d\Delta_i}{d\lambda}$ for operators $\mathcal{O}_i$ carrying mixed symmetry representations $\rho_i$ of $SO(d)$. This may not be too hard, and as in the $SO(2)$ and $SO(3)$ case one may hope to get flow equations which have the same form for any representation $\rho_i$. 
    \item How do errors propagate when one evolves approximate data with the flow equations? Are these ODEs chaotic? Can we find efficient and controlled numerical algorithms to solve them numerically in interacting examples? For preliminary work in this direction, see \cite{Xiao2026}.
    \item Is it possible to derive the ODE for the OPE coefficients in closed form? For AdS$_4$, it should read
    \begin{equation}
        \frac{dC_{ijk}^{(c)}}{d\lambda}=\sum_lb^{\hat\Phi}_l\sum_m\sum_{a,b}C_{lim}^{(a)}C_{mjk}^{(b)}\mathcal{K}^{(a,b,c)}_{ijk}(\Delta_l,\Delta_m,J_m)+\text{perms.}
        \label{eq:spinODE}
    \end{equation}
    where ``perms." stands for cyclic $(ijk)$ permutations, $\mathcal{K}$ would be integrated bulk-boundary-boundary-boundary blocks and $\mathcal{O}_i$, $\mathcal{O}_j$ and $\mathcal{O}_k$ would generically need to be spinning operators in order to close the system of equations. Computing $\mathcal{K}$ is quite difficult: the blocks on the 3D boundary are not known in closed form, deriving the bulk-boundary-boundary-boundary block and its integral over AdS thus seems like a very hard task. For AdS$_3$, instead, 2D boundary blocks are known in closed form for all spins \cite{Osborn:2012vt}. The main bottleneck, which also affects the 2D case \cite{Loparco:2026fki}, is the fact that one can at best only derive an integral representation of $\mathcal{K}$ rather than a closed form expression. This slows down the implementation of the flow equations greatly, as one has to evaluate multivariable numerical integrals for each summand in (\ref{eq:spinODE}) at each step in the integration of the flow equations.
    \item It should be possible to derive flow equations describing the QFT data on a defect inserted on the boundary of AdS. This setup is most naturally realized when studying a Yang-Mills flux tube stretched between Wilson lines inserted at the boundary of AdS \cite{Gabai:2025hwf,Gabai:2026myo}. In that case, there exists a displacement operator at the boundary, which has been recently proved to be protected under the bulk RG flow even if there is no stress tensor in the boundary theory \cite{Qiao:2026ijh,Bianchi:2026sax}.
\end{itemize}
\subsection{Possible applications}
There are many QFTs of interest which we would like to attack with the flow equations. Let us list a few examples in 3D and 4D 
\begin{itemize}
    \item Yang-Mills in 3D. Following the theory with Neumann boundary conditions, we could try to extract the masses of the lightest glueballs from the asymptotics of $\Delta_i(\lambda)$ and compare with the lattice results \cite{Athenodorou:2016ebg}.
    \item QCD in 3D. The masses of the resonances are again interesting targets. Moreover, one can study the conformal window and the values of the number of colors $N_c$ and fermions $N_f$ for which the transition between a gapped and gapless IR phase happens \cite{Appelquist:1989tc} (which would be signaled by the $\Delta_i(\lambda)$ growing vs. asymptoting to constants respectively).
    \item $\lambda\phi^4$ (and more generally $O(N)$ model) in 3D. In the strictly massless case, the IR fixed point is the 3D critical Ising (respectively, $O(N)$) model. Flat space perturbation theory fails in this case, but our flow equations should hold. Since the IR fixed point is a nontrivial BCFT, we would expect the curves $\Delta_i(\lambda)$ to asymptote to the values of the flat space BCFT \cite{Giombi:2020rmc,Giombi:2025pxx,Csipes:2026nyo}. In principle, one can then extract estimates for the bulk critical exponents by fitting the asymptotic behavior of the BOE coefficients $b_l^{\hat{\mathcal{O}}}$ for large $\Delta_l$. It would be great to compare them to the islands obtained through the conformal bootstrap \cite{Kos:2016ysd}. For recent perturbative and large $N$ work on this theory in AdS, see \cite{Carmi:2026spv,Dujava:2025php}.
\item Similarly to the $O(N)$ case, the 3D Gross-Neveu model has a strongly coupled gapless IR phase, which would be interesting to study with the flow equations. Its data could be compared to \cite{Iliesiu:2017nrv,Giombi:2017hpr,Diatlyk:2026eta}.
\item  QED in 3D. Depending on the number of fermions $N_f$, the IR of this theory is either gapped or flows to a CFT with global symmetry $SU(N_f)\times U(1)$ \cite{Appelquist:1986fd,DiPietro:2015taa,Karthik:2015sgq}. Determining the critical value of $N_f$ for which this transition happens, is an open problem. It would be great to study this issue with the flow equations, and observe a transition of the $\Delta_i(\lambda)$ curves from constant (in the BCFT phase) to growing (in the gapped phase). The BCFT data thus obtained could be matched with the upcoming $\epsilon$-expansion results from \cite{DeCesare:2026rup}.\footnote{For other recent studies of QED in AdS, see \cite{Ankur:2023lum,Ankur:2026ylr,DiPietro:QED3_WIP}.}

\end{itemize}
Other interesting QFTs, but which require treating the classically marginal relevant deformation as discussed in \ref{subsubsec:marginallyrelevant}, are
\begin{itemize}
    \item Yang-Mills and QCD in 4D. When the gauge fields in these theories have Dirichlet boundary conditions, the QFT data is expected to undergo merger-annihilation at some value of $R\Lambda_{\text{YM/QCD}}$ due to color confinement in the flat space theory \cite{Aharony:2012jf,Copetti:2023sya,Ciccone:2024guw,Ciccone:2025dqx,DiPietro:2025ozw}. Current estimates of the critical value of $R\Lambda_{\text{YM/QCD}}$ are obtained through perturbation theory. It would be interesting to study the explicit curves of the QFT data in these theories with the flow equations. With Neumann boundary conditions, we could in principle extract estimates of the masses of the resonances in both theories. The conformal window of QCD is also an interesting problem to study. 
\end{itemize}
\section*{Acknowledgments}
We would like to thank Tomas Reis, Lorenzo Quintavalle, Marco Serone, Jiaxin Qiao, Gregoire Mathys, Xiang Zhao, the participants and organizers of the conference ``QFT in AdS 2026" and especially Marco Meineri and Joao Penedones for useful conversations that helped overcome some of the challenges encountered in this work. We are grateful to Jiaxin Qiao, Xiang Zhao and Marco Meineri for detailed comments on an early draft. We also thank the anonymous referee for their very careful reading of our manuscript and for their detailed and constructive comments.

Our research is supported by the Italian Ministry of University and Research (MUR) under the FIS grant BootBeyond (CUP: D53C24005470001) and by the
INFN “Iniziativa Specifica” ST\&FI.
\appendix

\section{Derivation of blocks}
Here we derive the various conformal blocks introduced in the main text.
\subsection{Bulk-bulk}
\label{app:bulk-bulk}
AdS invariance forces the two-point function of bulk operators to take the form
\begin{equation}
    \langle\hat\Phi_1(X_1)\hat\Phi_2(X_2)\rangle=\mathcal{G}_{\hat\Phi_1\hat\Phi_2}(X_1\cdot X_2)\,.
    \label{eq:BBinvariant}
\end{equation}
At the same time, we can fix to the conformal frame
\begin{equation}
    (z_1,\mathbf{x}_1)=( z,0)\,,\qquad (z_2,\mathbf{x}_2)=(1,0)
    \label{eq:BBframe}
\end{equation}
and then perform the BOE (\ref{eq:BOE}) of $\hat\Phi_1$ 
\begin{equation}
\begin{aligned}
    \langle\hat{\Phi}_1( z,0)\hat{\Phi}_2(1,0)\rangle&=\sum_l b^{\hat{\Phi}_1}_l z^{\Delta_l}\sum_{n=0}^\infty\frac{(-1)^n}{n!2^{2n}(\Delta_l-\frac{d-2}{2})_n} z^{2n}\Box^n\langle\mathcal{O}_l(\mathbf{x})\hat{\Phi}_2(1,0)\rangle\Big|_{\mathbf{x}=0}\\
    &=\sum_l b^{\hat{\Phi}_1}_lb^{\hat{\Phi}_2}_l z^{\Delta_l}\sum_{n=0}^\infty\frac{(-1)^n}{n!2^{2n}(\Delta_l-\frac{d-2}{2})_n} z^{2n}\Box^n\left(\frac{1}{1+\mathbf{x}^2}\right)^{\Delta_l}\Big|_{\mathbf{x}=0}\\
    &=\sum_l b^{\hat{\Phi}_1}_lb^{\hat{\Phi}_2}_l z^{\Delta_l}\sum_{n=0}^\infty\frac{(\Delta_l)_n(\frac{d}{2})_n}{n!(\Delta_l-\frac{d-2}{2})_n} z^{2n}\\
    &=\sum_l b^{\hat{\Phi}_1}_lb^{\hat{\Phi}_2}_l z^{\Delta_l}\,_2F_1\left(\begin{matrix}
        \Delta_l, & \frac{d}{2}\\ \Delta_l-\frac{d-2}{2}, & 
    \end{matrix}; z^2\right)\\
    &=\sum_l b^{\hat{\Phi}_1}_lb^{\hat{\Phi}_2}_l\frac{ z^{\Delta_l}}{(1- z)^{2\Delta_l}}\,_2F_1\left(\begin{matrix}
        \Delta_l, & \Delta_l-\frac{d-1}{2},\\
       & 2\Delta_l-d+1
    \end{matrix};-\frac{4 z}{(1- z)^2}\right)
    \label{eq:bulkbulksteps}
\end{aligned}
\end{equation}
where in the last step we used Kummer's quadratic transformation identity
\begin{equation}
    \,_2F_1\left(\begin{matrix}
        a & b\\ &a-b+1 
    \end{matrix};z^2\right)=\frac{1}{(1-z)^{2a}}\,_2F_1\left(\begin{matrix}
        a & a-b+\frac{1}{2}\\
        & 2a-2b+1
    \end{matrix};-\frac{4z}{(1-z)^2}\right)
\end{equation}
Now consider the fact that, in frame (\ref{eq:BBframe}), $\sigma\equiv X_1\cdot X_2=-\frac{1+z^2}{2z}$. Inverting this relation and using (\ref{eq:BBinvariant}) we find 
\begin{equation}
    \langle\hat\Phi_1(X_1)\hat\Phi_2(X_2)\rangle=\sum_l b^{\hat{\Phi}_1}_lb^{\hat\Phi_2}_l\frac{1}{(-2-2\sigma)^{\Delta_l}}\,_2F_1\left(\begin{matrix}\Delta_l, & \Delta_l-\frac{d-1}{2},\\ & 2\Delta_l-d+1\end{matrix};\frac{2}{1+\sigma}\right)\,.
\end{equation}
\subsection{Bulk-boundary-boundary}
\label{app:bulk-boundary-boundary}
\paragraph{Scalars}
The three-point function of a bulk field with two boundary operators is also fixed up to a function of a single cross ratio
\begin{equation}
    \langle\mathcal{O}_i(P_1)\mathcal{O}_j(P_2)\hat\Phi(X)\rangle=\frac{1}{P_{12}^\frac{{\Delta_i+\Delta_j}}{2}}\left(\frac{P_1\cdot X}{P_2\cdot X}\right)^{\frac{\Delta_j-\Delta_i}{2}}\mathcal{G}_{\hat\Phi ij}(\chi)\,,
    \label{eq:genframeBbb}
\end{equation}
where we choose 
\begin{equation}
    \chi\equiv-\frac{1}{2}\frac{P_1\cdot P_2}{(P_1\cdot X)(P_2\cdot X)}\,.\label{eq:defchi}
\end{equation}
For simplicity, let us fix to the frame
\begin{equation}
    (z_1,\mathbf{x}_1)=(0,0)\,,\qquad  (z_2,\mathbf{x}_2)=(0,\infty)\,,\qquad (z_3,\mathbf{x}_3)=(z,\hat{\mathbf{x}}^1)\,,
    \label{eq:frameBbb}
\end{equation}
where $(z_3,\mathbf{x}_3)$ is $X$ in Poincar\'e coordinates, $\hat{\mathbf{x}}^1$ is the unit vector pointing in the $1$ direction and we take operators to infinity as follows
\begin{equation}
    \mathcal{O}_i(\infty)\equiv\lim_{\mathbf{x}^2\to\infty}(\mathbf{x}^2)^{\Delta_i}\mathcal{O}_i(\mathbf{x})
    \label{eq:opinfinity}
\end{equation}
Performing the BOE of $\hat\Phi$, and using (\ref{eq:scalar3pt}), we get
\begin{equation}
    \begin{aligned}\langle\mathcal{O}_i(0)\mathcal{O}_j(\infty)\hat\Phi(z,\hat{\mathbf{x}}^1)\rangle&=\sum_lb^{\hat\Phi}_lz^{\Delta_l}\sum_{n=0}^\infty\frac{(-1)^n}{n!2^{2n}(\Delta_l-\frac{d-2}{2})_n}z^{2n}\Box^n\langle\mathcal{O}_i(0)\mathcal{O}_j(\infty)\mathcal{O}_l(\mathbf{x})\rangle\Big|_{\mathbf{x}\to \hat{\mathbf{x}}^1}\\
    &=\sum_lb^{\hat\Phi}_lC_{lij}z^{\Delta_l}\sum_{n=0}^\infty\frac{(-1)^n(\frac{\Delta_{lij}}{2})_n(\frac{\Delta_{lij}}{2}-\frac{d-2}{2})_n}{n!(\Delta_l-\frac{d-2}{2})_n}z^{2n}\\
    &=\sum_lb^{\hat\Phi}_lC_{lij}z^{\Delta_l}\,_2F_1\left(\begin{matrix}\frac{\Delta_{lij}}{2} & \frac{\Delta_{lij}}{2}-\frac{d-2}{2}\\ & \Delta_l-\frac{d-2}{2}\end{matrix};-z^2\right)\\
    &=\sum_lb^{\hat\Phi}_lC_{lij}\frac{z^{\Delta_l}}{(z^2+1)^{\frac{\Delta_{lij}}{2}}}\,_2F_1\left(\begin{matrix}\frac{\Delta_{lij}}{2} & \frac{\Delta_{lij}}{2}\\ & \Delta_l-\frac{d-2}{2}\end{matrix};\frac{z^2}{z^2+1}\right)\,.
    \end{aligned}
\end{equation}
where in the last step we used the Pfaff transformation
\begin{equation}
    \,_2F_1\left(\begin{matrix}
        a & b\\ &c
    \end{matrix};z\right)=\frac{1}{(1-z)^{a}}\,_2F_1\left(\begin{matrix}
        a & c-b\\
        & c
    \end{matrix};\frac{z}{z-1}\right)
    \label{eq:pfaff}
\end{equation}
In frame (\ref{eq:frameBbb}), we have that $\chi=\frac{z^2}{z^2+1}$. Inverting this relation and matching with (\ref{eq:genframeBbb}), we can thus write
\begin{equation}
    \langle\mathcal{O}_i(P_1)\mathcal{O}_j(P_2)\hat\Phi(X)\rangle=\frac{1}{P_{12}^\frac{{\Delta_i+\Delta_j}}{2}}\left(\frac{P_1\cdot X}{P_2\cdot X}\right)^{\frac{\Delta_j-\Delta_i}{2}}\sum_lb^{\hat\Phi}_lC_{lij}G_{\text{Bbb}}^{\Delta_l,\Delta_i,\Delta_j}(\chi)
\end{equation}
with
\begin{equation}
   G_{\text{Bbb}}^{\Delta_l,\Delta_i,\Delta_j}(\chi)=\chi^\frac{\Delta_l}{2}\,_2F_1\left(\begin{matrix}
        \frac{\Delta_{lij}}{2}, & \frac{\Delta_{lji}}{2}\\
        &\Delta_l-\frac{d-2}{2}
    \end{matrix};\chi\right)\,.
    \label{eq:GBbb_app}
\end{equation}
\paragraph{AdS$_3$} 
When spin is involved in AdS$_3$, it is best to express boundary operators in the complex coordinates discussed in the main text. We will specifically need the case where the two boundary operators are identical, and have spin $J$.  

When we fix the position of the two boundary operators and the bulk operator in a non-degenerate way, the leftover stabilizer group is $SO(d-1)=SO(1)$, hence we have no independent spin cross-ratio in this case, and the three-point function can be fixed up to a function of a single cross-ratio
\begin{equation}
    \langle\mathcal{O}_i(\mathtt{z}_1,\bar{\mathtt{z}}_1)\mathcal{O}_i(\mathtt{z}_2,\bar{\mathtt{z}}_2)\hat\Phi(z,\mathtt{z}_3,\bar{\mathtt{z}}_3)\rangle=\frac{1}{\mathtt{z}_{12}^{2h_i}\bar{\mathtt{z}}_{12}^{2\bar h_i}}\mathcal{G}_{ii\hat\Phi}(\chi)
\end{equation}
where we take $\chi$ to be the same cross-ratio as in (\ref{eq:defchi}). We fix the frame 
\begin{equation}
    (\mathtt{z}_1,\bar{\mathtt{z}}_1)=(0,0)\,, \qquad (\mathtt{z}_2,\bar{\mathtt{z}}_2)=(\infty,\infty)\,,\qquad (\mathtt{z}_3,\bar{\mathtt{z}}_3)=(\mathtt{z},\bar{\mathtt{z}})\,,
\end{equation}
where taking an operator to infinity in these coordinates means
\begin{equation}
    \mathcal{O}_i(\infty,\infty)\equiv\lim_{\mathtt{z}\to\infty}\lim_{\bar{\mathtt{z}}\to\infty}\mathtt{z}^{2h}\bar{\mathtt{z}}^{2\bar h}\mathcal{O}_i(\mathtt{z},\bar{\mathtt{z}})\,.
\end{equation}
We expand $\hat\Phi$ to the boundary with the BOE (\ref{eq:BOE}). Using the form of the three-point function (\ref{eq:ads33pt}) and the laplacian in complex coordinates $\Box=2\partial_{\mathtt{z}}\partial_{\bar{\mathtt{z}}}$, we get
\begin{equation}
    \begin{aligned}
\langle\mathcal{O}_i(0,0)\mathcal{O}_i(\infty,\infty)\hat\Phi(z,\mathtt{z},\bar{\mathtt{z}})\rangle&=\sum_lb^{\hat\Phi}_lC_{lii}z^{\Delta_l}\sum_{n=0}^\infty\frac{(-1)^nz^{2n}}{n!2^{2n}(\Delta_l)_n}\Box^n\frac{1}{(\mathtt{z}\bar{\mathtt{z}})^{\frac{\Delta_{l}}{2}}}\\
    &=\sum_lb^{\hat\Phi}_lC_{lii}z^{\Delta_l}\frac{1}{(\mathtt{z}\bar{\mathtt{z}})^{\frac{\Delta_{l}}{2}}}\,_2F_1\left(\begin{matrix}
        \frac{\Delta_{l}}{2}, & \frac{\Delta_{l}}{2}\\
        & \Delta_l
    \end{matrix};-\frac{z^2}{\mathtt{z}\bar{\mathtt{z}}}\right)\\
    &=\sum_lb^{\hat\Phi}_lC_{lii}\left(\frac{z^2}{z^2+\mathtt{z}\bar{\mathtt{z}}}\right)^{\frac{\Delta_l}{2}}\,_2F_1\left(\begin{matrix}
        \frac{\Delta_{l}}{2}, & \frac{\Delta_{l}}{2}\\
        & \Delta_l
    \end{matrix};\frac{z^2}{z^2+\mathtt{z}\bar{\mathtt{z}}}\right)
    \end{aligned}
\end{equation}
where we used the Pfaff transform (\ref{eq:pfaff}). Now we recognize the cross ratio in this frame $\chi=\frac{z^2}{z^2+\mathtt{z}\bar{\mathtt{z}}}$. We thus can write
\begin{equation}
    \mathcal{G}_{ii\hat\Phi}(\chi)=\sum_lb^{\hat\Phi}_lC_{lii}\chi^{\frac{\Delta_l}{2}}\,_2F_1\left(\begin{matrix}
        \frac{\Delta_{l}}{2}, & \frac{\Delta_{l}}{2}\\
        & \Delta_l
    \end{matrix};\chi\right)\,,
\end{equation}
just like in the scalar case.
\paragraph{AdS$_4$}
 Now let us discuss the analogous case in 4D. In this case, fixing the position of all operators leaves the residual stabilizer group $SO(d-1)=SO(2)$ which allows for the three-point function to be fixed up to a function of a position cross-ratio and a spin cross-ratio
 \begin{equation}
     \langle\mathcal{O}_i^{(J)}(P_1,Z_1)\mathcal{O}_i^{(J)}(P_2,Z_2)\hat\Phi(X)\rangle=\frac{H_{1,2}^J}{P_{1,2}^{\Delta_i+J}}\mathcal{G}_{ii\hat\Phi}(\chi,v)
     \label{eq:Bbbansatzspin}
 \end{equation}
 where we choose $\chi$ as in (\ref{eq:defchi}) and 
\begin{equation}
    v\equiv\frac{1}{\chi-1}\frac{V_{1,23}V_{2,31}}{H_{1,2}}\,,\qquad P_3\to X
\end{equation}
where $H_{i,j}$ and $V_{i,jk}$ are defined in (\ref{eq:Hijdef}) and (\ref{eq:defV}) respectively. We choose the frame where one boundary operator is at the origin and one at infinity. In embedding notation, this corresponds to
\begin{equation}
\begin{aligned}
    P_1&=\left(\frac{1}{2},\mathbf{0},\frac{1}{2}\right),\quad P_2=\left(\frac{1}{2},\mathbf{0},-\frac{1}{2}\right),\quad Z_1=(0,\mathbf{z}_1,0),\quad Z_2=(0,\mathbf{z}_2,0)\\
    X&=\left(\frac{2+z^2}{2z},\frac{\hat{\mathbf{x}}^1}{z},-\frac{z}{2}\right)
\end{aligned}
\end{equation}
In this frame, the cross ratios take the form
\begin{equation}v=\frac{\mathbf{z}_1^1\mathbf{z}_2^1}{\mathbf{z}_1\cdot\mathbf{z}_2}\,,\qquad \chi=\frac{z^2}{1+z^2}\,.
\end{equation}
At the same time, the BOE of the bulk field gives
 \begin{equation}
 \begin{aligned}
     \langle\mathcal{O}_i^{(J)}&(0,\mathbf{z}_1)\mathcal{O}_i^{(J)}(\infty,\mathbf{z}_2)\hat\Phi(z,\hat{\mathbf{x}}^1)\rangle\\
     &=\sum_lb^{\hat\Phi}_lz^{\Delta_l}\sum_{n=0}^\infty\frac{(-1)^n}{n!2^{2n}(\Delta_l-\frac{1}{2})_n}z^{2n}\Box^n\langle\mathcal{O}_i^{(J)}(0,\mathbf{z}_1)\mathcal{O}_i^{(J)}(\infty,\mathbf{z}_2)\mathcal{O}_l(\mathbf{x})\rangle\Big|_{\mathbf{x}\to\hat{\mathbf{x}}^1}\\
     &=\sum_lb^{\hat\Phi}_lz^{\Delta_l}\sum_{m=0}^JC_{lii}^{(m)}\sum_{n=0}^\infty\frac{(-1)^n(\mathbf{z}_1\cdot\mathbf{z}_2)^J}{n!2^{2n}(\Delta_l-\frac{1}{2})_n}z^{2n}\Box^n\frac{\mathcal{H}_m\left(\frac{(\mathbf{z}_1\cdot\mathbf{x})(\mathbf{z}_2\cdot\mathbf{x})}{\mathbf{x}^2(\mathbf{z}_1\cdot\mathbf{z}_2)}\right)}{(\mathbf{x}^2)^{\frac{\Delta_l}{2}}}\Bigg|_{\mathbf{x}\to\hat{\mathbf{x}}^1}
\end{aligned}
 \end{equation}
 where we used the basis decomposition of the boundary three-point function (\ref{eq:JJO3ptalt}). 

To proceed, we use the following property of the $\mathcal{H}_m$ polynomials, the reason why we chose them as a basis of our tensor structures in the first place: the action of the Laplacian on the basis elements in this frame is diagonal
\begin{equation}
\Box \frac{1}{(\mathbf{x}^2)^{\frac{\Delta_l}{2}}}\mathcal{H}_m\left(\frac{(\mathbf{z}_1\cdot\mathbf{x})(\mathbf{z}_2\cdot\mathbf{x})}{\mathbf{x}^2(\mathbf{z}_1\cdot\mathbf{z}_2)}\right)=\frac{(\Delta_l+2m)(\Delta_l-2m-1)}{(\mathbf{x}^2)^{\frac{\Delta_l}{2}+1}}\mathcal{H}_m\left(\frac{(\mathbf{z}_1\cdot\mathbf{x})(\mathbf{z}_2\cdot\mathbf{x})}{\mathbf{x}^2(\mathbf{z}_1\cdot\mathbf{z}_2)}\right)
\end{equation}
Using this fact, we have
\begin{align}
     &\langle\mathcal{O}_i^{(J)}(0,\mathbf{z}_1)\mathcal{O}_i^{(J)}(\infty,\mathbf{z}_2)\hat\Phi(z,\hat{\mathbf{x}}^1)\rangle\nonumber\\
     &=(\mathbf{z}_1\cdot\mathbf{z}_2)^J\sum_lb^{\hat\Phi}_lz^{\Delta_l}\sum_{m=0}^JC_{lii}^{(m)}\mathcal{H}_m\left(\frac{\mathbf{z}^1_1\mathbf{z}_2^1}{\mathbf{z}_1\cdot\mathbf{z}_2}\right)\sum_{n=0}^\infty\frac{(-1)^nz^{2n}(\frac{\Delta_l}{2}+m)_n(\frac{\Delta_l-1-2m}{2})_n}{n!(\Delta_l-\frac{1}{2})_n}\nonumber\\
     &=(\mathbf{z}_1\cdot\mathbf{z}_2)^J\sum_lb^{\hat\Phi}_lz^{\Delta_l}\sum_{m=0}^JC_{lii}^{(m)}\mathcal{H}_m\left(\frac{\mathbf{z}^1_1\mathbf{z}_2^1}{\mathbf{z}_1\cdot\mathbf{z}_2}\right)\,_2F_1\left(\begin{matrix}
         \frac{\Delta_l}{2}+m, & \frac{\Delta_l-1-2m}{2},\\
         & \Delta_l-\frac{1}{2}
     \end{matrix};-z^2\right)\\
     &=(\mathbf{z}_1\cdot\mathbf{z}_2)^J\sum_lb^{\hat\Phi}_l\frac{z^{\Delta_l}}{(1+z^2)^{\frac{\Delta_l}{2}+m}}\sum_{m=0}^JC_{lii}^{(m)}\mathcal{H}_m\left(\frac{\mathbf{z}^1_1\mathbf{z}_2^1}{\mathbf{z}_1\cdot\mathbf{z}_2}\right)\,_2F_1\left(\begin{matrix}
         \frac{\Delta_l}{2}+m, & \frac{\Delta_l}{2}+m,\\
         & \Delta_l-\frac{1}{2}
     \end{matrix};\frac{z^2}{z^2+1}\right)\nonumber
\end{align}
Comparing with (\ref{eq:Bbbansatzspin}) and noticing that $H_{1,2}=\mathbf{z}_1\cdot\mathbf{z}_2$ in this frame, we get
\begin{equation}
    \mathcal{G}_{ii\hat\Phi}(\chi,v)=\sum_l b^{\hat\Phi}_l\sum_{n=0}^JC_{lii}^{(n)}\mathcal{H}_n(v)G_{\text{Bbb}}^{\Delta_l,n}(\chi)
\end{equation}
where
\begin{equation}
    G_{\text{Bbb}}^{\Delta_l,n}(\chi)=\chi^\frac{\Delta_l}{2}(1-\chi)^n\,_2F_1\left(\begin{matrix}
        \frac{\Delta_l}{2}+n, & \frac{\Delta_l}{2}+n\\
        & \Delta_l-\frac{1}{2}
    \end{matrix};\chi\right)\,
\end{equation}
Another interesting property of this basis decomposition is the behavior under cross-traces:
\begin{equation}
    (D_{\mathbf{z}_1}\cdot D_{\mathbf{z}_2})^m\left[(\mathbf{z}_1\cdot\mathbf{z}_2)^J\mathcal{H}_n\left(\frac{(\mathbf{z}_1\cdot\mathbf{x})(\mathbf{z}_2\cdot\mathbf{x})}{\mathbf{x}^2(\mathbf{z}_1\cdot\mathbf{z}_2)}\right)\right]=0\,,\qquad \forall m> J-n\,,
    \label{eq:vanishtraces}
\end{equation}
where $D_{\mathbf{z}}$ was defined in (\ref{eq:defza}). 
In other words, elements of this basis have a number of vanishing cross-traces which increases with $n$, and if in particular we take $m=J$, the only nonvanishing term is the one with $n=0$. 
\subsection{Bulk-bulk-boundary}
\label{app:bulk-bulk-boundary}
AdS invariance fixes bulk-bulk-boundary three-point functions up to two cross ratios
\begin{equation}
    \langle\hat\Phi_1(X_1)\hat\Phi_2(X_2)\mathcal{O}_i(P_3)\rangle=\frac{1}{(-2P_3\cdot X_2)^{\Delta_i}}\mathcal{G}_{\hat\Phi_1\hat\Phi_2 i}(\xi,\rho)\,.
\end{equation}
We choose
\begin{equation}
    \xi=\left(\frac{P_3\cdot X_1}{P_3\cdot X_2}\right)^2\,,\qquad \rho=-1-\xi-2(X_1\cdot X_2)\sqrt{\xi}
\end{equation}
At the same time, we can expand both bulk operators to the boundary. We do that in the following frame
\begin{equation}
    (z_1,\mathbf{x}_1)=(z,r\hat{\mathbf{x}}^1)\,,\qquad (z_2,\mathbf{x}_2)=(1,0)\,,\qquad (z_3,\mathbf{x}_3)=(0,\infty)\,.
    \label{eq:frameBBb2}
\end{equation}
We get
\begin{equation}
\begin{aligned}
    \langle&\hat\Phi_1(z,r\hat{\mathbf{x}}^1)\hat\Phi_2(1,0)\mathcal{O}_i(\infty)\rangle=\sum_{j}\sum_lb^{\hat\Phi_1}_l b^{\hat\Phi_2}_jz^{\Delta_l}\\
    &\sum_{n=0}^\infty\sum_{m=0}^\infty\frac{(-1)^{n+m}2^{-2(n+m)}z^{2n}}{n!m!(\Delta_l-\frac{d-2}{2})_n(\Delta_j-\frac{d-2}{2})_m}\Box^n\Box_2^m\langle\mathcal{O}_l(\mathbf{x})\mathcal{O}_j(\mathbf{x}_2)\mathcal{O}_i(\infty)\rangle\Big|_{\mathbf{x}_2\to0,\mathbf{x}\to r\hat{\mathbf{x}}^1}
    \label{eq:stepblockBBb}
\end{aligned}
\end{equation}
We will eventually want to match this to a block decomposition, which we expect to take the form
\begin{equation}
    \langle\hat\Phi_1(X_1)\hat\Phi_2(X_2)\mathcal{O}_i(P_3)\rangle=\frac{1}{(-2P_3\cdot X_2)^{\Delta_i}}\sum_{j}\sum_lb^{\hat\Phi_1}_l b^{\hat\Phi_2}_jC_{lij}G_{\text{BBb}}^{\Delta_l,\Delta_j,\Delta_i}(\xi,\rho)\,.
    \label{eq:BBbansatz}
\end{equation}
Moreover, in frame (\ref{eq:frameBBb2}), the cross ratios take the values $\xi=\frac{1}{z^2}$ and $\rho=\frac{r^2}{z^2}$ and the scale factor is $\frac{1}{(-2P_3\cdot X_2)^{\Delta_i}}\to\frac{1}{(\mathbf{x}_3^{2})^{\Delta_i}}$, which gets canceled when we take operator $\mathcal{O}_i$ to infinity as in (\ref{eq:opinfinity}).

Continuing from (\ref{eq:stepblockBBb}), we thus have
\begin{align}
   G_{\text{BBb}}^{\Delta_l,\Delta_j,\Delta_i}\left(\frac{1}{z^2},\frac{r^2}{z^2}\right)&= \sum_{n,m=0}^\infty\frac{(-1)^{n+m}(\frac{\Delta_{lji}}{2})_n(\frac{\Delta_{lji}}{2}-\frac{d-2}{2})_n(\frac{\Delta_{lji}}{2}+n)_m(\frac{\Delta_{lji}}{2}-\frac{d-2}{2}+n)_mz^{\Delta_l+2n}}{n!m!(\Delta_l-\frac{d-2}{2})_n(\Delta_j-\frac{d-2}{2})_mr^{\Delta_{lji}+2n+2m}}\nonumber\\
    &=\sum_{n,m=0}^\infty\frac{(-1)^{n+m}(\frac{\Delta_{lji}}{2})_{n+m}(\frac{\Delta_{lji}}{2}-\frac{d-2}{2})_{n+m}z^{\Delta_l+2n}}{n!m!(\Delta_l-\frac{d-2}{2})_n(\Delta_j-\frac{d-2}{2})_mr^{\Delta_{lji}+2n+2m}}\\
    &=\frac{z^{\Delta_l}}{r^{\Delta_{lji}}}F_4\left(\begin{matrix}\frac{\Delta_{lji}}{2} & \frac{\Delta_{lji}}{2}-\frac{d-2}{2}\\ \Delta_l-\frac{d-2}{2}, & \Delta_j-\frac{d-2}{2} \end{matrix}; -\frac{z^2}{r^2},-\frac{1}{r^2}\right)\nonumber
\end{align}
Matching with (\ref{eq:BBbansatz}), we get
\begin{equation}
    G_{\text{BBb}}^{\Delta_l,\Delta_j,\Delta_i}(\xi,\rho)=\frac{\xi^{\frac{\Delta_j-\Delta_i}{2}}}{\rho^{\frac{\Delta_{lji}}{2}}}F_4\left(\begin{matrix}\frac{\Delta_{lji}}{2} & \frac{\Delta_{lji}}{2}-\frac{d-2}{2}\\ \Delta_l-\frac{d-2}{2}, & \Delta_j-\frac{d-2}{2} \end{matrix}; -\frac{1}{\rho},-\frac{\xi}{\rho}\right)\,.
\end{equation}
This representation of the conformal block is not analytic for all possible values of the cross-ratios. If we instead expand the Appell $F_4$ as in (\ref{eq:appellF4def}) and resum over one of the variables, we can write
\begin{align}
     &G_{\text{BBb}}^{\Delta_l,\Delta_j,\Delta_i}(\xi,\rho)=\frac{\xi^{\frac{\Delta_j-\Delta_i}{2}}}{\rho^{\frac{\Delta_{lji}}{2}}}\sum_{m=0}^\infty\frac{(-1)^m(\frac{\Delta_{lji}}{2}-\frac{d-2}{2})_m(\frac{\Delta_{lji}}{2})_m\xi^m}{m!(\Delta_j-\frac{d-2}{2})_m \rho^m}\,_2F_1\left(\begin{matrix}
         \frac{\Delta_{lji}}{2}+m, & \frac{\Delta_{lji}}{2}-\frac{d-2}{2}+m\\\
         & \Delta_l-\frac{d-2}{2}
     \end{matrix};-\frac{1}{\rho}\right)\nonumber\\
    &=\frac{\xi^{\frac{\Delta_j-\Delta_i}{2}}}{(1+\rho)^{\frac{\Delta_{lji}}{2}}}\sum_{m=0}^\infty\frac{(-1)^m(\frac{\Delta_{lji}}{2}-\frac{d-2}{2})_m(\frac{\Delta_{lji}}{2})_m}{m!(\Delta_j-\frac{d-2}{2})_m (1+\rho)^m}\,_2F_1\left(\begin{matrix}
         \frac{\Delta_{lij}}{2}-m, & \frac{\Delta_{lji}}{2}+m\\\
         & \Delta_l-\frac{d-2}{2}
     \end{matrix};\frac{1}{1+\rho}\right)
\end{align}
The final step is to open up the hypergeometric function with its series representation and resum over $m$. We obtain
\begin{equation}
\begin{aligned}
    G_{\text{BBb}}^{\Delta_l,\Delta_j,\Delta_i}(\xi,\rho)=&\frac{\xi^{\frac{\Delta_j-\Delta_i}{2}}\Gamma(\Delta_j-\frac{d-2}{2})\Gamma(1-\frac{\Delta_{ilj}}{2})}{(1+\rho)^{\frac{\Delta_{lji}}{2}}}\sum_{n=0}^\infty\left(\frac{-1}{1+\rho}\right)^n\frac{(\frac{\Delta_{lji}}{2})_n}{n!(\Delta_l-\frac{d-2}{2})_n}\\
        &\times\,_3\tilde F_2\left(\begin{matrix}
            1-\frac{\Delta_{ilj}}{2}, & \frac{\Delta_{lji}}{2}-\frac{d-2}{2}, & \frac{\Delta_{lji}}{2}+n\\
            & 1-n-\frac{\Delta_{ilj}}{2}, & \Delta_j-\frac{d-2}{2}
        \end{matrix};-\frac{\xi}{1+\rho}\right)
        \label{eq:convergentBBb}
\end{aligned}
\end{equation}
The advantage of this representation is that it is analytic for all $\xi>0$ and $\rho>-1$, which in particular includes all configurations in AdS.

\section{Details about the derivation of the flow equations}

We present here some technical details used to derive the flow equations. 
\subsection{Flow of scaling dimensions}
\label{app:first_eq_details}
In this section, we derive the structural form of the integral
\begin{equation}
    I^{\Delta_l,\Delta_i,\Delta_j}_{\text{Bbb}}(\mathbf{x}_1,\mathbf{x}_2,\epsilon) = \int_{\mathcal{M}_\epsilon} \frac{dz d^d\mathbf{x}}{z^{d+1}} \frac{1}{|\mathbf{x}_{12}|^{\Delta_i+\Delta_j}} \left( \frac{|\mathbf{x}-\mathbf{x}_1|^2+z^2}{|\mathbf{x}-\mathbf{x}_2|^2+z^2} \right)^{\frac{\Delta_j-\Delta_i}{2}} G_{\text{Bbb}}^{\Delta_l,\Delta_i,\Delta_j}(\chi)\,.\label{eq:Ybis} 
\end{equation}
To evaluate this expression, we decompose the integral, separating the integration over the entire AdS space from the contributions of the subtracted regions near the boundary points, as in
\begin{equation}
    I^{\Delta_l,\Delta_i,\Delta_j}_{\text{Bbb}}(\mathbf{x}_1,\mathbf{x}_2,\epsilon) = I^{\Delta_l,\Delta_i,\Delta_j}_{\text{Bbb}}(\mathbf{x}_1, \mathbf{x}_2, 0) - I^{\Delta_l,\Delta_i,\Delta_j}_{\text{Bbb,in}}(\mathbf{x}_1, \mathbf{x}_2, \epsilon)\,,
\end{equation}
where $I^{\Delta_l,\Delta_i,\Delta_j}_{\text{Bbb,in}}$ represents the integral over the two subtracted half-balls, represented in figure \ref{fig:domain_first}.  This term is defined as
\begin{equation}
    I^{\Delta_l,\Delta_i,\Delta_j}_{\text{Bbb,in}}:= \int_{|\mathbf{x}-\mathbf{x}_k|^2+z^2 < \epsilon^2} \frac{dz d^d\mathbf{x}}{z^{d+1}} \frac{1}{|\mathbf{x}_{12}|^{\Delta_i+\Delta_j}} \left( \frac{|\mathbf{x}-\mathbf{x}_1|^2+z^2}{|\mathbf{x}-\mathbf{x}_2|^2+z^2} \right)^{\frac{\Delta_j-\Delta_i}{2}} G_{\text{Bbb}}^{\Delta_l,\Delta_i,\Delta_j}(\chi)\,.
    \label{eq:IBbb_in}
\end{equation}
We note that, if $I^{\Delta_l,\Delta_i,\Delta_j}_{\text{Bbb}}(\mathbf{x}_1,\mathbf{x}_2,0)$ were a convergent integral, conformal invariance would imply that it vanishes whenever $\Delta_i \neq \Delta_j$.
Consequently, one would obtain
\begin{equation}
I^{\Delta_l,\Delta_i,\Delta_j}_{\text{Bbb}}(\mathbf{x}_1, \mathbf{x}_2, \epsilon) =
-I^{\Delta_l,\Delta_i,\Delta_j}_{\text{Bbb,in}}(\mathbf{x}_1, \mathbf{x}_2, \epsilon) \, .
\label{eq:relIinout}
\end{equation}
Since the integral in the right-hand side is divergent for all values of the scaling dimensions, this relation must be intended as an analytic continuation, in the following sense: when considering the half-ball centered at $\mathbf{x}_1$, we take $\Delta_j > \Delta_i$, and the converse when integrating instead around $\mathbf{x}_2$. A justification of this extension is provided in appendix D of \cite{Loparco:2026fki}, and we will assume its validity in the following. We are thus left to evaluate the integration within the subtracted half-balls. Let us focus on the half-ball centered at $\mathbf{x}_1$, which is finite in the regime $\Delta_j > \Delta_i$. The one centered at $\mathbf{x}_2$ follows by exchanging $\Delta_i$ and $\Delta_j$. To compute it, we adopt spherical coordinates centered at $\mathbf{x}_1$ for the boundary directions, combined with polar coordinates for the half-plane spanned by the boundary radial distance $|\mathbf{r}|$ and the bulk coordinate $z$,
\begin{align}
 \mathbf{x}= \mathbf{x}_1+\mathbf{r}\,\quad\mathrm{with}\quad | \mathbf{r}|= R \cos\theta \, ,\qquad z = R \sin\theta\,,
\end{align}
where $R<\epsilon$, $\theta \in (0, \pi/2)$, and we call $\phi$ the angle between $\mathbf{r}$ and $\mathbf{x}_2-\mathbf{x}_1$. Let us now write the conformal block as a power series,
\begin{equation}
    G_{\text{Bbb}}^{\Delta_l,\Delta_i,\Delta_j}(\chi)=\sum_{n\geq0}g_n\,\chi^{\frac{\Delta_l}{2}+n}\,,\qquad g_n=\frac{\left(\frac{\Delta_{lij}}{2}\right)_n\left(\frac{\Delta_{lji}}{2}\right)_n}{n!\left(\Delta_l+1-\frac{d}{2}\right)_n}\,.
\end{equation}
For each power $\chi^k$, the integrand of \eqref{eq:IBbb_in} takes the simple form
\begin{equation}
    \frac{1}{|\mathbf{x}_{12}|^{\Delta_i+\Delta_j}}\left(\frac{|\mathbf{x}-\mathbf{x}_1|^2+z^2}{|\mathbf{x}-\mathbf{x}_2|^2+z^2}\right)^{\frac{\Delta_{ji}}{2}}\chi^k=\frac{\sin^{2k}\theta}{|\mathbf{x}_{12}|^{\Delta_i+\Delta_j}}\;t^{-\Delta_{ij}}\left(1-2t\cos\theta\cos\phi+t^2\right)^{\frac{\Delta_{ij}}{2}-k}\,,
\end{equation}
where $t\equiv R/|\mathbf{x}_{12}|$. The last factor is the generating function of the Gegenbauer polynomials,
\begin{equation}
    \left(1-2tx+t^2\right)^{-\nu}=\sum_{m\geq0}C^{(\nu)}_m(x)\,t^m\,.
\end{equation}
The contribution of the half-ball centered at $\mathbf{x}_1$ to \eqref{eq:IBbb_in} is therefore
\begin{equation}
\begin{split}
   I^{\Delta_l,\Delta_i,\Delta_j}_{\text{Bbb,in}}\Big|_{\mathbf{x}_1}={}&\frac{1}{|\mathbf{x}_{12}|^{\Delta_i+\Delta_j}}\sum_{n,m\geq0}g_n\int_0^{\epsilon}\frac{dR}{R}\left(\frac{R}{|\mathbf{x}_{12}|}\right)^{m-\Delta_{ij}}\\
   &\times\int d\Omega_d\int_0^{\pi/2}d\theta\,\cos^{d-1}\theta\,\sin^{\Delta_l+2n-d-1}\theta\;C^{(\frac{\Delta_{lji}}{2}+n)}_m(\cos\theta\cos\phi)\,.
\end{split}
\label{eq:half-ball}
\end{equation}
Since $C^{(\nu)}_m(-x)=(-1)^mC^{(\nu)}_m(x)$, the terms with odd $m$ vanish after integrating over the direction of $\mathbf{r}$, which can be flipped, $\cos\phi\to-\cos\phi$. For $m=2p$, the radial integral gives
\begin{equation}
    \int_0^{\epsilon} \frac{dR}{R} \left( \frac{R}{|\mathbf{x}_{12}|} \right)^{2p-\Delta_{ij}} = \frac{1}{2p-\Delta_{ij}} \left( \frac{\epsilon}{|\mathbf{x}_{12}|} \right)^{2p-\Delta_{ij}},
\end{equation}
which is finite for $\Delta_j>\Delta_i$, as anticipated. Adding the half-ball centered at $\mathbf{x}_2$ and using \eqref{eq:relIinout}, we obtain
\begin{equation}
   I^{\Delta_l,\Delta_i,\Delta_j}_{\text{Bbb}}(\mathbf{x}_1, \mathbf{x}_2, \epsilon) =  \frac{1}{|\mathbf{x}_{12}|^{\Delta_i+\Delta_j}} \sum_{p=0}^\infty \frac{a_{2p}(\Delta_{ij}, \Delta_l)}{\Delta_{ij} - 2p} \left( \frac{\epsilon}{|\mathbf{x}_{12}|} \right)^{2p-\Delta_{ij}} + (\Delta_i \leftrightarrow \Delta_j)\,,
   \label{eq:coeff_a}
\end{equation}
with 
\begin{equation}
    a_{2p}(\Delta_{ij},\Delta_l)=\sum_n g_n\,\int d\Omega_d\int_0^{\pi/2}d\theta\,\cos^{d-1}\theta\,\sin^{2n-d-1+\Delta_l}\theta\;C^{(n+\frac{\Delta_{lji}}{2})}_{2p}(\cos\theta\cos\phi)\,.
    \label{eq:a2pk_int}
\end{equation}
The angular integrals are elementary once we rewrite the Gegenbauer polynomials as
\begin{equation}
C^{(\nu)}_m(x)=\sum_{q=0}^{\lfloor m/2\rfloor}\frac{(-1)^q\,(\nu)_{m-q}}{q!\,(m-2q)!}\,(2x)^{m-2q}\,.
\end{equation}
Computing the integrals term by term and then summing over $q$ and $n$, we finally obtain
\begin{equation}
\begin{split}
    a_{2p}(\Delta_{ij},\Delta_l)={}&\frac{\pi^{\frac d2}\left(1-\frac{\Delta_{ij}}{2}\right)_p\left(\frac{\Delta_{lji}}{2}\right)_p\Gamma\left(\frac{\Delta_l-d}{2}\right)}{p!\;\Gamma\left(\frac{\Delta_l}{2}+p\right)}\\
    &\times{}_3F_2\left(\frac{\Delta_{lij}}{2},\frac{\Delta_{lji}}{2}+p,\frac{\Delta_l-d}{2};\Delta_l+1-\frac d2,\frac{\Delta_l}{2}+p;1\right)\,.
\end{split}
    \label{eq:a2p}
\end{equation}
The integrals in \eqref{eq:a2pk_int} converge for $\Delta_l>d$, and are otherwise defined by analytic continuation. For $p=0$, \eqref{eq:a2p} gives
\begin{equation}
    a_0(\Delta_{ij}, \Delta_l)=\frac{{\pi}^\frac{d}{2}\ \Gamma\left(\frac{\Delta_l-d}{2}\right)}{\Gamma\left(\frac{\Delta_l}{2}\right)} {}_3F_2\left( \frac{\Delta_{lij}}{2}, \frac{\Delta_{lji}}{2}, \frac{\Delta_l-d}{2} ; \Delta_l +1- \frac{d}{2}, \frac{\Delta_l}{2} ; 1 \right)\,.
\end{equation}
For the extraction of the flow equation, we require the value of $a_0$ when the boundary operators are identical, i.e., $\Delta_i = \Delta_j$. In this limit, the ${}_3F_2$ function reduces to a ${}_2F_1$ at unit argument, and Gauss's theorem yields
\begin{equation}
    a_0(0, \Delta_l) =-\frac{2 \pi ^{d/2} \Gamma (-\frac{d}{2}+\Delta_l+1)}{\Gamma (\frac{\Delta_l}{2})\Gamma (\frac{\Delta_l+2}{2}) (d-\Delta_l)}\,.
    \label{eq:a_coeff}
\end{equation}
Furthermore, since $a_0$ is even in $\Delta_{ij}$, its derivative with respect to $\Delta_{ij}$ vanishes at the origin:
\begin{equation}
    \left. \frac{d a_0(\Delta_{ij}, \Delta_l)}{d\Delta_{ij}} \right|_{\Delta_{ij}=0} = 0\,.
    \label{eq:a_der}
\end{equation}
Finally, because of the factor $\left(1-\frac{\Delta_{ij}}{2}\right)_p$, \eqref{eq:a2p} vanishes at $\Delta_{ij}=2,4,\dots,2p$. In particular, $a_{2p}(2p,\Delta_l)=0$ for every $\Delta_l$, in such a way that each term in the counterterm \eqref{eq:counterterm} is regular at $\Delta_{ij}=2p$. 

\subsection{Sign of local blocks}
\label{subsec:signlocal}
Here we prove the statement that local blocks (\ref{eq:Bbblocal}) with $\Delta_i=\Delta_j$ have a sign that is constant when we fix all parameters and let the cross ratio $\chi$ vary in the interval $(0,1)$. To begin, we use a Mellin Barnes representation, analogous to the one presented for AdS$_2$ in Appendix G of \cite{Meineri:2023mps}:
\begin{equation}
    G^{\Delta_l,\Delta_i,\Delta_i,\alpha}_{\text{Bbb}}(\chi)=\int_{\mathcal{C}}\frac{ds}{2\pi i}\,f_l(s)\,\chi^s\,\frac{\pi\sin\pi\big(\alpha-\frac{\Delta_l}{2}\big)}{\sin\pi\big(s-\frac{\Delta_l}{2}\big)\sin\pi(s-\alpha)}\,,
    \label{eq:GBbbalphaMB}
\end{equation}
where $\mathcal{C}$ is a vertical contour with $0<\text{Re}(s)<\min\big(\tfrac{\Delta_l}{2},\alpha\big)$, and
\begin{equation}
    f_l(s)\equiv\frac{\Gamma\big(\Delta_l-\frac{d-2}{2}\big)\Gamma(s)^2}{\Gamma\big(\frac{\Delta_l}{2}\big)^2\Gamma\big(1+s-\frac{\Delta_l}{2}\big)\Gamma\big(s+\frac{\Delta_l}{2}-\frac{d-2}{2}\big)}\,.
\end{equation}
This integral representation converges for $|\arg\chi|<2\pi$.

Using the identity 
\begin{equation}
\frac{-\pi\chi^{s-\alpha}}{\sin\pi(s-\alpha)}=\int_0^\infty dt\,\frac{t^{s-\alpha}}{t+\chi}\,,
\end{equation}
valid for $\alpha-1<\text{Re}(s)<\alpha$, and carrying out the integral over $s$ by recognizing the Mellin Barnes representation of the hypergeometric function, we find,  for $0<\alpha<\frac{\Delta_l}{2}+1$,
\begin{equation}
    G^{\Delta_l,\Delta_i,\Delta_i,\alpha}_{\text{Bbb}}(\chi)=\frac{\sin\pi\left(\alpha-\frac{\Delta_l}{2}\right)}{\pi}\,\chi^\alpha\int_0^\infty dt\,\frac{t^{\frac{\Delta_l}{2}-\alpha}}{t+\chi}\,{}_2F_1\left(\begin{matrix}\frac{\Delta_l}{2}, & \frac{\Delta_l}{2}\\ & \Delta_l-\frac{d-2}{2}\end{matrix};-t\right)\,.
    \label{eq:Galphasign}
\end{equation}
This is a rewriting of the dispersion relation defining the local block as an integral over the standard block (\ref{eq:dispBbb}). The purpose of the Mellin Barnes representation (\ref{eq:GBbbalphaMB}) is to provide a prescription to analytically continue beyond regime of convergence of this integral $\alpha<\frac{\Delta_l}{2}+1$. For $N-1<\alpha-\frac{\Delta_l}{2}<N+1$, with $N\geq1$ an integer, we can indeed move the contour $\mathcal{C}$ in (\ref{eq:GBbbalphaMB}) past the first $N$ poles at $s=\frac{\Delta_l}{2}+n$ to instead obtain
\begin{equation}
    G^{\Delta_l,\Delta_i,\Delta_i,\alpha}_{\text{Bbb}}(\chi)=\sum_{n=0}^{N-1}f_l\big(\tfrac{\Delta_l}{2}+n\big)\chi^{\frac{\Delta_l}{2}+n}+\frac{\sin\pi\left(\alpha-\frac{\Delta_l}{2}\right)}{\pi}\,\chi^\alpha\int_0^\infty dt\,\frac{t^{\frac{\Delta_l}{2}-\alpha}}{t+\chi}\,R_{l,N}(t)\,,
\end{equation}
where
\begin{equation}
    R_{l,N}(t)\equiv{}_2F_1\left(\begin{matrix}\frac{\Delta_l}{2}, & \frac{\Delta_l}{2}\\ & \Delta_l-\frac{d-2}{2}\end{matrix};-t\right)-\sum_{n=0}^{N-1}(-1)^n f_l\big(\tfrac{\Delta_l}{2}+n\big)\,t^n\,,
\end{equation}
is the hypergeometric function minus its Taylor polynomial of degree $N-1$. The range of $\alpha$ is the one where the integral over $t$ converges, since $R_{l,N}(t)\sim t^N$ for $t\to0$ and $R_{l,N}(t)\sim t^{N-1}$ for $t\to\infty$. To determine the sign of $R_{l,N}$, we use the Euler integral representation of the hypergeometric function, valid for $\Delta_l>d-2$ and thus in particular under our assumption $\Delta_l>d$,
\begin{equation}
    {}_2F_1\left(\begin{matrix}\frac{\Delta_l}{2}, & \frac{\Delta_l}{2}\\ & \Delta_l-\frac{d-2}{2}\end{matrix};-t\right)=\frac{\Gamma\big(\Delta_l-\frac{d-2}{2}\big)}{\Gamma\big(\frac{\Delta_l}{2}\big)\Gamma\big(\frac{\Delta_l-d+2}{2}\big)}\int_0^1du\,u^{\frac{\Delta_l}{2}-1}(1-u)^{\frac{\Delta_l-d}{2}}\,(1+tu)^{-\frac{\Delta_l}{2}}\,.
\end{equation}
Subtracting the Taylor polynomial amounts to replacing $(1+tu)^{-\frac{\Delta_l}{2}}$ by the remainder of its binomial series, which we write using Taylor's theorem with integral remainder:
\begin{equation}
g(x)-\sum_{k=0}^{N-1}\frac{g^{(k)}(0)}{k!}x^k=\frac{x^N}{(N-1)!}\int_0^1dv\,(1-v)^{N-1}g^{(N)}(xv)
\end{equation}
applied to $g(x)=(1+x)^{-\frac{\Delta_l}{2}}$, whose derivatives are $g^{(N)}(x)=(-1)^N\big(\frac{\Delta_l}{2}\big)_N(1+x)^{-\frac{\Delta_l}{2}-N}$:

\begin{equation}
    R_{l,N}(t)=\frac{(-1)^N\Gamma\big(\Delta_l-\frac{d-2}{2}\big)\big(\frac{\Delta_l}{2}\big)_Nt^N}{(N-1)!\,\Gamma\big(\frac{\Delta_l}{2}\big)\Gamma\big(\frac{\Delta_l-d+2}{2}\big)}\int_0^1du\int_0^1dv\,\frac{u^{\frac{\Delta_l}{2}+N-1}(1-u)^{\frac{\Delta_l-d}{2}}(1-v)^{N-1}}{(1+tuv)^{\frac{\Delta_l}{2}+N}}\,.
\end{equation}
The integrand is positive, so $(-1)^NR_{l,N}(t)>0$ for $t>0$, and likewise the hypergeometric function itself is positive. Moreover, we notice that $f_l(\frac{\Delta_l}{2}+n)=\frac{\big(\frac{\Delta_l}{2}\big)_n^2}{n!\,\big(\Delta_l-\frac{d-2}{2}\big)_n}>0\,.$

We can now read off the sign of the local block. If $\alpha>\frac{\Delta_l}{2}$, we choose $N=\lfloor\alpha-\frac{\Delta_l}{2}\rfloor$. Then $\sin\pi\big(\alpha-\frac{\Delta_l}{2}\big)$ has sign $(-1)^N$, like $R_{l,N}$, so the integral term is non-negative, while each term in the sum over $n$ is positive. For $N=0$ we use (\ref{eq:Galphasign}) instead, where both the sine and the hypergeometric function are positive. If $\alpha\leq\frac{\Delta_l}{2}$, (\ref{eq:Galphasign}) shows that the block has the sign of $\sin\pi\big(\alpha-\frac{\Delta_l}{2}\big)$. In summary, for all $\chi>0$, $G^{\Delta_l,\Delta_i,\Delta_i,\alpha}_{\text{Bbb}}(\chi)$ is positive for $\Delta_l<2\alpha$, vanishes for $\Delta_l\in2\alpha+2\mathbb{N}$, and otherwise has the sign of $\sin\pi\big(\alpha-\frac{\Delta_l}{2}\big)$.

 \subsection{Flow of BOE coefficients}
 \label{sec:sec_eq}
We want to compute the following quantity:
\begin{equation}
     I_{i}^{\hat{\Phi}\hat{\Phi}}(\mathbf{x}_1,\mathbf{x}_2, z_2, \epsilon)=\left(\frac{z_2}{\mathbf{x}_{12}^2+z_2^2}\right)^{\Delta_i} \sum_{j,l}b^{\hat\Phi}_l b^{\hat\Phi}_jC_{lij}I_\mathrm{BBb}^{\Delta_l,\Delta_j,\Delta_i}(\mathbf{x}_1,\mathbf{x}_2, z_2, \epsilon)\,,    \label{eq:I_second_eq2}
\end{equation}
by first computing the
integrated bulk-bulk-boundary block, defined as
\begin{equation}
    I_\mathrm{BBb}^{\Delta_l,\Delta_j,\Delta_i}(\mathbf{x}_1,\mathbf{x}_2, z_2, \epsilon)=\int_{|\mathbf{x}-\mathbf{x}_1|^2+z^2 \ge \epsilon^2} \frac{dz d^d\mathbf{x}}{z^{d+1}}G_{\text{BBb}}^{\Delta_l,\Delta_j,\Delta_i}(\xi,\rho)\,.
\end{equation}
 The above integral is finite for $\Delta_j>\Delta_i$, and can be re-expressed as the difference of two contributions, both finite in this regime: the integral in the whole AdS and the integral inside the subtracted half-ball around $\mathbf{x}_1$, as in
\begin{equation}
    I_\mathrm{BBb}^{\Delta_l,\Delta_j,\Delta_i}(\mathbf{x}_1,\mathbf{x}_2, z_2, \epsilon) = I_\mathrm{BBb}^{\Delta_l,\Delta_j,\Delta_i}(\mathbf{x}_1,\mathbf{x}_2, z_2, 0)-I_\mathrm{BBb,in}^{\Delta_l,\Delta_j,\Delta_i}(\mathbf{x}_1,\mathbf{x}_2, z_2, \epsilon) \,,
    \label{eq:IBBb_decomp}
\end{equation}
where
\begin{equation}
   I_\mathrm{BBb,in}^{\Delta_l,\Delta_j,\Delta_i}(\mathbf{x}_1,\mathbf{x}_2, z_2, \epsilon)   := \int_{|\mathbf{x}-\mathbf{x}_1|^2+z^2 < \epsilon^2} \frac{dz d^d\mathbf{x}}{z^{d+1}} G_{\text{BBb}}^{\Delta_l,\Delta_j,\Delta_i}(\xi,\rho)\,.
    \label{eq:IBBb_in}
\end{equation}
 Our strategy is to compute both terms separately, analytically continue them to arbitrary $\Delta_i$ and $\Delta_j$, and substitute the results back into \eqref{eq:I_second_eq2}. This gives
 \begin{equation}
     I_{i}^{\hat{\Phi}\hat{\Phi}}(\mathbf{x}_1,\mathbf{x}_2, z_2, \epsilon) =\left(\frac{z_2}{\mathbf{x}_{12}^2+z_2^2}\right)^{\Delta_i} \sum_{j,l}b^{\hat\Phi}_l b^{\hat\Phi}_jC_{lij}\left(I_\mathrm{BBb}^{\Delta_l,\Delta_j,\Delta_i}(\mathbf{x}_1,\mathbf{x}_2, z_2, 0) -I_\mathrm{BBb,in}^{\Delta_l,\Delta_j,\Delta_i}(\mathbf{x}_1,\mathbf{x}_2, z_2, \epsilon) \right)\,.
     \label{eq:IBBb_separated}
 \end{equation}
 We begin by evaluating $I_\mathrm{BBb}^{\Delta_l,\Delta_j,\Delta_i}(\mathbf{x}_1,\mathbf{x}_2, z_2, 0)$. Because this integral is independent of the configuration of points that we choose, it depends exclusively on the scaling dimensions $\Delta_i$, $\Delta_l$, and $\Delta_j$. We can therefore define a new quantity $\mathcal{J}_{\Delta_i}(\Delta_l,\Delta_j)$, that for convenience absorbs an overall minus sign,
 \begin{equation}
     \mathcal{J}_{\Delta_i}(\Delta_l,\Delta_j)\equiv -I_\mathrm{BBb}^{\Delta_l,\Delta_j,\Delta_i}(\mathbf{x}_1,\mathbf{x}_2, z_2, 0)\,.
 \end{equation}
  To compute $\mathcal{J}_{\Delta_i}(\Delta_l,\Delta_j)$, we select the specific points
\begin{equation}
   \mathbf{x}_1=\infty\,,\qquad (z_2,\mathbf{x}_2)=(1,0)\,,\qquad (z,\mathbf{x})=(z,r\hat{\mathbf{x}}^1)\,\,   ,
    \label{eq:frameBBb3}
\end{equation}
for which the cross ratios remarkably simplify:
$\xi={1}/{z^2}$, $\rho={r^2}/{z^2}$.
The integral then becomes
\begin{equation}
    \mathcal{J}_{\Delta_i}(\Delta_l,\Delta_j)=-{\text{Vol}(S^{d-1})} \int \frac{ dz dr\  r^{d-1}}{z^{d+1}}G_{\text{BBb}}^{\Delta_l,\Delta_j,\Delta_i}\left(-\frac{z^2}{r^2},-\frac{1}{r^2}\right)\,,
\end{equation}
which, using (\ref{eq:convergentBBb}), is
\begin{equation}
\begin{aligned}
    \mathcal{J}_{\Delta_i}(\Delta_l,&\Delta_j) = -\text{Vol}(S^{d-1}) \sum_{n=0}^\infty \frac{\left(\frac{\Delta_{ilj}}{2}\right)_n \left(\frac{\Delta_{lji}}{2}\right)_n}{n! \left(1-\frac{d}{2}+\Delta_l\right)_n} \int_0^\infty dr \int_0^\infty dz\, \frac{r^{d-1+\Delta_{ij}-\Delta_l}}{z^{d+1-2n-\Delta_l}} \\
    &\left(r^2+z^2\right)^{\frac{\Delta_{ij}-\Delta_l}{2}-n}  \,_3F_2\left(\begin{matrix} \frac{2-d}{2}+\frac{\Delta_{lji}}{2}, & n+\frac{\Delta_{lji}}{2}, & 1-\frac{\Delta_{ilj}}{2} \\ \frac{2-d}{2}+\Delta_j, & 1-n-\frac{\Delta_{ilj}}{2} \end{matrix}; -\frac{1}{r^2+z^2} \right)\,.
\end{aligned}
\end{equation}
This $_3F_2$ admits a representation as a finite sum of $_2F_1$, through the following identity
\begin{align}
     \,_2F_2\left(\begin{matrix}
        a_1, & a_2, & c\\ b, & c-n
     \end{matrix};z\right)
    =\sum_{l=0}^n\binom{n}{l}\frac{z^l}{(c-n)_l}\frac{(a_1)_l(a_2)_l}{(b)_l}
    \,_{2}F_{1}\left(a_1+l,a_2+l,b+l,z\right)
    \,.
\end{align}
Opening up the $_2F_1$ as well, we can carry out the integral over $z$. This results into
\begin{equation}
    \begin{aligned}
        \mathcal{J}_{\Delta_i}(\Delta_l,&\Delta_j) = -\text{Vol}(S^{d-1}) \sum_{m,n=0}^\infty \sum_{l=0}^n \frac{(-1)^{l+m} \binom{n}{l}}{2 m! n!} \frac{\Gamma\left(\frac{d+2l+2m-\Delta_{ij}}{2}\right) \Gamma\left(\frac{-d+2n+\Delta_l}{2}\right)\left(\frac{\Delta_{ilj}}{2}\right)_n }{\Gamma\left(l+m+n+\frac{\Delta_{lji}}{2}\right)\left(1-\frac{d}{2}+\Delta_l\right)_n} \\
        & \frac{\left(\frac{\Delta_{lji}}{2}\right)_n\left(\frac{2-d+\Delta_{lji}}{2}\right)_l \left(\frac{2n+\Delta_{lji}}{2}\right)_l\left(\frac{2-d+2l+\Delta_{lji}}{2}\right)_m \left(\frac{2l+2n+\Delta_{lji}}{2}\right)_m}{\left(1-\frac{d}{2}+\Delta_j\right)_l \left(1-n-\frac{\Delta_{ilj}}{2}\right)_l\left(1-\frac{d}{2}+l+\Delta_j\right)_m} \int_0^\infty dr\, r^{\Delta_{ij}-1-2l-2m}\,.
    \end{aligned}
\end{equation}
The remaining integral does not converge, but we can improve the situation by resumming over $ m$ and integrating over $r$ after that. 
The result reads
 \begin{equation}
   \mathcal{J}_{\Delta_i}(\Delta_l,\Delta_j)=-\frac{\pi ^{d/2} \Gamma \left(\frac{\Delta_{ji}}{2}\right) \Gamma \left(\Delta_l-\frac{d-2}{2}\right) \Gamma \left(\Delta_j-\frac{d-2}{2}\right) \Gamma \left(\frac{\Delta_l-d}{2}\right)}{\Delta_l \Gamma \left(\frac{\Delta_l}{2}\right) \Gamma \left(\frac{\Delta_{lji}}{2}\right) \Gamma \left(\frac{\Delta_i+\Delta_j-d+2}{2} \right) \Gamma \left(\frac{\Delta_{lji}-d+2}{2} \right)}\,.
   \label{eq:mathcalJ_app}
\end{equation}
We note that in the limit $\Delta_j\rightarrow\Delta_i$ this quantity diverges, and we get 
\begin{equation}
\mathcal{J}_{\Delta_i}(\Delta_l,\Delta_j)= \frac{2 \pi ^{d/2} \Gamma (-\frac{d}{2}+\Delta_l+1)}{\Delta_{ji}\Gamma (\frac{\Delta_l}{2})\Gamma (\frac{\Delta_l+2}{2}) (d-\Delta_l)} +[\mathcal{J}_{\Delta_i}(\Delta_l,\Delta_j)]_\mathrm{0}+\mathcal{O}(\Delta_{ji})\,,
\label{eq:J_delta_i}
\end{equation}
where
\begin{equation}
[\mathcal{J}_{\Delta_i}(\Delta_l,\Delta_j)]_\mathrm{0}=    \frac{2 \pi^{d/2} \Gamma\left(1 - \frac{d}{2} + \Delta_l\right) \left( \gamma_E - H_{-\frac{d}{2} + \Delta_i} + H_{\frac{\Delta_l - d}{2}} + \psi\left(\frac{\Delta_l}{2}\right) \right)}{\Delta_l (\Delta_l - d) \Gamma\left(\frac{\Delta_l}{2}\right)^2}\,.
\end{equation}
We are now left with the computation of $
     I_\mathrm{BBb,in}^{\Delta_l,\Delta_j,\Delta_i}$ in \eqref{eq:IBBb_in}. The first step is to 
     rewrite the bulk-bulk-boundary blocks in terms of bulk-boundary-boundary blocks. This can be done by starting from the three point function in \eqref{eq:BBb3point} and replacing one of the bulk  operator with its BOE expansion. Replacing then equation \eqref{eq:Bbbdec} and imposing that this decomposition has to match with the expansion in bulk-bulk-boundary blocks, one gets
     \begin{equation}
     \begin{aligned}
         G_{\text{BBb}}^{\Delta_l,\Delta_j,\Delta_i}(\xi,\rho)=&\left(\frac{\mathbf{x}_{12}^2+z_2^2}{z_2}\right)^{\Delta_i} z_2^{\Delta_j}\sum_{n} \frac{(-1)^n}{n!2^{2n}(\Delta_j-\frac{d-2}{2})_n}z_2^{2n}\\&\times\Box_{\mathbf{x}_2}^n\left(\frac{1}{|\mathbf{x}_{12}|^{\Delta_i+\Delta_j}} \left( \frac{|\mathbf{x}-\mathbf{x}_1|^2+z^2}{|\mathbf{x}-\mathbf{x}_2|^2+z^2} \right)^{\frac{\Delta_j-\Delta_i}{2}} G_{\text{Bbb}}^{\Delta_l,\Delta_i,\Delta_j}(\chi)\right)\,.
     \end{aligned}
    \end{equation}
     This implies that we can rewrite the integral in \eqref{eq:IBBb_in} in terms of  the bulk-boundary-boundary integrated block $I^{\Delta_l,\Delta_i,\Delta_j}_{\text{Bbb,in}}$ defined in \eqref{eq:IBbb_in}. We then obtain
    \begin{equation}
       I_\mathrm{BBb,in}^{\Delta_l,\Delta_j,\Delta_i}=\left(\frac{\mathbf{x}_{12}^2+z_2^2}{z_2}\right)^{\Delta_i}z_2^{\Delta_j}\sum_{n}\frac{(-1)^n}{n!2^{2n}(\Delta_j-\frac{d-2}{2})_n}z_2^{2n}\Box_{\mathbf{x}_2}^nI^{\Delta_l,\Delta_i,\Delta_j}_{\text{Bbb,in}}.
    \end{equation}
    Matching with \eqref{eq:coeff_a} and \eqref{eq:relIinout}, we get 

        \begin{equation}
        \begin{aligned}
I_\mathrm{BBb,in}^{\Delta_l,\Delta_j,\Delta_i}&=\left(\frac{\mathbf{x}_{12}^2+z_2^2}{z_2}\right)^{\Delta_i}\sum_{n,p} \frac{(-1)^{n+1}}{n!2^{2n}(\Delta_j-\frac{d-2}{2})_n}z_2^{2n+\Delta_j}\Box_{\mathbf{x}_2}^n    \frac{a_{2p}(\Delta_{ij}, \Delta_l)}{\Delta_{ij} - 2p}\frac{{\epsilon}^{2p-\Delta_{ij}}}{|\mathbf{x}_{12}|^{2\Delta_j+2p}} 
       \\ &=\left(\frac{\mathbf{x}_{12}^2+z_2^2}{z_2}\right)^{\Delta_i}\sum_{n,p} \frac{(-1)^{n+1} z_2^{2n+\Delta_j}a_{2p}(\Delta_{ij}, \Delta_l){\epsilon}^{2p-\Delta_{ij} }\Box_{\mathbf{x}_2}^n\Box_{\mathbf{x}_1}^p \langle \mathcal{O}_j(\mathbf{x}_1)\mathcal{O}_j(\mathbf{x}_2)\rangle}{n!2^{2n}(\Delta_j-\frac{d-2}{2})_n2^{2p}(\Delta_j-\frac{d-2}{2})_p(\Delta_j)_p(\Delta_{ij} - 2p)} \\&  =-\sum_{p} \frac{a_{2p}(\Delta_{ij}, \Delta_l){\epsilon}^{2p-\Delta_{ij} } }{2^{2p}(\Delta_j-\frac{d-2}{2})_p(\Delta_j)_p(\Delta_{ij} - 2p)}\Box_{\mathbf{x}_2}^p\left(\frac{\mathbf{x}_{12}^2+z_2^2}{z_2}\right)^{\Delta_j+\Delta_i}
        \end{aligned}
        \end{equation}
 Having in mind equation \eqref{eq:IBBb_separated}, we can insert this result in the sum
 \begin{equation}
 \begin{aligned}
     \left(\frac{z_2}{\mathbf{x}_{12}^2+z_2^2}\right)^{\Delta_i} &\sum_{j,l}b^{\hat\Phi}_l b^{\hat\Phi}_jC_{lij}I_\mathrm{BBb,in}^{\Delta_l,\Delta_j,\Delta_i}\\&=-\sum_{p,j,l} b^{\hat\Phi}_l b^{\hat\Phi}_jC_{lij}\frac{a_{2p}(\Delta_{ij}, \Delta_l){\epsilon}^{2p-\Delta_{ij} } }{2^{2p}(\Delta_i-\frac{d-2}{2})_p(\Delta_i)_p(\Delta_{ij} - 2p)}\Box_{\mathbf{x}_1}^p\left(\frac{\mathbf{x}_{12}^2+z_2^2}{z_2}\right)^{\Delta_j}\end{aligned}
 \end{equation}
First, consider the case $j\ne i$. Using the counterterm expressions found in \eqref{eq:counterterm}, we recover
 \begin{equation}
\sum_{j\ne i,p} b_j^{\hat{\Phi}}\frac{\delta\mathcal{Z}_{ij;p}}{\delta\lambda}\Box_{\mathbf{x}_1}^p \left(\frac{\mathbf{x}_{12}^2+z_2^2}{z_2}\right)^{\Delta_j}\,.
 \end{equation}
 To see what happens as $\Delta_{j}\rightarrow\Delta_{i}$ in the sum, we observe that the only non-vanishing contribution comes from the $p=0$ term, which reads:
    \begin{equation}\begin{aligned}
\left(\frac{\mathbf{x}_{12}^2+z_2^2}{z_2}\right)^{\Delta_i} b^{\hat{\Phi}}_i\sum_l b^{\hat{\Phi}}_l C_{jil} a_0(0, \Delta_l) \left( \frac{1}{\Delta_{ji}} + \log \epsilon + \log\left(\frac{z_2}{\mathbf{x}_{12}^2+z_2^2}\right) \right)  + \mathcal{O}(\Delta_{ji})\,,
\end{aligned}
\label{eq:I_BBb_in_Deltai}\end{equation}
where we used $\partial_{\Delta_{ji}}a_0(0, \Delta_l)=0$, as obtained in \eqref{eq:a_der}. Matching with the expression for $a_0$ found in \eqref{eq:a_coeff}, and replacing in \eqref{eq:IBBb_separated},  we find that the pole in $\Delta_{ji}$ cancels in the total result and we get 
\begin{equation}
\begin{aligned}
    I_{i}^{\hat{\Phi}\hat{\Phi}}(\mathbf{x}_1, \mathbf{x}_2, z_2,\epsilon)= -&\left(\frac{z_2}{\mathbf{x}_{12}^2+z_2^2}\right)^{\Delta_i} \sum_{j,l}b^{\hat\Phi}_l b^{\hat\Phi}_jC_{lij}[\mathcal{J}_{\Delta_i}(\Delta_l,\Delta_j)]_\mathrm{reg}\, \\&- b^{\hat{\Phi}}_i\left(\frac{z_2}{\mathbf{x}_{12}^2+z_2^2}\right)^{\Delta_i} \sum_l b^{\hat{\Phi}}_l C_{iil} 
\mathcal{I}( \Delta_l)  \log\left(\frac{z_2 \epsilon}{\mathbf{x}_{12}^2+z_2^2}\right)  \\&-\sum_{j\ne i} \sum_{n=0}^\infty b_j^{\hat{\Phi}}\frac{\delta\mathcal{Z}_{ij;n}}{\delta\lambda}\Box_{\mathbf{x}_2}^n \left(\frac{z_2}{\mathbf{x}_{12}^2+z_2^2}\right)^{\Delta_j(\lambda)} \,.
    \end{aligned}
\end{equation}

\subsubsection{Singularities at $\Delta_{ij}=2p$}
\label{app:resonances}
In this section, we consider the case $\Delta_{ij}=2p$ with $p\geq1$, i.e. when $\mathcal{O}_i$ is degenerate with the descendant $\Box^p\mathcal{O}_j$. Since the spectrum is non-degenerate, for given $i$ and $p$ there is at most one such $\mathcal{O}_j$. For this $j$, two terms in \eqref{eq:IBBb_separated} are singular. The first is $\mathcal{J}_{\Delta_i}(\Delta_l,\Delta_j)$, through the factor $\Gamma(\Delta_{ji}/2)$ in \eqref{eq:mathcalJ_app}, and it multiplies $\left(z_2/(\mathbf{x}_{12}^2+z_2^2)\right)^{\Delta_i}$. The second is the $p$-th term in the half-ball contribution, through the factor $a_{2p}(\Delta_{ij},\Delta_l)/(\Delta_{ij}-2p)$, and it multiplies $\epsilon^{2p-\Delta_{ij}}\,\Box^p_{\mathbf{x}_1}\left(z_2/(\mathbf{x}_{12}^2+z_2^2)\right)^{\Delta_j}$.

For $p=0$, i.e.\ for $\Delta_j\to\Delta_i$, the two poles multiply the same function and cancel block by block, as shown above. For $p\geq1$, instead, $\Box^p_{\mathbf{x}_1}\left(z_2/(\mathbf{x}_{12}^2+z_2^2)\right)^{\Delta_j}$ and $\left(z_2/(\mathbf{x}_{12}^2+z_2^2)\right)^{\Delta_i}$ are different functions of $\mathbf{x}_{12}$ and $z_2$, even at the singularity, so the two poles cannot cancel against each other. Since the regulated integral \eqref{eq:IBBb_def} is finite, each of them must cancel separately.

The half-ball contribution is in fact regular term by term, since $a_{2p}(2p,\Delta_l)=0$ for every $\Delta_l$, see \eqref{eq:a2p}. It reproduces the counterterm $\delta\mathcal{Z}_{ij;p}$ in \eqref{eq:counterterm}, exactly as for non-singular $j$. Its finite value is scheme dependent, but this is unimportant: as for all counterterms, it cancels against the corresponding term in \eqref{eq:Bbprop1} and does not enter the flow equation.

For $\mathcal{J}_{\Delta_i}$, instead, the finite part enters the flow equation \eqref{eq:second_eq_final}. We thus need to show that $\sum_l b_l^{\hat\Phi}C_{ijl}\,\mathcal{J}_{\Delta_i}(\Delta_l,\Delta_j)$ is finite at $\Delta_{ij}=2p$, and to compute its value. Since each term has a pole there, the sum is defined by continuing the terms away from $\Delta_{ij}=2p$, at fixed $b_l^{\hat\Phi}$, $C_{ijl}$ and $\Delta_l$, and then taking the limit. To proceed, we use local blocks. First, we note that since the standard blocks \eqref{eq:BBbdec} and the local blocks \eqref{eq:GBBbalpha1} decompose the same correlator, the subtraction terms must sum to zero:
\begin{equation}
     \sum_l b_l^{\hat\Phi}C_{ijl}\left(G_{\text{BBb}}^{\Delta_l,\Delta_j,\Delta_i,\alpha}(\xi,\rho)-G_{\text{BBb}}^{\Delta_l,\Delta_j,\Delta_i}(\xi,\rho)\right)=0\,.
     \label{eq:subBBb}
\end{equation}
Using the decomposition in \eqref{eq:GBBbalpha}, each subtraction term can be rewritten as
\begin{equation}
    G_{\text{BBb}}^{\Delta_l,\Delta_j,\Delta_i,\alpha}(\xi,\rho)-G_{\text{BBb}}^{\Delta_l,\Delta_j,\Delta_i}(\xi,\rho)=\sum_{m}\beta_m(\Delta_l,\Delta_{ij})\,\mathcal{T}_m(\xi,\rho)\,,
\end{equation}
where
\begin{equation}
    \beta_m(\Delta_l,\Delta_{ij})=\frac{\Gamma\left(\Delta_l-\frac{d-2}{2}\right)}{\Gamma\left(1+\alpha+m-\frac{\Delta_l}{2}\right)\Gamma\left(\alpha+m+\frac{\Delta_l}{2}-\frac{d-2}{2}\right)\Gamma\left(\frac{\Delta_l+\Delta_{ij}}{2}\right)\Gamma\left(\frac{\Delta_l-\Delta_{ij}}{2}\right)}\,,
    \label{eq:betam}
\end{equation}
and $\mathcal{T}_m$ depends on $\Delta_i$, $\Delta_j$ and $\alpha$, but not on $\Delta_l$. Since the $\mathcal{T}_m$ are independent functions of the cross ratios, \eqref{eq:subBBb} is equivalent to the sum rules
\begin{equation}
    \sum_{l} b_l^{\hat\Phi}C_{ijl}\,\beta_m(\Delta_l,\Delta_{ij})=0\,,\qquad m\in\mathbb{N}_0\,,
    \label{eq:betasumrule}
\end{equation}
which are a form of the locality sum rules \eqref{eq:linearsumrules}\footnote{The coefficients $\theta^{\alpha,n}_{\Delta_l\Delta_i\Delta_j}$ appearing in \eqref{eq:linearsumrules}, whose explicit expression can be found in \cite{Levine:2023ywq,Loparco:2025aag}, are related to $\beta_m(\Delta_l,\Delta_{ij})$ in \eqref{eq:betam} as follows
\begin{equation*}
    \beta_m(\Delta_l,\Delta_{ij})=\frac{1}{\Gamma\Big(\alpha+m+\frac{\Delta_{ij}}{2}\Big)\Gamma\Big(\alpha+m-\frac{\Delta_{ij}}{2}\Big)}\sum_{n=0}^{m}\frac{\big(\alpha+n+\frac{\Delta_{ij}}{2}\big)_{m-n}\big(\alpha+n-\frac{\Delta_{ij}}{2}\big)_{m-n}}{(m-n)!\,\big(2\alpha+2n+1-\frac{d}{2}\big)_{m-n}}\;\theta^{\alpha-1,n}_{\Delta_l\Delta_i\Delta_j}\,.
\end{equation*}
Thus, multiplying by $b^{\hat\Phi}_lC_{ijl}$ and summing over $l$, we see that the sum rules \eqref{eq:linearsumrules} imply \eqref{eq:betasumrule} for every $m\in\mathbb{N}_0$. Conversely, since the relation is triangular with non-vanishing diagonal, one can show that \eqref{eq:betasumrule} implies \eqref{eq:linearsumrules}.}. After integration over AdS, we can then write
\begin{equation}
    \mathcal{J}_{\Delta_i}(\Delta_l,\Delta_j)=\mathcal{J}^{\alpha}_{\Delta_i}(\Delta_l,\Delta_j)+\sum_m\beta_m(\Delta_l,\Delta_{ij})\,\widetilde{\mathcal{T}}_m(\Delta_i,\Delta_j)\,,
    \label{eq:JminusJalpha}
\end{equation}
where again $\widetilde{\mathcal{T}}_m$ does not depend on $\Delta_l$. For non-singular $j$, summing over $l$ with \eqref{eq:betasumrule} shows that standard and local blocks give the same result. At $\Delta_{ij}=2p$, however, $\widetilde{\mathcal{T}}_m$ has a pole, since $\mathcal{J}_{\Delta_i}$ does while $\mathcal{J}^{\alpha}_{\Delta_i}$ and $\beta_m$ do not, and this step requires more care.

The simplest option would be to discard the pole of each term, i.e.\ to apply the regularization \eqref{eq:regJ} around $\Delta_j=\Delta_i-2p$. Expanding $\widetilde{\mathcal{T}}_m=\widetilde{\mathcal{T}}_m^{(-1)}/(\Delta_{ij}-2p)+\widetilde{\mathcal{T}}_m^{(0)}+O(\Delta_{ij}-2p)$ in \eqref{eq:JminusJalpha}, this gives
\begin{equation}
    \sum_l b_l^{\hat\Phi}C_{ijl}\,[\mathcal{J}_{\Delta_i}(\Delta_l,\Delta_j)]_{\mathrm{reg}}=\sum_l b_l^{\hat\Phi}C_{ijl}\,\mathcal{J}^{\alpha}_{\Delta_i}(\Delta_l,\Delta_j)+\sum_m\widetilde{\mathcal{T}}_m^{(-1)}\sum_l b_l^{\hat\Phi}C_{ijl}\,\partial_{\Delta_{ij}}\beta_m(\Delta_l,2p)\,,
    \label{eq:naiveJ}
\end{equation}
where the terms with $\widetilde{\mathcal{T}}_m^{(0)}$ cancel by \eqref{eq:betasumrule}. The last term does not vanish. Indeed, discarding the poles amounts to continuing each term away from $\Delta_{ij}=2p$ at fixed $b_l^{\hat\Phi}$, $C_{ijl}$ and $\Delta_l$, and along this continuation the sum rules \eqref{eq:betasumrule} are violated. They hold at the physical value $\Delta_{ij}=2p$, and thus constrain the values of the $\beta_m$ there, but not their derivatives, and
\begin{equation}
    \partial_{\Delta_{ij}}\log\beta_m(\Delta_l,\Delta_{ij})=-\frac{1}{2}\left[\psi\left(\frac{\Delta_l+\Delta_{ij}}{2}\right)-\psi\left(\frac{\Delta_l-\Delta_{ij}}{2}\right)\right]
\end{equation}
depends on $\Delta_l$. The $O(\Delta_{ij}-2p)$ violation, multiplied by the pole of $\widetilde{\mathcal{T}}_m$, leaves the finite remainder in \eqref{eq:naiveJ}.

This remainder is spurious. The sum rules \eqref{eq:betasumrule} express the fact that standard and local blocks decompose the same correlator, and the local-block result, i.e.\ the first term on the right-hand side of \eqref{eq:naiveJ}, is unambiguous, since each $\mathcal{J}^{\alpha}_{\Delta_i}$ is finite at $\Delta_{ij}=2p$. A prescription for the sum over standard blocks is therefore correct only if it reproduces this result, i.e.\ only if the continuation away from $\Delta_{ij}=2p$ preserves the sum rules. This is achieved by rescaling each term by
\begin{equation}
    r_p(\Delta_l,\Delta_{ij})=\frac{\Gamma\left(\frac{\Delta_l+\Delta_{ij}}{2}\right)\Gamma\left(\frac{\Delta_l-\Delta_{ij}}{2}\right)}{\Gamma\left(\frac{\Delta_l}{2}+p\right)\Gamma\left(\frac{\Delta_l}{2}-p\right)}\,,
    \label{eq:rp}
\end{equation}
which equals one at $\Delta_{ij}=2p$, and is such that $r_p(\Delta_l,\Delta_{ij})\,\beta_m(\Delta_l,\Delta_{ij})=\beta_m(\Delta_l,2p)$ does not depend on $\Delta_{ij}$. Multiplying \eqref{eq:JminusJalpha} by $r_p$ and summing over $l$, the sum rules \eqref{eq:betasumrule} give
\begin{equation}
    \sum_l b_l^{\hat\Phi}C_{ijl}\,r_p(\Delta_l,\Delta_{ij})\,\mathcal{J}_{\Delta_i}(\Delta_l,\Delta_j)=\sum_l b_l^{\hat\Phi}C_{ijl}\,r_p(\Delta_l,\Delta_{ij})\,\mathcal{J}^{\alpha}_{\Delta_i}(\Delta_l,\Delta_j)
    \label{eq:JrpJalpha}
\end{equation}
for all $\Delta_j$ in a neighbourhood of $\Delta_i-2p$. The right-hand side is regular in this neighbourhood, because $\mathcal{J}^{\alpha}_{\Delta_i}$ in \eqref{eq:Jalpha} has no pole at $\Delta_{ij}=2p$, and the sum over $l$ converges as for non-singular $j$, since $r_p\to1$ for $\Delta_l\to\infty$. Hence the left-hand side is regular as well, which shows that the sum is finite. Its value at $\Delta_{ij}=2p$, where $r_p=1$, can be computed term by term with \eqref{eq:regJ}:
\begin{equation}
\begin{split}
    \sum_l b_l^{\hat{\Phi}}C_{ijl}\,\mathcal{J}_{\Delta_i}(\Delta_l,\Delta_j)\Big|_{\Delta_{ij}=2p}&=\sum_l b_l^{\hat{\Phi}}C_{ijl}\left[r_p(\Delta_l,\Delta_{ij})\,\mathcal{J}_{\Delta_i}(\Delta_l,\Delta_j)\right]_{\mathrm{reg}}\\
    &=\sum_l b_l^{\hat{\Phi}}C_{ijl}\bigg([\mathcal{J}_{\Delta_i}(\Delta_l,\Delta_j)]_{\mathrm{reg}}\\
    &\qquad\qquad-\sum_{k=-p}^{p-1}\frac{1}{\Delta_l+2k}\,\underset{\Delta_j=\Delta_i-2p}{\mathrm{Res}}\,\mathcal{J}_{\Delta_i}(\Delta_l,\Delta_j)\bigg)\,.
\end{split}
\label{eq:Jresonant_app}
\end{equation}
The second line shows that the correct value is obtained from the naive one by subtracting from each term a finite amount, proportional to its residue. Summed over $l$, these subtractions cancel the spurious remainder in \eqref{eq:naiveJ}.\footnote{We checked that the prescription in \eqref{eq:Jresonant_app} is the correct one in the free scalar theory, with $\mathcal{O}_j=\phi^2$ and $\mathcal{O}_i=[\phi^2]_p$, for $p=1,2$ in AdS$_2$ at $\lambda=4$. We used diagonal Pad\'e approximants for the sums over $l$, which diverge for standard blocks. In all cases, the right-hand side of \eqref{eq:Jresonant_app} agrees with the local-block sum to all computed digits, for several values of $\alpha$. }

\section{Some properties of $\frac{dC_{ijk}}{d\lambda}$}
\label{app:OPEflow}
In this work, we have not derived the higher dimensional analogue of the flow equation involving the derivative of the OPE coefficients \cite{Loparco:2026fki}, instead arguing that the crossing equation is a more efficient substitute. Nevertheless, we can prove certain properties of $\frac{dC_{ijk}}{d\lambda}$ without computing the required conformal integrals explicitly. These properties are important in the derivations of the merger-annihilation and level repulsion scenarios in section \ref{sec:mergerrepulsion}. We will focus on OPE coefficients of scalar operators for simplicity.

The variation of the OPE coefficients is captured by the correction to the boundary three-point function in perturbation theory. The renormalized operators have three-point function
\begin{equation}
    \langle\mathcal{O}^{(\text{ren})}_i(\mathbf{x}_1)\mathcal{O}^{(\text{ren})}_j(\mathbf{x}_2)\mathcal{O}^{(\text{ren})}_k(\mathbf{x}_3)\rangle_{\lambda+\delta\lambda}=\frac{C_{ijk}+\delta C_{ijk}}{(\mathbf{x}_{12}^2)^{\frac{\Delta_{ijk}}{2}+\frac{\delta\Delta_{ijk}}{2}}(\mathbf{x}_{23}^2)^{\frac{\Delta_{jki}}{2}+\frac{\delta\Delta_{jki}}{2}}(\mathbf{x}_{13}^2)^{\frac{\Delta_{ikj}}{2}+\frac{\delta\Delta_{ikj}}{2}}}
\end{equation}
Using (\ref{eq:barerenmatch}), the bare operators have instead
\begin{equation}
    \begin{aligned}\langle&\mathcal{O}^{(\text{bare})}_i(\mathbf{x}_1)\mathcal{O}^{(\text{bare})}_j(\mathbf{x}_2)\mathcal{O}^{(\text{bare})}_k(\mathbf{x}_3)\rangle_{\lambda+\delta\lambda}=\frac{C_{ijk}}{(\mathbf{x}_{12}^2)^{\frac{\Delta_{ijk}}{2}}(\mathbf{x}_{23}^2)^{\frac{\Delta_{jki}}{2}}(\mathbf{x}_{13}^2)^{\frac{\Delta_{ikj}}{2}}}\\
    &\times\Bigg[1+\frac{\delta C_{ijk}}{C_{ijk}}+\frac{\delta\Delta_i}{2}\log\left(\frac{\mathbf{x}_{23}^2}{\mathbf{x}_{12}^2\mathbf{x}_{13}^2}\right)+\frac{\delta\Delta_j}{2}\log\left(\frac{\mathbf{x}_{13}^2}{\mathbf{x}_{12}^2\mathbf{x}_{23}^2}\right)+\frac{\delta\Delta_k}{2}\log\left(\frac{\mathbf{x}_{12}^2}{\mathbf{x}_{23}^2\mathbf{x}_{13}^2}\right)\Bigg]\\
    &\qquad+\sum_m\sum_{p=0}^\infty\delta\mathcal{Z}_{im,p}\Box_1^p\frac{C_{mjk}}{(\mathbf{x}_{12}^2)^{\frac{\Delta_{mjk}}{2}}(\mathbf{x}_{23}^2)^{\frac{\Delta_{jkm}}{2}}(\mathbf{x}_{13}^2)^{\frac{\Delta_{mkj}}{2}}}\\
    &\qquad+\sum_m\sum_{p=0}^\infty\delta\mathcal{Z}_{jm,p}\Box_2^p\frac{C_{imk}}{(\mathbf{x}_{12}^2)^{\frac{\Delta_{imk}}{2}}(\mathbf{x}_{23}^2)^{\frac{\Delta_{mki}}{2}}(\mathbf{x}_{13}^2)^{\frac{\Delta_{ikm}}{2}}}\\
    &\qquad+\sum_m\sum_{p=0}^\infty\delta\mathcal{Z}_{km,p}\Box_3^p\frac{C_{ijm}}{(\mathbf{x}_{12}^2)^{\frac{\Delta_{ijm}}{2}}(\mathbf{x}_{23}^2)^{\frac{\Delta_{jmi}}{2}}(\mathbf{x}_{13}^2)^{\frac{\Delta_{imj}}{2}}}
    \label{eq:barebbb}
    \end{aligned}
\end{equation}
At the same time
\begin{align}
        \langle\mathcal{O}^{(\text{bare})}_i(\mathbf{x}_1)\mathcal{O}^{(\text{bare})}_j(\mathbf{x}_2)\mathcal{O}^{(\text{bare})}_k(\mathbf{x}_3)\rangle_{\lambda+\delta\lambda}=&\langle\mathcal{O}_i(\mathbf{x}_1)\mathcal{O}_j(\mathbf{x}_2)\mathcal{O}_k(\mathbf{x}_3)\rangle_{\lambda}\\
        &-\delta\lambda\int_\epsilon\frac{dzd^d\mathbf{x}}{z^{d+1}}\langle\hat\Phi(z,\mathbf{x})\mathcal{O}_i(\mathbf{x}_1)\mathcal{O}_j(\mathbf{x}_2)\mathcal{O}_k(\mathbf{x}_3)\rangle_{\lambda}\nonumber
\end{align}
where the subscript $\epsilon$ is the usual regularization in which we integrate everywhere in AdS except  semicircles of radius $\epsilon$ in half-plane coordinates centered around the insertions of the boundary operators. 

\begin{figure}
    \centering
    \includegraphics[width=0.5\linewidth]{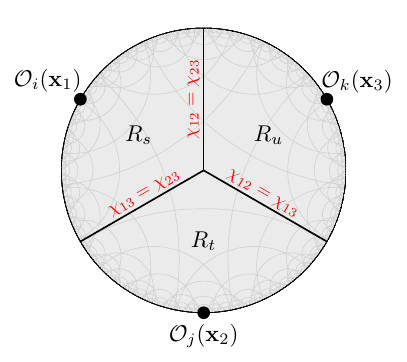}
    \caption{The three regions in which we split integral (\ref{eq:integralBbbb}). In region $R_s$ we use the $s$-channel OPE pairing $(jk)$, in region $R_t$ we pair $(ik)$ and in region $R_u$ we pair $(ij)$.}
    \label{fig:channels}
\end{figure}

Let us thus focus on 
\begin{equation}
    I_{Bbbb}^{\hat\Phi ijk}(\mathbf{x}_1,\mathbf{x}_2,\mathbf{x}_3;\epsilon)\equiv\int_{(\mathbf{x}-\mathbf{x}_i)^2+z^2\geq\epsilon^2}\frac{dzd^d\mathbf{x}}{z^{d+1}}\langle\hat\Phi(z,\mathbf{x})\mathcal{O}_i(\mathbf{x}_1)\mathcal{O}_j(\mathbf{x}_2)\mathcal{O}_k(\mathbf{x}_3)\rangle_{\lambda}\,.
    \label{eq:integralBbbb}
\end{equation}
We would like to carry out this integral universally in terms of the QFT data, hence to use a block decomposition. In this case, we would need to use the BOE of the bulk field $\hat\Phi$ and an OPE among the boundary operators. The crucial difference with the previous cases is that there is no OPE that converges in all AdS. Instead, we divide AdS in three regions which will be delimited by co-dimension 1 surfaces defined by the equations
\begin{equation}
    \chi_{12}=\chi_{23},\qquad \chi_{13}=\chi_{23},\qquad \chi_{13}=\chi_{12}\,, \qquad \chi_{ij}\equiv-\frac{1}{2}\frac{P_i\cdot P_j}{(X\cdot P_i)(X\cdot P_j)}\,.
    \label{eq:defchiij}
\end{equation}
We will split the integral in these three regions and use different OPEs in each. We give a pictorial representation in figure \ref{fig:channels}.
Let us focus on the $s$-channel region, corresponding to the OPE pairing $(jk)$. 
\begin{equation}
    I_{\text{Bbbb},s}^{\hat\Phi ijk}(\mathbf{x}_1,\mathbf{x}_2,\mathbf{x}_3;\epsilon)\equiv\int_{R_s|(\mathbf{x}-\mathbf{x}_1)^2+z^2\geq\epsilon^2}\frac{dzd^d\mathbf{x}}{z^{d+1}}\langle\hat\Phi(z,\mathbf{x})\mathcal{O}_i(\mathbf{x}_1)\mathcal{O}_j(\mathbf{x}_2)\mathcal{O}_k(\mathbf{x}_3)\rangle_{\lambda}\,.
    \label{eq:IBbbbs1}
\end{equation}
Now we perform the OPE $(jk)$ using (\ref{eq:defOPE})
\begin{equation}
    \langle\hat\Phi(z,\mathbf{x})\mathcal{O}_i(\mathbf{x}_1)\mathcal{O}_j(\mathbf{x}_2)\mathcal{O}_k(\mathbf{x}_3)\rangle=\sum_m\frac{C_{jkm}}{(\mathbf{x}_{23}^2)^\frac{\Delta_{jkm}}{2}}\mathcal{C}_{a_1\cdots a_{J_m}}(\mathbf{x}_{23},\partial_3)\langle\hat\Phi(z,\mathbf{x})\mathcal{O}_i(\mathbf{x}_1)\mathcal{O}_m^{a_1\cdots a_{J_m}}(\mathbf{x}_3)\rangle
\end{equation}
For the remaining bulk-boundary-boundary three-point function, we use the block decomposition (\ref{eq:JOPhidec})
\begin{align}\langle&\hat\Phi(z,\mathbf{x})\mathcal{O}_i(\mathbf{x}_1)\mathcal{O}_j(\mathbf{x}_2)\mathcal{O}_k(\mathbf{x}_3)\rangle\nonumber\\
    &=\sum_lb^{\hat\Phi}_l\sum_m\frac{C_{lim}C_{jkm}}{(\mathbf{x}_{23}^2)^\frac{\Delta_{jkm}}{2}}\mathcal{C}_{a_1\cdots a_{J_m}}(\mathbf{x}_{23},\partial_3)\frac{V_{3,1X}^{a_1\cdots a_{J_m}}}{(\mathbf{x}_{13}^2)^\frac{\Delta_i+\Delta_m+J_m}{2}}\left(\frac{z^2+(\mathbf{x}-\mathbf{x}_1)^2}{z^2+(\mathbf{x}-\mathbf{x}_3)^2}\right)^\frac{\Delta_m+J_m-\Delta_i}{2}G_{\text{Bbb},J_m}^{\Delta_l,\Delta_i,\Delta_m}(\chi_{13})\nonumber\\
    &\equiv\frac{\sum_lb^{\hat\Phi}_l\sum_mC_{lim}C_{jkm}G_{\text{Bbbb},s,J_m}^{\Delta_l,\Delta_m,\Delta_i,\Delta_j,\Delta_k}(\chi_{12},\chi_{13},\chi_{23})}{(\mathbf{x}_{12}^2)^{\frac{\Delta_{ijk}}{2}}(\mathbf{x}_{23}^2)^{\frac{\Delta_{jki}}{2}}(\mathbf{x}_{13}^2)^{\frac{\Delta_{ikj}}{2}}}
    \label{eq:defGBbbbJ}
\end{align}
where in the second line
\begin{equation}
    V_{3,1X}^{a_1\ldots a_{J}}\equiv \frac{1}{(\frac{d-2}{2})_JJ!}D_{\mathbf{z}_{a_1}}\cdots D_{\mathbf{z}_{a_J}}V_{3,12}\,,\qquad P_2\to X 
\end{equation}
and $V_{3,12}$ here is the pullback to local coordinates of (\ref{eq:defV}).
In the third line of (\ref{eq:defGBbbbJ}),
we used $\chi_{ij}$ defined in (\ref{eq:defchiij}) to form the three independent conformal cross ratios of this correlation function.

Using this in (\ref{eq:IBbbbs1}), we obtain
\begin{equation}
     I_{\text{Bbbb},s}^{\hat\Phi ijk}(\mathbf{x}_1,\mathbf{x}_2,\mathbf{x}_3;\epsilon)=\frac{\sum_lb^{\hat\Phi}_l\sum_mC_{lim}C_{mjk}I_{\text{Bbbb},s,J_m}^{\Delta_l\Delta_m\Delta_i\Delta_j\Delta_k}(\epsilon)}{(\mathbf{x}_{12}^2)^{\frac{\Delta_{ijk}}{2}}(\mathbf{x}_{23}^2)^{\frac{\Delta_{jki}}{2}}(\mathbf{x}_{13}^2)^{\frac{\Delta_{ikj}}{2}}}
     \label{eq:IBbbbphiijk}
\end{equation}
where
\begin{equation}
    I_{\text{Bbbb},s,J_m}^{\Delta_l\Delta_m\Delta_i\Delta_j\Delta_k}(\epsilon)\equiv\int_{R_s|(\mathbf{x}-\mathbf{x}_1)^2+z^2\geq\epsilon^2}\frac{dzd^d\mathbf{x}}{z^{d+1}}G_{\text{Bbbb},s,J_m}^{\Delta_l,\Delta_m,\Delta_i,\Delta_j,\Delta_k}(\chi_{12},\chi_{13},\chi_{23})\,.
    \label{eq:IBbbbs}
\end{equation}
This integral will generically diverge when $\epsilon\to0$. To study this divergence, let us consider the integral \textit{inside} the half-ball. For simplicity, we will only focus on the scalar contribution $J_m=0$. We will find that it is finite when $\Delta_m>\Delta_i$, while it diverges when  $\Delta_m\leq\Delta_i$. In the regime where it is finite, we find
\begin{align}
    &I_{\text{Bbbb},s,\text{in},0}^{\Delta_l\Delta_m\Delta_i\Delta_j\Delta_k}(\epsilon)=\int_{(\mathbf{x}-\mathbf{x}_1)^2+z^2<\epsilon^2}\frac{dzd^d\mathbf{x}}{z^{d+1}}G_{\text{Bbbb},s,0}^{\Delta_l,\Delta_m,\Delta_i,\Delta_j,\Delta_k}(\chi_{12},\chi_{13},\chi_{23})    \nonumber\\
    &=\frac{1}{(\mathbf{x}_{23}^2)^\frac{\Delta_{jkm}}{2}}\mathcal{C}(\mathbf{x}_{23},\partial_3)\frac{1}{(\mathbf{x}_{13}^2)^\frac{\Delta_i+\Delta_m}{2}}\int_{(\mathbf{x}-\mathbf{x}_1)^2+z^2<\epsilon^2}\frac{dzd^d\mathbf{x}}{z^{d+1}}\left(\frac{z^2+(\mathbf{x}-\mathbf{x}_1)^2}{z^2+(\mathbf{x}-\mathbf{x}_3)^2}\right)^\frac{\Delta_m-\Delta_i}{2}G_{\text{Bbb}}^{\Delta_l,\Delta_i,\Delta_m}(\chi_{13})\nonumber\\
&=-\frac{1}{(\mathbf{x}_{23}^2)^\frac{\Delta_{jkm}}{2}}\sum_{p=0}^\infty\frac{a_{2p}(\Delta_{im},\Delta_l)}{\Delta_{im}-2p}\frac{\epsilon^{2p-\Delta_{im}}}{2^{2p}(\Delta_{m})_p(\Delta_m-\frac{d-2}{2})_p}\mathcal{C}(\mathbf{x}_{23},\partial_3)\Box_1^p\frac{1}{(\mathbf{x}_{13}^2)^{\Delta_m}}\nonumber\\
&=-\sum_{p=0}^\infty\frac{a_{2p}(\Delta_{im},\Delta_l)}{\Delta_{im}-2p}\frac{\epsilon^{2p-\Delta_{im}}}{2^{2p}(\Delta_{m})_p(\Delta_m-\frac{d-2}{2})_p}\Box_1^p\frac{1}{(\mathbf{x}_{12}^2)^{\frac{\Delta_{mjk}}{2}}(\mathbf{x}_{23}^2)^{\frac{\Delta_{jkm}}{2}}(\mathbf{x}_{13}^2)^{\frac{\Delta_{mkj}}{2}}}\label{eq:IBbbbsinside}
\end{align}
where to go from the second to the third line we recognized (\ref{eq:IBbb_in}), and from the third to the fourth we used that
\begin{equation}
    \langle\mathcal{O}_m(\mathbf{x}_1)\mathcal{O}_j(\mathbf{x}_2)\mathcal{O}_k(\mathbf{x}_3)\rangle=\frac{C_{jkm}}{(\mathbf{x}_{23}^2)^{\frac{\Delta_{jkm}}{2}}}\mathcal{C}(\mathbf{x}_{23},\partial_3)\langle\mathcal{O}_m(\mathbf{x}_1)\mathcal{O}_m(\mathbf{x}_3)\rangle
\end{equation}
by virtue of the OPE (\ref{eq:defOPE}). In the final expression in (\ref{eq:IBbbbsinside}) we recognize the counterterms (\ref{eq:counterterm}). If we thus split the finite integral (\ref{eq:IBbbbs}) as 
\begin{equation}
    I_{\text{Bbbb},s,0}^{\Delta_l\Delta_m\Delta_i\Delta_j\Delta_k}(\epsilon)=\mathcal{K}_{\Delta_i\Delta_j\Delta_k}(\Delta_l,\Delta_m,0)-I_{\text{Bbbb},s,\text{in},0}^{\Delta_l\Delta_m\Delta_i\Delta_j\Delta_k}(\epsilon)
    \label{eq:inoutsubtract}
\end{equation}
where $\mathcal{K}$ is the integral over the full region $R_s$ with no subtraction, plugging things back into (\ref{eq:barebbb}) the counterterms and the logarithms when $\Delta_m=\Delta_i$ all cancel, leading to the following differential equation: 
\begin{equation}
    \frac{dC_{ijk}}{d\lambda}=\sum_lb^{\hat\Phi}_l\sum_mC_{lim}C_{mjk}\mathcal{K}_{\Delta_i\Delta_j\Delta_k}(\Delta_l,\Delta_m,0)+\text{perms.}+\text{spin}
\end{equation}
where ``perms." stands for the cyclic permutations of $i,j,k$ and ``spin" stands for contributions from exchanged spinning operators.


We do not determine the explicit form of $\mathcal{K}_{\Delta_i\Delta_j\Delta_k}(\Delta_l,\Delta_m,0)$. For the purposes of section \ref{sec:mergerrepulsion}, we need the following statements:
\paragraph{Pole at $\Delta_l=d$}
The block $G_{\text{Bbbb}}$ is obtained by performing the BOE of $\hat\Phi$, which goes as $\hat\Phi\sim b^{\hat\Phi}_lz^{\Delta_l}\mathcal{O}_l+\ldots$. As such, the integral near the boundary goes like
\begin{equation}
  \int_0\frac{dz}{z^{d+1}}G_{\text{Bbbb},s}^{\Delta_l,\Delta_m,\Delta_i,\Delta_j,\Delta_k}(\chi_{12},\chi_{13},\chi_{23})\sim\int_0\frac{dz}{z^{d+1}}z^{\Delta_l}\sim\frac{1}{(\Delta_l-d)}(\ldots)
\end{equation}
hence we expect
\begin{equation}
    \mathcal{K}_{\Delta_i\Delta_j\Delta_k}(\Delta_l,\Delta_m)\stackrel{\Delta_l\to d}{\sim}\frac{1}{\Delta_l-d}\times O(1)
\end{equation}
\paragraph{Pole at $\Delta_m=\Delta_i$}
Consider expression (\ref{eq:inoutsubtract}). 
The integral outside the half-sphere is finite even when $\Delta_m=\Delta_i$, while the integral inside diverges logarithmically in that case. This immediately tells us that $\mathcal{K}_{\Delta_i\Delta_j\Delta_k}(\Delta_l,\Delta_m)$ has a pole at $\Delta_m=\Delta_i$ with residue that has to match the one of (\ref{eq:IBbbbsinside}). This pole comes from the $p=0$ term in (\ref{eq:IBbbbsinside}). By explicitly matching the residue, we find
\begin{equation}
    \underset{\Delta_m=\Delta_i}{\text{Res}}\mathcal{K}_{\Delta_i\Delta_j\Delta_k}(\Delta_l,\Delta_m)=-a_0(0,\Delta_l)=-\mathcal{I}(\Delta_l)
\end{equation}
\section{Universal bounds on sums of QFT data}
\label{app:convergence}
Assuming that the QFT under consideration is unitary and is a relevant deformation of a UV CFT, in this section we prove bounds on truncated sums of BOE and OPE coefficients. These bounds are used in section \ref{sec:derivationflowequations} to study the convergence of the sums in the flow equations.
\subsection{Bounds on BOE coefficients}
Consider the two-point function of a relevant bulk operator $\hat\Phi$ with $\frac{d-1}{2}<\Delta_{\hat\Phi}^{\text{UV}}<d+1$ in the configuration
\begin{equation}
    (z_1,\mathbf{x}_1)=(z,\mathbf{0})\,,\qquad (z_2,\mathbf{x}_2)=(1,\mathbf{0})\,.
\end{equation}
In the coincident-points limit, $z\to1$, the two-point function is not sensitive to the AdS curvature or to the details of the RG flow, and as such it is dominated by its form in the UV CFT. In other words, there are constants $\kappa$ and $z_0$ for which, $\forall z\in(z_0,1)$,
\begin{equation}
    \langle\hat\Phi(z,\mathbf{0})\hat\Phi(1,\mathbf{0})\rangle\leq\kappa\left(\frac{z}{1-z}\right)^{2\Delta^{\text{UV}}_{\hat\Phi}}\,.
    \label{eq:coincidentsing}
\end{equation}
At the same time, the two-point function can be decomposed in blocks as in (\ref{eq:bulkbulksteps})
\begin{equation}
    \langle\hat\Phi(z,\mathbf{0})\hat\Phi(1,\mathbf{0})\rangle=\sum_l \left(b^{\hat\Phi}_l\right)^2z^{\Delta_l}\,_2F_1\left(\begin{matrix}\Delta_l &\frac{d}{2}\\ & \Delta_l-\frac{d-2}{2}\end{matrix};z^2\right)
\end{equation}
Notice that every term in this sum is non-negative: the assumption $\Delta_l>d$ ensures that each parameter of the hypergeometric function is positive, implying further that each coefficient in its Taylor expansion around $z=0$ is positive. This is expected by unitarity since individual terms in the Taylor series are associated to descendant states and the two-point function in this frame can be interpreted as computing a sum over their norms. Moreover, the hypergeometric function is greater than $1$, since that is the value of its leading Taylor coefficient.

These properties are enough to state that
\begin{equation}
    z^{\Delta_{l_\text{max}}}\sum_l^{l_\text{max}}\left(b^{\hat\Phi}_l\right)^2\leq\langle\hat\Phi(z,\mathbf{0})\hat\Phi(1,\mathbf{0})\rangle\leq\kappa\left(\frac{z}{1-z}\right)^{2\Delta^{\text{UV}}_{\hat\Phi}}
\end{equation}
In particular, if we choose $z=1-\frac{1}{\Delta_{l_{\text{max}}}}$, and we notice that $(1-\frac{1}{\Delta_{l_{\text{max}}}})^{-\Delta_{l_{\text{max}}}}\leq 4$ for $\Delta_{l_{\text{max}}}\geq2$,  we can state that there are constants $\Delta_0$ and $\kappa_b$ such that
\begin{equation}
    \sum_l^{l_\text{max}}\left(b^{\hat\Phi}_l\right)^2\leq\kappa_b\ \Delta_{l_{\text{max}}}^{2\Delta^{\text{UV}}_{\hat\Phi}}\,,\qquad \Delta_{l_{\text{max}}}\geq\Delta_0\,,
    \label{eq:boundb}
\end{equation}
where $\kappa_b=4\kappa$.
\subsection{Bounds on OPE coefficients}
Consider the  four-point function of identical boundary operators in AdS$_{d+1}$ placed on a line on the boundary and expressed in terms of the radial coordinate of \cite{Pappadopulo:2012jk,Hogervorst:2013sma}
\begin{equation}
    \langle\mathcal{O}_i(-1)\mathcal{O}_i(1)\mathcal{O}_i(\rho)\mathcal{O}_i(-\rho)\rangle\equiv\mathcal{G}(\rho)\,,\qquad 0<\rho<1\,,
    \label{eq:rhoframe}
\end{equation}
    where $\rho$ is related to the cross ratios $\eta$ and $\bar\eta$ defined in (\ref{eq:uvetabareta}) through $\eta=\bar \eta=\frac{4\rho}{(1+\rho)^2}$. Reflection positivity in radial quantization ensures this four-point function is positive: Hermitian conjugation maps the operators at $\rho$ and $-\rho$ to their reflections at $1$ and $-1$ through a sphere of radius $\sqrt{\rho}$.  Inserting a complete set of states of definite scaling dimension $\Delta$, we thus find a positive decomposition
\begin{equation}
    \mathcal{G}(\rho)=\frac{1}{(4\rho)^{2\Delta_i}}\sum_{\Delta}P_\Delta\,\rho^{\Delta}\,,\qquad P_\Delta\geq0\,,
\end{equation}
where each $P_\Delta$ is a sum of squared overlaps. The identity contributes $P_0=1$, and the primary state of each scalar $\mathcal{O}_l$ contributes
\begin{equation}
P_{\Delta_l}=\frac{(4\rho)^{2\Delta_i}}{\rho^{\Delta_l}}\langle0|\mathcal{O}_i(-1)\mathcal{O}_i(1)|\mathcal{O}_l\rangle\langle\mathcal{O}_l|\mathcal{O}_i(\rho)\mathcal{O}_i(-\rho)|0\rangle= C_{lii}^2\,4^{\Delta_l}\,.
\end{equation}
All other states, i.e.\ descendants and spinning primaries, give further non-negative contributions. Therefore, if we keep only primaries, we can state
\begin{equation}
    \frac{\rho^{\Delta_{l_\text{max}}}}{(4\rho)^{2\Delta_i}}\sum_l^{l_\text{max}}C_{lii}^2\,4^{\Delta_l}\leq\mathcal{G}(\rho)\,, \qquad 0<\rho<1
\end{equation}
where here and below the sum over $l$ only runs over scalar primary operators. In the limit $\rho\to1$, the operators collide pairwise, $\mathcal{O}_i(-\rho)$ with $\mathcal{O}_i(-1)$ and $\mathcal{O}_i(\rho)$ with $\mathcal{O}_i(1)$, and the correlator is dominated by the identity exchange in this channel
\begin{equation}
\mathcal{G}(\rho)\stackrel{\rho\to1}{\approx}
\frac{1}{(1-\rho)^{4\Delta_i}}\,.
\end{equation}
More precisely, there are constants $\kappa$ and $\rho_0$ for which, $\forall\rho\in(\rho_0,1)$,
\begin{equation}
    \frac{\rho^{\Delta_{l_\text{max}}}}{(4\rho)^{2\Delta_i}}\sum_l^{l_\text{max}}C_{lii}^2\,4^{\Delta_l}\leq\mathcal{G}(\rho)\leq\frac{\kappa}{(1-\rho)^{4\Delta_i}}\,.
\end{equation}
Choosing $\rho=1-\frac{1}{\Delta_{l_\text{max}}}$, we can state that there are constants $\Delta_0$ and $\kappa_C$ such that
\begin{equation}
    \sum_l^{l_\text{max}}C_{lii}^2\,4^{\Delta_l}\leq\kappa_C\ \Delta_{l_{\text{max}}}^{4\Delta_i}\,,\qquad \Delta_{l_{\text{max}}}\geq\Delta_0\,.
    \label{eq:boundCline}
\end{equation}
The same argument applies to OPE coefficients $C_{lij}$ with $i\neq j$. For the coefficients of the radial expansion to be non-negative, each operator in the inner pair must face an identical operator in the outer pair, at its position reflected through a sphere of radius $\sqrt{\rho}$:
\begin{equation}
    \langle\mathcal{O}_i(-1)\mathcal{O}_j(1)\mathcal{O}_j(\rho)\mathcal{O}_i(-\rho)\rangle=\mathcal{G}_{ij}(\rho)\,,\qquad 0<\rho<1\,.
\end{equation}
In this configuration we still have $\mathcal{G}_{ij}(\rho)=\frac{1}{(4\rho)^{\Delta_i+\Delta_j}}\sum_\Delta P_\Delta\,\rho^\Delta$ with $P_\Delta\geq0$. Each scalar primary $\mathcal{O}_l$ exchanged in the $\rho\to0$ OPE contributes 
\begin{equation}
    P_{\Delta_l}=\frac{(4\rho)^{\Delta_i+\Delta_j}}{\rho^{\Delta_l}}\langle0|\mathcal{O}_i(-1)\mathcal{O}_j(1)|\mathcal{O}_l\rangle\langle\mathcal{O}_l|\mathcal{O}_j(\rho)\mathcal{O}_i(-\rho)|0\rangle=C_{ijl}^24^{\Delta_l}\,.
\end{equation}
In the limit $\rho\to1$, $\mathcal{O}_i(-\rho)$ collides with $\mathcal{O}_i(-1)$ and $\mathcal{O}_j(\rho)$ with $\mathcal{O}_j(1)$, and the correlator is again dominated by the identity, $\mathcal{G}_{ij}\stackrel{\rho\to1}{\approx}(1-\rho)^{-2(\Delta_i+\Delta_j)}$. 
We can thus write, for some constants $\kappa$ and $\rho_0$ and  $\forall\rho\in(\rho_0,1)$
\begin{equation}
    \frac{\rho^{\Delta_{l_\text{max}}}}{(4\rho)^{\Delta_i+\Delta_j}}\sum_l^{l_\text{max}}C_{ijl}^2\,4^{\Delta_l}\leq\mathcal{G}_{ij}(\rho)\leq\frac{\kappa}{(1-\rho)^{2(\Delta_i+\Delta_j)}}\,.
\end{equation}
Analogously to the case of identical operators, we can thus conclude that there are constants $\Delta_0$ and $\kappa_C$ such that
\begin{equation}
    \sum_l^{l_\text{max}}C_{ijl}^2\,4^{\Delta_l}\leq\kappa_C\ \Delta_{l_{\text{max}}}^{2(\Delta_i+\Delta_j)}\,,\qquad \Delta_{l_{\text{max}}}\geq\Delta_0\,.
    \label{eq:boundCij}
\end{equation}
For $i=j$, this bound reduces to (\ref{eq:boundCline}).
\subsection{Bounds on weighted sums}
\label{subsec:weightedconvergence}
For the purposes of section \ref{subsec:flowdeltas}, we present here a modification of the bounds on BOE and OPE coefficients (\ref{eq:boundb}) and (\ref{eq:boundCij}). Both have the form
\begin{equation}
    f(\Delta_{l_{\text{max}}})\equiv\sum_{l|\Delta_l\leq\Delta_{l_{\text{max}}}}a_l\leq\kappa\Delta_{l_{\text{max}}}^\gamma\,, \qquad a_l\geq0\,, \qquad \Delta_{l_{\text{max}}}\geq\Delta_0\,,
    \label{eq:unibound}
\end{equation}
where we are here going to treat $\Delta_{l_{\text{max}}}$ as a continuous real number rather than literally the scaling dimension of some state at which we truncate.

The function $f(\Delta_{l_{\text{max}}})$ is a piecewise constant function that is non-decreasing in the truncation. What we want to achieve is a bound on the tail of the weighted sum 
\begin{equation}
    \sum_{l|\Delta_l>\Delta_{l_{\text{max}}}}a_l\Delta_l^{-\beta}\,,\qquad \beta>0\,.
\end{equation}
To do so, start by expressing the power as an integral:
\begin{equation}
    \Delta_l^{-\beta}=\beta\int_{\Delta_l}^\infty dx\ x^{-\beta-1}=\beta\int_{\Delta_{l_{\text{max}}}}^\infty dx\ x^{-\beta-1}\Theta(x-\Delta_l)
\end{equation}
where $\Theta$ is a Heaviside step function. This allows us to write
\begin{equation}
\begin{aligned}
    \sum_{l|\Delta_l>\Delta_{l_{\text{max}}}}a_l\Delta_l^{-\beta}&=\beta\int_{\Delta_{l_{\text{max}}}}^\infty dx\ x^{-\beta-1}\sum_{l|\Delta_l>\Delta_{l_{\text{max}}}}a_l\Theta(x-\Delta_l)\\
    &=\beta\int_{\Delta_{l_{\text{max}}}}^\infty dx\ x^{-\beta-1}\left(f(x)-f(\Delta_{l_{\text{max}}})\right)\,,
\end{aligned}
\end{equation}
where the exchange of sum and integral is here justified trivially due to the non-negativity of all terms.

This procedure is known as Abel summation, and is essentially the discrete analogue of integration by parts. Notice that we did not approximate the sum over operators with an integral; rather these manipulations are exact, and similar methods have been used in \cite{Pappadopulo:2012jk} to estimate the rate of convergence of the OPE.

Now we can bound the sum by dropping $f(\Delta_{l_{\text{max}}})$ and using (\ref{eq:unibound}):
\begin{equation}
    \sum_{l|\Delta_l>\Delta_{l_{\text{max}}}}a_l\Delta_l^{-\beta}\leq\beta\kappa\int_{\Delta_{l_{\text{max}}}}^\infty dx\ x^{\gamma-\beta-1}=\frac{\beta\kappa}{\beta-\gamma}\Delta_{l_{\text{max}}}^{\gamma-\beta}\,, \qquad \beta>\gamma\,.
\end{equation}
In the specific cases of the BOE and OPE coefficients this implies the bounds
\begin{equation}
    \sum_{l>l_{\text{max}}}\left(b^{\hat\Phi}_l\right)^2\Delta_l^{-\alpha_1}\leq K_b\Delta_{l_{\text{max}}}^{2\Delta^{\text{UV}}_{\hat\Phi}-\alpha_1}\,,\qquad \sum_{l>l_{\text{max}}}C_{lii}^24^{\Delta_l}\Delta_l^{-\alpha_2}\leq K_C\Delta_{l_{\text{max}}}^{4\Delta_i-\alpha_2}\,.
    \label{eq:weightedbounds}
\end{equation}
with $\alpha_1>2\Delta^{\text{UV}}_{\hat\Phi}$ and $\alpha_2>4\Delta_i$ and with $K_b$ and $K_C$ some unimportant constants.
\section{QFT data of free scalar and free tensor}
\label{app:QFTdata}
In section \ref{sec:checks}, we present checks of the flow equations in the theories of a free scalar and a free tensor. In such theories, the coupling of the deformation is the mass in units of the AdS radius, $\lambda=\frac{1}{2}m^2R^2$. Since the theories are free, we know the QFT data for any value of $\lambda$. For simplicity, we parametrize the QFT data in terms of the scaling dimension of the elementary boundary primary.
\subsection{Free scalar theory}
\subsubsection{Scaling dimensions}
For the free scalar, with action (\ref{eq:scalarAction}), the dimension of the elementary boundary primary $\phi$ is
\begin{equation}
    \Delta_\phi(\lambda)=\frac{d}{2}\pm\sqrt{\frac{d^2}{4}+2\lambda}
\end{equation}
and double trace operators have dimensions
\begin{equation}
    \Delta_{[\phi^2]_{n,l}}=2\Delta_\phi+2n+l
\end{equation}
\subsubsection{BOE coefficients}
We use two types of BOE coefficients in our checks: the one which appears in the expansion of the bulk elementary field $\hat\phi$ into the boundary primary $\phi$
\begin{equation}
    b^{\hat\phi}_\phi=\sqrt{\frac{\Gamma(\Delta_\phi)}{2\pi^\frac{d}{2}\Gamma(\Delta_\phi-\frac{d-2}{2})}}\,,
\end{equation}
which can be for example fixed by decomposing the canonically normalized massive bulk-bulk propagator into the bulk-bulk blocks (\ref{eq:BBdecompose}); and the coefficients appearing in the expansion of $\hat\phi^2$ into boundary scalar double trace primaries \cite{Fitzpatrick:2010zm}
\begin{equation}
    b^{\hat\phi^2}_{[\phi^2]_{n,0}}=\sqrt{\frac{(2(\Delta_\phi+n)-\frac{d}{2})\Gamma(\Delta_\phi+n)\Gamma(2(\Delta_\phi+n))\Gamma(\Delta_\phi-\frac{d-1}{2}+n)\Gamma(2\Delta_\phi-\frac{d}{2}+n)(\frac{d}{2})_n}{(2\pi)^dn!\Gamma(\Delta_\phi+\frac{1}{2}+n)\Gamma(\Delta_\phi-\frac{d-2}{2}+n)\Gamma(2\Delta_\phi-d+1+n)\Gamma(2(\Delta_\phi+n)-\frac{d-2}{2})}}\,,
    \label{eq:bphi2phi2}
\end{equation}
which can be extracted from the spectral decomposition of the bulk two-point function of $\hat\phi^2$.
\subsubsection{OPE coefficients}
Some OPE coefficients of free scalar theory are available in the literature. For example, \cite{Fitzpatrick:2011dm}
\begin{equation}
    C_{\phi\phi[\phi^2]_{n,s}}=\frac{(-1)^n2^{\frac{s+1}{2}}(\Delta_\phi)_{n+s}(\Delta_\phi-\frac{d-2}{2})_n}{\sqrt{n!s!(\frac{d}{2}+s)_n(2\Delta_\phi+n-d+1)_n(2\Delta_\phi-\frac{d}{2}+n+s)_n(2\Delta_\phi+2n+s-1)_s}}\,.
    \label{eq:Cphiphiphi2}
\end{equation}
In AdS$_3$, we are using the normalization of Osborn \cite{Osborn:2012vt}, hence we have
\begin{equation}
    \left(C_{\phi\phi[\phi^2]_{n,s}}^{(\text{AdS}_3)}\right)^2=\left(-\frac{1}{2}\right)^s\times(\text{\ref{eq:Cphiphiphi2}})^2\,.
\end{equation}
We also needed the OPE coefficients $C_{\phi^2[\phi^2]_n[\phi^2]_m}$. To compute them, we started from the three-point function $\langle\hat\phi^2(X_1)\hat\phi^2(X_2)\phi(P_3)\rangle$ in terms of Wick contractions and compared its expansion near the boundary to the expansion into bulk-bulk-boundary blocks (\ref{eq:BBbdec}). Using the knowledge of $b^{\hat\phi^2}_{[\phi^2]_n}$ (\ref{eq:bphi2phi2}), we obtained
\begin{equation}
\begin{aligned}
    &C_{\phi^2[\phi^2]_{n,0}[\phi^2]_{m,0}}\\
    &=\frac{\left(b^{\hat\phi^2}_{[\phi^2]_n}b^{\hat\phi^2}_{[\phi^2]_m}\right)^{-1}(-1)^{n+m}\sqrt{2}\Gamma(\Delta_\phi)^2(\Delta_\phi)_m(\Delta_\phi)_n(\Delta_\phi)_{n+m}(\Delta_\phi-\frac{d-2}{2})_{m+n}}{m!n!\pi^d\Gamma(\Delta_\phi-\frac{d-2}{2})^2(\Delta_\phi-\frac{d-2}{2})_m(\Delta_\phi-\frac{d-2}{2})_n(2\Delta_\phi-\frac{d}{2}+m)_m(2\Delta_\phi-\frac{d}{2}+n)_n}
    \label{eq:cphi2phi2phi2}
\end{aligned}
\end{equation}
Some more checks we performed in free theories involved $\frac{d\Delta_{\phi^m}}{d\lambda}$ for various values of $m\in\mathbb{N}$. The required OPE coefficients can be computed easily in terms of $C_{\phi\phi[\phi^2]_n}$ as follows: consider the three-point function
\begin{equation}
    \frac{1}{m!}\langle\phi^m(P_1)\phi^m(P_2)\hat\phi^2(X)\rangle=m \langle\phi(P_1)\phi(P_2)\rangle^{m-1}\langle\phi(P_1)\phi(P_2)\hat\phi^2(X)\rangle\,,
\end{equation}
where the $(m!)^{-1}$ is there to ensure proper normalization of the boundary operators, such that the two-point function is unit normalized
\begin{equation}
    \frac{1}{m!}\langle\phi^m(P_1)\phi^m(P_2)\rangle=\langle\phi(P_1)\phi(P_2)\rangle^m=\frac{1}{P_{12}^{m\Delta_\phi}}\,.
\end{equation}
Then, comparing the bulk-boundary-boundary expansion of the two sides (\ref{eq:Bbbdec}), we find
\begin{equation}
    C_{\phi^m\phi^m[\phi^2]_n}=mC_{\phi\phi[\phi^2]_n}\,.
\end{equation}
\subsection{Free tensor theory}
\subsubsection{Scaling dimensions}
In the free traceless symmetric tensor theory, with action (\ref{eq:actiontensor}), the boundary elementary field $h_{ab}$ has dimensions related to the mass $\lambda=\frac{1}{2}m^2R^2$ as
\begin{equation}
    \Delta_h(\lambda)=\frac{d}{2}\pm\sqrt{\frac{d^2}{4}+2\lambda}\,.
\end{equation}
The bulk field implementing the mass deformation, $\hat h^2\equiv \hat h^{\mu\nu}\hat h_{\mu\nu}$, has in its BOE an infinite set of scalar double trace primaries $[h^2]_{n,0}$ with dimensions
\begin{equation}
    \Delta_{[h^2]_{n,0}}=2\Delta_h+2n
\end{equation}
We will focus on the case of $d=3$.
\subsubsection{BOE$\times$OPE coefficients}
In this case we will only ever need a product of BOE coefficients and OPE coefficients, $b^{\hat h^2}_{[h^2]_{n,0}}C_{hh[h^2]_{n,0}}^{(m)}$ where $m=0,1,2$ labels the different tensor structures in the boundary three-point function $\langle h^{(2)}(P_1,Z_1)h^{(2)}(P_2,Z_2)[h^2]_{n,0}(P_3)\rangle$, where $h^{(2)}$ is the boundary elementary tensor with indices contracted by polarization tensors
\begin{equation}
    h^{(2)}(P,Z)\equiv Z_{A}Z_{B}h^{AB}(P)\,.
\end{equation}
To get this QFT data, we compute the three-point function $\langle h^{(2)}(P_1,Z_1)h^{(2)}(P_2,Z_2)\hat h^2(X)\rangle$ from Wick contractions in the special frame
\begin{equation}
\begin{aligned}
    P_1&=\left(\frac{1}{2},\mathbf{0},\frac{1}{2}\right),\quad P_2=\left(\frac{1}{2},\mathbf{0},-\frac{1}{2}\right),\quad Z_1=(0,\mathbf{z}_1,0),\quad Z_2=(0,\mathbf{z}_2,0)\\
    X&=\left(\frac{2+z^2}{2z},\frac{\hat{\mathbf{x}}^1}{z},-\frac{z}{2}\right)
\end{aligned}
\end{equation}
Using the bulk-boundary propagators of a free massive spin 2 field from \cite{Costa:2014kfa}, we get
\begin{equation}
\begin{aligned}
    \langle h^{(2)}(0,\mathbf{z}_1)&h^{(2)}(\infty,\mathbf{z}_2)\hat h^2(z,\hat{\mathbf{x}}^1)\rangle\\
    &=\frac{(\Delta_h+1)\Gamma(\Delta_h-1)}{\Gamma(\Delta_h-\frac{1}{2})\pi^\frac{3}{2}}\left(\frac{z^2}{1+z^2}\right)^{\Delta_h}\left(\mathbf{z}_1\cdot\mathbf{z}_2-2\frac{\mathbf{z}_1^1\mathbf{z}_2^1}{z^2+1}\right)^2
\end{aligned}
\end{equation}
We compare the $z\to0$ expansion of this expression to the same expansion applied to the basis decomposition (\ref{eq:BbbJ}) in the same frame, and extract the product of BOE and OPE coefficients:
\begin{equation}
    \begin{aligned}
        b^{\hat h^2}_{[h^2]_{n,0}}C_{hh[h^2]_{n,0}}^{(0)}&=\frac{f_n(\Delta_h)}{24(n+\Delta_h)^2(n+1+\Delta_h)^2}b^{\hat h^2}_{[h^2]_{n,0}}C_{hh[h^2]_{n,0}}^{(2)}\,,\\
        b^{\hat h^2}_{[h^2]_{n,0}}C_{hh[h^2]_{n,0}}^{(1)}&=\frac{5(3(\Delta_h+1)^2+n(6-8\Delta_h)-4n^2)}{6(\Delta_h+n+1)^2}b^{\hat h^2}_{[h^2]_{n,0}}C_{hh[h^2]_{n,0}}^{(2)}\,,\\
        b^{\hat h^2}_{[h^2]_{n,0}}C_{hh[h^2]_{n,0}}^{(2)}&=\frac{8(-1)^n\Gamma(\Delta_h-1)((\Delta_h+1)_{n+1})^2}{35\pi^\frac{3}{2}(\Delta_h+1)n!\Gamma(\Delta_h-\frac{1}{2})(2\Delta_h+n-\frac{3}{2})_n}\,,
        \label{eq:BCtensor}
    \end{aligned}
\end{equation}
where
\begin{equation}
\begin{aligned}
    f_n(\Delta_h)=7\Big(&8n^4+3\Delta_h^2(\Delta_h+1)^2+8n^3(4\Delta_h-3)\\
    &-2n\Delta_h(6+\Delta_h)(4\Delta_h-3)+2n^2(9+2\Delta_h(7\Delta_h-18))\Big)\,,
\end{aligned}
\end{equation}
and we isolated each contribution by taking traces using (\ref{eq:vanishtraces}).
\bibliography{bibliography}

@article{DiPietro:2025ozw,
    author = "Di Pietro, Lorenzo and Kousvos, Stefanos R. and Meineri, Marco and Piazza, Alessandro and Serone, Marco and Vichi, Alessandro",
    title = "{A Bootstrap Study of Confinement in AdS}",
    eprint = "2512.00150",
    archivePrefix = "arXiv",
    primaryClass = "hep-th",
    month = "11",
    year = "2025"
}

@article{DeCesare:2026rup,
    author = "De Cesare, Fabiana and Giombi, Simone",
    title = "{Conformal QED in AdS as a BCFT}",
    eprint = "2607.19464",
    archivePrefix = "arXiv",
    primaryClass = "hep-th",
    month = "7",
    year = "2026"
}

@article{Giombi:2017hpr,
    author = "Giombi, Simone and Sleight, Charlotte and Taronna, Massimo",
    title = "{Spinning AdS Loop Diagrams: Two Point Functions}",
    eprint = "1708.08404",
    archivePrefix = "arXiv",
    primaryClass = "hep-th",
    reportNumber = "PUPT-2540",
    doi = "10.1007/JHEP06(2018)030",
    journal = "JHEP",
    volume = "06",
    pages = "030",
    year = "2018"
}

@article{Csipes:2026nyo,
    author = "Csipes, Jozef and Va{\v{s}}ko, Petr",
    title = "{O(N) BCFT: new data from conformal partial wave expansions}",
    eprint = "2606.14733",
    archivePrefix = "arXiv",
    primaryClass = "hep-th",
    month = "6",
    year = "2026"
}

@article{Gorbenko:2018dtm,
    author = "Gorbenko, Victor and Rychkov, Slava and Zan, Bernardo",
    title = "{Walking, Weak first-order transitions, and Complex CFTs II. Two-dimensional Potts model at $Q>4$}",
    eprint = "1808.04380",
    archivePrefix = "arXiv",
    primaryClass = "hep-th",
    doi = "10.21468/SciPostPhys.5.5.050",
    journal = "SciPost Phys.",
    volume = "5",
    number = "5",
    pages = "050",
    year = "2018"
}

@unpublished{DiPietro:QED3_WIP,
  author = {Di Pietro, Lorenzo and Lanza, Stefano C. and Niro, Pierluigi},
  title  = {Symmetry Breaking in {QED}$_3$ from {Anti-de Sitter Space}},
  note   = {Work in progress},
  year   = {2026}
}

@article{Costa:2014kfa,
    author = "Costa, Miguel S. and Gon{\c{c}}alves, Vasco and Penedones, Jo{\~a}o",
    title = "{Spinning AdS Propagators}",
    eprint = "1404.5625",
    archivePrefix = "arXiv",
    primaryClass = "hep-th",
    doi = "10.1007/JHEP09(2014)064",
    journal = "JHEP",
    volume = "09",
    pages = "064",
    year = "2014"
}

@article{Fitzpatrick:2010zm,
    author = "Fitzpatrick, A. Liam and Katz, Emanuel and Poland, David and Simmons-Duffin, David",
    title = "{Effective Conformal Theory and the Flat-Space Limit of AdS}",
    eprint = "1007.2412",
    archivePrefix = "arXiv",
    primaryClass = "hep-th",
    reportNumber = "BUHET-07-14-10",
    doi = "10.1007/JHEP07(2011)023",
    journal = "JHEP",
    volume = "07",
    pages = "023",
    year = "2011"
}

@article{Fitzpatrick:2022dwq,
    author = "Fitzpatrick, A. Liam and Katz, Emanuel",
    title = "{Snowmass White Paper: Hamiltonian Truncation}",
    eprint = "2201.11696",
    archivePrefix = "arXiv",
    primaryClass = "hep-th",
    month = "1",
    year = "2022"
}

@article{Paley_Zygmund_1930, title={On some series of functions, (1)}, volume={26}, DOI={10.1017/S0305004100016078}, number={3}, journal={Mathematical Proceedings of the Cambridge Philosophical Society}, author={Paley, R. E. A. C. and Zygmund, A.}, year={1930}, pages={337–357}}

@article{Paley_Zygmund_1932, title={On some series of functions, (3)}, volume={28}, DOI={10.1017/S0305004100010860}, number={2}, journal={Mathematical Proceedings of the Cambridge Philosophical Society}, author={Paley, R. E. A. C. and Zygmund, A.}, year={1932}, pages={190–205}}

@article{Li:2017lmh,
    author = "Li, Daliang and Meltzer, David and Poland, David",
    title = "{Conformal Bootstrap in the Regge Limit}",
    eprint = "1705.03453",
    archivePrefix = "arXiv",
    primaryClass = "hep-th",
    doi = "10.1007/JHEP12(2017)013",
    journal = "JHEP",
    volume = "12",
    pages = "013",
    year = "2017"
}

@article{Deutsch:2018ulr,
    author = "Deutsch, Joshua M.",
    title = "{Eigenstate thermalization hypothesis}",
    eprint = "1805.01616",
    archivePrefix = "arXiv",
    primaryClass = "quant-ph",
    doi = "10.1088/1361-6633/aac9f1",
    journal = "Rept. Prog. Phys.",
    volume = "81",
    number = "8",
    pages = "082001",
    year = "2018"
}

@article{Poland:2018epd,
    author = "Poland, David and Rychkov, Slava and Vichi, Alessandro",
    title = "{The Conformal Bootstrap: Theory, Numerical Techniques, and Applications}",
    eprint = "1805.04405",
    archivePrefix = "arXiv",
    primaryClass = "hep-th",
    doi = "10.1103/RevModPhys.91.015002",
    journal = "Rev. Mod. Phys.",
    volume = "91",
    pages = "015002",
    year = "2019"
}

@article{Davoudi:2022Snowmass,
    author = "Davoudi, Zohreh and Neil, Ethan T. and others",
    title = "{Report of the Snowmass 2021 Topical Group on Lattice Gauge Theory}",
    eprint = "2209.10758",
    archivePrefix = "arXiv",
    primaryClass = "hep-lat",
    year = "2022"
}

@article{Maldacena:2015iua,
    author = "Maldacena, Juan and Simmons-Duffin, David and Zhiboedov, Alexander",
    title = "{Looking for a bulk point}",
    eprint = "1509.03612",
    archivePrefix = "arXiv",
    primaryClass = "hep-th",
    doi = "10.1007/JHEP01(2017)013",
    journal = "JHEP",
    volume = "01",
    pages = "013",
    year = "2017"
}

@article{Costa:2016xah,
    author = "Costa, Miguel S. and Hansen, Tobias and Penedones, Jo{\~a}o and Trevisani, Emilio",
    title = "{Radial expansion for spinning conformal blocks}",
    eprint = "1603.05552",
    archivePrefix = "arXiv",
    primaryClass = "hep-th",
    reportNumber = "CERN-TH-2016-079",
    doi = "10.1007/JHEP07(2016)057",
    journal = "JHEP",
    volume = "07",
    pages = "057",
    year = "2016"
}

@article{Erramilli:2019njx,
    author = "Erramilli, Rajeev S. and Iliesiu, Luca V. and Kravchuk, Petr",
    title = "{Recursion relation for general 3d blocks}",
    eprint = "1907.11247",
    archivePrefix = "arXiv",
    primaryClass = "hep-th",
    reportNumber = "PUPT-2593",
    doi = "10.1007/JHEP12(2019)116",
    journal = "JHEP",
    volume = "12",
    pages = "116",
    year = "2019"
}

@article{Erramilli:2020rlr,
    author = "Erramilli, Rajeev S. and Iliesiu, Luca V. and Kravchuk, Petr and Landry, Walter and Poland, David and Simmons-Duffin, David",
    title = "{blocks{\_}3d: software for general 3d conformal blocks}",
    eprint = "2011.01959",
    archivePrefix = "arXiv",
    primaryClass = "hep-th",
    reportNumber = "CALT-TH 2020-048",
    doi = "10.1007/JHEP11(2021)006",
    journal = "JHEP",
    volume = "11",
    pages = "006",
    year = "2021"
}

@article{Kravchuk:2016qvl,
    author = "Kravchuk, Petr and Simmons-Duffin, David",
    title = "{Counting Conformal Correlators}",
    eprint = "1612.08987",
    archivePrefix = "arXiv",
    primaryClass = "hep-th",
    reportNumber = "CALT-TH-2016-041",
    doi = "10.1007/JHEP02(2018)096",
    journal = "JHEP",
    volume = "02",
    pages = "096",
    year = "2018"
}

@article{Stahl1997TheCO,
  title={The convergence of Pad{\'e} approximants to functions with branch points},
  author={Herbert Stahl},
  journal={Journal of Approximation Theory},
  year={1997},
  volume={91},
  pages={139-204},
  url={https://api.semanticscholar.org/CorpusID:122298118}
}

@article{Stahl1986OrthogonalPW,
  title={Orthogonal polynomials with complex-valued weight function, I},
  author={Herbert Stahl},
  journal={Constructive Approximation},
  year={1986},
  volume={2},
  pages={225-240},
  url={https://api.semanticscholar.org/CorpusID:120504673}
}

@book{Baker_Graves-Morris_1996, place={Cambridge}, edition={2}, series={Encyclopedia of Mathematics and its Applications}, title={Padé Approximants}, publisher={Cambridge University Press}, author={Baker, George A. and Graves-Morris, Peter}, year={1996}, collection={Encyclopedia of Mathematics and its Applications}}

@article{Costa:2011mg,
    author = "Costa, Miguel S. and Penedones, Joao and Poland, David and Rychkov, Slava",
    title = "{Spinning Conformal Correlators}",
    eprint = "1107.3554",
    archivePrefix = "arXiv",
    primaryClass = "hep-th",
    reportNumber = "LPTENS-11-22, NSF-KITP-11-128",
    doi = "10.1007/JHEP11(2011)071",
    journal = "JHEP",
    volume = "11",
    pages = "071",
    year = "2011"
}

@article{Osborn:2012vt,
    author = "Osborn, H.",
    title = "{Conformal Blocks for Arbitrary Spins in Two Dimensions}",
    eprint = "1205.1941",
    archivePrefix = "arXiv",
    primaryClass = "hep-th",
    reportNumber = "DAMTP-12-37",
    doi = "10.1016/j.physletb.2012.09.045",
    journal = "Phys. Lett. B",
    volume = "718",
    pages = "169--172",
    year = "2012"
}

@article{Loparco:2026fki,
    author = "Loparco, Manuel and Mathys, Gr{\'e}goire and Penedones, Joao and Qiao, Jiaxin and Zhao, Xiang",
    title = "{QFT as a set of ODEs}",
    eprint = "2601.04310",
    archivePrefix = "arXiv",
    primaryClass = "hep-th",
    month = "1",
    year = "2026"
}

@article{Hogervorst:2013sma,
    author = "Hogervorst, Matthijs and Rychkov, Slava",
    title = "{Radial Coordinates for Conformal Blocks}",
    eprint = "1303.1111",
    archivePrefix = "arXiv",
    primaryClass = "hep-th",
    reportNumber = "CERN-PH-TH-2013-043, LPTENS-13-05",
    doi = "10.1103/PhysRevD.87.106004",
    journal = "Phys. Rev. D",
    volume = "87",
    pages = "106004",
    year = "2013"
}

@article{Gabai:2025hwf,
    author = "Gabai, Barak and Gorbenko, Victor and Qiao, Jiaxin",
    title = "{Yang-Mills Flux Tube in AdS}",
    eprint = "2508.08250",
    archivePrefix = "arXiv",
    primaryClass = "hep-th",
    month = "8",
    year = "2025"
}

@article{Gabai:2026myo,
    author = "Gabai, Barak and Gorbenko, Victor and Offertaler, Bendeguz",
    title = "{Yang-Mills Flux Tube in AdS II: Effective String Theory}",
    eprint = "2602.16694",
    archivePrefix = "arXiv",
    primaryClass = "hep-th",
    month = "2",
    year = "2026"
}

@ARTICLE{1929PhyZ...30..467V,
       author = {{von Neuman}, J. and {Wigner}, E.},
        title = "{Uber merkw{\"u}rdige diskrete Eigenwerte. Uber das Verhalten von Eigenwerten bei adiabatischen Prozessen}",
      journal = {Physikalische Zeitschrift},
         year = 1929,
        month = jan,
       volume = {30},
        pages = {467-470},
       adsurl = {https://ui.adsabs.harvard.edu/abs/1929PhyZ...30..467V}
}

@article{Xiao2026,
    author = "Xiao, Ruihao",
    title = "{Quantum Field Theory in Anti-de Sitter Spacetime}",
    year = "2026",
journal = "EPFL Master Thesis"
}

@article{Diatlyk:2026eta,
    author = "Diatlyk, Oleksandr and Giombi, Simone and Sun, Zimo",
    title = "{Boundary criticality in the Gross-Neveu-Yukawa model at higher orders}",
    eprint = "2606.07510",
    archivePrefix = "arXiv",
    primaryClass = "hep-th",
    month = "6",
    year = "2026"
}

@article{Giombi:2020rmc,
    author = "Giombi, Simone and Khanchandani, Himanshu",
    title = "{CFT in AdS and boundary RG flows}",
    eprint = "2007.04955",
    archivePrefix = "arXiv",
    primaryClass = "hep-th",
    doi = "10.1007/JHEP11(2020)118",
    journal = "JHEP",
    volume = "11",
    pages = "118",
    year = "2020"
}

@article{Giombi:2025pxx,
    author = "Giombi, Simone and Sun, Zimo",
    title = "{Higher loops in AdS: applications to boundary CFT}",
    eprint = "2506.14699",
    archivePrefix = "arXiv",
    primaryClass = "hep-th",
    doi = "10.1007/JHEP12(2025)011",
    journal = "JHEP",
    volume = "12",
    pages = "011",
    year = "2025"
}

@article{Ankur:2026ylr,
    author = "Ankur and Di Pietro, Lorenzo and Gorbenko, Victor and Komatsu, Shota and Sacchi, Veronica",
    title = "{Dressing and Screening in Anti-de Sitter}",
    eprint = "2601.04321",
    archivePrefix = "arXiv",
    primaryClass = "hep-th",
    month = "1",
    year = "2026"
}

@article{Qiao:2026ijh,
    author = "Qiao, Jiaxin",
    title = "{Protected operators in non-local defect CFTs from AdS}",
    eprint = "2605.13975",
    archivePrefix = "arXiv",
    primaryClass = "hep-th",
    month = "5",
    year = "2026"
}

@article{Bianchi:2026sax,
    author = "Bianchi, Lorenzo and de Sabbata, Elia and Meineri, Marco",
    title = "{Conformal defects and Goldstone bosons in Anti-de Sitter space}",
    eprint = "2605.13947",
    archivePrefix = "arXiv",
    primaryClass = "hep-th",
    month = "5",
    year = "2026"
}

@article{Bason:2025sxb,
    author = "Bason, Davide and Copetti, Christian and Di Pietro, Lorenzo and Ji, Ziming and Komatsu, Shota",
    title = "{F-theorem for Quantum Field Theories in Anti-de Sitter Space}",
    eprint = "2512.18392",
    archivePrefix = "arXiv",
    primaryClass = "hep-th",
    month = "12",
    year = "2025"
}

@article{He:2025fbk,
    author = "He, Yifei and Kruczenski, Martin",
    title = "{The Gauge Theory Bootstrap: Computing Pion amplitudes and low energy parameters from QCD}",
    doi = "10.22323/1.479.0040",
    journal = "PoS",
    volume = "CD2024",
    pages = "040",
    year = "2026"
}

@article{He:2025gws,
    author = "He, Yifei and Kruczenski, Martin",
    title = "{The Gauge Theory Bootstrap: Predicting pion dynamics from QCD}",
    eprint = "2505.19332",
    archivePrefix = "arXiv",
    primaryClass = "hep-th",
    month = "5",
    year = "2025"
}

@article{He:2024nwd,
    author = "He, Yifei and Kruczenski, Martin",
    title = "{Gauge Theory Bootstrap: Pion amplitudes and low energy parameters}",
    eprint = "2403.10772",
    archivePrefix = "arXiv",
    primaryClass = "hep-th",
    month = "3",
    year = "2024"
}

@article{He:2023lyy,
    author = "He, Yifei and Kruczenski, Martin",
    title = "{Bootstrapping gauge theories}",
    eprint = "2309.12402",
    archivePrefix = "arXiv",
    primaryClass = "hep-th",
    doi = "10.1103/PhysRevLett.133.191601",
    journal = "Phys. Rev. Lett.",
    volume = "133",
    pages = "191601",
    year = "2024"
}

@article{Guerrieri:2024jkn,
    author = {Guerrieri, Andrea and H{\"a}ring, Kelian and Su, Ning},
    title = "{From data to the analytic S-matrix: A Bootstrap fit of the pion scattering amplitude}",
    eprint = "2410.23333",
    archivePrefix = "arXiv",
    primaryClass = "hep-th",
    reportNumber = "CERN-TH-2024-177, CALT-TH-2024-038",
    doi = "10.21468/SciPostPhys.20.2.034",
    journal = "SciPost Phys.",
    volume = "20",
    number = "2",
    pages = "034",
    year = "2026"
}

@article{Albert:2026xyz,
    author = "Albert, Jan and Kosva, Dilara and Rastelli, Leonardo",
    title = "{Bootstrapping Pion Form Factors at Large $N$}",
    eprint = "2606.19420",
    archivePrefix = "arXiv",
    primaryClass = "hep-th",
    month = "6",
    year = "2026"
}

@article{Albert:2022oes,
    author = "Albert, Jan and Rastelli, Leonardo",
    title = "{Bootstrapping pions at large N}",
    eprint = "2203.11950",
    archivePrefix = "arXiv",
    primaryClass = "hep-th",
    reportNumber = "YITP-SB-2022-07",
    doi = "10.1007/JHEP08(2022)151",
    journal = "JHEP",
    volume = "08",
    pages = "151",
    year = "2022"
}

@article{Dujava:2025php,
    author = "Dujava, Jon{\'a}{\v{s}} and Va{\v{s}}ko, Petr",
    title = "{Finite-coupling spectrum of O(N) model in AdS}",
    eprint = "2503.16345",
    archivePrefix = "arXiv",
    primaryClass = "hep-th",
    doi = "10.1007/JHEP12(2025)036",
    journal = "JHEP",
    volume = "12",
    pages = "036",
    year = "2025"
}

@article{Albert:2023jtd,
    author = "Albert, Jan and Rastelli, Leonardo",
    title = "{Bootstrapping pions at large N. Part II. Background gauge fields and the chiral anomaly}",
    eprint = "2307.01246",
    archivePrefix = "arXiv",
    primaryClass = "hep-th",
    reportNumber = "YITP-SB-2023-15",
    doi = "10.1007/JHEP09(2024)039",
    journal = "JHEP",
    volume = "09",
    pages = "039",
    year = "2024"
}

@article{Carmi:2026spv,
    author = "Carmi, Dean and Ciccone, Riccardo and Sukholuski, Shimon",
    title = "{Closing the loop on $\Phi^4$ in AdS$_3$}",
    eprint = "2606.06589",
    archivePrefix = "arXiv",
    primaryClass = "hep-th",
    reportNumber = "CERN-TH-2026-121",
    month = "6",
    year = "2026"
}

@article{Polchinski:1999ry,
    author = "Polchinski, Joseph",
    title = "{S matrices from AdS space-time}",
    eprint = "hep-th/9901076",
    archivePrefix = "arXiv",
    reportNumber = "NST-ITP-99-02",
    month = "1",
    year = "1999"
}

@article{Susskind:1998vk,
    author = "Susskind, Leonard",
    editor = "Burgess, C. P. and Myers, Robert C.",
    title = "{Holography in the flat space limit}",
    eprint = "hep-th/9901079",
    archivePrefix = "arXiv",
    doi = "10.1063/1.1301570",
    journal = "AIP Conf. Proc.",
    volume = "493",
    number = "1",
    pages = "98--112",
    year = "1999"
}

@article{Loparco:2025aag,
    author = "Loparco, Manuel and Mathys, Gr{\'e}goire and Penedones, Jo{\~a}o and Qiao, Jiaxin and Zhao, Xiang",
    title = "{Locality constraints in AdS$_2$ without parity}",
    eprint = "2511.20749",
    archivePrefix = "arXiv",
    primaryClass = "hep-th",
    month = "11",
    year = "2025"
}

@article{Kaplan:2009kr,
    author = "Kaplan, David B. and Lee, Jong-Wan and Son, Dam T. and Stephanov, Mikhail A.",
    title = "{Conformality Lost}",
    eprint = "0905.4752",
    archivePrefix = "arXiv",
    primaryClass = "hep-th",
    reportNumber = "INT-PUB-09-020",
    doi = "10.1103/PhysRevD.80.125005",
    journal = "Phys. Rev. D",
    volume = "80",
    pages = "125005",
    year = "2009"
}

@article{Meineri:2023mps,
    author = "Meineri, Marco and Penedones, Joao and Spirig, Taro",
    title = "{Renormalization group flows in AdS and the bootstrap program}",
    eprint = "2305.11209",
    archivePrefix = "arXiv",
    primaryClass = "hep-th",
    month = "5",
    year = "2023"
}

@article{Paulos:2016fap,
    author = "Paulos, Miguel F. and Penedones, Joao and Toledo, Jonathan and van Rees, Balt C. and Vieira, Pedro",
    title = "{The S-matrix bootstrap. Part I: QFT in AdS}",
    eprint = "1607.06109",
    archivePrefix = "arXiv",
    primaryClass = "hep-th",
    reportNumber = "CERN-TH-2016-162",
    doi = "10.1007/JHEP11(2017)133",
    journal = "JHEP",
    volume = "11",
    pages = "133",
    year = "2017"
}

@article{Hogervorst:2021spa,
    author = "Hogervorst, Matthijs and Meineri, Marco and Penedones, Joao and Vaziri, Kamran Salehi",
    title = "{Hamiltonian truncation in Anti-de Sitter spacetime}",
    eprint = "2104.10689",
    archivePrefix = "arXiv",
    primaryClass = "hep-th",
    reportNumber = "CERN-TH-2021-065",
    doi = "10.1007/JHEP08(2021)063",
    journal = "JHEP",
    volume = "08",
    pages = "063",
    year = "2021"
}

@article{Mazac:2018mdx,
    author = "Mazac, Dalimil and Paulos, Miguel F.",
    title = "{The analytic functional bootstrap. Part I: 1D CFTs and 2D S-matrices}",
    eprint = "1803.10233",
    archivePrefix = "arXiv",
    primaryClass = "hep-th",
    reportNumber = "LPTENS/18/05, LPTENS-18-05",
    doi = "10.1007/JHEP02(2019)162",
    journal = "JHEP",
    volume = "02",
    pages = "162",
    year = "2019"
}

@article{Cordova:2022pbl,
    author = "C{\'o}rdova, Luc{\'\i}a and He, Yifei and Paulos, Miguel F.",
    title = "{From conformal correlators to analytic S-matrices: CFT$_{1}$/QFT$_{2}$}",
    eprint = "2203.10840",
    archivePrefix = "arXiv",
    primaryClass = "hep-th",
    doi = "10.1007/JHEP08(2022)186",
    journal = "JHEP",
    volume = "08",
    pages = "186",
    year = "2022"
}

@article{vanRees:2022zmr,
    author = "van Rees, Balt C. and Zhao, Xiang",
    title = "{Quantum Field Theory in AdS Space instead of Lehmann-Symanzik-Zimmerman Axioms}",
    eprint = "2210.15683",
    archivePrefix = "arXiv",
    primaryClass = "hep-th",
    doi = "10.1103/PhysRevLett.130.191601",
    journal = "Phys. Rev. Lett.",
    volume = "130",
    number = "19",
    pages = "191601",
    year = "2023"
}

@article{Ciccone:2024guw,
    author = "Ciccone, Riccardo and De Cesare, Fabiana and Di Pietro, Lorenzo and Serone, Marco",
    title = "{Exploring confinement in Anti-de Sitter space}",
    eprint = "2407.06268",
    archivePrefix = "arXiv",
    primaryClass = "hep-th",
    doi = "10.1007/JHEP12(2024)218",
    journal = "JHEP",
    volume = "12",
    pages = "218",
    year = "2024",
    note = "[Erratum: JHEP 06, 037 (2025)]"
}

@article{Aharony:2012jf,
    author = "Aharony, Ofer and Berkooz, Micha and Tong, David and Yankielowicz, Shimon",
    title = "{Confinement in Anti-de Sitter Space}",
    eprint = "1210.5195",
    archivePrefix = "arXiv",
    primaryClass = "hep-th",
    reportNumber = "WIS-14-12-AUG-DPPA",
    doi = "10.1007/JHEP02(2013)076",
    journal = "JHEP",
    volume = "02",
    pages = "076",
    year = "2013"
}

@article{Callan:1989em,
    author = "Callan, Jr., Curtis G. and Wilczek, Frank",
    title = "{INFRARED BEHAVIOR AT NEGATIVE CURVATURE}",
    reportNumber = "IASSNS-HEP-90-4, PUPT-1168",
    doi = "10.1016/0550-3213(90)90451-I",
    journal = "Nucl. Phys. B",
    volume = "340",
    pages = "366--386",
    year = "1990"
}

@article{Copetti:2023sya,
    author = "Copetti, Christian and Di Pietro, Lorenzo and Ji, Ziming and Komatsu, Shota",
    title = "{Taming Mass Gaps with Anti{\textendash}de Sitter Space}",
    eprint = "2312.09277",
    archivePrefix = "arXiv",
    primaryClass = "hep-th",
    doi = "10.1103/PhysRevLett.133.081601",
    journal = "Phys. Rev. Lett.",
    volume = "133",
    number = "8",
    pages = "081601",
    year = "2024"
}

@article{Gorbenko:2018ncu,
    author = "Gorbenko, Victor and Rychkov, Slava and Zan, Bernardo",
    title = "{Walking, Weak first-order transitions, and Complex CFTs}",
    eprint = "1807.11512",
    archivePrefix = "arXiv",
    primaryClass = "hep-th",
    doi = "10.1007/JHEP10(2018)108",
    journal = "JHEP",
    volume = "10",
    pages = "108",
    year = "2018"
}

@article{Fitzpatrick:2011dm,
    author = "Fitzpatrick, A. Liam and Kaplan, Jared",
    title = "{Unitarity and the Holographic S-Matrix}",
    eprint = "1112.4845",
    archivePrefix = "arXiv",
    primaryClass = "hep-th",
    reportNumber = "SLAC-PUB-14979",
    doi = "10.1007/JHEP10(2012)032",
    journal = "JHEP",
    volume = "10",
    pages = "032",
    year = "2012"
}

@article{Kruczenski:2022lot,
    author = "Kruczenski, Martin and Penedones, Joao and van Rees, Balt C.",
    title = "{Snowmass White Paper: S-matrix Bootstrap}",
    eprint = "2203.02421",
    archivePrefix = "arXiv",
    primaryClass = "hep-th",
    month = "3",
    year = "2022"
}

@article{Levine:2023ywq,
    author = "Levine, Nat and Paulos, Miguel F.",
    title = "{Bootstrapping bulk locality. Part I: Sum rules for AdS form factors}",
    eprint = "2305.07078",
    archivePrefix = "arXiv",
    primaryClass = "hep-th",
    doi = "10.1007/JHEP01(2024)049",
    journal = "JHEP",
    volume = "01",
    pages = "049",
    year = "2024"
}

@article{Carmi:2018qzm,
    author = "Carmi, Dean and Di Pietro, Lorenzo and Komatsu, Shota",
    title = "{A Study of Quantum Field Theories in AdS at Finite Coupling}",
    eprint = "1810.04185",
    archivePrefix = "arXiv",
    primaryClass = "hep-th",
    doi = "10.1007/JHEP01(2019)200",
    journal = "JHEP",
    volume = "01",
    pages = "200",
    year = "2019"
}

@article{Antunes:2021abs,
    author = "Antunes, Ant{\'o}nio and Costa, Miguel S. and Penedones, Jo{\~a}o and Salgarkar, Aaditya and van Rees, Balt C.",
    title = "{Towards bootstrapping RG flows: sine-Gordon in AdS}",
    eprint = "2109.13261",
    archivePrefix = "arXiv",
    primaryClass = "hep-th",
    doi = "10.1007/JHEP12(2021)094",
    journal = "JHEP",
    volume = "12",
    pages = "094",
    year = "2021"
}

@article{Lauria:2023uca,
    author = "Lauria, Edoardo and Milam, Michael N. and van Rees, Balt C.",
    title = "{Perturbative RG flows in AdS. An {\'e}tude}",
    eprint = "2309.10031",
    archivePrefix = "arXiv",
    primaryClass = "hep-th",
    doi = "10.1007/JHEP03(2024)005",
    journal = "JHEP",
    volume = "03",
    pages = "005",
    year = "2024"
}

@article{Behan:2017mwi,
    author = "Behan, Connor",
    title = "{Conformal manifolds: ODEs from OPEs}",
    eprint = "1709.03967",
    archivePrefix = "arXiv",
    primaryClass = "hep-th",
    reportNumber = "YITP-SB-17-34",
    doi = "10.1007/JHEP03(2018)127",
    journal = "JHEP",
    volume = "03",
    pages = "127",
    year = "2018"
}

@article{Penedones:2015aga,
    author = "Penedones, Jo{\~a}o and Trevisani, Emilio and Yamazaki, Masahito",
    title = "{Recursion Relations for Conformal Blocks}",
    eprint = "1509.00428",
    archivePrefix = "arXiv",
    primaryClass = "hep-th",
    reportNumber = "IPMU15-0139",
    doi = "10.1007/JHEP09(2016)070",
    journal = "JHEP",
    volume = "09",
    pages = "070",
    year = "2016"
}

@article{Athenodorou:2016ebg,
    author = "Athenodorou, Andreas and Teper, Michael",
    title = "{SU(N) gauge theories in 2+1 dimensions: glueball spectra and k-string tensions}",
    eprint = "1609.03873",
    archivePrefix = "arXiv",
    primaryClass = "hep-lat",
    doi = "10.1007/JHEP02(2017)015",
    journal = "JHEP",
    volume = "02",
    pages = "015",
    year = "2017"
}

@article{Appelquist:1989tc,
    author = "Appelquist, Thomas and Nash, Daniel",
    title = "{Critical Behavior in (2+1)-dimensional {QCD}}",
    reportNumber = "YCTP-P15-89",
    doi = "10.1103/PhysRevLett.64.721",
    journal = "Phys. Rev. Lett.",
    volume = "64",
    pages = "721",
    year = "1990"
}

@article{Kos:2016ysd,
    author = "Kos, Filip and Poland, David and Simmons-Duffin, David and Vichi, Alessandro",
    title = "{Precision Islands in the Ising and $O(N)$ Models}",
    eprint = "1603.04436",
    archivePrefix = "arXiv",
    primaryClass = "hep-th",
    reportNumber = "CERN-TH-2016-050",
    doi = "10.1007/JHEP08(2016)036",
    journal = "JHEP",
    volume = "08",
    pages = "036",
    year = "2016"
}

@article{Iliesiu:2017nrv,
    author = "Iliesiu, Luca and Kos, Filip and Poland, David and Pufu, Silviu S. and Simmons-Duffin, David",
    title = "{Bootstrapping 3D Fermions with Global Symmetries}",
    eprint = "1705.03484",
    archivePrefix = "arXiv",
    primaryClass = "hep-th",
    reportNumber = "PUPT-2524",
    doi = "10.1007/JHEP01(2018)036",
    journal = "JHEP",
    volume = "01",
    pages = "036",
    year = "2018"
}

@article{Appelquist:1986fd,
    author = "Appelquist, Thomas W. and Bowick, Mark J. and Karabali, Dimitra and Wijewardhana, L. C. R.",
    title = "{Spontaneous Chiral Symmetry Breaking in Three-Dimensional QED}",
    reportNumber = "YTP-85-26",
    doi = "10.1103/PhysRevD.33.3704",
    journal = "Phys. Rev. D",
    volume = "33",
    pages = "3704",
    year = "1986"
}

@article{DiPietro:2015taa,
    author = "Di Pietro, Lorenzo and Komargodski, Zohar and Shamir, Itamar and Stamou, Emmanuel",
    title = "{Quantum Electrodynamics in d=3 from the {\ensuremath{\varepsilon}} Expansion}",
    eprint = "1508.06278",
    archivePrefix = "arXiv",
    primaryClass = "hep-th",
    doi = "10.1103/PhysRevLett.116.131601",
    journal = "Phys. Rev. Lett.",
    volume = "116",
    number = "13",
    pages = "131601",
    year = "2016"
}

@article{Karthik:2015sgq,
    author = "Karthik, Nikhil and Narayanan, Rajamani",
    title = "{No evidence for bilinear condensate in parity-invariant three-dimensional QED with massless fermions}",
    eprint = "1512.02993",
    archivePrefix = "arXiv",
    primaryClass = "hep-lat",
    doi = "10.1103/PhysRevD.93.045020",
    journal = "Phys. Rev. D",
    volume = "93",
    number = "4",
    pages = "045020",
    year = "2016"
}

@article{Penedones:2010ue,
    author = "Penedones, Joao",
    title = "{Writing CFT correlation functions as AdS scattering amplitudes}",
    eprint = "1011.1485",
    archivePrefix = "arXiv",
    primaryClass = "hep-th",
    doi = "10.1007/JHEP03(2011)025",
    journal = "JHEP",
    volume = "03",
    pages = "025",
    year = "2011"
}

@article{Hollands:2023txn,
    author = "Hollands, Stefan and Wald, Robert M.",
    title = "{The Operator Product Expansion in Quantum Field Theory}",
    eprint = "2312.01096",
    archivePrefix = "arXiv",
    primaryClass = "hep-th",
    month = "12",
    year = "2023"
}

@article{Levine:2024wqn,
    author = "Levine, Nat and Paulos, Miguel F.",
    title = "{Bootstrapping bulk locality. Part II: Interacting functionals}",
    eprint = "2408.00572",
    archivePrefix = "arXiv",
    primaryClass = "hep-th",
    month = "8",
    year = "2024"
}

@article{Ankur:2023lum,
    author = "Ankur and Carmi, Dean and Di Pietro, Lorenzo",
    title = "{Scalar QED in AdS}",
    eprint = "2306.05551",
    archivePrefix = "arXiv",
    primaryClass = "hep-th",
    doi = "10.1007/JHEP10(2023)089",
    journal = "JHEP",
    volume = "10",
    pages = "089",
    year = "2023"
}

@article{Antunes:2024hrt,
    author = "Antunes, Ant{\'o}nio and Lauria, Edoardo and van Rees, Balt C.",
    title = "{A bootstrap study of minimal model deformations}",
    eprint = "2401.06818",
    archivePrefix = "arXiv",
    primaryClass = "hep-th",
    reportNumber = "DESY-23-141",
    doi = "10.1007/JHEP05(2024)027",
    journal = "JHEP",
    volume = "05",
    pages = "027",
    year = "2024"
}

@article{Komatsu:2020sag,
    author = "Komatsu, Shota and Paulos, Miguel F. and Van Rees, Balt C. and Zhao, Xiang",
    title = "{Landau diagrams in AdS and S-matrices from conformal correlators}",
    eprint = "2007.13745",
    archivePrefix = "arXiv",
    primaryClass = "hep-th",
    reportNumber = "CPHT-RR119.122020",
    doi = "10.1007/JHEP11(2020)046",
    journal = "JHEP",
    volume = "11",
    pages = "046",
    year = "2020"
}

@article{vanRees:2023fcf,
    author = "van Rees, Balt C. and Zhao, Xiang",
    title = "{Flat-space Partial Waves From Conformal OPE Densities}",
    eprint = "2312.02273",
    archivePrefix = "arXiv",
    primaryClass = "hep-th",
    month = "12",
    year = "2023"
}

@article{Hijano:2019qmi,
    author = "Hijano, Eliot",
    title = "{Flat space physics from AdS/CFT}",
    eprint = "1905.02729",
    archivePrefix = "arXiv",
    primaryClass = "hep-th",
    doi = "10.1007/JHEP07(2019)132",
    journal = "JHEP",
    volume = "07",
    pages = "132",
    year = "2019"
}

@article{Li:2021snj,
    author = "Li, Yue-Zhou",
    title = "{Notes on flat-space limit of AdS/CFT}",
    eprint = "2106.04606",
    archivePrefix = "arXiv",
    primaryClass = "hep-th",
    doi = "10.1007/JHEP09(2021)027",
    journal = "JHEP",
    volume = "09",
    pages = "027",
    year = "2021"
}

@article{Antunes:2025iaw,
    author = "Antunes, Ant{\'o}nio and Levine, Nat and Meineri, Marco",
    title = "{Demystifying integrable QFTs in AdS: No-go theorems for higher-spin charges}",
    eprint = "2502.06937",
    archivePrefix = "arXiv",
    primaryClass = "hep-th",
    month = "2",
    year = "2025"
}

@article{Pappadopulo:2012jk,
    author = "Pappadopulo, Duccio and Rychkov, Slava and Espin, Johnny and Rattazzi, Riccardo",
    title = "{OPE Convergence in Conformal Field Theory}",
    eprint = "1208.6449",
    archivePrefix = "arXiv",
    primaryClass = "hep-th",
    reportNumber = "LPTENS-12-31",
    doi = "10.1103/PhysRevD.86.105043",
    journal = "Phys. Rev. D",
    volume = "86",
    pages = "105043",
    year = "2012"
}

@article{Ciccone:2025dqx,
    author = "Ciccone, Riccardo and De Cesare, Fabiana and Di Pietro, Lorenzo and Serone, Marco",
    title = "{QCD in AdS}",
    eprint = "2511.04752",
    archivePrefix = "arXiv",
    primaryClass = "hep-th",
    month = "11",
    year = "2025"
}
\bibliographystyle{utphys}
\end{document}